\documentclass[a4paper,UKenglish,cleveref, autoref, thm-restate]{lipics-v2021}

\hideLIPIcs

\usepackage{graphicx} % Required for inserting images

\usepackage{multirow}
\usepackage{comment}

\usepackage{tabularx}
\usepackage{tikz}
\usetikzlibrary{arrows}
\usepackage{tikz-qtree}

\usepackage{amsmath}
\usepackage{amssymb}
\usepackage{amsfonts}
\usepackage{amsthm}

\usepackage{thmtools}
\usepackage{thm-restate}

\usepackage{algorithm}
\usepackage{algpseudocode}

\usepackage{blindtext}
\usepackage{hyperref}
\usepackage{cleveref}
\usepackage{xspace}

\usepackage[dvipsnames]{xcolor}

\usepackage{todonotes}

\newenvironment{proofsketch}{\begin{proof}[Proof Sketch]}{\end{proof}}

\newcommand{\T}{\ensuremath{\mathcal{T}}\xspace}
\newcommand{\R}{\ensuremath{\mathbb{R}}\xspace}
\newcommand{\Rcon}{\ensuremath{\overline{\mathbb{R}}}\xspace}
\newcommand{\rrcon}{\ensuremath{\overline{\mathcal{R}}}\xspace}
\newcommand{\Q}{\ensuremath{\mathbb{Q}\xspace}}
\newcommand{\N}{\ensuremath{\mathbb{N}}\xspace}
\newcommand{\Z}{\ensuremath{\mathbb{Z}}\xspace}

\newcommand{\Voc}{\ensuremath{\mathsf{Voc}}\xspace}
\newcommand{\MVoc}{\ensuremath{\mathsf{MVoc}}\xspace}
\newcommand{\Var}{\ensuremath{\mathsf{Var}}\xspace}
\newcommand{\FV}{\ensuremath{\mathsf{FV}}\xspace}
\newcommand{\VAR}{\ensuremath{\mathsf{VAR}}\xspace}
\newcommand{\conf}{\ensuremath{\mathsf{conf}}\xspace}
\newcommand{\A}{\ensuremath{\mathcal{A}}\xspace}
\newcommand{\D}{\ensuremath{\mathcal{D}}\xspace}
\newcommand{\rr}{\ensuremath{\mathcal{R}}\xspace}
\newcommand{\W}{\ensuremath{\mathbf{F}}\xspace}

\newcommand{\B}{\ensuremath{\mathcal{B}}\xspace}

\newcommand{\M}{\ensuremath{M}\xspace}

\newcommand{\program}{\ensuremath{\mathcal{P}}\xspace}
\renewcommand{\:}{\colon}
\newcommand{\satisfies}{\ensuremath{\vDash}}
\newcommand{\fromto}{\ensuremath{\leftrightarrow}}

\newcommand{\Con}{\ensuremath{\mathsf{Con}}}
\newcommand{\Fun}{\ensuremath{\mathsf{Fun}}}
\newcommand{\Rel}{\ensuremath{\mathsf{Rel}}}

\newcommand{\Th}{\ensuremath{\mathsf{Th}}}
\newcommand{\BTh}{\ensuremath{\mathsf{BTh}}}

\newcommand{\scalar}[1]{\ensuremath{\mu_{#1}}\xspace}

\renewcommand{\P}{\textup{P}} % P 
\newcommand{\NP}{\ensuremath{\textup{NP}}\xspace} % NP
\newcommand{\coNP}{\ensuremath{\textup{coNP}}\xspace} % coNP
\newcommand{\BP}{\ensuremath{\textup{BP}}\xspace}

\newcommand{\PH}{\ensuremath{\textup{PH}}\xspace} % PH
\newcommand{\BH}{\ensuremath{\textup{BH}}\xspace} % BH
\newcommand{\SAT}{\ensuremath{\textup{SAT}}\xspace} % SAT
\newcommand{\FO}{\ensuremath{\textup{FO}}\xspace} % FO
\newcommand{\SO}{\ensuremath{\textup{SO}}\xspace} % SO
\newcommand{\monus}{\ensuremath{\mathbin{\ooalign{\hss\raise1ex\hbox{.}\hss\cr
  \mathsurround=0pt$-$}}}}

\newcommand{\lbr}{\ooalign{\ensuremath{\lfloor}\cr\ensuremath{\lceil}}}
\newcommand{\rbr}{\ooalign{\ensuremath{\rfloor}\cr\ensuremath{\rceil}}}

\newcommand{\start}{\textup{\texttt{start}}}
\newcommand{\update}{\textup{\texttt{update}}}
\newcommand{\accept}{\textup{\texttt{accept}}}
\newcommand{\Instruction}{\texttt{instruction}}
\newcommand{\Index}{\textup{\texttt{index}}}
\newcommand{\Input}{\textup{\texttt{input}}}
\newcommand{\sumform}{\textup{\texttt{summation}}}
\newcommand{\prodform}{\textup{\texttt{product}}}
\newcommand{\orderform}{\textup{\texttt{ordering}}}
\newcommand{\valid}{\textup{\texttt{valid}}}
\newcommand{\satisfied}{\textup{\texttt{satisfied}}}

\newcommand{\bin}{\ensuremath{\textup{bin}}\xspace}

\newcommand{\dom}{\textup{dom}}
\newcommand{\code}{\textup{code}}
\newcommand{\evaluate}{\ensuremath{\textup{\texttt{evaluate}}}\xspace}
\newcommand{\primary}{\ensuremath{\mathsf{P}}}
\newcommand{\secondary}{\ensuremath{\mathsf{S}}}
\newcommand{\weight}{\ensuremath{\mathsf{W}}}
\newcommand{\struct}{\ensuremath{\textup{Struct}}\xspace}
\newcommand{\Bstruct}{\ensuremath{\textup{BStruct}}\xspace}
\newcommand{\ordering}{\ensuremath{\leq\!}}
\newcommand{\ifthenbranch}[3]{\ensuremath{\mathsf{if} \ #1 \ \mathsf{then} \ \mathsf{goto} \ #2 \ \mathsf{else} \ \mathsf{goto} \ #3}}

\newcommand{\partialto}{\rightharpoonup} %partial function notation

\title{On Equivalent Characterizations of the Polynomial\\
Hierarchy in Abstract Models of Computation}

\titlerunning{On Equivalent Characterizations of $\PH$ in Abstract Models of Computation}
\author{Jeremy C. Kirn}{Utrecht University}{j.c.kirn@uu.nl}{https://orcid.org/0009-0009-2903-3846}{generously supported by the Netherlands Organisation for Scientific Research (NWO) under project no. VI.Vidi.213.150.}

\author{Lucas Meijer}{Utrecht University}{l.meijer2@uu.nl}{https://orcid.org/0009-0002-4901-5249}{generously supported by the Netherlands Organisation for Scientific Research (NWO) under project no. VI.Vidi.213.150.}

\author{Tillmann Miltzow}{Utrecht University}{t.miltzow@uu.nl}{https://orcid.org/0000-0003-4563-2864}{generously supported by the Netherlands Organisation for Scientific Research (NWO) under project no. VI.Vidi.213.150.}

\author{Hans L. Bodlaender}{Utrecht University}{h.l.bodlaender@uu.nl}{https://orcid.org/0000-0002-9297-3330}{}

\authorrunning{Kirn, Meijer, Miltzow, Bodlaender} 

\Copyright{Jane Open Access and Joan R. Public} %TODO mandatory, please use full first names. LIPIcs license is ''CC-BY'';  http://creativecommons.org/licenses/by/3.0/

 %TODO mandatory: Please choose ACM 2012 classifications from https://dl.acm.org/ccs/ccs_flat.cfm 

 %TODO mandatory; please add comma-separated list of keywords

\category{} %optional, e.g. invited paper

\relatedversion{} 

\acknowledgements{The first author gratefully acknowledges conversations with Nima Motamed. }

\relatedversion{A short version of this paper appears in the Proceedings of the 51st International Symposium on Mathematical Foundations of Computer Science (MFCS 2026) \cite{kirn_et_al:LIPIcs.MFCS.2026.63}. This full version contains supplementary proofs and details omitted from the conference version due to space constraints.}

\hideLIPIcs

\begin{document}
%%%%%%%%%%%%%%%%%%%%%%%%%%%%%%%%%%%%%%%%%

\newpage
\maketitle

\begin{abstract}

    %\jeremy{Highlight contributions to Boolean parts, higher levels of the hierarchy, and oracle result?}
    
    We investigate machine models similar to Turing machines that are augmented with the operations of a first-order structure $\rr$, and we show that under weak conditions on $\rr$, the complexity class $\Sigma_k\rr$ may be characterized in four equivalent ways: (1) by polynomial-time algorithms implemented on $\rr$-machines together with witness strings, (2) by the $\Sigma_k\rr$-complete problem $\Sigma_k\SAT(\rr)$, (3) by the $k$th existential fragment of second-order metafinite logic over $\rr$ via descriptive complexity, and (4) via oracles. By characterizing $\Sigma_k\rr$ in these four ways, we extend previous work and embed it in one coherent framework. In addition, we derive similar results for $\exists_k\rr$, the constant-free Boolean part of $\Sigma_k\rr$, by showing that $\exists_k\rr$ may be characterized in four analogous ways. 
    
    Some conditions on $\rr$ must be assumed in order to achieve the above quaternity because there are infinite-vocabulary structures for which $\NP(\rr) = \Sigma_1 \rr$ does not have a complete problem. Surprisingly, even in these cases, we show that $\NP(\rr)$ \textit{does} have a characterization in terms of existential second-order metafinite logic, suggesting that descriptive complexity theory is well suited to working with infinite-vocabulary structures, such as real vector spaces. 
\end{abstract}

\newpage

\tableofcontents

%%%%%%%%%%%%%%%%%%%%%%%%%%%%%%%%%%%%%%
\newpage
\section{Introduction}
%%%%%%%%%%%%%%%%%%%%%%%%%%%%%%%%%%%%%%

%\jeremy{Could be compacted}

Algorithms are often presented in terms of primitive operations, and the complexity of these algorithms is often measured by counting how many times the primitive operations are applied. Ignoring the implementation details of these primitive operations is an abstraction that simplifies and enhances algorithm analysis. This abstraction is common in scientific computing, wherein algorithms are presented in terms of operations on the real numbers such as arithmetic and trigonometric functions, the exponential function, and others \cite{blum2004computing}. It is also common in algebraic dynamic programming, wherein algorithms are presented in terms of operations that are interpreted in different semirings to solve different problems \cite{giegerich2004discipline}. Already at this level of abstraction, it is infeasible to solve some algorithmic problems because they are complete for some higher level, $\Sigma_k$ or $\Pi_k$, of a version of the polynomial hierarchy that depends on these operations. We describe an example of this phenomenon in a way that motivates our work, and we then provide an overview of our contribution.

\subsection{Motivation}
\label{sec:motivation}
%%%%%%%%%%%%%%%%%%%%%%%%%%%%%%%%%

The polynomial hierarchy $\PH$ contains some of the most important complexity classes in theoretical computer science. We may characterize the $k$th existential level $\Sigma_k$ in at least four different ways, resulting in a quaternity:
\begin{enumerate}
    \item The class $\Sigma_k$ consists of decision problems $L \subseteq \{0,1\}^*$ for which there exists a polynomial-time Turing machine $M$ and a polynomial $q$ such that for all $v \in \{0,1\}^n$,
    $$v \in L \iff (Q_1 w_1 \in R^{q(n)})  \dots (Q_k w_k \in R^{q(n)}) \M(v, w_1, \dots, w_k)=1,$$ where the quantifiers $Q_i \in \{\exists, \forall\}$ alternate, starting with $Q_1 = \exists$. This is usually taken as the definition of $\Sigma_k$ \cite{Arora_Barak_2009}. Although one may define $\Sigma_k$ equivalently in terms of polynomial-time nondeterministic Turing machines, we work with the above definition of $\Sigma_k$ because it generalizes in a more straightforward manner to our setting of interest. 
    \item The class $\Sigma_k$ consists of decision problems $L \subseteq \{0,1\}^*$ that can be reduced by a polynomial-time Turing machine to $\Sigma_k\SAT \in \Sigma_k$, where $\Sigma_k\SAT$ is the problem of deciding if a quantified Boolean formula, with $k-1$ quantifier alternations starting with $\exists$, is satisfied. 
    The Cook-Levin theorem establishes this characterization of $\Sigma_1 = \NP$, which is expressed by saying that $\SAT$ is $\NP$-complete \cite{cook1971complexity}\cite{levin1973universal}, and the $\Sigma_k$-completeness of $\Sigma_k\SAT$ was established in \cite{stockmeyer1976polynomial}.
    % \item The class $\NP$ consists of decision problems $L \subseteq \{0,1\}^*$ that can be reduced by a polynomial-time Turing machine to the decision problem $\SAT \in \NP$, where \SAT is the problem of deciding if a Boolean formula is satisfiable. 
    % This characterization of $\NP$ is established by the Cook-Levin theorem \cite{cook1971complexity}\cite{levin1973universal}. This is expressed by saying that $\SAT$ is $\NP$-complete.
    \item The class $\Sigma_k$ consists of decision problems $L \subseteq \{0,1\}^*$ for which there is a sentence $\Phi$ of the $k$th existential fragment of second-order logic $\Sigma_k\SO$ describing $L.$ Fagin's theorem establishes this characterization of $\Sigma_1 = \NP$, which is expressed by saying that $\Sigma_1\SO$ captures $\NP$ \cite{fagin1974generalized}, and the fact that $\Sigma_k\SO$ captures $\Sigma_k$ was first established in \cite{stockmeyer1976polynomial}.
    % \item The class $\NP$ consists of decision problems $L \subseteq \{0,1\}^*$ for which there is a sentence $\Phi$ of existential second-order logic describing $L.$ This characterization is established by Fagin's theorem \cite{fagin1974generalized}. We express this by saying that $\Sigma_1\SO$ captures $\NP$.
    \item The class $\Sigma_k$ consists of decision problems $L \subseteq \{0,1\}^*$ for which there is a polynomial-time Turing machine $M$ with access to an oracle $Q \in \Sigma_{k-1}$ and a polynomial $q$ such that for all $v \in \{0,1\}^n$, $v \in L \iff (\exists w \in \{0,1\}^{q(n)}) \ M^Q(v,w) = 1$. We express this by saying that $L \in \NP^{\Sigma_{k-1}}.$ The fact that $\Sigma_k = \NP^{\Sigma_{k-1}}$ was first established in \cite{stockmeyer1976polynomial}.
\end{enumerate}

% \hans{There is a well known fourth characterization of NP, namely by a non-deterministic Turing Machine using polytime. This is probably an even more used definition. What does that mean for our story?}
% \till{I wrote a pragraph below.}

In part motivated by the success of the theory of $\NP$-completeness, Blum, Shub, and Smale initiated the study of computational complexity over the real numbers in \cite{blum1989theory}. They did this by investigating machines over the real numbers that have the ability to write any real number into memory, to evaluate infinite-precision subtraction, addition, and multiplication of real numbers, as well as to test the order relation between two real numbers and branch depending on the answer. Performing each of these operations takes one time step. In our terminology, this is a machine over the structure $\Rcon = (\R,(r : r \in \R),-,+,\cdot,\leq)$, which we obtain from the ordered ring of real numbers $\R = (\R,0,1,-,+,\cdot,\leq)$ by expanding it with a constant for each $r \in \R$. 
% Using machines over $\Rcon$, one may recreate the above quaternity for the analogous complexity class $\NP(\Rcon)$ by characterizing it in three ways:

% Using machines over $\Rcon$, one may 
% %investigate the complexity of decision problems over $\Rcon$, which are sets $L \subseteq R^*$ of strings of real numbers. 
% recreate the above quaternity for the analogous complexity class $\NP(\Rcon)$ by making the following modifications. A decision problem is now a set $L \subseteq R^*$ of strings of real numbers. Witness strings are now strings of real numbers $w \in R^*$. The decision problem $\SAT(\Rcon)$...

Using machines over $\Rcon$, one may recreate the above quaternity for $\NP(\Rcon)$. By definition, $\NP(\Rcon)$ consists of \textit{real} decision problems $L \subseteq \R^*$ for which there exists a polynomial-time verification algorithm, using \textit{real} witness strings $w \in \R^*$, implemented by an $\Rcon$-machine. One may then show that $\SAT(\Rcon)$ is $\NP(\Rcon)$-complete under polynomial-time $\Rcon$-machine reductions, where $\SAT(\Rcon)$ is the problem of deciding if a quantifier-free formula using the operations of $\Rcon$ is satisfiable in $\Rcon$. This Cook-Levin theorem over $\Rcon$ is implied by results in \cite{blum1998complexity}. One may also show that $\Sigma_1\SO(\Rcon)$ captures $\NP(\Rcon)$, where $\Sigma_1\SO(\Rcon)$ is existential second-order metafinite logic over $\Rcon$. This Fagin's theorem over $\Rcon$ is shown in \cite{gradel1995descriptive}. Finally, the oracle characterization is trivial at this base level of the hierarchy.

More recently, Schaefer and \v{S}tefankovi\v{c}
observed in \cite{schaefer2017fixed} that many Boolean decision problems $L \subseteq \{0,1\}^*$ in computational geometry and beyond may be reduced to the problem $\exists\SAT(\R) = \text{ETR}$ of deciding if a quantifier-free formula using the operations of the ordered ring of reals $\R = (\R,0,1,-,+,\cdot,\leq)$ is satisfiable in $\R$. Notably, such a formula may only contain the constants 0 and 1, and its satisfiability may be decided by a Turing machine \cite{T48}. The above observation prompted the introduction of the complexity class $\exists \R$ consisting of those problems reducible to $\exists\SAT(\R)$ with a polynomial-time Turing machine. Already more than 200 problems have been identified as $\exists\R$-complete, attesting to the importance of the complexity class $\exists\R$ \cite{SCM24}. 

Using machines over $\R$, which are just $\Rcon$-machines restricted to write only the constants 0 and 1 into memory, we may recreate the above quaternity for $\exists\R$. The class $\exists\R$ consists of \textit{Boolean} decision problems $L \subseteq \{0,1\}^*$ for which there exists a polynomial-time verification algorithm, using \textit{real} witness strings $w \in \R^*$, implemented by an $\R$-machine. This machine characterization of $\exists\R$ is shown in \cite{BC06,EvdHM20}, and it is expressed by saying that $\exists\R$ is the constant-free Boolean part of $\NP(\Rcon)$, or $\exists \R = \BP^0(\NP(\overline{\R})).$ By definition, $\exists\SAT(\R)$ is $\exists\R$-complete under polynomial-time Turing machine reductions. We will show in this work that $\exists\SO(\R)$ captures $\exists\R$, where $\exists\SO(\R)$ is the restriction of $\Sigma_1\SO(\Rcon)$ to use only the constants 0 and 1, while a similar result was recently (and independently of our work) established in \cite{meer2025some} by restricting $\Sigma_1\SO(\Rcon)$ to use only rational constants. Finally, the oracle characterization is again trivial at this base level of the hierarchy.

Since it is possible to recreate the above quaternity for both $\NP(\Rcon)$ and $\exists\R$, these complexity classes faithfully resemble $\Sigma_1 = \NP$. As we noted above, algorithms are often analyzed over different domains according to different sets of primitive operations that depend on the application area. Given the importance of $\NP$ and $\Sigma_k$ more generally, a fundamental question in the complexity of algorithms is the following: In which domains and with which sets of primitive operations can we recreate the above quaternity for $\Sigma_k$? 

We formalize this question with first-order structures $\rr$, 
which consist of a set $R$ together with constants $c \in R$, functions $f \: R^k \to R$, and relations $P \subseteq R^k$. 
Using register machines over $\rr$, we refine our fundamental question as follows: Under what conditions on $\rr$ is it possible to recreate the above quaternity for both $\Sigma_k\rr$ and $\exists_k\rr$ by characterizing them in terms of machine models, complete problems, descriptive complexity, and oracles?

\subsection{Overview of Our Contribution}
\label{sub:OurContribution}
%%%%%%%%%%%%%%%%%%%%%%%%%%%%%%%%%%%%%%%%%%%%%%%

% In this work, under weak assumptions on a first-order structure $\rr$, we generalize the above quaternity by defining $\NP(\rr)$ and $\exists\rr$ in terms of register machines over $\rr$ and characterizing $\NP(\rr)$ and $\exists\rr$ in terms of complete problems and descriptive complexity. Furthermore, we generalize these characterizations to each level of the polynomial hierarchy over $\rr$ and the Boolean hierarchy over $\rr$, as depicted in \Cref{fig:PolynomialAndBooleanHierarchyOverR}, and we give additional characterizations of these hierarchies using oracles.

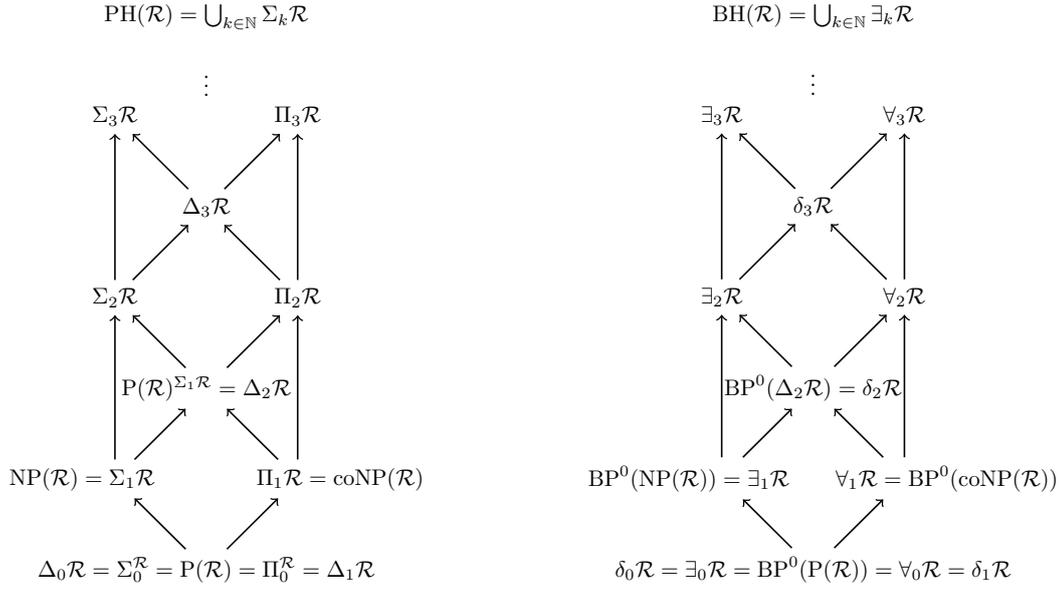
\begin{figure}
%\centering % This centers the pair of diagrams on the page
%--- FIRST DIAGRAM ---
\begin{tikzpicture}[->, node distance=2cm, semithick, scale=.85, transform shape]
    % Define all the nodes (complexity classes) first
    \node (P)      {$\Delta_0\rr =\Sigma_0^\rr = \P(\rr) = \Pi_0^\rr = \Delta_1 \rr $};
    \node (Sigma1) [above left of=P]       {$\NP(\rr) = \Sigma_1\rr$ \hspace*{0.9cm}};
    \node (Pi1)    [above right of=P]      {\hspace*{1.2cm} $\Pi_1\rr = \coNP(\rr)$};
    \node (Delta2) [above left of=Pi1]     {$\P(\rr)^{\Sigma_1\rr} = \Delta_2\rr$};
    \node (Sigma2) [above left of=Delta2]  {$\Sigma_2\rr$};
    \node (Pi2)    [above right of=Delta2] {$\Pi_2\rr$};
    \node (Delta3) [above left of=Pi2]     {$\Delta_3\rr$};
    \node (Sigma3) [above left of=Delta3]  {$\Sigma_3\rr$};
    \node (Pi3)    [above right of=Delta3] {$\Pi_3\rr$};
    \node (dots)   [above of=Delta3]       {\vdots};
    \node (PH)     [above=0.25cm of dots]       {$\PH(\rr) = \bigcup_{k \in \N}\Sigma_k\rr$};
    % Draw the arrows between the nodes
    \draw (P)      -> (Sigma1);
    \draw (P)      -> (Pi1);
    \draw (Sigma1) -> (Sigma2);
    \draw (Sigma1) -> (Delta2);
    \draw (Pi1)    -> (Pi2);
    \draw (Pi1)    -> (Delta2);
    \draw (Delta2) -> (Sigma2);
    \draw (Delta2) -> (Pi2);
    \draw (Sigma2) -> (Sigma3);
    \draw (Sigma2) -> (Delta3);
    \draw (Pi2)    -> (Pi3);
    \draw (Pi2)    -> (Delta3);
    \draw (Delta3) -> (Sigma3);
    \draw (Delta3) -> (Pi3);
\end{tikzpicture}
\hfill % This command adds flexible horizontal space, pushing the diagrams apart
%--- SECOND DIAGRAM ---
\begin{tikzpicture}[->, node distance=2cm, semithick, scale=.85, transform shape]
    % Define all the nodes (complexity classes) first
    \node (P)      {$\delta_0\rr =\exists_0\rr = \BP^0(\P(\rr)) = \forall_0\rr = \delta_1\rr$};
    \node (Sigma1) [above left of=P]      {$\BP^0(\NP(\rr)) = \exists_1\rr$ \hspace*{0.9cm}};
    \node (Pi1)    [above right of=P]     {\hspace*{1.2cm} $\forall_1\rr = \BP^0(\coNP(\rr))$};
    \node (Delta2) [above left of=Pi1]    {$\BP^0(\Delta_2\rr) = \delta_2\rr$};
    \node (Sigma2) [above left of=Delta2]  {$\exists_2\rr$};
    \node (Pi2)    [above right of=Delta2] {$\forall_2\rr$};
    \node (Delta3) [above left of=Pi2]    {$\delta_3\rr$};
    \node (Sigma3) [above left of=Delta3]  {$\exists_3\rr$};
    \node (Pi3)    [above right of=Delta3] {$\forall_3\rr$};
    \node (dots)   [above of=Delta3]      {\vdots};
    \node (BH)     [above=0.25cm of dots]        {$\BH(\rr) = \bigcup_{k \in \N}\exists_k\rr$};
    % Draw the arrows between the nodes
    \draw (P)      -> (Sigma1);
    \draw (P)      -> (Pi1);
    \draw (Sigma1) -> (Sigma2);
    \draw (Sigma1) -> (Delta2);
    \draw (Pi1)    -> (Pi2);
    \draw (Pi1)    -> (Delta2);
    \draw (Delta2) -> (Sigma2);
    \draw (Delta2) -> (Pi2);
    \draw (Sigma2) -> (Sigma3);
    \draw (Sigma2) -> (Delta3);
    \draw (Pi2)    -> (Pi3);
    \draw (Pi2)    -> (Delta3);
    \draw (Delta3) -> (Sigma3);
    \draw (Delta3) -> (Pi3);
\end{tikzpicture}
\caption{The polynomial hierarchy and the Boolean hierarchy over a structure $\rr$.}
\label{fig:PolynomialAndBooleanHierarchyOverR}
\end{figure}

In this work, we define $\Sigma_k\rr$ and $\exists_k\rr$ in terms of register machines over $\rr$, resulting in the polynomial and Boolean hierarchies over a structure, as depicted in \Cref{fig:PolynomialAndBooleanHierarchyOverR}. We then identify mild conditions on $\rr$ which allow us to generalize the above quaternity by characterizing $\Sigma_k\rr$ and $\exists_k\rr$ in terms of complete problems, descriptive complexity, and oracles.

Each of our results requires us to assume that $\rr$ is a \textit{bipointed} structure, by which we mean that $\rr$ has at least two distinct constants $0 \neq 1$. Our completeness results require us to assume further that $\rr$ is of \textit{finite type}, by which we mean that $\rr$ has only finitely many functions and relations. Finally, for completeness with respect to non-Boolean decision problems, we must assume that $\rr$ \textit{has all constants}, by which we mean that $\rr$ has a unique constant $r$ for each element in its universe $R$.

We begin with complete problems. 
%Let $\SAT(\rr)$ be the set of quantifier-free formulas using the operations of $\rr$ that are satisfiable in $\rr$. Similarly, let $\exists\SAT(\rr)$ be the subset of such sentences containing only the constants 0 and 1. 
It is already known from \cite{goode1994accessible} that if $\rr$ is a bipointed structure of finite type with all constants, then $\SAT(\rr)$ is $\NP(\rr)$-complete under polynomial-time $\rr$-machine reductions. Furthermore, it is folklore that, under the same conditions on $\rr$, $\Sigma_k\SAT(\rr)$ is $\Sigma_k\rr$-complete under polynomial-time $\rr$-machine reductions \cite{CK99,de2004completeness}. This folklore result is sometimes attributed to \cite{Poizat_1995}, which is a foundational text that is no longer in wide circulation, preventing us from accessing it. In \Cref{thm:HigherCookLevin}, we offer a detailed proof that if $\rr$ is a bipointed structure of finite type with all constants, then $\Sigma_k\SAT(\rr)$ is $\Sigma_k\rr$-complete under polynomial-time $\rr$-machine reductions. We offer this proof not only for completeness, but also because analyzing this proof allows us to explain why a slight weakening of the assumed conditions on $\rr$ are probably necessary. Furthermore, our analysis of this proof allows us to derive \Cref{thm:HigherBooleanCookLevin}, in which we show that if $\rr$ is a bipointed structure of finite type, then $\exists_k\SAT(\rr)$ is $\exists_k\rr$-complete under polynomial-time Turing machine reductions.

%that under the same conditions $\Sigma_k\SAT(\rr)$ is $\Sigma_k\rr$-complete, a fact we recall as \Cref{thm:HigherCookLevin}. We analyze this proof in detail and explain why we cannot expect to weaken any of the assumptions. Furthermore, our analysis allows us to derive \Cref{thm:HigherBooleanCookLevin}, showing that $\exists_k\SAT(\rr)$ is $\exists_k\rr$-complete under polynomial-time Turing machine reductions.

We next turn to descriptive complexity theory. In \Cref{thm:HigherFagin}, we show that if $\rr$ is a bipointed structure, then $\Sigma_k\SO(\rr)$ captures $\Sigma_k\rr$ on metafinite structures over $\rr$. Note that we are not required to assume that $\rr$ is of finite type, nor that $\rr$ has all constants, in contrast to our completeness results. Since there is a structure $\mathcal{Z}$ of infinite type for which there is no $\NP(\mathcal{Z})$-complete problem \cite{gassner1997np}, \Cref{thm:HigherFagin} is strictly more general than any possible generalization of \Cref{thm:HigherCookLevin}. This suggests that descriptive complexity theory is well suited for dealing with structures of infinite type, such as real vector spaces. Additionally, from \Cref{thm:HigherFagin} we derive \Cref{thm:HigherBooleanFagin} in which we show that if $\rr$ is a bipointed structure, then $\exists_k\SO(\rr)$ captures $\exists_k\rr$ on Boolean metafinite structures over $\rr$.

We finally turn to oracle characterizations of these hierarchies. In \Cref{thm:OraclePolynomialHierarchyCharacterization}, we show that if $\rr$ is a bipointed structure, then $\Sigma_{k+1}\rr = \NP(\rr)^{\Sigma_k\rr}$.
Additionally, from \Cref{thm:OraclePolynomialHierarchyCharacterization} we derive \Cref{thm:OracleBooleanHierarchyCharacterization} in which we show that if $\rr$ is a bipointed structure, then $\exists_{k+1}\rr = \exists_k\rr^{\Sigma_k\rr^0}$. The class $\Sigma_k\rr^0$ is obtained by restricting $\Sigma_k\rr$ to those decision problems whose verification algorithms only involve the machine constants 0 and 1, while still allowing non-Boolean inputs. From this corollary, we can see that it is unlikely that we can replace the non-Boolean oracle class $\Sigma_k\rr^0$ with the Boolean oracle class $\exists_k\rr \subseteq \Sigma_k\rr^0$. In particular, it is unlikely that $\exists\R^{\exists\R} = \exists_2\R$. 

We organize our results as follows: In \Cref{sub:Preliminaries} we present preliminary definitions, in \Cref{sub:CookLevinOverStructure} we present our results about complete problems, in \Cref{sub:FaginOverStructure} we present our results about descriptive complexity, and in \Cref{sub:OraclesAndHierarchies} we present our results about oracles. In each section, we provide proof sketches of our results, as well as a discussion of related work. Full proofs appear in the appendices of this paper. There is also an extended preliminaries appendix to make our results more accessible. Each appendix of this paper may be seen as an extended version of the corresponding section in the main body of our paper.

\section{Preliminaries}
\label{sub:Preliminaries}
%%%%%%%%%%%%%%%%%%%%%%%%%%%%%%%%%%%%%%%%%%%%%%

We present a selection of the preliminaries that are most important for understanding our results. The reader who desires more formal detail or more intuition may find both of these in \Cref{sec:ExtendedPreliminaries}.

\subsection{Register Machines Over a Structure}
\label{sub:RegisterMachinesOverStructure}
%%%%%%%%%%%%%%%%%%%%%%%%%%%%%%%%%%%%%%%%%%%%%%

A vocabulary $\Voc$ specifies a set of constant symbols $c$ and, for each arity $k \geq 1$, a set function symbols $f$ and a set of relation symbols $P$. A (first-order) $\Voc$-structure $\rr$ specifies an interpretation of each of these symbols in an underlying set $R$, which we call the universe of $\rr$. Conflating a symbol with its interpretation, we may say that $\rr$ has constants $c \in R$, functions $f : R^k \to R$, and relations $P \subseteq R^k$.

Recall that we say $\rr$ is \textit{bipointed} if it has at least two constants $0 \neq 1$, that $\rr$ is of \textit{finite type} if it has only finitely many functions and relations, and that $\rr$ \textit{has all constants} if it has a unique constant $r$ for each element of its universe $R$. Additionally, we treat equality $=$ as a logical relation \cite{van1994logic}, so we always assume that the equality relation is implicitly present in every structure. This appears to be necessary for most of our results to hold in generality, though equality can be omitted in some particular contexts, as is often done when comparing the complexity of different syntactic fragments of the theory of $\R$, viewed as decision problems in $\exists\R$ and related complexity classes \cite{SS17,SS22}.

\begin{figure}[h]
    \centering
    \includegraphics{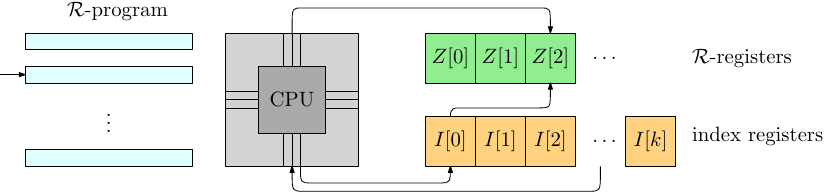}
    \caption{Visualization of an $\rr$-machine adapted from~\cite{gassner2019introduction}.}
    \label{fig:RegisterMachine}
\end{figure}

The symbols of a vocabulary $\Voc$ constitute the basic ingredients of a programming language in which we may specify computations over a $\Voc$-structure $\rr$. These computations are carried out by a register machine over $\rr$, also called an $\rr$-machine, as depicted in \Cref{fig:RegisterMachine}.

Such machines have an infinite sequence $Z \lbr 0 \rbr, Z \lbr 1 \rbr, Z \lbr 2 \rbr, \dots$ of $\rr$-registers, each of which may store a single element $r$ from the universe $R$ of $\rr$, and a finite sequence $I \lbr 0 \rbr, \dots , I \lbr k \rbr$ of index registers, each of which may store a single natural number $n \in \N$. The primary role of index registers $I 
\lbr j \rbr$ is to keep track of indexing and counting during computations, as well as to point to $\rr$-registers $Z \lbr I \lbr j \rbr \rbr$.

%\jeremy{Cut detailed examples of instructions?}

% Such machines have an infinite sequence $Z \lbr 0 \rbr, Z \lbr 1 \rbr, Z \lbr 2 \rbr, \dots$ of $\rr$-registers, each of which may store a single element $r$ from the universe $R$ of $\rr$. These are idealized memory registers that may hold infinite objects. For example, when the universe of $\rr$ is $\R$, an $\rr$-register may hold an infinite-precision real number.

% Register machines over $\rr$ also have a finite sequence $I \lbr 0 \rbr, \dots , I \lbr k \rbr$ of index registers, each of which may store a single natural number $n \in \N$. We may think of index registers $I \lbr j \rbr$ as pointers to $\rr$-registers $Z \lbr I \lbr j \rbr \rbr$ or as a memory location for discrete data. Their primary role is to keep track of indexing and counting during computations.

An $\rr$-machine manipulates the contents of its registers based on its $\rr$-program $\program$, which is a finite enumerated sequence of $\rr$-instructions. 
Importantly, there are $\rr$-instructions directing $\rr$-machines to evaluate each operation of the structure $\rr$ in the following way:

\begin{enumerate}
    \item the instruction $Z \lbr j_0 \rbr := c$ directs a machine to store the constant $c$ in register $Z \lbr j_0 \rbr$,  
    \item the instruction $Z\lbr j_0 \rbr := f( Z \lbr j_1 \rbr, \dots, Z \lbr j_k \rbr)$ directs a machine to evaluate the function $f(Z \lbr j_1 \rbr, \dots, Z \lbr j_k \rbr)$ and store the result in register $Z \lbr j_0 \rbr$,
    \item the instruction $\ifthenbranch{P( Z \lbr j_1 \rbr, \dots, Z \lbr j_k \rbr)}{\ell_1}{\ell_2}$ directs a machine to test the relation $P( Z \lbr j_1 \rbr, \dots, Z \lbr j_k \rbr)$ and go to instruction $\ell_1$ or $\ell_2$ depending on the result.
\end{enumerate}
In addition to the above, $\rr$-instructions may also direct an $\rr$-machine to branch based on a test of equality between $\rr$-registers or index registers, to copy the contents of one $\rr$-register to another, and to increment or decrement the contents of an index register. Finally, when we would like to equip an $\rr$-machine with an oracle $Q \subseteq R^*$, we allow $\rr$-instructions directing $\rr$-machines to query this oracle.  A full list of $\rr$-instructions is presented in \Cref{tab:RInstructions}.
% \begin{enumerate}
% \setcounter{enumi}{3}
%     \item The instruction $\ifthenbranch{(Z \lbr 0 \rbr ,\dots, Z\lbr I \lbr 0 \rbr \rbr) \in Q}{\ell_1}{\ell_2}$ directs a machine to make the query $(Z \lbr 0 \rbr ,\dots, Z\lbr I \lbr 0 \rbr \rbr) \in Q$ and go to step $\ell_1$ or $\ell_2$ of its $\rr$-program depending on the result.
% \end{enumerate}

In the formal definition of $\rr$-machines below, we include the structure $\rr$ as a component of $\rr$-machines. We do this so that the $\rr$-machine knows how to interpret the symbols in its program, and in this sense we may view the structure $\rr$ as the CPU from \Cref{fig:RegisterMachine}, while the other components of $\rr$-machines correspond in the expected way to this picture.

\begin{definition}[$\rr$-Machines]
\label{def:RMachine}
    A \textit{machine over $\rr$}, also called an \textit{$\rr$-machine}, is a tuple $\M = (\program,\rr,\N^{k_\program},R^\N)$ specifying the following information:
    \begin{enumerate}
        \item an $\rr$-program $\program$,
        \item the structure $\rr$,
        \item a space of index states $\N^{k_\program}$, 
        \item a space of memory states $R^\N$.
    \end{enumerate}
    We say that the constants of $\rr$, other than $0$ and $1$, appearing in an instruction of $\program$ are the \textit{machine constants} of $\M$, and that $\M$ is \textit{constant free} if it has no machine constants. The \textit{input} and \textit{output space} of $\M$ are $R^* = \bigcup_{n=0}^\infty R^n.$ 
\end{definition}

Our definition of $\rr$-machines is taken with minimal modification from \cite{hemmerling1996computability, gassner2019introduction}, and it is similar to the original definition of BSS machines given in \cite{blum1989theory}, which closely resemble register machines, a precursor to random access machines introduced in \cite{shepherdson1963computability}. Because our $\rr$-instructions allow only sequential access to memory by incrementing or decrementing an index register, we use the name register machine over $\rr$.

A significant difference between BSS machines over a ring $\rr$ and $\rr$-machines is that BSS machines over $\rr$ may write an arbitrary element $r \in R$ into memory, while $\rr$-machines may write an element $r$ into memory only if $r$ is a constant of the structure $\rr$, or if $r$ can be computed from an input using the operations of $\rr$. We choose to limit $\rr$-machines in this way so that the power of $\rr$-machines more closely reflects the operations present in the structure $\rr$, including the constant operations. 

% However, we may reconcile this difference between BSS-machines over $\rr$ and $\rr$-machines by passing to the structure $\overline{\rr}$ that has been expanded with a constant for each $r \in R$. An $\overline{\rr}$-program is an $\rr$-quasiprogram in the terminology of \cite{hemmerling1996computability}.

An element $r \in R$ usually plays the role of an \textit{input} to an $\rr$-machine. When we turn $r$ into a \textit{constant}, we assign it a very different role. As a constant, $r$ may encode an infinite amount of information into $\rr$-programs. For example, consider the ordered ring of reals expanded with all constants $\Rcon$ and suppose $r_\textup{halt} \in \R$ encodes the characteristic function of the diagonal halting problem. Upon input $n \in \N$ encoded in binary, an $\Rcon$-machine with $r_\textup{halt}$ as a machine constant can extract the $n$th bit of $r_\textup{halt}$, as described in \cite{koiran1994computing}, to decide if the $n$th Turing machine halts on input $n$, thereby deciding the diagonal halting problem. This means that $\Rcon$-machines with $r_\textup{halt}$ as a machine constant are strictly more powerful than Turing machines, even when we restrict these $\Rcon$-machines to Boolean inputs, whereas constant-free Boolean $\Rcon$-machines may be simulated by Turing machines. 

%However, a constant-free $\R$-machine restricted to Boolean inputs can be simulated by a Turing machine. Thus, such machines create a bridge between the nonBoolean world of $\R$-machines and the Boolean world of Turing machines. Similarly, constant-free $\rr$-machines restricted to Boolean inputs form a bridge between the computation in $\rr$ and Boolean computation, although we note that Turing machines cannot simulate constant-free Boolean $\rr$-machines in general.

Another subtlety concerning constants arises when one regards constants as functions of arity zero. We refrain from taking this view because it would clash with the definition of structures of finite type, which may have infinitely many constants but only finitely many functions. Assuming that $\rr$ is of finite type is useful to us because this guarantees that $\rr$ has an \textit{effective encoding}, which is an encoding the symbols in the vocabulary of $\rr$ as strings in $R^*$ in such a way that an $\rr$-machine can interpret that code and then evaluate the corresponding operation. In general, we will use an effective encoding of $\rr$ when we show membership of satisfiability problems over $\rr$ in complexity classes over $\rr$. Note that there is an effective encoding for $\rr$ if and only if there is a universal $\rr$-machine \cite{hemmerling1996computability}.

Structures with infinitely many constants may still have an effective encoding because a constant may always stand as a code for itself, whereas there are structures with infinitely many unary functions, such as the linear reals $\R_\textup{lin} = (\R,0,1,-,+,(\scalar{r} : r \in \R))$, for which there is no effective encoding.  The linear reals split multiplication into infinitely many scalar multiplication operations $\scalar{r} : x \mapsto r \cdot x$ in accordance with the standard way to turn vector spaces into first-order structures, as explained in \Cref{rem:ScalarMultiplication}.

To get an intuition for why $\R_\textup{lin}$ cannot have an effective encoding, consider why there can be no universal $\R_\textup{lin}$-machine. Any $\R_\textup{lin}$-machine can have only finitely many scalar multiplication functions $\scalar{r}$ in its program, and it is not possible to express each function in $\{\scalar{r} : r \in \R\}$ in terms of a finite subset of these functions. Thus, for any given $\R_\textup{lin}$-machine $M$, there will be an operation $\scalar{r^*}$ that $M$ cannot evaluate, preventing $M$ from simulating any other $\R_\textup{lin}$-machine with $\scalar{r^*}$ in its program.
%The intuition behind this phenomenon is that an $\rr$-machine can only have finitely many functions in its program, and it may not be possible to simulate. Thus, there are structures $\rr$ for which there is no universal $\rr$-machine.

There is a long history capturing computations over a first-order structure with machines and methods similar to ours \cite{feferman2013computingreals}. We refer the reader to the introduction of \cite{hemmerling1996computability} for further references into this history. There are multiple incomparable ways to generalize computation to the real numbers \cite{papayannopoulos2021unrealistic}. Machines over $\R$ given an algebraic generalization, while computable analysis and type two effectivity give a topological generalization \cite{neumann2018topological}. The algebraic approach provides fertile ground for complexity theory. For an introductory survey of early results in the complexity of computation over $\R$ and other structures, see \cite{MM97}.

\subsection{Complexity Classes}
\label{sub:ComplexityClasses}
%%%%%%%%%%%%%%%%%%%%%%%%%%%%%%%%%%%%%%%%%%%%%%

We view each subset $L \subseteq R^*=\bigcup_{n =0}^\infty R^n$ as a decision problem over a structure $\rr$. Under a suitable encoding, the main decision problems we will work with consist of satisfied sentences of a specified grammatical form. A sentence $\Phi$ is in prenex normal form when it is a string of quantifiers followed by a quantifier free formula $\varphi$ as in $Q_1 x_1 \dots Q_n \varphi(x_1,\dots,x_n),$ where $Q_i \in \{\exists, \forall\}$. Supposing $\Phi$ has $k-1$ quantifier alternations, we say that $\Phi$ is a $\Sigma_k$ sentence when $Q_1 = \exists$, and we say that $\Phi$ is a $\Pi_k$ sentence when $Q_1 = \forall$.

% \begin{definition}[Decision Problems of Satisfied Sentences]
%     The $k$th existential fragment of the theory of $\rr$ is the set $\Sigma_k\SAT(\rr)$ of $\Sigma_k$ sentences that are satisfied in $\rr$, while the $k$th universal fragment of the theory of $\rr$ is the set of $\Pi_k$ sentences satisfied in $\rr$. 
% \end{definition}

\begin{definition}[Decision Problems of Satisfied Sentences]
    The theory of $\rr$ is the set $\Th(\rr)$ of sentences $\Phi$ in the vocabulary of $\rr$ that are satisfied in $\rr$, and the Boolean theory of $\rr$ is the subset $\BTh(\rr) \subseteq \Th(\rr)$ of sentences $\Phi$ such that all constants in $\Phi$ are 0 or 1. We identify the following fragments of these theories that we refer to as the $k$th existential or universal fragment of the theory or Boolean theory of $\rr$, respectively:
    
    \begin{enumerate}
        \item $\Sigma_k\SAT(\rr)  =  \{\Phi : \Phi \text{ is a $\Sigma_k$ sentence and } \rr \satisfies \Phi\}$,
        \item $\Pi_k\SAT(\rr)  =  \{\Phi : \Phi \text{ is a $\Pi_k$ sentence and } \rr \satisfies \Phi\}$,
        \item $\exists_k\SAT(\rr)  =  \{ \Phi \in \Sigma_k\SAT(\rr) : \text{all constants in $\Phi$ are 0 or 1}\}$,
        \item $\forall_k\SAT(\rr)  =  \{ \Phi \in \Pi_k\SAT(\rr) : \text{all constants in $\Phi$ are 0 or 1}\}$.
    \end{enumerate}
    
    % $$
    % {
    % \renewcommand{\arraystretch}{1.5}
    % \begin{array}{rcl}
    %     \Sigma_k\SAT(\rr) & = & \{\Phi : \Phi \text{ is a $\Sigma_k$ sentence and } \rr \satisfies \Phi\}  \\
    %     \Pi_k\SAT(\rr) & = & \{\Phi : \Phi \text{ is a $\Pi_k$ sentence and } \rr \satisfies \Phi\} \\
    %     \exists_k\SAT(\rr) & = & \{ \Phi \in \Sigma_k\SAT(\rr) : \text{all constants in $\Phi$ are 0 or 1}\} \\
    %     \forall_k\SAT(\rr) & = & \{ \Phi \in \Pi_k\SAT(\rr) : \text{all constants in $\Phi$ are 0 or 1}\}
    % \end{array}
    % }
    % $$
\end{definition}

A complexity class $\mathcal{C} \subseteq 2^{R^*}$ is any set of decision problems $L \subseteq R^*$. The polynomial hierarchy $\PH(\rr)$ of complexity classes over a structure $\rr$ generalize the classical polynomial hierarchy, and these classes are our main focus.

\begin{definition}[Polynomial Hierarchy Over $\rr$]
\label{def:PolynomialHierarchyR}
    We define two sequences of complexity classes $\Sigma_k\rr$ and $\Pi_k\rr$ for $k \in \N$ as follows. A decision problem $L \subseteq R^*$ is in $\Sigma_k\rr$ if and only if there is a polynomial $q$ and a polynomial-time $\rr$-machine $\M$ such that for all $n \in \N$ and all $v \in R^n$ $$v \in L \iff (Q_1 w_1 \in R^{q(n)})  \dots (Q_k w_k \in R^{q(n)}) \M(v, w_1, \dots, w_k)=1,$$ where the quantifiers $Q_i \in \{\exists, \forall\}$ alternate, starting with $Q_1 = \exists$. The class $\Pi_k\rr$ is defined by the same condition except the first quantifier is $Q_1 = \forall$. Define the \textit{polynomial hierarchy over $\rr$} as the class $\PH(\rr) = \bigcup_{k\in\N}\Sigma_k\rr.$
\end{definition}

We define $\P(\rr) = \Sigma_0\rr$, and we note that this is the class of decision problems that can be decided by a polynomial-time $\rr$-machine. Furthermore, we define $\NP(\rr) = \Sigma_1\rr$ and $\coNP(\rr) = \Pi_1\rr$, and we note that these classes generalize classical $\NP$ and $\coNP$. We may relate a complexity class $\mathcal{C}$ over a structure to classical complexity classes by taking the constant-free Boolean part of $\mathcal{C}.$

\begin{definition}[Constant-Free Boolean Part]
\label{def:ConstantFreeBooleanPart}
    The \textit{Boolean part} of a decision problem $L$ is the set $\BP(L) = L \cap \{0,1\}^*$. The Boolean part of a complexity class $\mathcal{C}$ is the set $\BP(\mathcal{C}) = \{\BP(L) : L \in \mathcal{C}\}.$ If $\mathcal{C}$ is defined in terms of $\rr$-machines, then $\mathcal{C}^0$ is the class obtained by restricting to $\rr$-machines with no machine constants. We write $\BP^0(\mathcal{C})$ for $\BP(\mathcal{C}^0)$, and we say that $\BP^0(\mathcal{C})$ is the \textit{constant-free Boolean part} of $\mathcal{C}$.
\end{definition}

By taking the constant-free Boolean part of each level of $\PH(\rr)$, we obtain the Boolean hierarchy over a structure $\rr$, which is a hierarchy of Boolean complexity classes.

\begin{definition}[Boolean Hierarchy Over $\rr$]
\label{def:BooleanHierarchyR}
    We define the complexity classes $\exists_k\rr$ and $\forall_k\rr$ as the constant-free Boolean parts of $\Sigma_k\rr$ and $\Pi_k\rr$, respectively, for $k \in \N$. That is, $\exists_k \rr= \BP^0(\Sigma_k\rr) \textup{ and } \forall_k \rr = \BP^0(\Pi_k\rr).$ The Boolean hierarchy over $\rr$ is the complexity class $\BH(\rr) = \bigcup_{k \in \N} \exists_k\rr$. 
    %\jeremy{Need to add the definition of $\delta_k\rr$ here in order to accord with the definition, where $\delta_k\rr = \BP^0(\Delta_k\rr)$}
\end{definition}

% When a complexity class $\mathcal{C}$ is defined in terms of $\rr$-machines and $\mathcal{D}$ is another complexity class, the class $\mathcal{C}^\mathcal{D}$ results from allowing $\rr$-machines with oracles from $\mathcal{D}$ in the definition of $\mathcal{C}$.

As we will see, it is quite fruitful to analyze the polynomial and Boolean hierarchies over $\rr$ in terms of $\rr$-machines with oracles.

\begin{definition}[Oracle Complexity Classes]
    Let $\mathcal{D}$ be a complexity class over $\rr$. For each of level of the polynomial and Boolean hierarchies over $\rr$, the complexity classes $\Sigma_k\rr^\mathcal{D}$, $\Pi_k\rr^\mathcal{D}$, $\exists_k\rr^\mathcal{D}$, and $\forall_k\rr^\mathcal{D}$ are obtained by modifying the definition of the corresponding class to allow $\rr$-machines with oracles from $\mathcal{D}$.
\end{definition}

\subsection{Metafinite Structures}
\label{sub:MetafiniteStructures}
%%%%%%%%%%%%%%%%%%%%%%%%%%%%%%%%%%%%%%%%%%%%%%

Metafinite structures may be thought of as finite structures $\A$ that are given finite access to a potentially infinite structure $\rr$ by a finite set $\W$ of weight functions $W \: A^{k} \to R$. 
They were first used to generalize Fagin's theorem from $\NP$ to $\NP(\Rcon)$ in \cite{gradel1995descriptive}, and their general theory was developed in~\cite{graedel1998metafinite}. 
Just as $\rr$-machines separate the finite and discrete operations of index registers from the potentially infinite and continuous operations of $\rr$-registers, metafinite structures separate the finite and discrete aspect of a structure from the potentially infinite and continuous aspects of a structure. 

%Good examples to hold in mind as we delve into the specifics are graphs with real number weights on the edges, relational databases with a numerical domain that supports arithmetic operations on that domain, or more generally any kind finite structure with numerical attributes.  

In order to talk about metafinite structures, we need a notion of metafinite vocabularies, which
%Just as a metafinite structure has a clean separation between a finite structure $\A$ and a potentially infinite structure $\rr$ that interact via a finite set of weight functions $\W$, a metafinite vocabulary 
will consist of three separate vocabularies $\MVoc = (\primary(\MVoc),\secondary(\MVoc),\weight(\MVoc))$. The \textit{primary vocabulary} $\primary(\MVoc)$ is a finite vocabulary, the \textit{secondary vocabulary} $\secondary(\MVoc)$ is a potentially infinite vocabulary, and the \textit{weight vocabulary} $\weight(\MVoc)$ is a finite set of function symbols $F$, each with a given arity. We say that $\MVoc$ is a \textit{metafinite vocabulary over $\rr$} whenever the secondary vocabulary of $\MVoc$ is the vocabulary of $\rr$.

\begin{definition}
\label{def:MetafiniteRStructure}
    A \textit{metafinite structure} for a metafinite vocabulary $\MVoc$ is a triple $\D = (\A,\rr,\W)$ where
    \begin{enumerate}
        \item the \textit{primary part} of $\D$ is a finite structure $\A$ for vocabulary $\primary(\MVoc)$,
        \item the \textit{secondary part} of $\D$ is a potentially infinite structure $\rr$ for vocabulary $\secondary(\MVoc)$,
        \item the set of \textit{weight functions} of $\D$ is a finite set $\W$ of functions $F \: A^{k_F} \to R$.  
    \end{enumerate}
    We say that $\D$ is a \textit{Boolean} metafinite structure if the image of each weight function $F \in \W$ is contained in $\{0,1\}$, so that $F \:A^{k_F} \to \{0,1\}\subseteq R$.
\end{definition}

The \textit{size} of a metafinite structure $\D$, denoted by $|\D|$, is the cardinality of the universe $A$ of its primary part $\A$. In practice, we will conflate the function $F \: A^{k_F} \to R$ with the symbol for $F$. Whenever the secondary part of $\D$ is $\rr$, then we say that $\D$ is an \textit{$\rr$-structure}. For a given metafinite vocabulary $\MVoc$ over $\rr$, let $\struct(\MVoc)$ be the class of metafinite structures in this vocabulary.

\begin{example}
    A simple class of metafinite structures is obtained by letting $\A = (V,E)$ be a finite graph, letting $\rr = \R$ be the ordered ring of real numbers, and letting $\W$ contain a single binary function $F \: V^2 \to \R$ that assigns $0$ to all pairs $(v_1,v_2)$ that are not related by $E$. The resulting $\R$-structure is a finite graph with edges that have real number weights. Conversely, every graph with real weights can be interpreted as an \R-structure.
\end{example}

% \begin{example}
%     Another simple class of $\rr$-structures is obtained by letting $\A = (\{1, \dots, n\})$ be an initial segment of the positive integers and letting $\W$ contain a single unary function $F \: \{1,\dots,n\} \to R$. The resulting $\rr$-structure induces the string
%     $(F_1, \dots, F_n) \in R^n$ such that $F_i = F(i)$. 
%     Conversely, every string $(F_1,\dots,F_n) \in R^n$ induces an $\rr$-structure in this same class. From this we can see that elements of $R^*$ are in bijection with $\rr$-structures of a certain metafinite vocabulary.
% \end{example}

An important restriction of metafinite logic is that first-order variables, those from $\Var = \{x_1,x_2,x_3,\dots \}$, range only over the primary part $\A$ of an $\MVoc$-structure $\D = (\A,\rr,\W)$. Second-order variables of positive arity $k$, those from $\VAR_k = \{X_1^k,X_2^k,X_3^k,\dots\}$, range over all possible functions $W \: A^k \to R$, even those not in $\W$.  First-order metafinite logic allows only quantification over first-order variables, while second-order metafinite logic allows quantification over second-order variables as well. Note that second-order quantification subsumes quantification over the secondary structure by accessing a second-order variable $X \: A \to R$ at a specific point $X(a_0)$.

% An important restriction of metafinite logic is that first-order variables, those from $\Var = \{x_1,x_2,x_3,\dots \}$, range only over the primary part $\A$ of an $\MVoc$-structure $\D = (\A,\rr,\W)$. Using these variables, as well as the constants and functions from $\primary(\Voc)$, we can build \textit{point terms} that range over the primary part of $\MVoc$-structures. Using point terms, weight functions, and the constants and functions from $\secondary(\MVoc)$, we can build \textit{weight terms} that range over the secondary part of $\MVoc$-structures. First-order metafinite formulas are built from $\MVoc$ using first-order variables $\Var$, the equality symbol $=$, connectives from $\{\wedge,\vee,\neg\}$, and quantifiers from $\{\exists,\forall\}$ as one would expect.

% Second-order metafinite logic expands first-order metafinite logic with second-order variables of positive arity $k$, those from $\VAR_k = \{X_1^k,X_2^k,X_3^k,\dots\}$. These variables range over all possible functions $W \: A^k \to R$ on an $\MVoc$-structure $\D=(\A,\rr,\W)$. First-order metafinite logic allows only quantification over first-order variables, while second-order metafinite logic allows quantification over second-order variables as well. Note that even second-order metafinite logic does not allow quantification over the secondary structure $\rr$.

A sentence $\Phi$ of second-order metafinite logic is in prenex normal form when it is a string of second-order quanifiers followed by a first-order sentence $\varphi$, as in $QX_1 \dots Q X_N \varphi(X_1,\dots,X_N)$, where $Q_i \in \{\exists,\forall\}$. Supposing $\Phi$ has $k-1$ quantifier alternations, we say that $\Phi$ is a $\Sigma_k$ sentence when $Q_1 = \exists$, and we say that $\Phi$ is a $\Pi_k$ sentence when $Q_1 = \forall$.

\begin{definition}[Fragments of Second-Order Metafinite Logic]
    Let $\SO(\rr)$ be the set of all second-order sentences in any metafinite vocabulary $\MVoc$ over $\rr$. We identify the following fragments of metafinite second order logic over $\rr$ that we refer to as the $k$th existential or universal fragment or Boolean fragment of $\SO(\rr)$, respectively:
    \begin{enumerate}
        \item $\Sigma_k\SO(\rr) = \{\Phi \in \SO(\rr) : \Phi \text{ is a $\Sigma_k$ sentence} \}$,
        \item $\Pi_k\SO(\rr) = \{\Phi \in \SO(\rr) : \Phi \text{ is a $\Pi_k$ sentence} \}$,
        \item $\exists_k\SO(\rr) \, = \{\Phi \in \Sigma_k\SO(\rr) :  \text{all constants from $\rr$ in $\Phi$ are 0 or 1} \}$,
        \item $\forall_k\SO(\rr) \, = \{\Phi \in \Pi_k\SO(\rr) :  \text{all constants from $\rr$ in $\Phi$ are 0 or 1} \}$.
    \end{enumerate}
\end{definition}

\section{Cook-Levin Theorem Over a Structure}
\label{sub:CookLevinOverStructure}
%%%%%%%%%%%%%%%%%%%%%%%%%%%%%%%%%%%%%%%%%

%\jeremy{This has to be too long.}

The idea behind our hardness results is a generalization of the idea behind the classical Cook-Levin theorem: It is possible to encode the computation of a machine in a logical formula. In the classical setting, the machine is a polynomial-time Turing machine and the formula is a Boolean formula. In our setting, the machine is a polynomial-time $\rr$-machine, and the formula is a first-order formula in the vocabulary of $\rr$.

\begin{table}[h!]
\centering

\renewcommand{\arraystretch}{1}

\begin{tabularx}{\textwidth}{|c|| *{7}{>{\centering\arraybackslash}X|}}
\hline

& \multicolumn{7}{c|}{Registers} \\
\hline
% --- HEADER ROW ---
% The top-left cell now labels the first column as ''Time''.
Time & $L \lbr 0 \rbr$ & $I \lbr 0 \rbr $ &  $\dots$ & $I \lbr k_\program-1 \rbr$ & $Z \lbr 0 \rbr $ &  $\dots$ & $Z \lbr n^m \rbr$ \\
\hline \hline % A double horizontal line to clearly separate the header from the data rows.

% --- DATA ROWS ---
% The structure remains the same as before.
0 & $\ell_0 $ & $\nu_{0,0} $ & $\dots$ & $\nu_{k_\program-1,0}$ & $\rho_{0,0}$  & $\dots$ & $\rho_{n^m,0}$ \\
\hline
\vdots & & & & & & & \\ % \vdots creates vertical dots, useful for showing a sequence.
\hline
$t$ & $\ell_t $ & $\nu_{0,t} $ & $\dots$ & $\nu_{k_\program-1,t}$ & $\rho_{0,t}$ & $\dots$ & $\rho_{n^m,t}$ \\
\hline
\vdots & & & & & & & \\
\hline
$n^m$ &  $\ell_{n^m} $ & $\nu_{0,n^m} $ & $\dots$ & $\nu_{k_\program-1,n^m}$ & $\rho_{0,n^m}$ & $\dots$ & $\rho_{n^m,n^m}$ \\
\hline
\end{tabularx}
\caption{Computation table of $\M$ running in time $n^m$ on input $v \in R^n$.}
\label{tab:ConfigurationSequenceTable}
\end{table}

\begin{restatable}[Encoding Computations in First-Order Formulas]{lemma}{EncodingComputationFOFormula}
\label{lemma:EncodingComputationFOFormula}
    Let $\rr$ be a bipointed structure with all constants, $\M$ be a polynomial-time $\rr$-machine, $q$ be a polynomial, and $v \in R^*$ be a string. Then for all $k \in \N$, there is a first-order formula $\exists\overline{y}\varphi_{v,k}(\overline{x}_1, \dots, \overline{x}_k,\overline{y})$ in the vocabulary of $\rr$, with $\varphi_{v,k}$ quantifier free, such that for all strings $w_1,\dots,w_k \in R^{q(|v|)}$, $$\M(v,w_1,\dots,w_k)=1 \text{ if and only if } \rr \satisfies \exists \overline{y} \ \varphi_{v,k}(w_1, \dots, w_k,\overline{y}).$$ 
\end{restatable}

\begin{proofsketch}
    The time bound of $M$ allows us to represent the computation of $M$ on input $v$ in a finite computation table, as depicted in \Cref{tab:ConfigurationSequenceTable}. Since the program of $\M$ is built from the operations of $\rr$, we may use a formula $\varphi_{v,k} \equiv \start_v \wedge \update \wedge \accept$ in the vocabulary of $\rr$ to describe this table. We need to assume that $\rr$ has all constants because the subformula $\start_{v}$ uses constants to for each $v_i$ in $v = (v_1,\dots,v_n) \in R^n$ to express the zeroth row of the table.
    The subformula $\update$ expresses that row $t+1$ of the table results from modifying row $t$ of the table according to the program of $\M$, 
    while the subformula $\accept$ expresses that $\M$ returns 1 in the last row of the table. The computation table of $\M$ on input $(v,w_1,\dots,w_k)$ is then captured as a sequence $\overline{z} \in R^*$ such that $\M(v,w_1,\dots,w_k)=1$ if and only if $\rr \satisfies \varphi_{v,k}(w_1,\dots,w_k,\overline{z})$.
\end{proofsketch}

\begin{restatable}[Higher Cook-Levin Analogue]{theorem}{HigherCookLevin}
\label{thm:HigherCookLevin}
    Let $\rr$ be a bipointed structure of finite type with all constants. Then for all $k \in \N$ with $k \geq 1$, $\Sigma_k\SAT(\rr)$ is $\Sigma_k\rr$-complete and $\Pi_k\SAT(\rr)$ is $\Pi_k\rr$-complete under polynomial-time $\rr$-machine reductions.
\end{restatable}

\begin{proofsketch}
        To show that $\Sigma_k\SAT(\rr)$ is in $\Sigma_k(\rr)$, it suffices to show that a polynomial-time $\rr$-machine can decide if $\rr \satisfies \varphi(\overline{r})$ for any quantifier-free $\Voc(\rr)$-formula $\varphi(x_1,\dots,x_m)$ and any $\overline{r} = (r_1,\dots,r_m) \in R^m$. We show this by arguing that a polynomial-time $\rr$-machine can recursively decide if $\rr \satisfies \varphi(\overline{r})$ based on the syntactic structure of $\varphi$. In order to implement the base case of this algorithm, we need $\rr$ to have an effective encoding so that an $\rr$-machine can take the encoding of a symbol in the vocabulary of $\rr$ and use it to evaluate the corresponding operation. Our assumption that $\rr$ is of finite type guarantees the existence of an effective encoding that may, moreover, be decoded in constant time. Hence, this recursive algorithm may be implemented on a polynomial-time $\rr$-machine.

        To show hardness, suppose $L \in \Sigma_k\rr$. For any $v \in R^n$, let $Q\overline{y}\varphi_{v,k}$ be the formula depending on the machine associated with $L$ and $v$ defined above, or its negation if $k$ is even, and let $\Phi_{v,k}$ be the $\Sigma_k$ sentence $\Phi_{v,k} \equiv  \exists \overline{x}_1 \forall \overline{x}_2 \dots Q \overline{x}_k Q \overline{y} \  \varphi_{v,k}(\overline{x}_1,\overline{x}_2,\dots,\overline{x}_k,\overline{y})$. Then by \Cref{lemma:EncodingComputationFOFormula} together with the logical replacement theorem of \cite{kleene2002mathematical}, which allows us to prepend the appropriate string of quantifiers, $v \in L$ if and only if $\rr \satisfies \Phi_{v,k}.$ Since the number of atomic subformulas in $\varphi_{v,k}$ is polynomial in $|v|,$ an $\rr$-machine can construct $\Phi_{v,k}$ in polynomial time from $v$ and $L$. Thus $v \mapsto \Phi_{v,k}$ is a polynomial-time reduction from $L$ to $\Sigma_k\SAT(\rr).$
\end{proofsketch}

As a corollary, we see that $\SAT(\rr)$ is $\NP(\rr)$-complete, as was shown in \cite{goode1994accessible} with special cases shown already in \cite{BSS89,BCSS98}. \Cref{thm:HigherCookLevin} was claimed already in \cite{de2004completeness} and attributed to \cite{Poizat_1995}, which is a foundational text that is no longer in wide circulation, preventing us from accessing it. We offer the present proof not only for completeness, but also because we slightly modify the proof to obtain \Cref{thm:HigherBooleanCookLevin} and because we analyze the proof in order to demonstrate that a slight weakening of the assumptions of the theorem are probably necessary.

We know that the assumption that $\rr$ is of finite type is not strictly necessary because there are structures $\rr$ of infinite type for which $\SAT(\rr)$ is $\NP(\rr)$ complete, as shown in \cite{megiddo1990general,Hemmerling98a}. However, we have a strong intuition that the existence of an effective encoding for $\rr$ is a necessary condition for $\SAT(\rr)$ to be in $\NP(\rr)$ because if there is an operation of $\rr$ that $\rr$-machines cannot decode, then $\rr$-machines should not be able to test the truth of formulas with this operation. Given the fact that $\rr$ has an effective encoding if and only if there is a universal $\rr$-machine \cite{Hemmerling98a}, we conjecture that $\SAT(\rr)$ is in $\NP(\rr)$ if and only if there is a universal $\rr$-machine that runs with polynomial-overhead.

% We also have a strong intuition that the presence of all constants in $\rr$ is a necessary condition for $\SAT(\rr)$ to be $\NP(\rr)$-hard. Constants for each $v_i$ in $v = (v_1,\dots,v_n) \in R^n$ are used to hard-code $v$ into the formula $\varphi_{v,1}(\overline{x},\overline{y})$. Since this formula expresses the computation table of $\M$ on inputs of the form $(v,w)$ for some $w$, it seems necessary to specify $v$ in this formula, and it is not clear how to do this without constants while retaining the membership of $\SAT(\rr)$ in $\NP(\rr)$. 

Previously, most research into the polynomial hierarchy over $\rr$ focused on the cases when $\rr$ is either $\R_\textup{add}=(\R,0,1,-,+)$ or $\R_\textup{lin}$ or a number of related structures \cite{CK95}. Many of the results for the additive reals remain true when $\rr$ is a ring, and in this case we already knew already that a $k$-fold quantified version of polynomial feasibility is complete for each level of the polynomial hierarchy over $\rr$ \cite{BCSS98, BC09}. 

Exotic quantifiers are used in \cite{BC09} to extend the polynomial hierarchy over $\R$ in order to achieve completeness results for problems from semialgebraic geometry within these newly defined classes. When restricting to the constant-free Boolean part of these newly defined classes, the exotic quantifiers can be eliminated, thereby yielding completeness results in the Boolean hierarchy over $\R$, which is treated in our next corollary.

\begin{restatable}[Higher Boolean Cook-Levin Analogue]{corollary}{HigherBooleanCookLevin}
\label{thm:HigherBooleanCookLevin}
    Let $\rr$ be a bipointed structure of finite type. Then for all $k \in \N$ with $k \geq 1$, $\exists_k\SAT(\rr)$ is $\exists_k\rr$-complete and $\forall_k\SAT(\rr)$ is $\forall_k\rr$-complete under polynomial-time Turing machine reductions.
\end{restatable}

\begin{proofsketch}
    All constants in a sentence $\Phi$ of  $\exists_k\SAT(\rr)$ are either 0 or 1, implying that the encoding of $\Phi$ is a Boolean string, so $\exists_k\SAT(\rr)$ is in $\exists\rr$ by definition. The reduction in \Cref{thm:HigherCookLevin} from $L \in \Sigma_k\rr$ to $\Sigma_k\SAT(\rr)$ restricts to a reduction from $L \in \exists_k\rr$ to $\exists_k\SAT(\rr)$. Finally, the polynomial-time $\rr$-machine implementing this reduction takes only Boolean inputs and need not execute operations of $\rr$ other than writing 0 and 1 and testing for equality, so this reduction may be implemented on a polynomial-time Turing machine.
\end{proofsketch}

As a corollary we get that $\exists\SAT(\rr)$ is $\exists\rr$-complete. Indeed, this same sort of argument is used in \cite{BC06} to show that $\exists\SAT(\rr)$ is $\exists\rr$-complete when $\rr$ is a ring. When comparing \Cref{thm:HigherBooleanCookLevin} with \Cref{thm:HigherCookLevin}, we see that the assumption that $\rr$ has all constants has been dropped. This is possible because, even though $\exists\rr$ is defined in terms of $\rr$-machines, it is a Boolean complexity class, by which we mean that $L \subseteq \{0,1\}^*$ for every $L \in \exists\rr$.

To the best of our knowledge, the first results about higher levels of the Boolean hierarchy over a structure $\rr$ appear in \cite{CK95}, wherein $\rr$ is taken to be $\R_\textup{add}$ or $\R_\textup{lin}$, among other structures. In these cases, the authors show that $\exists\R_\textup{add} = \NP = \exists\R_\textup{lin}$, and furthermore that the Boolean hierarchy over these structures coincides with the classical polynomial hierarchy. 

The authors of \cite{BC09} introduce the Boolean hierarchy over $\R$ in terms of machine models and show completeness for every level of this hierarchy of a problem that is essentially the circuit version of $\exists_k\SAT(\R)$ and $\forall_k\SAT(\R)$. This hierarchy also is introduced in \cite{SS23} as the real hierarchy by \textit{defining} $\exists_k\R$ as the closure of a problem essentially equivalent to $\exists_k\SAT(\R)$ under polynomial-time Turing machine reductions. \Cref{thm:HigherBooleanCookLevin} offers another proof that their definition in terms of complete problems coincides the machine-based definition found here and in \cite{BC09,OracleSeparationRPH}. 

% Finally, this hierarchy is introduced as the real polynomial hierarchy in \cite{OracleSeparationRPH}, wherein it is defined using real Turing machines. These are easily seen to be equivalent to $\R$-machines, establishing that their hierarchy is identical to Boolean hierarchy over $\R$. 

A number of problems have already been identified as complete for classes in the second level of this hierarchy, including area-universality~\cite{DKMR18}, Hausdorff-distance~\cite{JKM22}, escape-games~\cite{DCLNOW21}, surjectivity~\cite{BC09,SS24}, and image density~\cite{BC09, JJ23}.

\section{Fagin's Theorem Over a Structure}
\label{sub:FaginOverStructure}
%%%%%%%%%%%%%%%%%%%%%%%%%%%%%%%%%%%%%%%%%

The idea behind our results in this section is a generalization of the idea behind Fagin's theorem: It is possible to encode the computation of a machine in second-order logical formula. In the setting of Fagin's theorem, the machine is a polynomial-time Turing machine, and the formula is a second-order formula that is interpreted in finite structures. In our setting, the machine is a polynomial-time $\rr$-machine, and the formula is a second-order metafinite formula that is interpreted in metafinite structures over $\rr$, which we will abbreviate to $\rr$-structures.

\begin{restatable}[Encoding Computations in Second-Order Formulas]{lemma}{EncodingComputationSOFormula}
\label{lemma:EncodingComputationSOFormula}
    Let $\rr$ be a bipointed structure, $\M$ be a polynomial-time $\rr$-machine, $q$ be a positive integer, and $\MVoc$ be a metafinite vocabulary over $\rr$. Then for all $k \in \N$, there is a second-order formula $\exists \overline{Y}\varphi_{\M,k}(X_1,\dots,X_k,\overline{Y})$ in the vocabulary $\MVoc$, with $\varphi_{\M,k}$ first-order, such that for all $\MVoc$-structures $\D$ and all weight functions $W_1,\dots,W_k \: A^q \to R$, $$\M(\D,W_1,\dots,W_k) = 1 \text{ if and only if } \D \satisfies \exists \overline{Y}\varphi_{\M,k}(W_1,\dots,W_k,\overline{Y}).$$ 
\end{restatable}

\begin{proofsketch}
    %In more detail, let $\M$ be an $\rr$-machine running in time $n^m$, and let $\D$ be an $\rr$-structure for a metafinite vocabulary $\MVoc$. 
    The time bound of $\M$ allows us to represent the computation of $\M$ on input $\D$ in a finite computation table, as depicted in \Cref{tab:ConfigurationSequenceTable}, which we may describe with the metafinite formula $\varphi_{\M,k} \equiv \start_\M \wedge \update \wedge \accept$ whose subformulas have the same intended interpretation as those of $\varphi_{v,k}$ in \Cref{lemma:EncodingComputationFOFormula}. 
    
    However, while $\varphi_{v,k}$ describes an accepting computation of $\M$ on a \textit{fixed} input $v$, the formula $\varphi_{\M,k}$ describes an accepting computation of $\M$ on input $\D$ for \textit{any} $\MVoc$-structure $\D$. In order to achieve this, we construct a subformula $\Input_\M$ of $\start_\M$ such that $\Input_\M$ expresses the encoding of $\D$ as a string in $R^*$ when it is interpreted in $\D$. The computation table of $\M$ on input $(\D,W_1,\dots,W_k)$ is then captured as a sequence $\overline{Z}$ of weight functions on $\D$ such that $\M(\D,W_1,\dots,W_k)=1$ if and only if $\D \satisfies \varphi_{\M,k}(W_1,\dots,W_k,\overline{Z}).$
    %We use second-order variables $\overline{Y}$ ranging over all weight functions $Z \: A^m \to R$ to encode the values occurring in the computation table of $\M$ on input $\D = (\A,\rr,\W)$. Instead of representing $t$, $\ell_t$, and $\nu_{i,t}$ as binary strings, as we did in \Cref{lemma:EncodingComputationFOFormula}, we represent them as tuples in $A^{m}$ using their $n$-ary representation when $|A| = |\{0,\dots,n-1\}|$. We do this because metafinite formulas only have access to the secondary structure $\rr$ through weight functions $Z \: A^m \to R$, so we need to represent $t$ as some tuple $\overline{t} \in A^m$ if we want to reference the value $Z(\overline{t}) \in R$ stored in an $\rr$-register at time $t$. 
    % By encoding the time $t$, the label $\ell_t$, and the index $\nu_{i,t}$ as tuples in $A^m$, we capture the computation table of $\M$ on input $\D = (\A,\rr,\W)$ with a finite sequence $\overline{Z}$ of weight functions. The formula $\varphi_\M(\overline{Y})$ is constructed so that $\overline{Z}$ correctly represents an accepting computation table of $\M$ on input $\D$ if and only if $\D \satisfies \varphi_\M(\overline{Z})$. Since $\M(\D) = 1$ if and only if there is such a $\overline{Z}$, we see that $\M(\D) = 1 \text{ if and only if } \D \satisfies \exists\overline{Y} \varphi_\M(\overline{Y}).$ By allowing some variables in $\start_\M(\overline{Y})$ to remain unconstrained, we make space for witness weight functions $W$ in the input to $\M$, which yields the following lemma. 
\end{proofsketch}

\begin{restatable}[Higher Fagin's Analogue]{theorem}{HigherFagin}
\label{thm:HigherFagin}
    Let $\rr$ be a bipointed structure. Then for all $k \in \N$ with $k \geq 1$, $\Sigma_k\SO(\rr)$ captures $\Sigma_k\rr$ and $\Pi_k\SO(\rr)$ captures $\Pi_k\rr$ on $\rr$-structures
\end{restatable}

\begin{proofsketch}
    We first show that the data complexity of $\Sigma_1\SO(\rr)$ is $\NP(\rr)$. Suppose $\Phi$ is a $\Sigma_1\SO(\rr)$ sentence, which means that $\Phi$ is of the form $\Phi \equiv \exists X_1 \dots \exists X_N \varphi(X_1,\dots,X_N),$ where 
    %$X_i$ is a second-order variable of arity $k_i$, and
    $\varphi(X_1,\dots,X_N)$ is a first-order formula in some metafinite vocabulary $\MVoc$ over $\rr$. To show that $\{\D \in \struct(\MVoc) : \D \satisfies \Phi\} \in \Sigma_k\rr$, it suffices to show that given an $\MVoc$-structure $\D = (\A,\rr,\W)$ and weight functions $W_1,\dots,W_N$ with $W_i \: A^{k_i} \to R$, an $\rr$-machine can decide in polynomial time if $\D \satisfies \varphi(W_1,\dots,W_N).$ This is essentially a matter of showing that an $\rr$-machine requires only polynomial-time overhead when simulating the operations of the primary part $\A$ of an $\rr$-structure $\D = (\A,\rr,\W)$ that it receives as input, which may be done by analyzing the encoding of the structure $\D$ as a string in $R^*$ according to our knowledge of the fixed metafinite vocabulary $\MVoc$ of $\D$.

    We next show that if $L \in \Sigma_k\rr$ is a decision problem of $\MVoc$-structures, then there is a sentence $\Phi$ of $\Sigma_k\SO(\MVoc)$ describing $L$. Each such $L \in \Sigma_k\rr$ is of the form $$L = \{\D \in \struct(\MVoc) : (\exists W_1 : A^q \to R)\dots(Q W_N : A^q \to R) \ \M(\D,W_1,\dots,W_N) = 1\}$$ for some positive integer $q$ and some polynomial-time $\rr$-machine $\M$. Let $Q\overline{Y}\varphi_{M,k}$ be the formula associated with the machine $M$ defined above, or its negation if $k$ is even, and let $\Phi_{M,k}$ be the $\Sigma_k\SO(\MVoc)$ sentence $\Phi_{M,k} \equiv \exists Z_1 \dots Q Z_N Q\overline{Y} \varphi_{M,k}(Z_1,\dots,Z_N,\overline{Y}).$ Then by \Cref{lemma:EncodingComputationSOFormula} together with the logical replacement theorem \cite{kleene2002mathematical}, which allows us to prepend the appropriate string of quantifiers, we see that $L = \{\D \in \struct(\MVoc) : \D \satisfies \Phi_{M,k}\}.$ We conclude that $\Sigma_k\SO(\rr)$ captures $\Sigma_k\rr$ on $\rr$-structures.
\end{proofsketch}

While \Cref{thm:HigherCookLevin} assumes that $\rr$ is of finite type with all constants, \Cref{thm:HigherFagin} makes neither of these assumptions. This is due mainly to the way in which descriptive complexity theory approaches complexity classes. We describe in more detail why we can do without each of these assumptions below. 

The assumption that $\rr$ is of finite type is used \Cref{thm:HigherCookLevin} to guarantee that $\rr$ has an effective encoding, which we require when showing that $\SAT(\rr)$ is in $\NP(\rr)$. The corresponding part of \Cref{thm:HigherFagin} shows that the data complexity of $\Sigma_k\SO(\rr)$ is $\Sigma_k\rr$, which we do by constructing a machine that takes input $(\D,\overline{W})$ and decides, for a fixed formula $\varphi$, if $\D \satisfies \varphi(\overline{W})$. Since $\varphi$ is fixed and not part of the input, we do not need to encode $\varphi$ in a way that a machine can understand. Hence, we do not need the assumption that $\rr$ has an effective encoding, let alone the assumption that $\rr$ is of finite type. Notably, this means that \Cref{thm:HigherFagin} applies to the structure $\R_\textup{lin}$, as well as structures $\mathcal{Z}$ of infinite type for which we know that $\NP(\mathcal{Z})$ has no complete problems \cite{gassner1997np}. 

The assumption that $\rr$ has all constants is used in \Cref{thm:HigherCookLevin} when showing that $\SAT(\rr)$ is $\NP(\rr)$-hard in order to build the formula $\varphi_{v,k}$, which hard-codes the fixed input $v$. The corresponding part of \Cref{thm:HigherFagin} builds the formula $\varphi_{M,k}$, which does describe a fixed input, so no constants other than 0 or 1 or the machine constants of $\M$ are necessary, and these are a subset of the constants of the structure $\rr$. 

%shows that for each $L \in \NP(\rr)$ there is some $\Sigma_1\SO(\rr)$ sentence $\exists X \exists \overline{Y} \varphi_{\M,1}(X,\overline{Y})$ that describes the machine $\M$ that verifies $L$. Since the formula $\varphi_{\M,1}(X,\overline{Y})$ describes the computation of $\M$ on inputs of the form $(\D,W)$ for \textit{any} $\rr$-structure $\D$ when it is interpreted in $\D$, we do not need to use constants to encode a specific $\D$ into $\varphi_{\M,1}(X,\overline{Y})$. Hence, we do not need the assumption that $\rr$ has all constants. 

Our proof strategy generalizes that found in \cite{gradel1995descriptive}, which shows that $\Sigma_1\SO(\Rcon)$ captures $\NP(\Rcon)$. A different proof strategy, relying on a prior descriptive characterization of $\P(\rr)$, is used in \cite{de2004completeness} to show that $\Sigma_1\SO(\rr)$ captures $\NP(\rr)$ on structures of finite type.

A generalization of Fagin's theorem to a class of structures on the natural numbers, called arithmetical structures, is already found in \cite{graedel1998metafinite}, though the authors use a different measure of the size of an input, so their results are incomparable with ours. In more detail, arithmetical structures are metafinite structures $\D = (\A,\mathcal{N},\W)$ where $\mathcal{N}$ resembles $\mathcal{N}_0 = (\N,0,1,+,\cdot,\max,\sum,\prod)$ in a technical sense. In an effort to capture $\NP$ over arithmetical structures, the authors define the size of such a structure as the length of its binary encoding. In contrast, we define its size as the length of its encoding as a string of natural numbers because our aim is to capture $\NP(\mathcal{N})$ rather than $\NP$. Due to differing definitions of the size of an input, our results are not directly comparable.

Another result that is similar but incomparable to ours appears in \cite{hannula2020descriptive} in which the authors introduce separate-branching machines over $\R$ and prove a Fagin-type theorem for the corresponding class $\textup{S-NP(\R)}$. Their result is incomparable to ours because there is no structure $\rr$ for which $\rr$-machines correspond to separate-branching machines over $\R$. The authors also identify the constant-free Boolean part of $\textup{S-NP}(\R)$ with a natural fragment $\exists[0,1]^\leq$  of $\exists\R$, and we expect that a small modification of their Fagin-type theorem for $\textup{S-NP}(\R)$ leads to a characterization of $\exists[0,1]^\leq$ in terms of existential second-order metafinite logic, similar to the characterization of $\exists\rr$ we give in the following corollary. 

We mention one last result that is similar but incomparable to ours that has recently been established in \cite{barlag2025logical}. Herein, the authors derive a Fagin's type theorem over semirings using semiring semantics, rather than the metafinite semantics that we employ here. These logics appear to be intimately related, though, because they may both be used to capture some of the same complexity classes.

In a result that is very similar to our own (but derived independently of our work), the authors of \cite{meer2025some} show that by restricting $\Sigma_1\SO(\Rcon)$ to those formulas with only rational constants, one can capture $\exists\R$ over discrete $\Rcon$-structures, whose definition makes use of the fact that $\N \subseteq \R$. Since we work with structures $\rr$ whose universe may contain neither $\Q$ nor $\N$, we restrict $\Sigma_k\SO(\rr)$ to those formulas with only $0$ and $1$ as constants, and we capture $\exists_k\rr$ over Boolean $\rr$-structures.

% \begin{proofsketch}
%     The argument showing that $\Sigma_k\SO(\rr)$ captures $\Sigma_k\rr$ on $\rr$-structures follows from \Cref{lemma:EncodingComputationSOFormula} in essentially the same way as the argument showing that $\Sigma_1\SO(\rr)$ captures $\NP(\rr)$ on $\rr$-structures, with small modifications and an additional step to account for the alternating string of quantifiers in the definition of $\Sigma_k\SO(\rr)$ and $\Sigma_k\rr$.
% \end{proofsketch}

In the same way that we modified the proof that $\Sigma_1\SO(\rr)$ captures $\NP(\rr)$ on $\rr$-structures to show that $\exists\SO(\rr)$ captures $\exists\rr$ on Boolean $\rr$-structures, we may modify the proof that $\Sigma_k\SO(\rr)$ captures $\Sigma_k\rr$ on $\rr$-structures to show that $\exists_k\SO(\rr)$ captures $\exists_k\rr$ on Boolean $\rr$-structures.

\begin{restatable}[Higher Boolean Fagin's Analogue]{corollary}{HigherBooleanFagin}
\label{thm:HigherBooleanFagin}
    Let $\rr$ be a bipointed structure. Then for all $k \in \N$ with $k \geq 1$, $\exists_k\SO(\rr)$ captures $\exists_k\rr$ and $\forall_k\SO(\rr)$ captures $\forall_k\rr$ on Boolean metafinite structures over $\rr$.
\end{restatable}

\begin{proofsketch}
    %We first show that the data complexity of $\exists\SO(\rr)$ is $\exists\rr$. 
    Suppose $\Phi \equiv \exists \overline X_1 \dots Q \overline X_N \varphi(X_1, \dots, X_N)$ is a sentence of $\exists_k\SO(\MVoc)$, and let $L = \{\D \in \Bstruct(\MVoc) : \D \satisfies \Phi\}.$ Since all constants from $\rr$ occurring in $\varphi$ are either 0 or 1, the $\rr$-machine from \Cref{thm:HigherFagin} that decides if $\D \satisfies \varphi(\overline{W})$ is constant free. Moreover, since $L$ is a decision problem of Boolean $\rr$-structures, we see that $L \in \BP(\NP^0(\rr)) = \exists\rr$. From this we conclude that the data complexity of $\exists\SO(\rr)$ is $\exists\rr$.
    
    Each $L \in \exists_k\rr$ is of the form specified in \Cref{thm:HigherFagin} where $\M$ is a polynomial-time constant-free $\rr$-machine. Since $\M$ is constant free, all constants from $\rr$ occurring in $\varphi_{\M,k}$ are 0 or 1. Thus, $\exists X \exists\overline{Y} \varphi_{\M,1}(X,\overline{Y})$ is an $\exists\SO(\rr)$ sentence, and from this we conclude that $\exists\SO(\rr)$ captures $\exists\rr$ on Boolean $\rr$-structures.
\end{proofsketch}

An advantage of using the class of Boolean $\rr$-structures is that they may be identified with the class of finite structures that is used in classical descriptive complexity theory. Indeed, this is the perspective  taken by the authors of \cite{hannula2022tractability} when they show that $\exists\SO(\R_\textup{add})$ captures $\NP$ on finite structures, using the fact that $\exists\R_\textup{add} = \NP$ established in \cite{CK95}. Given the similar result that $\exists\R_\textup{lin} = \NP$, one implication of \Cref{thm:HigherBooleanFagin} is that $\exists\SO(\R_\textup{lin})$ captures $\NP$ on Boolean $\R_\textup{lin}$-structures. In this way, we can see that Boolean $\rr$-structures have a natural place in descriptive complexity theory.

\section{Oracles and Hierarchies}
\label{sub:OraclesAndHierarchies}
%%%%%%%%%%%%%%%%%%%%%%%%%%%%%%%%%%%%%%%%%

% \jeremy{Need to include the discussion about why we choose to use $\NP(\rr)^{\Sigma_k\rr}$ instead of $\NP(\rr)^{\Sigma_k\SAT(\rr)}$. Should say that everything is relatively standard argument, except we choose this more general definition using oracle classes rather than just problems so that it applies to all structures and not just structures for which $\Sigma_k\SAT(\rr)$ is $\Sigma_k\rr$-complete.}

% \jeremy{I need to introduce the $\Sigma$ and $\Pi$ notation because it is really helpful to explain the proof below.}

% \jeremy{I wonder if the notation $\Sigma$ instead of $\exists$ as a complexity class operator is well chosen, because apparently the $\exists$ notation is well-established already, and I don't really see a reason to prefer $\Sigma$ over $\exists$, since both are asking for a string $w \in R^*$, rather than a string $w \in \{0,1\}^*$... Perhaps a good way to disambiguate could be to say $\exists^\mathsf{p}$ or $\exists^{\mathsf{poly}}$}

% \jeremy{Ahhhhh, actually the $\Sigma$ notation might actually be misleading because it suggests that $\Sigma \Delta_k\rr = \Sigma_{k+1}\rr$. Is this true? It would be surprising because we know that $\Sigma \Pi_k\rr = \Sigma_{k+1}\rr$, but perhaps that is the result of the statement that $\Sigma_{k+1}\rr = \NP(\rr)^{\Sigma_k\rr} = \Sigma(\P(\rr)^{\Sigma_k\rr}) = \Sigma\Delta_k\rr.$}

The classical polynomial hierarchy was originally defined in terms of oracle classes \cite{stockmeyer1976polynomial}. Since the polynomial hierarchy is now usually defined in terms of machine models and witness strings, characterizing each level of this hierarchy in terms of oracles requires proof. The proof strategy we use below is a generalization of that found in \cite{Arora_Barak_2009}, for the classical polynomial hierarchy, and \cite{CK95}, for the polynomial hierarchy over $\R_\textup{add}$ and a number of related structures. The latter proof was already generalized to rings in \cite{BCSS98}. 

\begin{restatable}[Oracle Polynomial Hierarchy Characterization]{theorem}{OraclePolynomialHierarchyCharacterization}
\label{thm:OraclePolynomialHierarchyCharacterization}
    Let $\rr$ be a bipointed structure. Then for all $k \in \N$, $\Sigma_{k+1}\rr = \NP(\rr)^{\Sigma_k\rr}$ and $\Pi_{k+1}\rr = \coNP(\rr)^{\Sigma_k\rr}.$  
\end{restatable}

\begin{proofsketch}
    We proceed by induction on $k \in \N$. The base case holds immediately, and in the inductive step establishing that $\Sigma_{k+1}\rr = \NP(\rr)^{\Sigma_k\rr}$, the difficult inclusion to show is $\NP(\rr)^{\Sigma_k\rr} \subseteq \Sigma_{k+1}\rr$. Our inductive hypothesis implies that it is sufficient to show that 
    a polynomial-time $\rr$-machine $\M$ with a $Q \in \Sigma_{k}\rr$ oracle may be simulated by a polynomial-time $\rr$-machine $N$ with an $S \in \Pi_{k-1}\rr$ oracle, together with the help of witness strings that are provided in addition to the input. 
    This is possible because for each $Q \in \Sigma_k\rr$ there is some $S \in \Pi_{k-1}\rr$ such that $u \in Q$ if and only if there is some $w$ such that $(u,w) \in S$, and $u \not\in Q$ if and only if $(u,w') \not\in S$ for all $w'.$ Given the answers $b_1, \dots, b_p \in \{0,1\}$ to the queries $u_1,\dots,u_p$ that $\M$ makes on input $v$, as well as the witness strings $w_1,\dots,w_p$ and $w'_1, \dots,w'_p$, the machine $N$ can simulate $\M$ by branching according to the answers $b_1,\dots,b_p$. Furthermore, $N$ can check that these answers are correct by querying $(u_i,w_i) \in S$ if $b_i=1$ and querying $(u_i,w'_i) \in S$ if $b_i=0.$ Thus, $\Sigma_{k+1}\rr = \NP(\rr)^{\Sigma_k\rr}$, and this implies $\Pi_{k+1}\rr = \coNP(\rr)^{\Sigma_k\rr}$, thereby completing our induction.
    % We do this by programming $N$ to compute everything in the square brackets below:
    % \[
    % {
    % \renewcommand{\arraystretch}{1.2}
    % \begin{array}{rcl}
    %     \M(v) = 1 & \iff & \exists (p,(b_1,\dots,b_p),(w_1,\dots,w_p)) \forall(w'_1,\dots,w'_p)  \\
    %      & & [\text{On input $v$, if $\M$ branches according to answers $b_1,\dots,b_p,$} \\
    %      & & \text{then $\M$ will make queries $u_1,\dots,u_p$ before accepting $v$}] \wedge \\
    %      & & [\bigwedge_{i=1}^p (b_i=1 \implies (u_i,w_i) \not\in \overline{S}] \wedge \\
    %      & & [\bigwedge_{i=1}^p (b_i=0 \implies (u_i,w'_i) \in \overline{S}].
    % \end{array}
    % }
    % \]
    %Thus, $\Sigma_{k+1}\rr = \NP(\rr)^{\Sigma_k\rr}$, and this implies $\Pi_{k+1}\rr = \coNP(\rr)^{\Sigma_k\rr}$.
\end{proofsketch}

One usually states the corresponding theorem for the classical polynomial hierarchy as $\Sigma_{k+1} = \NP^{\Sigma_k\SAT}$, where $\Sigma_k\SAT$ is the problem of deciding if a quantified Boolean formula of the appropriate form is true. This is equivalent to saying that $\Sigma_{k+1} = \NP^{\Sigma_k}$ because $\Sigma_k\SAT$ is $\Sigma_k$-complete. We use the oracle class $\Sigma_k\rr$, rather than the single oracle $\Sigma_k\SAT(\rr)$, so that \Cref{thm:OraclePolynomialHierarchyCharacterization} applies to structures for which $\Sigma_k\SAT(\rr)$ is not $\Sigma_k\rr$-complete. Modifying the proof of \Cref{thm:OraclePolynomialHierarchyCharacterization}, we obtain the following corollary.

\begin{restatable}[Oracle Boolean Hierarchy Characterization]{corollary}{OracleBooleanHierarchyCharacterization}
\label{thm:OracleBooleanHierarchyCharacterization}
    Let $\rr$ be a bipointed structure. Then for all $k \in \N$, $\exists_{k+1}\rr = \exists \rr^{\Sigma_k\rr^0}$ and $\forall_{k+1}\rr = \forall\rr^{\Sigma_k\rr^0}.$    
\end{restatable} 
% \begin{proofsketch}
%     By \Cref{thm:OraclePolynomialHierarchyCharacterization}, we have the equalities $\exists\rr^{\Sigma_k\rr} = \BP^0(\NP(\rr))^{\Sigma_k\rr} = \BP^0(\NP(\rr)^{\Sigma_k\rr}) = \BP^0(\Sigma_{k+1}\rr) = \exists_{k+1}\rr.$ Similarly, $\forall_k\rr^{\Sigma_k\rr} = \forall_{k+1}\rr$. 
% \end{proofsketch}

This corollary tells us that it is unlikely that we can replace the non-Boolean oracle class $\Sigma_k\rr^0$ with the Boolean oracle class $\exists_k\rr$, as one might expect from a straightforward analogy with the classical polynomial hierarchy. For example, while $\NP^{\NP} = \Sigma_2$ and $\NP(\rr)^{\NP(\rr)} = \Sigma_2\rr$, \Cref{thm:OracleBooleanHierarchyCharacterization} tells us that $\exists\rr^{\NP(\rr)^0} = \exists_2\rr$, so it is unlikely that $\exists\rr^{\exists\rr} = \exists_2\rr$. It is instructive to reflect on why our proof requires the constant-free, non-Boolean oracle class $\NP(\rr)^0$ when showing that $\exists\rr^{\NP(\rr)^0} = \exists_2\rr$.

% It is instructive to reflect on why our proof requires the constant-free, non-Boolean oracle class $\Sigma_k\rr^0$. Already at $k=0$, in order to show that $\exists\rr^{\Sigma_0\rr^0} = \exists\rr$, we need to use a constant-free machine to simulate a constant-free machine with an oracle $Q \in \Sigma_0\rr^0=\P(\rr)^0$. This is possible only because $Q$ is itself decidable with a constant-free machine. Otherwise, we could hide a constant in the oracle $Q$, making this simulation impossible. Thus, we need to take the constant-free part of $\Sigma_k\rr$ as an oracle class.

% We cannot take the Boolean part of $\Sigma_k\rr^0$ as an oracle class because witness strings are non-Boolean. In order to show the inclusion $\exists_2\rr \subseteq \exists\rr^{\NP(\rr)^0}$, we need to 

% It is instructive to reflect on why our proof requires the constant-free, non-Boolean oracle class $\NP(\rr)^0$ when showing that $\exists\rr^{\NP(\rr)^0} = \exists_2\rr$. In order to show the inclusion $\exists\rr^{\NP(\rr)^0} \subseteq \exists_2\rr$, we proceed as in \Cref{thm:OraclePolynomialHierarchyCharacterization} by using a constant free machine $N$, together with witness strings, to simulate a constant-free machine $\M$ with an oracle $Q \in \NP(\rr)^0$. This is only possible because we assume that $Q$ has a constant-free verification algorithm. 

In order to show the inclusion $\exists\rr^{\NP(\rr)^0} \subseteq \exists_2\rr$, we use a constant-free machine $N$ and witness strings to simulate a constant-free machine $M$ with an oracle $Q \in \NP(\rr)^0$. If $Q$ were not in the constant-free part of $\NP(\rr)$, we could hide a constant in the oracle $Q$, making a constant-free simulation impossible. In order to show the inclusion $\exists_2\rr \subseteq \exists\rr^{\NP(\rr)^0}$, we replace witness strings $w \in R^*$ with an oracle $Q \in \NP(\rr)^0$. Since witness strings can be non-Boolean, it should generally be impossible to replace them with an oracle in the Boolean part of $\NP(\rr)^0$. Thus, we take $\NP(\rr)^0$ as our oracle class because we need to take the constant-free part of $\NP(\rr)$, but we cannot take the Boolean part of $\NP(\rr)^0$.

%%%%%%%%%%%%%%%%%%%%%%%%%%%%%%%%%%%%%%%%%
\section{Conclusion}
\label{sec:Conclusion}
%%%%%%%%%%%%%%%%%%%%%%%%%%%%%%%%%%%%%%%%%

We have characterized the polynomial and Boolean hierarchies over a first-order structure $\rr$ in terms of machine models, complete problems, descriptive complexity, and oracles. Additionally, we have shown that descriptive complexity is well suited to working with algorithms over infinite vocabulary structures, such as real vector spaces. In this way, we have answered fundamental questions about the complexity of algorithms presented in terms of the primitive operations present in a first-order structure.

\newpage

\bibliography{ER, references}

% \printbibliography

\appendix

%%%%%%%%%%%%%%%%%%%%%%%%%%
\newpage
\section{Extended Preliminaries}
\label{sec:ExtendedPreliminaries}
%%%%%%%%%%%%%%%%%%%%%%%%%%

The section provides introductory material for readers who are familiar with theoretical computer science or mathematics, but who may not be familiar with first-order logic, machine models over first-order structures, or the descriptive complexity of metafinite structures. We assume that many readers will use this section only as a form of reference to look up certain details of our definitions when necessary.

% We will start with first-order logic and structures in \Cref{sub:FirstOrderStructures}.
% One can typically think of the reals as the primary example of interest.
% Then we explain our model of computation with access to possibly infinite structures in \Cref{sub:MachinesOverR}.
% We then proceed to metafinite structures in \Cref{sub:MetafiniteLogicandStructures}. 
% \hans{Some reference is missing here}
% Formally, metafinite structures are finite structures with finite access to infinite structures. 
% At last, we explain how oracles-access can be added to our machines.

%%%%%%%%%%%%%%%%%%%%%%%%%%%%%%%%%%%%%%%%%
\subsection{First-Order Structures}
\label{sub:FirstOrderStructures}
%%%%%%%%%%%%%%%%%%%%%%%%%%%%%%%%%%%%%%%

In many fields of mathematics and computer science, one studies structures $\rr$ composed of a set $R$, called the universe or domain of the structure, together with some elements $c \in R$, called constants, some functions $f:R^k \to R$, and some relations $P \subseteq R^k,$ where $k$ can be any positive integer. These are called first-order structures because the language of first-order logic is designed to speak about them. In this section, we formally introduce first-order structures, leaving the syntax and semantics of first-order logic to \Cref{sub:FirstOrderLogic}. 

The symbols $c$, $f$, and $P$ that we use to refer to the constants, functions, and relations of a structure form a vocabulary with which we can speak about the structure. Formalizing these vocabularies is the first step towards formalizing the relationship between language and structure via first-order logic and first-order structures.

\begin{definition}[Vocabularies]
\label{def:Vocabulary}
    A \textit{vocabulary} $\Voc$ consists of three disjoint sets:
    \begin{enumerate}
        \item a set $\Con$ of constant symbols $c$,
        \item a set $\Fun$ of function symbols $f$, each with a positive integer $k$ called the \textit{arity} of  $f$.
        \item a set $\Rel$ of relation symbols $P$, each with a positive integer $k$ called the \textit{arity} of $P$. 
    \end{enumerate}
    In general, we will write vocabularies as $\Voc = (\Con,\Fun,\Rel)$. If $\Voc$ is finite, we may write $$\Voc = (c_1,\dots c_{|\Con|}, f_1,\dots,f_{|\Fun|}, P_1,\dots,P_{|\Rel|}).$$
\end{definition}

\begin{remark}
    We make no assumption about the nature of symbols. They can be the elements of any set, including elements such as real numbers that we can never write down on paper in their entirety. Nonetheless, just as we can refer to a real number $r \in \R$ with the variable ``$r$'', we can refer to these infinite symbols with the variables ``$c$'', ``$f$'', and ``$P$''. Technically, ``$c$'', ``$f$'', and ``$P$'' are metavariables of our metalanguage, which is the language that we use to speak about the formal languages of first-order logic built from such symbols. In this case, our metalanguage is the language of mathematics.
    %which is the language that we use to speak about sentences composed of the symbols $c$, $f$, and $P$.
\end{remark}

Symbols have no inherent meaning. Though we think of a constant symbol $c$ as something that can refer to an element of some set, vocabularies do not specify a fixed interpretation of $c$. Similarly, there is no fixed function that a function symbol $f$ refers to, nor a fixed relation that a relation symbol $P$ refers to. Providing an interpretation of these symbols as constants, functions, and relations in a given set results in a \textit{first-order structure}, which we will often simply call a structure.

\begin{definition}[Structures]
\label{def:Structure}
    A \textit{(first-order) structure} $\rr$ for a vocabulary $\Voc = (\Con,\Fun,\Rel)$, also called a \textit{$\Voc$-structure}, provides an interpretation of the symbols in $\Voc$ by specifying the following:
    %\till{Do we really need to use super scripts here? }
    %\jeremy{I think the distinction between symbols and their interpretations is important conceptually. This is the standard notation to express this distinction. In practice we will often we will drop the superscripts.}
    \begin{enumerate}
        \item a nonempty set $R$ called the \textit{universe}, \textit{domain}, or \textit{underlying set} of $\rr$,
        \item an element $c^\rr \in R$ for each $c \in \Con$,
        \item a function $f^\rr \: R^{k} \to R$ for each $f\in \Fun$ with arity $k$,
        \item a relation $P^\rr \subseteq R^{k}$ for each $P \in \Rel$, with arity $k$.
    \end{enumerate}
    A \textit{structure} is a structure for some vocabulary $\Voc$. This means that each structure is equipped with a particular vocabulary. Given a structure $\rr$, the vocabulary of $\rr$ is  $\Voc(\rr)$. We often write structures as $$\rr = (R,(c^\rr : c \in \Con),(f^\rr : f \in \Fun),(P^\rr : P \in \Rel)),$$ where the universe of $\rr$ is often denoted with the same character, possibly in a different font. We use sequence notation rather than set notation to emphasize that, for example, $(c^\rr : c \in \Con)$ is the function that takes $c$ to $c^\rr$, rather than the image of this function. If $\Voc(\rr)$ is finite, we may write $$\rr = (R,c^\rr_1,\dots c^\rr_{|\Con|}, f^\rr_1,\dots,f^\rr_{|\Fun|}, P^\rr_1,\dots,P^\rr_{|\Rel|}).$$ In practice, we will often conflate a symbol of $\Voc$ with its interpretation in a $\Voc$-structure $\rr$ by dropping the superscript $\rr$, although it will sometimes be useful to maintain this distinction.
\end{definition}

\begin{remark}
    We could have defined structures without reference to vocabularies, but as soon as we want to talk about the relationship between a structure $\rr$ and some language, we need to choose a vocabulary $\Voc$, and we thereby arrive at $\Voc$-structures anyways. For this reason, structures come equipped with a choice of vocabulary, and different choices result in different structures. This may unsettle the reader who notices that there are many different but essentially equivalent choices of vocabulary for a given mathematical structure. We assure this reader that it is possible to translate between different choices of vocabulary, as explained Section 5.3, especially Remark 1, of \cite{Hodges_1993}.
\end{remark}

\begin{example}[Booleans]
    Boolean values are often represented with the set $\{0,1\}$ where $0$ represents false and $1$ represents true. We may regard Booleans as a structure for the vocabulary $\Voc_\textup{bool} = (0,1)$ containing two constant symbols $0$ and $1$ by setting $\B = (\{0,1\},0^\B,1^\B)$ where $0^\B = 0$ and $1^\B = 1$. Since the interpretation of $0$ and $1$ is clear in $\B$, we may drop the superscript $\B$ and write $\B = (\{0,1\},0,1)$. 
    %Another $\Voc_\textup{bool}$-structure is $\mathcal{C} = (\{0,1\},0^\mathcal{C},1^\mathcal{C})$ where $0^\mathcal{C} = 1$ and $1^\mathcal{C}=0$.
\end{example}

\begin{example}[Graphs]
    A directed graph $(V,R)$ is a set of vertices $V$ together with a binary edge relation $R \subseteq V^2$. We may regard a directed graph as a structure for the vocabulary $\Voc_\textup{graph} = (E)$ containing only the binary relation symbol $E$ by setting $\mathcal{G} = (V,E^\mathcal{G})$ where $E^\mathcal{G} = R$. If $E^\mathcal{G}$ is a symmetric relation, then we may view $\mathcal{G}$ as an undirected graph. For example, $\mathcal{K}_2 = (\{0,1\},\{(0,1),(1,0)\})$ is a $\Voc_\textup{graph}$-structure representing the complete undirected graph on two vertices.
\end{example}

\begin{example}[Partially Ordered Sets]
    A partially ordered set $(X,R)$ is a set $X$ together with a binary relation $R \subseteq X^2$ such that $R$ is reflexive, antisymmetric, and transitive. We may regard a partially ordered set as a structure for the vocabulary $\Voc_\textup{pos} = (\leq)$ containing the binary relation symbol $\leq$ by setting $\mathcal{X} = (X,\leq^\mathcal{X})$ where $\leq^\mathcal{X} = R$. For example, $\mathcal{N} = (\N,\leq)$ is a $\Voc_\textup{pos}$-structure where $\leq^\mathcal{N} = \leq$ is the usual order on the natural numbers. Similarly, $\mathcal{P}_n = (2^{\{1,\dots,n\}},\subseteq)$ is a $\Voc_\textup{pos}$-structure where $\leq^{\mathcal{P}_n} = \subseteq$ is the subset inclusion relation.
\end{example}

\begin{example}[Monoids]
    A monoid $(M,0,+)$ is a set $M$ together with an associative binary operation $+ : M^2 \to M$ and a constant $0 \in M$ that is an identity for the $+$ operation. We may regard a monoid as a structure for the vocabulary $\Voc_\textup{mon} = (0,+)$ by setting $\M = (M,0^\M,+^\M)$ where $0^\M = 0$ and $+^\M = +$. For example, $\mathcal{N} = (\N,0,+)$ is $\Voc_\textup{mon}$-structure where $0^\mathcal{N} = 0 \in \N$ and $+^\mathcal{N} = +$ is addition. Similarly, $\mathcal{Z}=(\Z,1,\cdot)$ is a $\Voc_\textup{mon}$-structure where $0^\mathcal{Z} = 1 \in \Z$ and $+^\mathcal{Z} = \cdot$ is multiplication.
\end{example}

\begin{example}[Groups]
    A group $(G,0,-,+)$ is a monoid $(G,0,+)$ together with an operation $- \: G \to G$ that maps each element $x$ of $G$ to its inverse $-x$ under the $+$ operation. If  $x + y = y + x$ for all $x$ and $y$ in $G$, then we say that $\mathcal{G}$ is a commutative group. We may regard a group as a structure for the vocabulary $\Voc_\textup{group} = (0,-,+)$. For example, $\mathcal{Z} = (\Z,0,-,+)$ is a $\Voc_\textup{group}$-structure where each of the symbols $0$, $-$, and $+$ receive their expected interpretation. Similarly, $\mathcal{Q} = (\Q\setminus \{0\},1,^{-1},\cdot)$ is a $\Voc_\textup{group}$-structure where $0^\mathcal{Q} = 1 \in \Q$, the function $-^\mathcal{Q} = ^{-1}$ sends $x$ to its multiplicative inverse $x^{-1}$, and $+^\mathcal{Q} = \cdot$ is multiplication. 
    
    %Furthermore, both of these groups are commutative. 
    
    %For an example of a group that is not commutative, consider the set $S_3$ of permutations of the set $\{1,2,3\}$. If $e$ is the permutation that does nothing, $^{-1} \: S_3 \to S_3$ is the operation that sends a permutation to its inverse as a function, and $\circ \: S_3^2 \to S_3$ composes two permutations, then the structure $\mathcal{S}_3 = (S_3,e,^{-1},\circ)$ is a non-commutative group that we may interpret as the symmetries of a triangle.
\end{example}

\begin{example}[Rings]
    A ring $(R,0,1,-,+,\cdot)$ is a commutative group $(R,0,-,+)$ together with a monoid $(R,1,\cdot)$ that is compatible in the sense that the $\cdot$ operation distributes over the $+$ operation. We may regard each ring as a structure for the vocabulary $\Voc_\textup{ring} = (0,1,-,+,\cdot)$. For example, the usual definitions of addition and multiplication on $\Z$, $\Q$, $\R$, and $\mathbb{C}$ give rise to a ring structure on each of these sets where each symbol of $\Voc_\textup{ring}$ receives its expected interpretation. Additionally, for any natural number $m \in \N$, $\mathcal{Z}_m = (\{0,\dots,m-1\},0,1,-,+,\cdot)$ is a $\Voc_\textup{ring}$-structure where the symbols $-$, $+$, and $\cdot$ are interpreted as the corresponding operation modulo $m$. 
\end{example}

\begin{example}[Ordered Rings]
    An ordered ring $(R,0,1,-,+,\cdot,\leq)$ is a ring $(R,0,1,-,+,\cdot)$ together with a linear ordering $\leq$ that is compatible with the ring structure on $R$ in the sense that for all $x,y,z \in R$, if $x \leq y$, then $x+z \leq y + z$, and if $0 \leq x$ and $0 \leq y$, then $0 \leq x \cdot y$. We may regard each ordered ring as a structure for the vocabulary $\Voc_\textup{oring} = (0,1,-,+,\cdot,\leq)$. For example, the ordered ring of real numbers $\R = (\R,0,1,-,+,\cdot,\leq)$ is a $\Voc_\textup{oring}$-structure.
\end{example}

\begin{example}[Real Vector Spaces]
\label{example:VectorSpacesOverReals}
    A vector space over $\R$ is a commutative group $(V,0,-,+)$ together with a scalar multiplication operation $\cdot \: \R \times V \to V$ such that for all $r,s \in \R$ and all $x,y \in V$,
    \begin{enumerate}
        \item $1 \cdot x = x$,
        \item $r \cdot (x + y) = r \cdot x + r \cdot y$,
        \item $(r+s)\cdot x = r\cdot x + s \cdot x$,
        \item $(rs) \cdot x = r \cdot (s \cdot x)$.
    \end{enumerate}
    We may regard each real vector space as a structure for the vocabulary $\Voc_\textup{lin} = (0,-,+,(\scalar{r} : r \in \R))$ by setting $\mathcal{V} = (V,0^\mathcal{V},-^\mathcal{V},+^\mathcal{V},(\scalar{r}^\mathcal{V} : r \in \R))$ where each of $0$, $-$, and $+$ receives their expected interpretation, and $\scalar{r}^\mathcal{V} \: V \to V$ is the function that sends the vector $x$ to the vector $r \cdot x$. For example, for any fixed $n \in \N$, $\R^n = (\R^n,0,-,+,(\scalar{r} : r \in \R))$ is a $\Voc_\textup{lin}$-structure where $\scalar{r}(x_1,\dots,x_n) = (r \cdot x_1, \dots, r \cdot x_n)$. In particular, the linear reals $\R_\textup{lin} = (\R,0,-,+,(\scalar{r} : r \in \R))$ are a $\Voc_\textup{lin}$-structure with $\scalar{r}(x) = r \cdot x$.
\end{example}

\begin{example}[Modules over a Ring $\rr$]
    If we replace $\R$ in the definition of vector spaces with another ring $\rr$, then we obtain the definition of a module over $\rr$. We may regard each module over $\rr$ as a structure for $\Voc_\textup{mod} = (0,-,+,(\scalar{r} : r \in R))$. For example, any commutative group $\mathcal{G}$ may be expanded to a module over the ring of integers $\mathcal{Z}$ by adding the functions $(\mu_z : z \in \Z)$ to the structure $\mathcal{G}$, where $\mu_z \: G \to G$ sends $x$ to $\mu_z(x) = \sum_{i=1}^z x$ if $z$ is nonnegative and to $\mu_z(x) = -\left(\sum_{i=1}^z x\right)$ if $z$ is negative.
\end{example}

\begin{remark}
\label{rem:ScalarMultiplication}
    The scalar multiplication operation $\cdot \: \R \times V \to V$ of a vector space $V$ cannot in general be viewed as an $m$-ary operation on $V$. To fit the operation $\cdot$ into the formalism of single-sorted first-order logic, we split the operation $\cdot$ into uncountably many operations $(\scalar{r} : r \in \R)$. Another option is to work in the more general system of many-sorted logic, with a sort for scalars from $\R$ and a sort for vectors from $V$, in which $\cdot\:\R\times V \to V$ is a valid operation. This example shows that there are mathematical objects that do not fit cleanly into the formalism of single-sorted first-order logic. Topological spaces are another example of this phenomenon. Despite this, we work in single-sorted first-order logic for its simplicity and wide applicability.
\end{remark}

\subsubsection{Bipointed Structures of Finite Type with All Constants.} In the following sections, we will often encode discrete objects, such as natural numbers, into formulas built from the vocabularies of structures. For this reason, we will often assume that structures have two constants $0$ and $1$ that we can use to encode discrete objects in binary. We call these structures \textit{bipointed} because they have two distinguished points denoted by $0$ and $1$. 

\begin{definition}[Bipointed Structure]
\label{def:BipointedStructure}
    A structure $\rr$ is \textit{bipointed} if $\Voc(\rr)$ has two constant symbols $0$ and $1$ that $\rr$ interprets as different elements of its universe. That is, $0^\rr \neq 1^\rr$. 
\end{definition}

Many structures are not bipointed. As long as a structure $\rr$ has at least two distinct elements $a$ and $b$ in its universe, we may expand the vocabulary of $\rr$ to include two constant symbols $0$ and $1$, and we may expand the structure $\rr$ by setting $0^\rr = a$ and $1^\rr = b$. The expanded structure $\rr'$ is strictly different from the original structure $\rr$, but it is not essentially different.

\begin{example}[Bipointed Graphs]
    The complete undirected graph on 2 vertices $\mathcal{K}_2 = (\{0,1\},\{(0,1),(1,0)\})$ has vocabulary $\Voc(\mathcal{K}_2) = (E)$. We may expand this to the vocabulary $\Voc' = (E,0,1)$ and we may expand $\mathcal{K}_2$ to the structure $\mathcal{K}'_2 = (\{0,1\},0,1,\{(0,1),(1,0)\})$ where $0^{\mathcal{K}'_2} = 0$ and $1^{\mathcal{K}'_2} = 1$.
\end{example}

\begin{example}[Bipointed Groups]
    The group $\rr = (\R,0,+)$ has vocabulary $\Voc(\rr) = (0,+)$. We may expand this to the vocabulary $\Voc' = (0,1,+)$ and we may expand $\rr$ to the structure $\rr' = (\R,0,1,+)$ by setting $1^{\rr'} = 1$.
\end{example}

We will often make two further assumptions about structures that are essential to \Cref{sec:cook-levin}. One of the main contribution of this paper is the observation that these assumptions are not needed for the results in \Cref{sec:OracleMachinePolynomialHierarchy} and \Cref{sec:Fagin'sTheorem}. The first assumption is that structures are of finite type. This assumption is essential to the results in \Cref{sub:SATR-NPR-Membership} establishing that $\SAT(\rr)$ is in $\NP(\rr)$.

\begin{definition}[Structures of Finite Type]
\label{def:StructuresOfFiniteType}
    A structure $\rr$ with vocabulary $\Voc(\rr) = (\Con,\Fun,\Rel)$ is of \textit{finite type} if both $\Fun$ and $\Rel$ are finite sets of symbols. Note that $\Con$ may be infinite.
\end{definition}

The second assumption is that structures have all constants, by which we mean that for all elements of the universe, there is a unique constant naming that element. This assumption is essential to the results in \Cref{sub:SATRIsNPR-hard} establishing that $\SAT(\rr)$ is $\NP(\rr)$-hard.

\begin{definition}[Structures with All Constants]
\label{def:StructuresWithAllConstants}
    A structure $\rr$ with vocabulary $\Voc(\rr) = (\Con,\Fun,\Rel)$ \textit{has all constants} if there is a bijective correspondence between the universe $R$ of $\rr$ and the set $\{c^\rr : c \in \Con\}$ of constants in $\rr$.
\end{definition}

If $\rr$ has all constants, then for every $r$ in the universe $R$ of $\rr$, there is a unique constant symbol $c_r$ such that $c_r^\rr = r$. Since we often conflate a symbol with its interpretation in a structure, we will often write $r$ for both $c_r$ and $c_r^\rr$. Similarly, we will often write a structure with all constants as $$\rr = (R,R,(f^\rr : f \in \Fun),(P^\rr : P \in \Rel)).$$ If $\rr$ is a structure without all constants, then $\rrcon$ denotes the structure in which $(c^\rr : c \in \Con)$ has been replaced by $(c_r^\rr : r \in R)$. Note that it is possible for $\rr$ to have all constants, even uncountably many, but still be of finite type.

\begin{example}[Ordered Rings with All Constants]
    The structure $\R = (\R,0,1,-,+,\cdot,\leq)$ does not have all constants, but the structure $\Rcon = (\R,(r : r \in \R),-,+,\cdot,\leq)$ does, and $\Rcon$ remains bipointed because it has both constants $0$ and $1$ interpreted in their natural way in $\R$. Note that even though $\Rcon$ has uncountably many constants, it is still of finite type.
\end{example}

\subsubsection{Discussion.} 
In most fields that investigate first-order structures, emphasizing the symbols used to speak about a structure is not common practice, so this emphasis on symbols requires some justification.
To a graph theorist, a directed graph $\mathcal{G} = (V,E)$ is simply a set $V$ together with a binary relation $E \subseteq V^2$, and there is no need to mention a symbol ``$E$'' used to refer to the relation $E$. Similarly, to a group theorist, a group $\mathcal{X} = (X,0,-,+)$ is simply a set $X$ together with an associative binary operation $+ \: X^2 \to X$, a unary operation that maps $x$ to its inverse $-x$ for the $+$ operation, as well as an element $0 \in X$ that is an identity element for the $+$ operation. It is just as valid to represent this group in multiplicative notation as $\mathcal{X} = (X,1,^{-1},\cdot)$, so emphasizing particular symbols seems superfluous at best.

Our motivation for specifying the symbols in our language is that this is the first step towards formalizing the relationship between language and structure. For example, let us regard $E$ as a symbol, and let $E^\mathcal{G}$ denote the edge relation of the graph $\mathcal{G}$. That is, $E^\mathcal{G}$ is the interpretation of the symbol $E$ in the graph $\mathcal{G}$. As we will see in the following section, using the symbol $E$, we can write a sentence $\forall x \forall y(E(x,y) \to E(y,x))$ specifying the condition that $\mathcal{G}$ must satisfy in order for $E^\mathcal{G}$ to be symmetric. If $E^\mathcal{G}$ is symmetric, then we say that $\mathcal{G}$ satisfies $\forall x \forall y(E(x,y) \to E(y,x))$ and we write $\mathcal{G} \satisfies \forall x \forall y(E(x,y) \to E(y,x))$. 

Formalizing the relationship between language and structure 
%using the satisfaction relation $\satisfies$ 
is surprisingly fruitful. Doing so leads to complete problems for important complexity classes including $\NP$ and $\exists\R$ \cite{Arora_Barak_2009,schaefer2024existentialdubbel}, enables machine-independent characterizations of many complexity classes via descriptive complexity theory \cite{libkin2004elements}, and provides new perspectives on algebraic geometry and analysis, as well as new methods to solve problems therein \cite{Pillay2000ModelTheory}.

% Deciding if the Boolean structure $\B = (\{0,1\},0,1)$ satisfies sentences of a particular syntactic form gives us complete problems for important complexity classes such as $\NP$, every level of the polynomial hierarchy $\PH$, and \textup{PSPACE} \cite{Arora_Barak_2009}. Furthermore, the complexity of many problems in geometry, probability, artificial intelligence and beyond may be reduced to the problem of deciding if the structure $\R = (\R,0,1,-,+,\leq)$ satisfies certain sentences \cite{schaefer2024existentialdubbel} In a different way, we may capture many complexity classes in a machine-independent way by phrasing membership in these classes in terms of satisfaction of sentences in finite structures, as is done in descriptive complexity theory \cite{libkin2004elements}.

% Take, for example, the Boolean structure $\B = (\{0,1\},0,1)$ that has constants $0$ and $1$ encoding false and true. If $\varphi$ is a Boolean combination of variable assignments to the values $0$ and $1$, then $\varphi$ represents a theoremal formula, and deciding if $\varphi$ is satisfiable is the same thing as deciding if $\B \satisfies \exists x_1 \dots \exists x_n \varphi$. This is the canonical $\NP$-complete problem $\SAT$ of satisfiability. Furthermore, allowing different strings of quantifiers before the formula $\varphi$ gives us complete problems for every level of the polynomial hierarchy $\PH$, as well as a complete problem for \textup{PSPACE}. 

%%%%%%%%%%%%%%%%%%%%%%%%%%%%%%%%%%%%%%%
\subsection{First-Order Logic}
\label{sub:FirstOrderLogic}
%%%%%%%%%%%%%%%%%%%%%%%%%%%%%%%%%%%%%%%

The relationship between language and structure is the subject of model theory, a branch of mathematical logic. We will focus on first-order logic because it is both widely applicable and well behaved. (Specifically, we will focus on Boolean, single-sorted first-order logic with equality, although we will also treat first and second-order metafinite logic, which is a variant of two-sorted logic.) In the rest of this section, we formally introduce the syntax and semantics of first-order logic.

Symbols of a vocabulary are basic ingredients that first-order sentences are made of. Another kind of basic ingredient is variables.

\begin{definition}
    A \textit{(first-order) variable} is an element of the set $\Var = \{x_1,x_2,x_3, \dots\}$. We sometimes use other characters, such as $x,y,z,\dots$, to denote variables.
\end{definition}

Using constant symbols, function symbols, and variables, we may build \textit{terms} that refer to elements of a structure.

\begin{definition}[Terms]
     The set of \textit{terms} for a vocabulary $\Voc = (\Con,\Fun,\Rel)$, also called the set of $\Voc$-terms, is the smallest set of expressions such that: 
     \begin{enumerate}
        \item $x$ is a term whenever $x \in \Var$,
        \item $c$ is a term whenever $c \in \Con$,
        %\item all variables are terms,
        %\item all constants are terms,
        %\item if $f \in \Fun$ has arity $m$ and $t_1, \dots, t_m$ are terms, then $f(t_1,\dots,t_m)$ is a term.
        \item $f(t_1,\dots,t_k)$ is a term whenever $f \in \Fun$ has arity $k$ and $t_1,\dots,t_{k}$ are terms.
     \end{enumerate}
    For any term $t$, the set $\FV(t) \subseteq \Var$ of \textit{free variables} of $t$ is defined as follows:
    \begin{enumerate}
        \item $\FV(x) = \{x\}$, 
        %where $x$ is a variable,
        \item $\FV(c) = \emptyset$, 
        %where $c$ is a constant,
        \item $\FV(f(t_1,\dots,t_k)) = \bigcup_{i=1}^k \FV(t_i)$.
    \end{enumerate}
    We may write a term $t$ as $t(x_1,\dots,x_n)$ to emphasize that $\FV(t) \subseteq \{x_1,\dots,x_n\}.$ For any $\Voc$-structure $\rr$ and any $n \in \N$, each term $t(x_1,\dots,x_n)$ induces a function $t^\rr \: R^n \to R$ that sends $\overline{r} = (r_1,\dots,r_n) \in R^n$ to $t^\rr(\overline{r})$, which is defined as follows:
    \begin{enumerate}
        \item $r_i$ whenever $t$ is the variable $x_i$,
        \item $c^\rr$ whenever $t$ is the constant $c$,
        \item $f^\rr(t_1^\rr(\overline{r}),\dots,t_n^\rr(\overline{r}))$ whenever $t$ is $f(t_1,\dots,t_n).$
    \end{enumerate}
\end{definition}

\begin{definition}[Formulas and Sentences]
    The set of formulas for a vocabulary $\Voc = (\Con,\Fun,\Rel)$, also called the set of $\Voc$-formulas, is the smallest set of expressions such that:
    \begin{enumerate}
        \item $t_1 = t_2$ is a formula whenever $t_1$ and $t_2$ are terms,
        \item $P(t_1,\dots,t_k)$ is a formula whenever $P \in \Rel$ has arity $k$ and $t_1,\dots,t_k$ are terms,
        \item $\neg (\varphi)$ is a formula whenever $\varphi$ is a formula,
        \item $(\varphi \wedge \psi)$ is a formula whenever $\varphi$ and $\psi$ are formulas, 
        \item $(\varphi \vee \psi)$ is a formula whenever $\varphi$ and $\psi$ are formulas, 
        \item $\forall x (\varphi)$ is a formula whenever $x$ is a variable and $\varphi$ is a formula
        \item $\exists x (\varphi)$ is a formula whenever $x$ is a variable and $\varphi$ is a formula.
    \end{enumerate}
    If a formula $\varphi$ is of the type specified in line 1 or 2, then $\varphi$ is an \textit{atomic formula}. We may drop the parentheses if this enhances clarity, and we may write $\varphi \to \psi$ to abbreviate $\neg \varphi \vee \psi$ and $\varphi \fromto \psi$ to abbreviate $(\varphi \to \psi) \wedge (\psi \to \varphi)$. We use the symbol $\equiv$ to denote equality between formulas so as to avoid confusion with the symbol $=$, which may occur in formulas. For any formula $\varphi$, the set $\FV(\varphi) \subseteq \Var$ of \textit{free variables} of $\varphi$ is defined as follows:
    \begin{enumerate}
        \item $\FV(t_1=t_2) = \FV(t_1) \cup \FV(t_2),$
        \item $\FV(R(t_1,\dots,t_m)) = \bigcup_{i=1}^m \FV(t_i),$
        \item $\FV(\neg\varphi) = \FV(\varphi)$,
        \item $\FV(\varphi \wedge \psi) = \FV(\varphi) \cup \FV(\psi)$, 
        \item $\FV(\varphi \vee \psi) = \FV(\varphi) \cup \FV(\psi)$,
        \item $\FV(\forall x \varphi) = \FV(\varphi) \setminus \{x\}$,
        \item $\FV(\exists x \varphi) = \FV(\varphi) \setminus \{x\}$.
    \end{enumerate}
    A variable occurring in a formula $\varphi$ immediately after a quantifier is \textit{bound} by that quantifier within its scope, which is the smallest subformula of $\varphi$ containing that quantifier.
    %We assume, for each formula $\varphi$ that the set of free variables of $\varphi$ is disjoint from the set of bound variables of $\varphi$. 
    If a formula $\varphi$ has no free variables, then we say that $\varphi$ is a \textit{sentence}. We may write a formula $\varphi$ as $\varphi(x_1,\dots,x_n)$ to emphasize that $\FV(\varphi) \subseteq \{x_1,\dots,x_n\}$, and we sometimes abbreviate this to $\varphi(\overline{x})$. For any $\Voc$-structure $\rr$, any $n \in \N$, any formula $\varphi(x_1,\dots,x_n)$, and any $\overline{r} = (r_1,\dots,r_n) \in R^n$, we define the satisfaction relation $\satisfies$ as follows:
    %each formula $\varphi(x_1,\dots,x_n)$ defines a set $\{\overline{r} \in R^n : \rr \satisfies \varphi(\overline{r})\}$, which is defined in terms of the satisfaction relation $\satisfies$ as follows:
    \begin{enumerate}
        \item $\rr \satisfies (t_1 = t_2)(\overline{r})$ whenever $t_1^\rr(\overline{r}) = t_2^\rr(\overline{r}),$
        \item $\rr \satisfies P(t_1,\dots,t_m)(\overline{r})$ whenever $(t_1^\rr(\overline{r}),\dots,t_m^\rr(\overline{r})) \in P^\rr$,
        \item $\rr \satisfies \neg \varphi(\overline{r})$ whenever $\rr \not\satisfies \varphi(\overline{r})$,
        \item $\rr \satisfies (\varphi \wedge \psi)(\overline{r})$ whenever $\rr \satisfies \varphi(\overline{r})$ and $\rr \satisfies \psi(\overline{r})$,
        \item $\rr \satisfies (\varphi \vee \psi)(\overline{r})$ whenever $\rr \satisfies \varphi(\overline{r})$ or $\rr \satisfies \psi(\overline{r})$,
        \item $\rr \satisfies \forall x \varphi(\overline{r},x)$ whenever $\rr \satisfies \varphi(\overline{r},r)$ for all $r \in R$,
        \item $\rr \satisfies \exists x \varphi(\overline{r},x)$ whenever $\rr \satisfies \varphi(\overline{r},r)$ for some $r \in R$.
    \end{enumerate}
    If $\rr \satisfies \varphi(\overline{r})$, then we say that $\rr$ \textit{satisfies} $\varphi(\overline{r})$, or that $\rr$ is a \textit{model} of $\varphi(\overline{r})$, or that $\varphi(\overline{r})$ is \textit{true} in $\rr$.
\end{definition}

\begin{example}[Booleans]
    The only terms in the vocabulary of Booleans $\Voc_\textup{bool} = (0,1)$ are variables $x$ and the constants $0$ and $1$. Atomic formulas include $x=y$, which we can make into a sentence $\forall x \exists y(x=y)$ by quantifying the variables $x$ and $y$. Since $0^\B = 0 \neq 1 = 1^\B$ in the Boolean structure $\B = (\{0,1\},0,1)$, we see that $\B$ satisfies the sentence expressing that $0$ is not equal to $1$. That is, $$\B \satisfies \neg(0=1).$$ On the other hand, for the tuple $\overline{r} = (0,0) \in \{0,1\}^2$, the terms $x_1$ and $x_2$ induce functions $x_1^\B$ and $x_2^\B$ for which $x_1^\B(\overline{r}) = 0$ and $x_2^\B(\overline{r})=0$, so $$\B \satisfies (x_1 = x_2)(\overline{r}).$$
\end{example}

\begin{example}[Graphs]
    The only terms in the vocabulary of graphs $\Voc_\textup{graphs} = (E)$ are variables. Atomic formulas include both $x=y$ and $E(x,y)$, and sentences include $\forall x \forall y E(x,y)$. Since $(0,1)$ is an edge of the graph $\mathcal{K}_2  = (\{0,1\},\{(0,1),(1,0)\})$, we see that $\mathcal{K}_2 \satisfies E(0,1)$. Since the edge relation of the graph $\mathcal{K}_2$ is symmetric, we see that $$\mathcal{K}_2 \satisfies \forall x \forall y ( E(x,y) \to E(y,x)).$$ However, since $(0,0)$ is not in the edge relation $E^{\mathcal{K}_2} = \{(0,1),(1,0)\}$, we see that $$\mathcal{K}_2 \satisfies \neg E(x_1,x_2)(0,0).$$ 
\end{example}

\begin{example}[Partially Ordered Sets]
    The only terms in the vocabulary of partially ordered sets $\Voc_\textup{pos}=(\leq)$ are variables. Since the ordering relation of any partially ordered set $\mathcal{P} = (P,\leq)$ is transitive, we see that $$\mathcal{P} \satisfies \forall x \forall y \forall z ( (x \leq y \wedge y \leq z) \to x \leq z).$$ Since $\mathcal{N} = (\N,\leq)$ is linearly ordered, we see that $$\mathcal{N} \satisfies \forall x \forall y (x \leq y \vee y \leq x).$$ Furthermore, $\mathcal{P}_2 = (2^{\{1,2\}},\subseteq)$ is not linearly ordered because $\{1\} \not\subseteq \{2\}$ and $\{2\} \not\subseteq \{1\}$, so we see that $$\mathcal{P}_2 \satisfies (\neg(x_1 \leq x_2) \wedge \neg(x_2 \leq x_1))(\{1\},\{2\}).$$
\end{example}

\begin{example}[Monoids]
    Terms in the vocabulary of monoids $\Voc_\textup{mon}=(0,+)$ include expressions recursively built from variables $x$ and the constant $0$ using the $+$ symbol. For example, each of $x$, $y$, $0$, $+(x,y)$, and $+(+(x,y),0)$ are terms. When it is conventional, we will use infix notation, in which the last two terms become $x+y$ and $(x+y)+0$, respectively. Atomic formulas include $x+x = 0$ and $x+y=y+x$. Since addition in $\mathcal{N} = (\N,0,+)$ is associative, we see that $$\mathcal{N} \satisfies \forall x \forall y \forall z ((x+y)+z) = (x+(y+z)).$$ However, there is no additive inverse of $1 \in \N$, so we see that $$\mathcal{N} \not\satisfies \exists x_2 (x_1+x_2 = 0)(1).$$
\end{example}

\begin{example}[Groups]
    Terms in the vocabulary of groups $\Voc_\textup{group}=(0,-,+)$ include all terms of $\Voc_\textup{mon}$, as well as expressions such as $-x$,$-x + y$, $-(x+y)$, and $-x+-y$. Atomic formulas include $-(x+y) = -x+-y$. Since every element of $\mathcal{Z} = (\Z,0,-,+)$ has an additive inverse, we see that $$\mathcal{Z} \satisfies \forall x \exists y (x+y=0).$$  Furthermore, based on our knowledge of addition, we see that $$\mathcal{Z} \satisfies (x_1+-(x_2+x_3) = 0)(5,2,3,4).$$
\end{example}

\begin{example}[Ordered Rings] 
    In the vocabulary of ordered rings $\Voc_\textup{oring}=(0,1,+,\cdot,\leq)$, terms include $0 \cdot (x+y)$ and $1+(0\cdot(x+y))$, as well as any polynomial with coefficients built from the constants $0$ and $1$ using addition and multiplication: we may see the polynomial $x^2 + 2x+2$ as the term $$(x \cdot x) + ((1+1)\cdot x) + (1+1).$$ Atomic formulas include $0 \cdot x = 0$ and $0 \cdot (x + y) \leq 1 + (0 \cdot (x+y))$. In the ordered ring of real numbers $\R = (\R,0,1,-,+,\cdot,\leq)$, multiplying some real number $r$ by $-1$ results in $-r$, so we see that $$\R \satisfies \forall r (-1 \cdot r = -r).$$ Furthermore, since $\sqrt{2}$ is a real number, we see that $$\R \satisfies ( x_1 \cdot x_1 = x_2)(\sqrt{2},2).$$ 
    The ordering of $\R$ is compatible with addition and multiplication in $\R$ in the sense that addition is a monotone function and multiplication of positives results in a positive, from which we see that
    $$\R \satisfies \forall x \forall y \forall z (( x \leq y \to x+z \leq y+z) \wedge  ((0 \leq x \wedge 0 \leq y) \to 0 \leq x \cdot y)).$$
\end{example}

\begin{example}[Real Vector Spaces]
    In the vocabulary of real vector spaces $\Voc_\textup{lin} = (0,-,+,(\scalar{r} : r \in \R))$ terms include $-0+\scalar{\sqrt{2}}(x) + \scalar{1/3}(y)$, as well as any polynomial of degree 1, with coefficients in $\R$ whose zero-degree term is zero. We may express the system of linear equations $\pi x+1/2 y=0$ and $-x+\sqrt{2}x + z = 0$ with the formula $$(\scalar{\pi}(x) + \scalar{1/2}(y) = 0) \wedge (-x + \scalar{\sqrt{2}}(x)+z = 0).$$ Because $\R$ itself is a real vector space, we see that $\R_\textup{lin}=(\R,0,-,+,(\scalar{r} : r \in \R))$ satisfies the usual axioms of real vector spaces described in \cref{example:VectorSpacesOverReals}. In particular, for all $r$ and $s$ in $\R$, we see that
    \begin{enumerate}
        \item $\R_\textup{lin} \satisfies \forall x (\scalar{1}(x) = x)$,
        \item $\R_\textup{lin} \satisfies \forall x \forall y (\scalar{r}(x+y) = \scalar{r}(x) + \scalar{r}(y))$,
        \item $\R_\textup{lin} \satisfies \forall x (\scalar{r+s}(x) = \scalar{r}(x)+\scalar{s}(x))$,
        \item $\R_\textup{lin} \satisfies \forall x (\scalar{rs}(x) = \scalar{r}(\scalar{s}(x)))$.
    \end{enumerate}
\end{example}

\subsubsection{Equality.} In first-order logic, the equality symbol $=$ is usually treated as a logical symbol that is interpreted in every structure as the equality relation on the universe of that structure. This uniform interpretation justifies treating the equality symbol as a logical symbol, similar to $\neg$, $\wedge$, $\vee$, $\forall$, and $\exists$, rather than treating it as a relation symbol, whose interpretation may change from structure to structure. 

Because the equality symbol is logical rather than relational, equality is not usually included as a relation of structures, even though the formulas of first-order logic may use equality to speak about structures. This is not problematic because for every underlying set of a structure, there is a uniquely determined equality relation on that set that the equality symbol refers to. We highlight the importance of equality here because the main arguments of this paper use equality in an essential way, and it is not clear if our results hold without equality.

To emphasize the convention that equality is considered a logical symbol, one may say first-order logic \textit{with equality}. There are contexts in which the equality-free fragment of first-order logic becomes relevant, such as when comparing the computational complexity of deciding whether or not a structure satisfies a sentence taken from different fragments of first-order logic. This is especially common in research related to the complexity class $\exists\R$ \cite{SS17,SS22}. In this context, one usually includes equality as a relation of structures to emphasize its presence or absence. To translate our results into a form that agrees with this latter convention, one need only add the equality relation to every structure.

%%%%%%%%%%%%%%%%%%%%%%%%%%%%%%%%%%%%
\subsection{Fragments of Theories: $\Sigma_k\SAT(\rr)$ and $\exists_k\SAT(\rr)$}
\label{sub:DefinitionSatisfiabilityInR}
%%%%%%%%%%%%%%%%%%%%%%%%%%%%%%%%%%%%%

Euclidean geometry, as developed in Euclid's Elements, is the oldest known mathematical theory formulated in terms of basic assumptions called axioms. Early physical theories were modeled after Euclidean geometry, and their basic assumptions were called laws, such as the laws of motion or the laws of thermodynamics. Explicitly stated basic assumptions remain an important component of modern scientific theories, although philosophers of science would now include other components as well \cite{sep-structure-scientific-theories}. Nonetheless, the syntactic view of a theory as a set of basic assumptions has left its mark on the terminology of first-order logic and model theory.

\begin{definition}[Theories and Models]
    Given a vocabulary $\Voc$, a \textit{$\Voc$-theory} $T$ is simply a set of $\Voc$-sentences. 
    %A \textit{theory} is a $\Voc$-theory for some vocabulary $\Voc$. 
    Whenever a $\Voc$-structure $\rr$ satisfies each sentence $\varphi$ in $T$, we say that $\rr$ is a \textit{model} of $T$ or that $\rr$ \textit{models} $T$, and we write $\rr \satisfies T$.
\end{definition}

\begin{example}[Theory of Partially Ordered Sets]
    In the vocabulary of partially ordered sets $\Voc_\textup{pos} = (\leq)$, the theory of partially ordered sets $T_\textup{pos}$ is composed of the usual axioms for partially ordered sets stating that $\leq$ is reflexive, antisymmetric, and transitive. That is, $T_\textup{pos}$ is composed of the following sentences:
    \begin{enumerate}
        \item $\forall x ( x \leq x) $,
        \item $\forall x \forall y ((x \leq y \wedge y \leq x)\to x = y)$,
        \item $\forall x \forall y \forall z((x \leq y \wedge y \leq z) \to x \leq z).$
    \end{enumerate}
    Since $\mathcal{P}_n = (2^{\{1,\dots,n\}},\subseteq)$ is a partially ordered set, we see that $\mathcal{P}_n \satisfies T_\textup{pos}$. In general, a $\Voc_\textup{pos}$-structure $\mathcal{P}$ is a partially ordered set if and only if $\mathcal{P} \satisfies T_\textup{pos}$.
\end{example} 

\begin{example}[Theory of Groups]
    In the vocabulary of groups $\Voc_\textup{group} = (0,-,+)$, the theory of groups $T_\textup{group}$ is composed of the usual axioms for groups stating that $0$ is an additive identity, $-x$ is the additive inverse of $x$, and $+$ is associative. That is, $T_\textup{group}$ contains the following sentences:
    \begin{enumerate}
        \item $\forall x(x+0 = x \wedge 0+x = x)$,
        \item $\forall x(x+(-x) = 0 \wedge (-x)+x = 0)$,
        \item $\forall x \forall y \forall z ( x+(y+z) = (x+y)+z)$.
    \end{enumerate}
    To obtain the theory of commutative groups $T_\textup{cgroup}$, we add the following sentence:
    \begin{enumerate}[resume]
        \item $\forall x \forall y(x+y = y+x)$.
    \end{enumerate}
    Since $\mathcal{Z} = (\Z,0,-,+)$ is a commutative group, we see that $\mathcal{Z} \satisfies T_\textup{cgroup}$. In general, a $\Voc_\textup{group}$ structure $\mathcal{G}$ is a commutative group if and only if $\mathcal{G} \satisfies T_\textup{cgroup}.$
\end{example}

\begin{example}[Theory of Real Vector Spaces]
    In the vocabulary of real vector spaces $\Voc_\textup{lin} = (0,-,+,(\scalar{r} : r \in \R))$, the theory of real vector spaces $T_\textup{lin}$ is composed of the usual axioms for real vector spaces stating that the vector space is a commutative group and that scalar multiplication behaves as expected. That is $T_\textup{lin}$ is the union of the following sets of sentences:
    \begin{enumerate}
        \item $T_\textup{cgroup}$,
        \item $\{ \forall x (\scalar{1}(x) = x) \}$,
        \item $\{ \forall x \forall y (\scalar{r}(x+y) = \scalar{r}(x) + \scalar{r}(y)) : r \in \R\}$,
        \item $\{ \forall x (\scalar{r+s}(x) = \scalar{r}(x)+\scalar{s}(x)) : r,s \in \R\}$,
        \item $\{ \forall x (\scalar{rs}(x) = \scalar{r}(\scalar{s}(x))) : r,s \in \R\}$.
    \end{enumerate}
    Since $\R_\textup{lin} = (\R,0,-,+,(\scalar{r}:r \in \R))$ is a real vector space, $\R_\textup{lin} \satisfies T_\textup{lin}$. In general, a $\Voc_\textup{lin}$-structure $\mathcal{V}$ is a real vector space if and only if $\mathcal{V} \satisfies T_\textup{lin}$.
\end{example}

While many of the usual theories are finite, such as the theory of partially ordered sets and the theory of groups, the example of real vector spaces shows that there are natural examples of infinite theories. Another natural example is the theory $\Th(\rr)$ of a fixed structure $\rr$. This theory is composed of all sentences that are true in $\rr$. If we restrict the theory of a structure to those sentences in which the only constants that occur are $0$ and $1$, then we obtain the Boolean theory of a structure.

\begin{definition}[Theory of a Structure]
\label{def:TheoryOfAStructure}
    The \textit{theory} of a structure $\rr$ is the set $\Th(\rr)$ of sentences in the vocabulary of $\rr$ that are satisfied in $\rr$. That is, $$\Th(\rr) = \{\varphi : \text{$\varphi$ is a $\Voc(\rr)$-sentence such that } \rr \satisfies \varphi\}.$$ The \textit{Boolean theory} of a structure $\rr$ is the set $\BTh(\rr)$ of $\varphi \in \Th(\rr)$ such that every constant occurring in $\varphi$ is either $0$ or $1$. That is, $$\BTh(\rr)=\{\varphi \in \Th(\rr) : \text{every constant in  $\varphi$ is 0 or 1} \}.$$
\end{definition}

We will call a subset of a theory a \textit{fragment} of that theory. In this way, the Boolean theory of a structure is a fragment of the theory of that same structure. In the following sections, we will show that for many complexity classes $\mathcal{C}$ over a structure $\rr$, there are fragments of $\Th(\rr)$ that are complete decision problems for $\mathcal{C}$. In order to define these fragments, we first need to define the prenex normal form of formulas, as well as several syntactic gradations within formulas of this form.

\begin{definition}[Prenex Normal Form]
    A formula $\Phi$ is in \textit{prenex normal form} if $\Phi$ is a (possibly empty) string of quantifiers followed by a quantifier-free formula $\varphi$. In other words, $\Phi$ is of the form $$Q_1 x_1 \dots Q_m x_m \varphi(x_1, \dots, x_m, y_1, \dots y_n),$$ where the variables $x_i$ are bound by the quantifiers $Q_i \in \{\exists, \forall\}$, the variables $y_j$ are free, and $\varphi$ is a quantifier-free formula. 
    
    %If each $Q_i$ is existential, then we say that $\Phi$ is a $\Sigma_1$ formula, and if each $Q_i$ is universal, then we say that $\Phi$ is a $\Pi_1$ formula. More generally, we say that $\Phi$ is a $\Sigma_k$ formula if $\Phi$ is of the form $$Q_1 \overline{x}_1 \dots Q_k \overline{x}_k \varphi(\overline{x}_1,\dots,\overline{x}_k,\overline{y}),$$ where the quantifier blocks $Q_i \overline{x}_i = Q_{i1}x_{i1}\dots Q_{im_i}x_{im_i}$ are composed of the same quantifier, starting with an existential block $Q_1 \overline{x}_1$ and alternating quantifiers between adjacent blocks until reaching the quantifier-free formula $\varphi$. Similarly, we say that $\Phi$ is a $\Pi_k$ formula if the same condition holds, except the first quantifier block $Q_1 \overline{x}_1$ is universal.
    
    %Similarly, we say that a $\Sigma_k$ formula is one in prenex normal form, starting with an existential quantifier and alternative quantifiers $k-1$ times before reaching a quantifier-free formula. In more detail, $\Phi$ is a $\Sigma_k$ formula if $\Phi$ is of the form $$Q_1 \overline{x}_1 \dots Q_k \overline{x}_k \varphi(\overline{x}_1,\dots,\overline{x}_k,\overline{y})$$
\end{definition}

\begin{remark}
    We say that a $\Voc$-formula $\varphi(x_1,\dots,x_n)$ is \textit{logically equivalent} to a $\Voc$-formula $\varphi'(x_1,\dots,x_n)$ if, for any $\Voc$-structure $\rr$ and any tuple $(r_1,\dots,r_n) \in R^n$, $\rr \satisfies \varphi(r_1,\dots,r_n)$ if and only if $\rr \satisfies \varphi'(r_1,\dots,r_n)$. In every vocabulary $\Voc$, every $\Voc$-formula $\varphi$ is logically equivalent to a $\Voc$-formula $\varphi'$ in prenex normal form such that $\varphi'$ has the same free and bound variables as $\varphi$ \cite{van1994logic}. In particular, every sentence is logically equivalent to a sentence in prenex normal form. 
\end{remark}

\begin{definition}[$\Sigma_k$ and $\Pi_k$ Formulas]
\label{def:SigmaKFormulasandPiKFormulas}
    We say that a formula $\Phi$ in prenex normal form is a \textit{$\Sigma_1$ formula} if each quantifier $Q_i$ occurring in $\Phi$ is existential, and we say it is a \textit{$\Pi_1$ formula} if each quantifier $Q_i$ occurring in $\Phi$ is universal. More generally, we say that $\Phi$ is a \textit{$\Sigma_k$ formula} if $\Phi$ is of the form $$Q_1 \overline{x}_1 \dots Q_k \overline{x}_k \varphi(\overline{x}_1,\dots,\overline{x}_k,\overline{y}),$$ where the quantifier blocks $Q_i \overline{x}_i = Q_{i1}x_{i1}\dots Q_{im_i}x_{im_i}$ are composed of the same quantifier, starting with an existential block $Q_1 \overline{x}_1$ and alternating quantifiers $k-1$ times between adjacent blocks until reaching the quantifier-free formula $\varphi$. Similarly, we say that $\Phi$ is a $\Pi_k$ formula if the same condition holds, except the first quantifier block $Q_1 \overline{x}_1$ is universal.
\end{definition}

If $\varphi(x_1,\dots,x_n)$ is a quantifier-free formula in the vocabulary of a structure $\rr$, then we say that $\varphi(x_1,\dots,x_n)$ is \textit{satisfiable in $\rr$} if there are elements $r_1,\dots,r_n \in R$ such that $\rr \satisfies \varphi(r_1,\dots,r_n)$. We can see from the definition of the satisfaction relation $\satisfies$ that $\varphi(x_1,\dots,x_n)$ is satisfiable in $\rr$ if and only if $\exists x_1 \dots \exists x_n \varphi(x_1,\dots,x_n) \in \Th(\rr).$ In this way, we may identify the set of satisfiable quantifier-free formulas $\SAT(\rr)$ with the first existential fragment $\Sigma_1\SAT(\rr)$ of $\Th(\rr)$ that results from existentially quantifying over quantifier-free formulas. More generally, we define the $k$th existential and universal fragments of $\Th(\rr)$ as follows.

\begin{definition}
\label{def:SigmaKSATRandPiKSATR}
    The \textit{$k$th existential fragment} of the theory of $\rr$ is the set $\Sigma_k\SAT(\rr)$ of $\Sigma_k$ sentences $\Phi$ in the vocabulary of $\rr$ that are satisfied in $\rr$. That is, $$\Sigma_k\SAT(\rr) = \{\Phi : \Phi \text{ is a $\Sigma_k$ sentence and } \rr \satisfies \Phi\}.$$ In particular, we define $\SAT(\rr) = \Sigma_1\SAT(\rr)$. Similarly, the \textit{$k$th universal fragment} of the theory of $\rr$ is the set $\Pi_k\SAT(\rr)$ of $\Pi_k$ sentences $\Phi$ in the vocabulary of $\rr$ that are satisfied in $\rr$. That is, $$\Pi_k\SAT(\rr) = \{\Phi : \Phi \text{ is a $\Pi_k$ sentence and } \rr \satisfies \Phi\}.$$
\end{definition}

% \till{The name ''satisfiability'' refers to there being an existential quantifier. So $\forall \SAT(\rr)$ is somewhat contradictory. Maybe $\forall \rr $ and $\forall_k \rr$ would make more sense. 
% Also $\SAT(\rr) = \Sigma_1(\rr)$ would make more sense to me. To be discussed.} 

% \jeremy{Perhaps the remark below sufficiently justifies our notation?}

% \lucas{Might be worth mentioning it's called the tautology problem for the Boolean case?}

% \jeremy{This is now mentioned in the first example below.}

\begin{remark}
    The reader versed in computational complexity theory may be troubled by the notation $\Sigma_k\SAT(\rr)$ and $\Pi_k\SAT(\rr)$ when they interpret $\SAT$ as an abbreviation for ``satisfiable'', as then $\SAT$ would only appropriately describe $\Sigma_1$ sentences. We hope this cognitive dissonance may be resolved by reinterpreting $\SAT$ as an abbreviation for ``satisfied'' 
    %or as a reference to the satisfaction relation $\satisfies$. 
    so that one may pronounce $\Sigma_k\SAT(\rr)$ as ``the $\Sigma_k$ sentences satisfied in $\rr$''. We choose this notation to remain consistent with the standard notation $\Sigma_k\SAT$ and $\Pi_k\SAT$ for the decision problems that generalize $\SAT$ and provide complete problems for each level of the classical polynomial hierarchy, as in \cite{Arora_Barak_2009}.
\end{remark}    

For various structures $\rr$, the fragments $\Sigma_k\SAT(\rr)$ and $\Pi_k\SAT(\rr)$ are equivalent to well-known decision problems in the Boolean Turing machine setting and the nonBoolean setting BSS-machine setting. We give a few examples here.

\begin{example}
    Over the Boolean structure $\mathcal{B} = (\{0,1\}, 0,1)$, $\text{SAT}(\mathcal{B})$ is equivalent to the Boolean problem $\SAT$ of deciding the satisfiability of Boolean formulas. This is because each quantifier-free formula $\varphi(x_1,\dots,x_n)$ in the vocabulary of $\mathcal{B}$ is logically equivalent to a Boolean combination of variable assignments $x_i = 0$ or $x_i = 1$, which we may think of as setting variable $x_i$ to false or true, respectively. Similarly, $\Pi_1\SAT(\mathcal{B})$ is equivalent to the Boolean problem $\Pi_1\SAT$ of deciding if a Boolean formula is a tautology. More generally, $\Sigma_k\SAT(\mathcal{B})$ is equivalent to $\Sigma_k\SAT$, and $\Pi_k\SAT(\mathcal{B})$ is equivalent to $\Pi_k\SAT$, as defined in \cite{Arora_Barak_2009}. 
    % \till{Did we ever introduce $\Sigma_k\SAT$?
    % Also maybe it deserves some attention that $\Pi_1 \SAT$ has really nothing to do with the origin of the word satisfiability. }
    % \jeremy{I now mention $\Sigma_k\SAT$ in the remark above, and I give a reference for the reader to read further about it, but it would take us too far afield to define it formally because it deals with propositional formulas rather than first-order formulas, so I assume it as background knowledge that is safe for the reader to ignore.}
    
    % \jeremy{The use of the $\SAT$ notation is, hopefully, addressed in the remark above. Would you agree?}
\end{example}

\begin{example}
    Over the ordered ring of real numbers with only $0$ and $1$ as constants $\R = (\R,0,1,-,+,\cdot,\leq)$, a quantifier-free formula defines a semialgebraic set presented as a Boolean combination of integer-coefficient polynomial equalities and inequalities. Such a formula is satisfiable in $\R$ if and only if this semialgebraic set is nonempty, so $\SAT(\R) = \textup{ETR}$, where $\textup{ETR}$ is the existential theory of the reals decision problem \cite{schaefer2017fixed,SCM24}.
\end{example}

\begin{example}
    Over the ordered ring of real numbers with all constants $\Rcon = (\R,(r : r \in \R),-,+,\cdot,\leq)$, a quantifier-free formula in the vocabulary of $\Rcon$ defines a semialgebraic set presented as a Boolean combination of real-coefficient polynomial equalities and inequalities. Such a formula is satisfiable over $\Rcon$ if and only if this semialgebraic set is nonempty, so $\SAT(\Rcon)$ is the semialgebraic feasibility problem SA-FEAS of \cite{blum1998complexity}. 
\end{example}

As we will see in \Cref{sub:ComplexityClasses}, $\SAT(\R)$ is a Boolean decision problem because $\SAT(\R)$ may be encoded as a set of binary strings $\SAT(\R) \subseteq \{0,1\}^*$. On the other hand, we will see that $\SAT(\Rcon)$ is a nonBoolean decision problem because the presence of real number constants in sentences means there is no binary encoding of $\SAT(\Rcon)$, which is instead encoded as a set of strings of real numbers $\SAT(\Rcon) \subseteq \R^*$. In order to derive a Boolean decision problem from a structure including constants other than 0 and 1, we introduce the $k$th existential and universal Boolean fragments of a theory.

\begin{definition}
\label{def:ExistsKSATRandForallKSATR}
    The \textit{$k$th existential Boolean fragment} of the theory of $\rr$ is the set $\exists_k\SAT(\rr)$ of sentences $\Phi$ in $\Sigma_k\SAT(\rr)$ such that every constant occurring in $\Phi$ is either $0$ or $1$. That is, $$\exists_k\SAT(\rr) = \{\Phi \in \Sigma_k\SAT(\rr) : \text{every constant in $\Phi$ is 0 or 1} \}.$$ In particular, we define $\exists\SAT(\rr) = \exists_1\SAT(\rr)$. Similarly, the \textit{$k$th universal Boolean fragment} of $\BTh(\rr)$ is the set $\forall_k\SAT(\rr)$ of sentences $\Phi$ in $\Pi_k\SAT(\rr)$ such that every constant occurring in $\Phi$ is either $0$ or $1$. That is, $$\forall_k\SAT(\rr) = \{\Phi \in \Pi_k\SAT(\rr) : \text{every constant in $\Phi$ is 0 or 1} \}.$$
\end{definition}

\begin{example}
    The only constants in the structure $\R = (\R,0,1,-,+,\cdot,\leq)$ are 0 and 1, so we have $\exists\SAT(\R) = \SAT(\R)$. Though the structure $\Rcon = (\R,(r : r \in \R),-,+,\cdot,\leq)$ has all constants, $\exists\SAT(\Rcon)$ restricts to sentences whose only constants are among 0 and 1, so we have $\exists\SAT(\Rcon) = \exists_1\SAT(\Rcon) = \Sigma_1\SAT(\R) = \SAT(\R)$. 
\end{example}

With this background knowledge in first-order logic, we are now ready to introduce machines over first-order structures. We will first give an extended example of a machine over the structure $\R = (\R,0,1,-,+,\cdot,\leq)$ to provide some intuition for these abstract devices.

%%%%%%%%%%%%%%%%%%%%%%%%%%%%%%%%%%%%%%%%%%%%%%%%%%%%
\subsection{Extended Example of a Machine over $\R$}
\label{sub:ExtendedExampleRMachine}
%%%%%%%%%%%%%%%%%%%%%%%%%%%%%%%%%%%%%%%%%%%%%%%%%%%%

We provide an extended example of a machine over $\R$ that computes the sum of a string of real numbers of variable length. This problem illustrates the role of index registers and $\R$-registers, as well as how $\R$-instructions form the syntax of a basic programming language.

\subsubsection{The Algorithm.} Suppose we are working in the structure $\R=(\R,0,1,-,+,\cdot,\leq)$. If we would like to compute the sum of a string of real numbers $v=(r_1,\dots,r_n)$, then the following simple algorithm suggests itself: set a variable $i$ to have value $1$, and set a variable $x$ to have value $r_i$ before iteratively incrementing $i$ and adding $r_i$ to the value of $x$ until and including when $i$ reaches $n$. 

This algorithm may be expressed in low-level programming language based on the operations of $\R$, as we have done in Table \ref{table:ExampleR-Program}. The result is a finite sequence of instructions, such as $Z \lbr 0 \rbr := Z \lbr 0 \rbr + Z \lbr 1 \rbr$ to add two real numbers $r_1$ and $r_2$ stored in registers $Z \lbr 0 \rbr$ and $Z \lbr 1 \rbr$, as well as indexing instructions, such as $I \lbr 2 \rbr := I \lbr 2 \rbr + 1$ to increment the natural number $i-1$ stored in the register $I \lbr 2 \rbr$.

\subsubsection{$\R$-Registers and Index Registers.} In order to store the real number $r_i$ in memory with infinite precision, we use an idealized machine with $\R$-registers $Z \lbr j \rbr$ that are capable of storing a single real number. This machine will have an unbounded sequence of $\R$-registers $Z \lbr 0 \rbr, Z \lbr 1 \rbr, Z \lbr 2 \rbr, \dots$ where each $Z \lbr j \rbr$ for $j \in \N$ can store a single real number $r \in \R$. On input $v=(r_1,\dots,r_n) \in R^n$, $\R$-registers $Z \lbr i-1 \rbr$ are initialized to hold the real number $r_i$, for $1 \leq i \leq n$, while all other $\R$-registers are initialized to hold the value $0$. For technical reasons explained in the discussion at the end of \Cref{sub:MachinesOverR}, we start indexing $\R$-registers with $0$, which causes $r_i$ to be stored in $Z \lbr i-1 \rbr$.

In order to increment the index $i$ in the above algorithm, we also need to store the natural number $i$ in memory. Our idealized machine will do this with a bounded sequence of index registers $I \lbr 0 \rbr, \dots, I \lbr k-1 \rbr$ where each $I \lbr j \rbr$ for $0 \leq j \leq k-1$ is capable of storing a single natural number. On input $v = (r_1,\dots,r_n) \in \R^n$, index register $I \lbr 0 \rbr$ is initialized to hold the value $n$, while all other index registers are initialized to hold the value $0$. The $\R$-registers and the index registers, together with an idealized central processing unit capable of manipulating these registers, form an $\R$-machine as in \Cref{fig:RegisterMachine}.

Since $i$ and $n$ are natural numbers, they could have been stored in the $\R$-registers. We prefer to use dedicated index registers for two reasons. First, in algorithms based on the structure $\R$, the role of indexing operations is fundamentally different than the role of operations in $\R$. In a sense, indexing operations happen ``outside'' of the structure $\R$. Second, we would like to generalize this approach to structures $\rr$ whose universe may not contain the natural numbers.

\subsubsection{Instructions, Programs, and Machines.} The instructions in the program of \Cref{table:ExampleR-Program} are examples of $\R$-instructions. The full range of possible $\R$-instructions is detailed in \Cref{tab:RInstructions}. These include instructions to print each constant $0$ and $1$ of $\R$, to evaluate each function $+$ and $\cdot$ of $\R$, to branch based on testing the relation $\leq$ of $\R$, and to branch on  equality $=$, as well as basic indexing instructions. A sequence of $\R$-instructions that ends with the $\mathsf{stop}$ instruction forms an $\R$-program. \Cref{table:ExampleR-Program} implements the above algorithm for computing the sum of a string of real numbers $v = (r_1,\dots,r_n)$ as an $\R$-program $\program$.

The central processing unit of an $\R$-machine interprets and executes the instructions of an $\R$-program until the $\mathsf{stop}$ instruction is reached, at which point the machine returns the string $v' = (r'_1,\dots,r'_{n'}) \in \R^{n'}$ that is contained in registers $Z \lbr 0 \rbr, \dots, Z \lbr n' - 1 \rbr$, where $n'$ is contained in $I \lbr 0 \rbr$. So just as $I \lbr 0 \rbr$ is initialized with the length of the input when the machine starts, $I \lbr 0 \rbr$ determines the length of the output when the machine stops. 

This completes the intuitive picture of $\R$-machines presented in \Cref{fig:RegisterMachine}. While it is important to keep this picture in mind, it is also important to note that, formally, an $\R$-machine is just a tuple $\M = (\program,\R, \N^k,\R^\N)$ where $\program$ is an $\R$-program, $\R$ is the structure, $\N^k$ is a space of index states, and $\R^\N$ is a space of memory states. For example, if $\program$ is the $\R$-program of \Cref{table:ExampleR-Program}, then $\M = (\program,\R,\N^4,\R^\N)$ is an $\R$-machine that computes the sum of a string of real numbers $v = (r_1,\dots,r_n) \in \R^n$ for any $n \in \N$, meaning that this string can have any length.

\subsubsection{Computation.} We now describe how this $\R$-machine works. We use register $I \lbr 1 \rbr$ to store the value $1$ and we use register $I \lbr 2 \rbr$ to store the value $i-1$ indexing the register $Z \lbr i-1 \rbr$ storing the real number $r_i$, for some $1 \leq i \leq n$. We also use register $I \lbr 3 \rbr$ to store the value $0$. Steps 0 through 2 check if the length of the input is 0 or 1, in which case the $\R$-machine stops and returns the value in $Z \lbr 0 \rbr$, which has been initialized to either $0$ or $r_1$, respectively. 

Step 3 sets the value of $I \lbr 0 \rbr$ to $n \monus 1$, which is the number of times addition needs to be applied. We use the $\monus$ operation, called monus or truncated subtraction, so that indices do not become negative. Starting with the value $i-1=0$, step 4 increments the value $i-1$ of $Z \lbr i-1 \rbr$ containing $r_i$. Step 5 copies $r_i$ from $Z \lbr i-1 \rbr$ into $Z \lbr 1 \rbr$, and step 6 adds the contents of $Z \lbr 0 \rbr$ and $Z \lbr 1 \rbr$, storing the result in $Z \lbr 0 \rbr$. This results in $r_1 + \dots + r_i$ being stored in $Z \lbr 0 \rbr$ after $i-1$ executions of steps 4 through 7, which means the program should stop looping when $i-1 = n-1$. The value of index register $I \lbr 0 \rbr$ at the end of the execution of a program is the length of the output. Steps 8 through 10 set the value of $I \lbr 0 \rbr$ to 1. Finally, step 12 stops the $\R$-machine and returns the value in register $Z \lbr 0 \rbr$, which is $r_1 + \dots + r_n$.
\begin{table}[h!]
\fbox{\begin{minipage}{\textwidth}
On input $(r_1,\dots,r_n) \in \R^n$, initialize the registers as follows: 
$$
\begin{array}{cccccccc}
    Z \lbr 0 \rbr := r_1 & Z \lbr 1 \rbr := r_2 & Z \lbr 2 \rbr := r_3 & Z \lbr 3 \rbr := r_4 & \dots & Z \lbr n-1 \rbr := r_n & Z \lbr n \rbr := 0 & \dots \\
    & & & & & & & \\
    I \lbr 0 \rbr := n, & I \lbr 1 \rbr := 0 & I \lbr 2 \rbr := 0  & I \lbr 3 \rbr := 0  & & & &
\end{array}
$$
$$
{\renewcommand{\arraystretch}{1.5} % Increase row spacing by 50%
\begin{array}{ccll}
    0 & : & \ifthenbranch{I \lbr 0 \rbr = I \lbr 1 \rbr}{10}{1} & \textit{\# check if the input has length 0}\\
    1 & : & I \lbr 1 \rbr := I \lbr 1 \rbr +1 
    & \textit{\# increment $I \lbr 1 \rbr$ to 1} \\
    2 & : & \ifthenbranch{I \lbr 0 \rbr = I \lbr 1 \rbr}{11}{3} & \textit{\# check if the input has length 1}\\
    3 & : & I \lbr 0 \rbr := I \lbr 0 \rbr \monus 1 & \textit{\# decrement $I \lbr 0 \rbr$ to $n \monus 1$}\\
    4 & : & I \lbr 2 \rbr := I \lbr 2 \rbr + 1 & \textit{\# increment the index $i-1$ that points to $r_i$}\\
    5 & : & Z \lbr I \lbr 1 \rbr \rbr := Z \lbr I \lbr 2 \rbr \rbr & \textit{\# copy $r_i$ into $Z \lbr 1 \rbr$}\\
    6 & : & Z \lbr 0 \rbr := Z \lbr 0 \rbr + Z \lbr 1 \rbr & \textit{\# add $r_1+\dots+r_{i-1}$ in $Z \lbr 0 \rbr$ to $r_i$ in $Z \lbr 1 \rbr$} \\
    7 & : & \ifthenbranch{I \lbr 2 \rbr = I \lbr 0 \rbr}{8}{4} & \textit{\# check if $i-1 = n-1$}\\
    8 & : & I \lbr 0 \rbr := I \lbr 0 \rbr \monus 1 & \textit{\# decrement $I \lbr 0 \rbr$}\\
    9 & : & \ifthenbranch{I \lbr 0 \rbr = I \lbr 3 \rbr}{10}{8} & \textit{\# check if $I \lbr 0 \rbr$ contains 0}\\
    10 & : & I \lbr 0 \rbr := I \lbr 0 \rbr + 1 & \textit{\# store $1$ in $I \lbr 0 \rbr$ denoting length of output} \\
    11 & : & \mathsf{stop} & \textit{\# stop and return $Z \lbr 0 \rbr$}\\
\end{array}}
$$
\end{minipage}}
\caption{An $\R$-program that computes the sum of a finite sequence of real numbers of variable length.}
\label{table:ExampleR-Program}
\end{table}

\subsubsection{Structures and Operations.} By allowing each operation of the structure $\R = (\R,0,1,-,+,\cdot,\leq)$ as a type of $\R$-instruction in $\R$-programs, we capture algorithms and computations over $\R$ in the form of $\R$-programs and $\R$-machines. The power of this basic programming language of $\R$-programs is parameterized by the strength of the operations present in the structure $\R$. We can consider additional operations, such as the exponential function $\exp$, as primitive operations by adding them as instructions to our programming language and equipping our machines with the ability to execute these operations. This corresponds to passing to the expanded structure $\R_{\exp} = (\R,0,1,+,\cdot, \exp,\leq).$

Similarly, we capture algorithms and computations over the complex numbers by modifying our machines in two ways. First, we must change their registers $Z \lbr 0 \rbr, Z \lbr 1 \rbr, Z \lbr 2 \rbr, \dots$ from $\R$-registers to $\mathbb{C}$-registers that can hold a single complex number. Second, we must modify the way these machines interpret the constants $0$ and $1$ to agree with their interpretation in $\mathbb{C}$, and we must change the way these machines execute the operations $+$, $\cdot$, and $\exp$ to coincide with their interpretation in $\mathbb{C}$. This corresponds to passing to the structure $\mathbb{C}_{\exp} = (\mathbb{C},0,1,+,\cdot, \exp).$

%%%%%%%%%%%%%%%%%%%%%%%%%%%%%%%%%%%%%%%%%%%%%%%%%%%%%%%%%%%%%%%%%
\subsection{Register Machines over a Structure $\rr$}
\label{sub:MachinesOverR}
%%%%%%%%%%%%%%%%%%%%%%%%%%%%%%%%%%%%%%%%%%%%%%%%%%%%%%%%%%%%%%%%%

We can see from the discussion in the previous section that any structure $\rr$ determines a class of $\rr$-programs and a class of $\rr$-machines. The vocabulary of $\rr$ determines the class of $\rr$-instructions, the universe of $\rr$ determines the type of data $\rr$-machines manipulate, and the interpretation of the vocabulary of $\rr$ in the universe of $\rr$ determines the meaning of $\rr$-programs when they are executed on $\rr$-machines. In this way, we capture algorithms and computations over $\rr$ in the form of $\rr$-programs and $\rr$-machines. 

\subsubsection{Instructions and Programs.} We now formalize this idea for an arbitrary structure $\rr$. The definitions of $\rr$-program and $\rr$-machine that we provide are slight modifications of the definitions found in \cite{gassner2019introduction} and \cite{Hemmerling98a}, which in turn share characteristics with the definitions of machines given in \cite{blum1989theory} and \cite{gradel1995descriptive}, as well as many others. A discussion comparing our model with other machine models is deferred until the end of this section. 

Our formalization begins by defining the basic components of $\rr$-programs, namely $\rr$-instructions. These instructions are syntactic objects built from the vocabulary of $\rr$, and so we first define $\Voc$-instructions for an arbitrary vocabulary $\Voc$. Working with arbitrary vocabularies at this stage will allow a $\Voc$-program to be interpreted in any $\Voc$-structure, which can be useful for transferring computational complexity results between structures of the same vocabulary, as is done in \cite{michaux1994p}. 

We also include a query instruction for an unspecified oracle $Q$ that is meant to be interpreted as some $Q \subseteq R^*$. Note that query instructions are not allowed in $\rr$-programs unless an $\rr$-machine has been equipped with an oracle $Q \subseteq R^*$.

\begin{definition}[$\Voc$-Instructions and $\rr$-Instructions]
\label{def:RInstruction}
    For a vocabulary $\Voc = (\Con,\Fun,\Rel)$, a \textit{$\Voc$-instruction} is an instruction of the form (1)-(9) in Table \ref{tab:RInstructions} such that:
    \begin{enumerate}
        \item $j_0,j_1,j_2,\dots,j_k,\ell_1,\ell_2$ are any nonnegative integers,
        \item $c \in \Con$ is a constant symbol,
        \item $f \in \Fun$ is an $k$-ary function symbol, for any positive integer $k$,
        \item $P \in \Rel$ is an $k$-ary relation symbol, for any positive integer $k$.
    \end{enumerate} 
    For a structure $\rr$, an \textit{$\rr$-instruction} is a $\Voc(\rr)$-instruction.
    %An \textit{$\rr$-instruction} is a $\Voc(\rr)$-instruction, that is, an instruction in the vocabulary of $\rr$.
\end{definition}
\begin{table}[h!]
\fbox{\begin{minipage}{\textwidth}
\textit{Computation instructions:}
\begin{enumerate}
    \item $Z \lbr j_0 \rbr := c$  
    \item $Z\lbr j_0 \rbr := f( Z \lbr j_1 \rbr, \dots, Z \lbr j_k \rbr)$
\end{enumerate}
\textit{Branching instructions:}
\begin{enumerate}
\setcounter{enumi}{2}
    \item $\ifthenbranch{Z \lbr j_1 \rbr = Z \lbr j_2 \rbr}{\ell_1}{\ell_2}$
    \item $\ifthenbranch{P( Z \lbr j_1 \rbr, \dots, Z \lbr j_k \rbr)}{\ell_1}{\ell_2}$
\end{enumerate}
\textit{Copy instructions:}
\begin{enumerate}
\setcounter{enumi}{4}
    \item $Z \lbr I \lbr j_0 \rbr \rbr := Z \lbr I \lbr j_1 \rbr \rbr$ 
\end{enumerate}
\textit{Index instructions:}
\begin{enumerate}
\setcounter{enumi}{5}
    \item $I \lbr j_0 \rbr := I \lbr j_0 \rbr + 1$
    \item $I \lbr j_0 \rbr := I \lbr j_0 \rbr \monus 1$
    \item $\ifthenbranch{I \lbr j_1 \rbr = I \lbr j_2 \rbr}{\ell_1}{\ell_2}$
\end{enumerate}
\textit{Stop instruction:}
\begin{enumerate}
\setcounter{enumi}{8}
    \item $\mathsf{stop}$
\end{enumerate}
\textit{Query instruction for an oracle $Q$}
\begin{enumerate}
\setcounter{enumi}{9}
    \item $\ifthenbranch{(Z \lbr 0 \rbr ,\dots, Z\lbr I \lbr 0 \rbr \rbr) \in Q}{\ell_1}{\ell_2}$
\end{enumerate}
\end{minipage}}
\caption{$\Voc$\textbf{-instructions}}
\label{tab:RInstructions}
\end{table}

%Recalling our assumption that $0,1 \in \Con(\rr)$ and $= \hspace{.1cm} \in \Rel(\rr)$ for all $\rr$ in our consideration, we see that $\rr$-instructions always allow writing $0$ and $1$ into $\rr$-registers, as well as comparing the contents of two $\rr$-registers for equality. 
As we saw in the previous section, we may think of index registers $I \lbr j \rbr$ as pointers to $\rr$-registers $Z \lbr I \lbr j \rbr \rbr$ or as a memory location for discrete data. Index instructions of the form $(6)$ and $(7)$ allow incrementing and decrementing the contents of index registers, which corresponds to moving this pointer left or right along the unbounded sequence $Z\lbr 0 \rbr, Z \lbr 1 \rbr, Z \lbr 2 \rbr, \dots$ of $\rr$-registers. Since this sequence is one-way infinite, decrement instructions of the form $(7)$ use truncated subtraction, or monus, defined as $n \monus m = n - m$ if $n \geq m$, and $n \monus m = 0$ if $n < m$. 

\begin{definition}[$\Voc$-Programs and $\rr$-Programs]
\label{def:RProgram}
    For a vocabulary $\Voc$, a \textit{$\Voc$-program $\program$} is a finite sequence of labeled $\Voc$-instructions of the form $$0: \mathsf{instruction}_0; \ 1: \mathsf{instruction}_1; \ \dots ; \ \ell_\program-1: \mathsf{instruction}_{l-1}; \ \ell_\program : \mathsf{stop}.$$ The positive integer $\ell_\program+1$ is the length of the program $\program$. In each $\Voc$-program $\program$, the $\mathsf{stop}$ instruction occurs exactly once, as the last instruction in the sequence. The nonnegative integer $k_\program-1$ is the largest $j$ such that $I \lbr j \rbr$ occurs in an instruction of $\program$. For a structure $\rr$, an $\rr$-program is a $\Voc(\rr)$-program. 
\end{definition}

% \begin{definition}
% \label{def:RProgram}
%     An \textit{$\rr$-program $\program$} is a finite sequence of labeled $\rr$-instructions of the form $$0: \mathsf{instruction}_0; \ 1: \mathsf{instruction}_1; \ \dots ; \ \ell_\program-1: \mathsf{instruction}_{l-1}; \ \ell_\program : \mathsf{stop}.$$ The positive integer $\ell_\program+1$ is the length of the program $\program$. In each $\rr$-program $\program$, the $\mathsf{stop}$ instruction occurs exactly once, as the last instruction in the sequence. The nonnegative integer $k_\program-1$ is the largest $j$ such that $I \lbr j \rbr$ occurs in an instruction of $\program$. 
% \end{definition}

%Without an interpretation of the symbols in a vocabulary $\Voc$, there is no inherent meaning of a $\Voc$-program $\program$. A $\Voc$-structure $\rr$ provides this meaning by interpreting the symbols in $\program$. The $\Voc$-program $\program$, together with the $\Voc$-structure $\rr$, the index registers, and $\rr$-registers form an $\rr$-machine as depicted in \Cref{fig:RegisterMachine}, wherein we may think of $\rr$ as part of the CPU interpreting the program.   

\subsubsection{Machines.} Similar to Turing machines, register machines, and random-access machines, each $\rr$-machine will contain a fixed $\rr$-program. We maintain a distinction between $\rr$-programs and $\rr$-machines as advocated by Scott in \cite{scott1967some} and endorsed by Ga\ss ner in \cite{gassner2019introduction}. Programs are syntactic objects built from the vocabulary of $\rr$, and machines give meaning to programs by specifying how the instructions are to be interpreted.

\begin{definition}[$\rr$-Machines]
\label{def:RMachine}
    A \textit{machine over $\rr$}, also called an \textit{$\rr$-machine}, is a tuple $\M = (\program,\rr,\N^{k_\program},R^\N)$ specifying the following information:
    \begin{enumerate}
        \item an $\rr$-program $\program$,
        \item the structure $\rr$,
        \item a space of index states $\N^{k_\program}$, 
        \item a space of memory states $R^\N$.
    \end{enumerate}
    We say that the constants of $\rr$, other than $0$ and $1$, appearing in an instruction of $\program$ are the \textit{machine constants} of $\M$. The \textit{input} and \textit{output space} of $\M$ are $R^* = \bigcup_{n=0}^\infty R^n.$ 
\end{definition}

In the above definition, we include the structure $\rr$ as a component of $\rr$-machines. We do this so that the $\rr$-machine knows how to interpret the symbols in its program, and in this sense we may view the structure $\rr$ as the CPU from \Cref{fig:RegisterMachine}, while the other components of $\rr$-machines correspond in the expected way to this picture.

Note that if $\rr$ and $\mathcal{Q}$ are both $\Voc$-structures, then every $\rr$-program $\program$ is a $\mathcal{Q}$-program and vice versa. This means that $\program$ induces both the $\rr$-machine $\M = (\program,\rr,\N^{k_\program},R^\N)$ and the $\mathcal{Q}$-machine $\mathcal{N} = (\program, \mathcal{Q},\N^{k_\program},Q^\N)$. Despite the fact that these machines have the same program, in general they will interpret this program differently. That is, the machines $\M$ and $\mathcal{N}$ will compute different functions, a notion that we formalize in the following section.

\subsubsection{Discussion.}

Our definition of $\rr$-machines is a slight modification of the definition of BSS RAM's presented by Ga\ss ner in \cite{gassner2019introduction}. It is a generalization of the original definition of BSS-machines given in \cite{blum1989theory}, which resemble register machines because they have high-level instructions from which to build programs. Note that the later definition of BSS-machines given in \cite{blum1998complexity} was modified to more closely resemble a Turing machine by replacing indirect access instructions with instructions that shift a single head left or right one step at a time. There is a long history of research into computation and complexity over structures. For a brief discussion of this history with extensive references, see the introduction of \cite{hemmerling1996computability}.

We have modified the definition of BSS RAM's found in \cite{gassner2019introduction} in several ways. We allow machines to accept the empty string as an input, and in order to encode the length of this input in an index register---as we do for all inputs---we allow index registers to contain 0, in contrast to the original definition in which only positive integers are allowed. This means we must have an $\rr$-register indexed by 0. For uniformity, we also have an index register indexed by $0$. 

More significantly, we always allow branching on equality tests, in line with our use of first order logic with equality in which equality is always present by default. (See the note on equality at the end of \Cref{sub:FirstOrderLogic}.) Furthermore, we do not include input and output maps since these are fixed in a standard way for each machine, and we do not include the set of labels of $\program$, nor a finite-vocabulary structure $\mathcal{\rr'}$ that is a reduct of $\rr$ with sufficient structure to interpret $\program$ since both of these may be determined by the program $\program$. 

Finally, we avoid calling our machines RAM's because our programs have no instructions that allow \textit{random} access to memory, even though the copy instruction $Z \lbr I \lbr j_0 \rbr \rbr := Z \lbr I \lbr j_1 \rbr \rbr$ does allow \textit{indirect} access to memory. When paired with indirect access to memory, an instruction of the form $I \lbr j \rbr := n$, for any $n \in \N$, would allow random access to memory, as in the paper originally introducing random access machines \cite{cook1972time}. Without such an instruction, our machines are required to sequentially increment $I \lbr j \rbr$ until it reaches $n$, so our machines have only sequential access to memory. If one needs to differentiate our machine model from others, we suggest the name \textit{register machines over a structure}.

% \jeremy{Should these last paragraphs be included?}
% A machine over the Boolean structure $\mathcal{B} = (\{0,1\},0,1)$ is equivalent to a Turing machine, up to polynomial-time simulations.
% %, a fact we establish in Theorem \ref{}. 

% On the other hand, a machine over the structure $(\R,+,-,\times,/,<)$ is equivalent to a BSS-machine over the real numbers, up to simulations with constant-factor overhead in time. From this we can see that machines over structures faithfully generalize both Turing machines and BSS-machines.

%%%%%%%%%%%%%%%%%%%%%%%%%%%%%%%%%%%%%%%
\subsection{Computations Over $\rr$}
\label{sub:ComputationsOverR}
%%%%%%%%%%%%%%%%%%%%%%%%%%%%%%%%%%%%%%%

\subsubsection{Configurations and Transitions.} In order to define precisely how an $\rr$-machine $\M$ interprets its program $\program$, we must define the notion of a configuration of $\M$. A configuration of $\M$ is a tuple that encodes a snapshot of $\M$ at one point in time. This tuple specifies the instruction of $\program$ that $\M$ is about to execute, the current contents of the index registers $I \lbr 0 \rbr,\dots , I \lbr k_\program - 1 \rbr$ and the current contents of the $\rr$-registers $Z \lbr 0 \rbr, Z \lbr 1 \rbr, Z \lbr 2 \rbr,\dots$.

\begin{definition}[Configurations]
    A \textit{configuration} of an $\rr$-machine $\M = (\program,\rr,\N^{k_\program},R^\N)$ is a tuple $$\conf = (\ell, \nu, \rho) \in \{0,\dots,\ell_\program\} \times \N^{k_\program} \times R^\N.$$
    Here $\ell$ is the label of the instruction currently being executed, $\nu \in \N^{k_\program}$ describes the contents of the index registers, and $\rho \in R^\N$ describes the contents of the $\rr$-registers.
    If $\ell = 0$, then $\conf$ is an \textit{initial configuration}, and if $\ell = \ell_\program$, then $\conf$ is a   \textit{stop configuration}. The \textit{initial configuration on $v = (r_1,\dots,r_n) \in R^n$} is $$\conf_0(v) = (0,(n,0, \dots,0),(r_1,\dots,r_n,0,\dots)).$$ The \textit{state space} of $\M$ is the set $\mathcal{S}_\M = \{0,\dots, \ell_\program\} \times \N^{k_\program} \times R^\N$ of all possible configurations for $\M$. 
    We simply write $\mathcal{S}$, instead of $\mathcal{S}_\M$, whenever \M is clear from the context.
\end{definition}

We now define the transition relation of $\M$, which is a binary relation $\to_\M$ on the state space of $\M$. 
Again, we write $\to$, instead of $\to_\M$, whenever \M is clear from the context.
The relation $\to$ describes how one configuration of $\M$ transitions into another as $\M$ alters the contents of its registers based on the instructions in its program. We may think of this as the \textit{operational semantics} of $\rr$-programs.

%\till{I am a bit confused that you call it transition-relation instead of transition-function. But I guess there is not much of a difference.}\till{I would drop the index \M for the following definition. I think it will reduce clutter.}

\begin{definition}[Transition Relation]
\label{def:TransitionRelationOfRMachines}
    The \textit{transition relation} $\to \ \subseteq \mathcal{S}^2$ of an $\rr$-machine $\M$
    % $= (\program,\rr,\N^{k_\program},R^\N)$ 
    is determined as follows. 
    Suppose $\conf = (\ell,\nu,\rho) \in \{0,\dots,\ell_\program\} \times \N^{k_\program} \times R^\N$ is a configuration of $\M$.
    \begin{enumerate}
        \item If the $\ell$th instruction of $\program$ is $\ell : Z \lbr j_0 \rbr := c$, then $\conf \to (\ell',\nu',\rho')$ where
        \begin{itemize}
            \item $\ell' = \ell+1$,
            \item for all $m \in \{0, \dots, k_\program-1\}$, $\nu'(m) = \nu(m)$,
            \item for all $n \in \N$, $\rho'(n) = \begin{cases}
                c^\rr & \text{ if } n = j_0 \\
                \rho(n) & \text{ if } n \neq j_0.
            \end{cases}$
        \end{itemize}
        \item If the $\ell$th instruction of $\program$ is $\ell: Z \lbr j_0 \rbr := f(Z \lbr j_1 \rbr, \dots, Z \lbr j_k \rbr )$, then $\conf \to (\ell',\nu',\rho')$ where
        \begin{itemize}
            \item $\ell' = \ell+1$,
            \item for all $m \in \{0, \dots, k_\program-1\}$, $\nu'(m) = \nu(m)$,
            \item for all $n \in \N$, $\rho'(n) = \begin{cases}
                f^\rr(\rho({j_1}), \dots, \rho({j_k})) & \text{ if } n = j_0 \\
                \rho(n) & \text{ if } n \neq j_0.
            \end{cases}$
        \end{itemize}
        \item If the $\ell$th instruction of $\program$ is $\ifthenbranch{Z \lbr j_1 \rbr = Z \lbr j_2 \rbr}{\ell_1}{\ell_2}$, then $\conf \to (\ell',\nu',\rho')$ where
        \begin{itemize}
            \item $\ell' = \begin{cases}
                \ell_1 & \text{ if } \rho({j_1}) = \rho({j_2}) \\
                \ell_2 & \text{ if } \rho({j_1}) \neq \rho({j_2})
            \end{cases}$
            \item for all $m \in \{0, \dots, k_\program-1\}$, $\nu'(m) = \nu(m)$,
            \item for all $n \in \N$, $\rho'(n) = \rho(n)$.
        \end{itemize}
        \item If the $\ell$th instruction of $\program$ is $\ifthenbranch{P( Z \lbr j_1 \rbr, \dots, Z \lbr j_k \rbr)}{\ell_1}{\ell_2}$, then $\conf \to (\ell',\nu',\rho')$ where
        \begin{itemize}
            \item $\ell' = \begin{cases}
                \ell_1 & \text{ if } (\rho(j_1), \dots, \rho(j_k)) \in P^\rr \\
                \ell_2 & \text{ if } (\rho(j_1), \dots, \rho(j_k)) \not\in P^\rr
            \end{cases}$
            \item for all $m \in \{0, \dots, k_\program-1\}$, $\nu'(m) = \nu(m)$,
            \item for all $n \in \N$, $\rho'(n) = \rho(n)$.
        \end{itemize}
        \item If the $\ell$th instruction of $\program$ is $\ell : Z \lbr I \lbr j_0 \rbr \rbr := Z \lbr I \lbr j_1 \rbr \rbr$, then $\conf \to (\ell',\nu',\rho')$ where
        \begin{itemize}
            \item $\ell' = \ell+1$
            \item for all $m \in \{0, \dots, k_\program-1\}$, $\nu'(m) = \nu(m)$,
            \item for all $n \in \N$, $\rho'(n) = \begin{cases}
                \rho(\nu(j_1)) & \text{ if } n = \nu(j_0) \\
                \rho(n) & \text{ if } n \neq \nu(j_0).
            \end{cases}$
        \end{itemize}
        \item If the $\ell$th instruction of $\program$ is $\ell : I \lbr j_0 \rbr := I \lbr j_0 \rbr + 1$, then $\conf \to (\ell',\nu',\rho')$ where
        \begin{itemize}
            \item $\ell' = \ell+1,$
            \item for all $0 \leq j \leq k_\program$, $\nu'(m) = \begin{cases}
                \nu(m)+1 & \text{ if } m = j_0 \\
                \nu(m) & \text{ if } m \neq j_0,
            \end{cases}$
            \item for all $n \in \N$, $\rho'(n) = \rho(n).$
        \end{itemize}
        \item If the $\ell$th instruction of $\program$ is $\ell : I \lbr j_0 \rbr := I \lbr j_0 \rbr \monus 1$, then $\conf \to (\ell',\nu',\rho')$ where
        \begin{itemize}
            \item $\ell' = \ell+1,$
            \item for all $0 \leq j \leq k_\program$, $\nu'(m) = \begin{cases}
                \nu(m) \monus 1 & \text{ if } m = j_0 \\
                \nu(m) & \text{ if } m \neq j_0,
            \end{cases}$
            \item for all $n \in \N$, $\rho'(n) = \rho(n).$
        \end{itemize}
        \item If the $\ell$th instruction of $\program$ is $\ell : \ifthenbranch{I \lbr j_1 \rbr = I \lbr j_2 \rbr}{\ell_1}{\ell_2}$, then $\conf \to (\ell',\nu',\rho')$ where
        \begin{itemize}
            \item $\ell' = \begin{cases}
                \ell_1 & \text{ if } \nu(j_1) = \nu(j_2) \\
                \ell_2 & \text{ if } \nu(j_1) \neq \nu(j_2),
            \end{cases}$
            \item for all $m \in \{0, \dots, k_\program-1\}$, $\nu'(m) = \nu(m)$,
            \item for all $n \in \N$, $\rho'(n) = \rho(n)$.
        \end{itemize}
        \item If the $\ell$th instruction of $\program$ is $\ell_\program : \mathsf{stop}$, then $\conf \to \conf'$ does not hold for any $\conf' \in \mathcal{S}_\M$.
    \end{enumerate}
We work only with deterministic machines, which means that the transition relation for any $\rr$-machine $\M$ is a partial function $(\to) \: \mathcal{S}_\M \partialto \mathcal{S}_\M$, analogous to the transition function of Turing machines or the computing endomorphism of BSS-machines. Since $(\to)$ is a partial function, may write $(\to)(\conf) = \conf'$ instead of $\conf \to \conf'$ whenever $(\conf,\conf') \in \hspace{.1cm} \to$.
\end{definition}

\subsubsection{Computations.} The inputs that we can give to an $\rr$-machine are finite-length strings $v = (r_1,\dots,r_n)$ from the universe of $\rr$. As such, the set of strings over the universe of $\rr$ plays an important role as a domain for computation with $\rr$-machines.

\begin{definition}[Strings]
    Given a set $R$, the \textit{set of strings} over $R$ is the set $R^* = \bigcup_{n=0}^\infty R^n$. Elements $v = (r_1,\dots,r_n)$ of $R^*$ are called \textit{strings}. The \textit{size} of a string is its length $|v| = n \in \N$. We will typically use $u,v,w,\dots,v_1,v_2,v_3,\dots$ to denote strings in $R^*$.
\end{definition}

Whenever an $\rr$-machine $\M$ is given an input $v = (r_1,\dots,r_n) \in R^*$, its registers are initialized so that $v$ is stored in its $\rr$-registers, and $|v|=n$ is stored in its index registers, in the following manner:
$$
\begin{array}{cccccccc}
    Z \lbr 0 \rbr := r_1 & Z \lbr 1 \rbr := r_2 & Z \lbr 2 \rbr := r_3 & Z \lbr 3 \rbr := r_4 & \dots & Z \lbr n-1 \rbr := r_n & Z \lbr n \rbr := 0 & \dots \\
    & & & & & & & \\
    I \lbr 0 \rbr := n, & I \lbr 1 \rbr := 0 & I \lbr 2 \rbr := 0  & I \lbr 3 \rbr := 0  & \dots  & I \lbr k_\program - 1 \rbr := 0. & &
\end{array}
$$
The component $r_i$ of $v$ is stored in $\rr$-register $Z \lbr i-1 \rbr$ because we begin indexing $\rr$-registers with 0, for reasons explained in the discussion at the end of \Cref{sub:MachinesOverR}, but we begin indexing $v$ with 1, as is traditional for vectors. 

Since we work only with deterministic machines, the input $v \in R^*$ induces a single, potentially infinite sequence of configurations such that $$\conf_0 = \conf_0(v) \text{ and } \conf_{t+1} = (\to)(\conf_t)$$ for all $t \in \N$ for which $(\to)$ is defined on $\conf_t$. This sequence is the \textit{computation} of $\M$ on $v$.  

\begin{definition}[Computation and Running Time]
    The \textit{computation} of $\M$ on input $v \in R^*$ is the (potentially infinite) sequence of configurations $$\conf_0(v) \to \conf_1 \to \ \dots \to \conf_n \to \dots$$ that cannot be extended. We say that $\M$ \textit{halts} on input $v$ whenever there is a finite sequence of configurations $$\conf_0(v) \to \conf_1 \to \dots \to \conf_n$$ such that $\conf_n$ is a stopping configuration. In this case, we call the sequence $(\conf_0(v), \dots, \conf_n)$ a \textit{halting computation}, and we say that $\M$ halts on input $v$ in time $T_\M(v)=n$. If $\M$ does not halt on input $v$, then we say that $\M$ \textit{diverges} on input $v$, and we set $T_\M(v) = \infty$. The value $T_\M(v)$ is the \textit{running time} of $\M$ on input $v$.
\end{definition}

If $\M$ halts on input $v$ in configuration $(\ell_\program,(n_0,\dots,n_{k_\program-1}),(r'_0,\dots,r'_{n_0-1},\dots))$, then the \textit{output} of $\M$ on input $v$ is $\M(v) = (r'_0,\dots,r'_{n_0-1}) \in R^{n_0}$. The value $n_0$ stored in the zeroth index register $I \lbr 0 \rbr$ when $\M$ halts determines the length of the output, just $I \lbr 0 \rbr$ is initialized to hold the length of the input when the machine starts. In this way, $\M$ computes a partial function $\M \: R^* \partialto R^*$ that sends $v$ to $\M(v)$ whenever $v$ is in the domain $\dom(\M) = \{v \in R^* : \M \text{ halts on input } v \}$.

\begin{definition}[Computing a Function with a Time Bound]
    Given a partial function $F \: R^* \partialto R^*$, we say that $\M$ \textit{computes} $F$ whenever $\M(v) = F(v)$ for all $v \in \dom(\M) = \dom(F) \subseteq R^*$. Furthermore, if $F$ is a total function, then given a total function $T \: \N \to \N$, we say that \textit{$\M$ computes $F$ in time $T$} whenever $\M$ computes $F$ and $T_\M(v) \leq T(|v|)$ for all $v \in R^*$.
\end{definition}

\subsubsection{Efficiently Simulating Turing Machines.} 

We will often want $\rr$-machines to implement standard algorithms on binary inputs in a reasonable amount of time. 
For example, we claim in Subsection \ref{sub:SATR-NPR-Membership} that $\rr$-machines can implement standard techniques used to parse a context-free grammar in polynomial time, and we claim in Subsection \ref{sub:DataComplexityESO(R)} that $\rr$-machines can perform binary search in logarithmic time. 

Both of these claims are justified by our knowledge that the same can be done by Turing machines, and by the fact that $\rr$-machines can simulate Turing machines with constant-factor overhead. Intuitively, the contents of an index register may be thought of as the position of one head of a Turing machine with one tape and multiple heads. 
We make this analogy precise in Theorem \ref{thm:RMachinesSimulateBooleanOneTapeTuringMachinesWithConstantOverhead}.

\begin{theorem}
\label{thm:RMachinesSimulateBooleanOneTapeTuringMachinesWithConstantOverhead}
    Let \rr be any bipointed structure and \T be any single-tape Turing machine.
    Then there exists an \rr-machine \M that simulates $\T$ with constant-factor overhead.
\end{theorem}
\begin{proof}
    We may assume without loss of generality that the alphabet of \T is $\{0,1\}$ and that \T uses a single one-directional tape.
    We describe one for one all possible actions that \T can make and how they can be simulated by \M{}.
    \begin{itemize}
        \item Writing a $0$ or $1$ to the tape of $\T$ may be simulated by an $\rr$-machine $\M$ using computation instructions that set an $\rr$-register to $0$ or $1$, respectively.
        \item Changing the state of $\T$ based on the contents of the cell currently being read by $\T$ may be simulated by $\M$ using branching instructions that check if a register is currently set to $0$ or $1$ using the equality relation. This may be done with constant-factor overhead. 
        \item Changing the position of the tape head can be simulated by changing the value of the index registers.
    \end{itemize}
    Indeed, the entirety of the transition function of $\T$ may be encoded into the program of \M, and each transition of \T may be simulated by \M in constant time.
\end{proof}

Evoking this correspondence is the reason we have used decrement instructions of the form $(7)$ instead of the reset instruction of \cite{gassner2019introduction}, which set the contents of an index register to the smallest possible value in one time step. These instruction sets can be simulated by one another, so they are of equivalent power, but the decrement instruction brings the time complexity analysis of $\rr$-machines closer to that of Turing machines.

\subsubsection{Discussion.}

The way we have defined the running time of an $\rr$-machine uses \textit{unit cost} for operations, in which each operation in the vocabulary of $\rr$ can be performed in one time step. Others have considered cost functions in which a basic operation incurs a cost proportional its size, as is done in Chapter 20 of \cite{blum1998complexity} and \cite{K97b}. 

We have defined the \textit{size} of an input $v \in R^*$ to be its length. This means that the size of an element of the universe $r \in R$ is $|r| = 1$, even if $r$ is an infinite object such as a real number. There are other measures of size that one may use to suit one's purpose. For example, one may use bit size, in which the size of an integer $z \in \Z$ is the length of the binary encoding of $z$. This size function is often used when analyzing BSS-machines over the integers, as is done in Chapter 6 of \cite{blum1998complexity}.

%%%%%%%%%%%%%%%%%%%%%%%%%%%%%%%%%%%%%%
\subsection{Complexity Classes Over $\rr$}
\label{sub:ComplexityClasses}
%%%%%%%%%%%%%%%%%%%%%%%%%%%%%%%%%%%%%%

\begin{table}[h!]
\renewcommand{\arraystretch}{1.5}
    \centering
    \begin{tabularx}{\textwidth}{
    | >{\raggedright\arraybackslash}X
    | >{\raggedright\arraybackslash}X
    | >{\raggedright\arraybackslash}X | }
    \hline
    \bf{Term} & \bf{Definition} & \bf{Example} \\
    \hline
    universe & $R$ & $\R$ \\
    element & $r \in R$ & $r = \sqrt{2}$ \\
    strings over $R$ & $R^* = \bigcup_{n \in \N} R^n$ & $\R^* = \bigcup_{n \in \N} \R^n$ \\
    string & $v = (r_1,\dots,r_n) \in R^*$ & $v = (0,0,1.2,\sqrt{2},4) \in \R^*$ \\
    size & $|v| = n \in \N$ & $|v| = |(0,0,1.2,\sqrt{2},4)| = 5 \in \N$ \\
    powerset of strings over $R$ & $2^{R^*} = \{S : S \subseteq R^*\}$ & $2^{\R^*} = \{S : S \subseteq \R^*\}$ \\
    decision problem & $S \subseteq R^* $ &  $S_\ell = \{v \in R^* : \sum_{i=1}^{|v|} r_i = \ell\}$\\ 
    complexity class & $\mathcal{C} \subseteq 2^{R^*}$ & $\mathcal{C} = \{S_\ell : \ell \in \N\}$ \\
    \hline
    \end{tabularx}
    \caption{Complexity Theory Terminology}
    \label{tab:ComplexityTheoryTerminology}
\end{table}

One of the most basic tasks that we may devise an $\rr$-machine to carry out is to decide if some string $v$ lies in some subset $S \subseteq R^*$. An $\rr$-machine may decide this for us by computing the characteristic function $\chi_S \: R^* \to \{0,1\}$ of $S$, for which 
%$\chi_S(v) = 1$ if $v \in S$ and $\chi_S(v) = 0$ if $v \not\in S$. 
$$\chi_S(v) = \begin{cases}
        1 & \text{ if } v \in S \\
        0 & \text{ if } v \not\in S.
    \end{cases}$$ 
In this way, we may view each subset $S \subseteq R^*$ as a \textit{decision problem}, and if an $\rr$-machine $\M$ \textit{decides} $S$, we may view $\M$ as a \textit{decider}.
%and we may view an $\rr$-machine deciding $S$ as a \textit{decider}.

\begin{definition}[Decision Problems and Deciders]
\label{def:DecisionProblem}
    A \textit{decision problem} over $\rr$ is a set $S \subseteq R^*$. We say that an $\rr$-machine $\M$ \textit{decides} $S$ when $\M$ computes the characteristic function $\chi_S$ of $S$. In this case, $\M$ is an example of a \textit{decider}, which is an $\rr$-machine that returns $1$ or $0$ on every input $v \in R^*$. We say that $\M$ \textit{accepts} $v$ if $\M(v)=1$, and we say that $\M$ \textit{rejects} $v$ if $\M(v)=0$. 
\end{definition}

\begin{definition}
\label{def:PRandNPR}
    The complexity class $\P(\rr)$ is the set of decision problems that can be decided by a polynomial-time $\rr$-machine. The complexity class $\NP(\rr)$ is the set of decision problems $L$ such that there is a decision problem $S \in \P(\rr)$ and some polynomial $q$ such that $$ L = \{v \in R^* : (\exists w \in R^{q(|v|)})  (v,w) \in S\}.$$ In case $(v,w) \in S,$ we call $w$ a \textit{witness} for $v$. If $\M$ is a polynomial time $\rr$-machine deciding $S$, then we may present $L$ equivalently as $$L = \{v \in R^* : (\exists w \in R^{q(|v|)})  \M(v,w)=1\}.$$ In this case, we may view $\M$ as a \textit{verifier}, and we say that $\M$ \textit{verifies} $L$. Similarly, the complexity class $\coNP(\rr)$ is the set of decision problems $L$ such that there is a polynomial $q$ and a polynomial-time machine $\M$ such that $$L = \{v \in R^* : (\forall w \in R^{q(|v|)}) \M(v,w) = 1\}.$$ 
\end{definition}

% \jeremy{I should update the definition of $\P(\rr)$ and $\NP(\rr)$ to be also in terms of $\M(v)=1$ and $\M(v,w)=1$, as well as explaining what a ``witness'' is, as well as redifining $\NP(\rr)$ in terms of witnesses of length exactly $q(n)$. }

In addition to analyzing complexity classes $\mathcal{C}$ over $\rr$, we will relate them to Boolean complexity classes over the domain $\{0,1\}$. Towards this end, we define the Boolean part of decision problems $S$ as the set of binary strings in $S$. If $\M$ is an $\rr$-machine that decides $S$, then taking the Boolean part of $S$ corresponds to restricting the inputs of $\M$ to binary strings, or modifying the program of $\M$ to reject all non-Boolean strings.

Just as we may restrict the inputs to $\rr$-machines, may also restrict machines by only allowing the use of the constants $0$ and $1$. If $\M$ uses only these constants, then we say that $\M$ has no \textit{machine constants.} This becomes relevant if the domain of $\rr$ is a set like $\R$ in which a single element can encode infinite information. Restricting the use of constants brings $\rr$-machines closer to Boolean Turing machines. These modifications are summarized in the next definition. 

\begin{definition}
\label{def:ConstantFreeBooleanPart}
    The \textit{Boolean part} of a decision problem $S$ is the set $\BP(S) = S \cap \{0,1\}^*$. The Boolean part of a complexity class $\mathcal{C}$ is the set $\BP(\mathcal{C}) = \{\BP(S) : S \in \mathcal{C}\}.$ If $\mathcal{C}$ is defined in terms of $\rr$-machines, then $\mathcal{C}^0$ is the class obtained by restricting to $\rr$-machines with no machine constants. We write $\BP^0(\mathcal{C})$ for $\BP(\mathcal{C}^0)$, and we say that $\BP^0(\mathcal{C})$ is the \textit{constant-free Boolean part} of $\mathcal{C}$.
\end{definition}

\subsubsection*{Warning.} While the Boolean part operator $\BP(\cdot)$ acts directly on complexity classes, the constant-free operator $(\cdot)^0$ acts on \textit{definitions} of complexity classes by replacing machines in the definition of a complexity class $\mathcal{C}$ with constant-free machines to arrive at a well-defined complexity class $\mathcal{C}^0$. This means that even if there is an equality of complexity classes $\mathcal{C} = \mathcal{D}$, it might be that $\mathcal{C}^0 \neq \mathcal{D}^0$ because the definitions of $\mathcal{C}$ and $\mathcal{D}$ use machines in different ways. 
%For an example of this, see the discussion at the end of \Cref{sec:OracleMachinePolynomialHierarchy}.

For example, consider the structure $\Rcon_\textup{add}^\leq = (\R,(r : r \in \R),0,1,-,+,\leq)$. We will show in \Cref{thm:OraclePolynomialHierarchyCharacterization} that $ \NP(\Rcon_\textup{add}^\leq)^{\NP(\Rcon_\textup{add}^\leq)} = \Sigma_2 \Rcon_\textup{add}^\leq$. However, we can derive a contradiction from the assumption that $ (\NP(\Rcon_\textup{add}^\leq)^{\NP(\Rcon_\textup{add}^\leq)})^0 = (\Sigma_2\Rcon_\textup{add}^\leq)^0$ in the following way.  

We know from \cite{CK95} that $\BP(\NP(\Rcon_\textup{add}^\leq) = \NP / \textup{poly}$, the classical nonuniform complexity class defined by machines with polynomial advice, and that $\BP((\Sigma_2\Rcon_\textup{add}^\leq)^0) = \Sigma_2$, the second level of the classical polynomial hierarchy. Since restricting to constant-free oracle machines does not affect the oracles these machines have access to, we see that $(\NP(\Rcon_\textup{add}^\leq)^{\NP(\Rcon_\textup{add}^\leq)})^0 = (\NP(\Rcon_\textup{add}^\leq)^0)^{\NP(\Rcon_\textup{add}^\leq)}$. It is clear that $\NP(\Rcon_\textup{add}^\leq) \subseteq (\NP(\Rcon_\textup{add}^\leq)^0)^{\NP(\Rcon_\textup{add}^\leq)}$. Thus, if we assume that $(\NP(\Rcon_\textup{add}^\leq)^{\NP(\Rcon_\textup{add}^\leq)})^0 = (\Sigma_2\Rcon_\textup{add}^\leq)^0$, then we have that $$\NP / \textup{poly} = \BP(\NP(\Rcon_\textup{add}^\leq)) \subseteq \BP((\NP(\Rcon_\textup{add}^\leq)^{\NP(\Rcon_\textup{add}^\leq)})^0) = \BP((\Sigma_2\Rcon_\textup{add}^\leq)^0) = \Sigma_2,$$ which is a contradiction because $\NP/\textup{poly}$ contains undecidable problems.

\begin{definition}
\label{def:ExistsR}
    For any structure $\rr$, we define $\exists \rr = \BP^0(\NP(\rr))$ as the constant-free Boolean part of $\NP(\rr).$ Similarly, we define $\forall\rr = \BP^0(\coNP(\rr))$, and we define $\delta\rr = \BP^0(\P(\rr)).$ 
\end{definition}

\begin{remark}
    The notation $\delta\rr$ is meant to evoke a similarity with the notation $\Delta_0=\P$ of the polynomial hierarchy. We will see that $\delta\rr \subseteq \exists\rr \cap \forall\rr$, which resembles $\Delta_0 \subseteq \Sigma_1 \cap \Pi_1$ where $\Sigma_1 = \NP$ and $\Pi_1 = \coNP$.
\end{remark}

\begin{definition}[Polynomial Hierarchy Over $\rr$]
\label{def:PolynomialHierarchyR}
    We define two sequences of complexity classes $\Sigma_k\rr$ and $\Pi_k\rr$ for $k \in \N$ as follows. A decision problem $L \subseteq R^*$ is in $\Sigma_k\rr$ if and only if there is a polynomial $q$ and a polynomial-time $\rr$-machine $\M$ such that for all $n \in \N$ and all $v \in R^n$ $$v \in L \text{ if and only if } (Q_1 w_1 \in R^{q(n)})  \dots (Q_k w_k \in R^{q(n)}) \M(v, w_1, \dots, w_k)=1,$$ where the quantifiers $Q_i \in \{\exists, \forall\}$ alternate, starting with $Q_1 = \exists$. The class $\Pi_k\rr$ is defined by the same condition except the first quantifier is $Q_1 = \forall$. Define the \textit{polynomial hierarchy over $\rr$} as the class $\PH(\rr) = \bigcup_{k\in\N}\Sigma_k\rr.$ 
\end{definition}

\begin{definition}[Boolean Hierarchy Over $\rr$]
\label{def:BooleanHierarchyR}
    We define the complexity classes $\exists_k\rr$ and $\forall_k\rr$ as the constant-free Boolean parts of $\Sigma_k\rr$ and $\Pi_k\rr$, respectively, for $k \in \N$. That is, $$\exists_k \rr= \BP^0(\Sigma_k\rr) \textup{ and } \forall_k \rr = \BP^0(\Pi_k\rr).$$ The Boolean hierarchy over $\rr$ is the complexity class $\BH(\rr) = \bigcup_{k \in \N} \exists_k\rr$. 
\end{definition}

\begin{example}
    If $\mathcal{B} = (\{0,1\},0,1)$ is the Boolean structure, then $\Sigma \mathcal{B} = \exists \mathcal{B} = \NP$ and $\Pi\mathcal{B} = \forall\mathcal{B} = \coNP$.
\end{example}

\begin{example}
    If $\R = (\R,(r : r \in \R),+,-,\times,/,<)$, then $\Sigma_1 \R = \NP(\R)$ as defined in \cite{blum1998complexity}, and $\exists_1 \R = \exists \R$ as defined in \cite{SCM24}.
\end{example}

\begin{example}
    If $\R = (\R,0,1,-,+,\cdot,\leq)$ then $\exists \R = \BP^0(\NP(\R))$ is the complexity class whose defining hard problem is the existential theory of the reals as defined in \cite{SCM24}. 
\end{example}

\begin{definition}
\label{def:SigmaComplexityClassJumpAndExistsComplexityClass}
    For any complexity class $\mathcal{C}$ over $\rr$, the complexity class $\Sigma \mathcal{C}$ contains a decision problem $L \subseteq R^*$ if and only if there is a polynomial $q$ and a decision problem $S \in \mathcal{C}$ such that for all $v \in R^*$ $$v \in L \textup{ if and only if } \exists w \in R^{q(|v|)} (v,w) \in S.$$ Similarly, the complexity class $\Pi \mathcal{C}$ contains a decision problem $L \subseteq R^*$ if and only if there is a polynomial $q$ and a decision problem $S$ such that for all $v \in R^*$ $$v \in L \textup{ if and only if } \forall w \in R^{q(|v|)} (v,w) \in S.$$ 
    %The complexity class $\exists \mathcal{C} = \BP^0(\Sigma \mathcal{C})$ is the constant-free Boolean part of $\Sigma\mathcal{C}$, and $\forall \mathcal{C} = \BP^0(\Pi\mathcal{C})$ is the constant-free Boolean part of $\Pi\mathcal{C}$. 
\end{definition}

% \begin{lemma}
%     For any structure $\rr$, define the complexity classes $\Sigma_k \rr$ and $\Pi_k \rr$ over $\rr$, for any $k \in \N$, as follows. First, we define $\Sigma_0\rr = \Pi_0\rr = \P(\rr).$ Next, we define $$\Sigma_{k+1} \rr = \Sigma \Pi_k \rr \textup{ and } \Pi_{k+1}\rr = \Pi \Sigma_k \rr.$$ The polynomial hierarchy over $\rr$ is the complexity class $\PH(\rr) = \bigcup_{k \in \N} \Sigma_k\rr$. 
% \end{lemma}

\begin{lemma}
    For all $k \in \N,$ $\Sigma_{k+1} \rr = \Sigma \left(\Pi_k \rr \right)$ and $\Pi_{k+1}\rr = \Pi \left( \Sigma_k \rr\right).$ 
\end{lemma}
\begin{proof}
    Simply pad each witness string so that they are all of the same length.
\end{proof}

\subsubsection{Oracle Complexity Classes}

Oracles are perhaps the central concept in computability theory because they introduce a notion of relative computability \cite{soare2009turing}, and they have played a significant role in computational complexity theory.

\begin{definition}
\label{def:OracleMachines}
    An $\rr$-machine with a $Q \subseteq R^*$ oracle is an $\rr$-machine whose program may contain instructions of the form $$\ifthenbranch{(Z \lbr 0 \rbr ,\dots, Z\lbr I \lbr 0 \rbr \rbr) \in Q}{\ell_1}{\ell_2}.$$ 
\end{definition}

\begin{definition}
    If $\mathcal{C}$ is a complexity class, then $\P(\rr)^\mathcal{C}$ is the complexity class of decision problems $L \subseteq R^*$ for which there is a decision problem $Q \in \mathcal{C}$ and some polynomial-time $\rr$-machine $\M$ such that $\M$ with oracle access to $Q$ decides $L$.
\end{definition}

\begin{example}
    The class $\NP(\rr)^{\NP(\rr)}$ is the class $(\Sigma\P(\rr))^{\NP(\rr)} = \Sigma\left(\P(\rr)^{\NP(\rr)}\right).$
\end{example}

Usually, complexity classes defined in terms of oracle machines are defined using a single fixed oracle. For example, in the classical polynomial hierarchy one may show that $\Sigma_2 = \NP^\SAT$. Using the fact that $\SAT$ is $\NP$-complete, one may show that $\NP^\SAT = \NP^\NP$. Defining the comlexity class $\P(\rr)^\mathcal{C}$ in terms of the entire class $\mathcal{C}$, rather than a complete problem for $\mathcal{C}$, extends our results to classes $\mathcal{C}$ that may not have a complete problem. In particular, we thereby extend our results to structures $\rr$ of infinite type for which we do not know if $\NP(\rr)$ has a complete problem.

Within the classical polynomial hierarchy, we define the sequence of complexity classes $\Delta_k$ as $\Delta_0=\P$, and $\Delta_{k+1} = \P^{\Sigma_k}$. We extend this definition to arbitrary structures $\rr$ in the next definition.

\begin{definition}
\label{def:DeltaKR}
    For any structure $\rr$, we define a sequence of complexity classes $\Delta_k\rr$ as follows. Define $\Delta_0\rr = \P(\rr)$, and define $\Delta_{k+1}\rr = \P(\rr)^{\Sigma_k\rr}$. Furthermore, define $\delta_k\rr$ as the constant-free Boolean part of $\Delta_k\rr$. That is, $$\delta_k\rr = \BP^0(\Delta_k\rr).$$
\end{definition}

As we will see, it is quite fruitful to analyze the polynomial and Boolean hierarchies over $\rr$ in terms of $\rr$-machines with oracles.

\begin{definition}[Oracle Complexity Classes]
    Let $\mathcal{D}$ be a complexity class over $\rr$. For each of level of the polynomial and Boolean hierarchies over $\rr$, the complexity classes $\Sigma_k\rr^\mathcal{D}$, $\Pi_k\rr^\mathcal{D}$, $\exists_k\rr^\mathcal{D}$, and $\forall_k\rr^\mathcal{D}$ are obtained by modifying the definition of the corresponding class to allow $\rr$-machines with oracles from $\mathcal{D}$.
\end{definition}

\subsubsection{Reductions}

Throughout this paper we use the standard notion of a polynomial-time computable many-one reduction executed either by an $\rr$-machine, when in the $\Sigma_k\rr$ setting, or by a Turing machine, when in the $\exists_k \rr$ setting.

\begin{definition}[$\rr$-Machine Reductions]
    An $\rr$-machine reduction from a decision problem $A \subseteq R^*$ to a decision problem $B \subseteq R^*$ is a function $f$ such that
    \begin{enumerate}
        \item for all $x \in R^*$, $x \in A$ if and only if $f(x) \in B$
        \item the function $f$ is computable in polynomial time on an $\rr$-machine.
    \end{enumerate}
\end{definition}

When we work in the constant-free Boolean part of complexity classes, such as $\exists_k\rr$, we switch to Turing machine reductions in order to relate these classes to classical complexity classes.

\begin{definition}[Turing-Machine reductions]
    An $\rr$-machine reduction from a decision problem $A \subseteq \{0,1\}^*$ to a decision problem $B \subseteq \{0,1\}^*$ is a function $f$ such that
    \begin{enumerate}
        \item for all $x \in \{0,1\}^*$, $x \in A$ if and only if $f(x) \in B$
        \item the function $f$ is computable in polynomial time on an Turing machine.
    \end{enumerate}
\end{definition}

%%%%%%%%%%%%%%%%%%%%%%%%%%%%%%%%%%%%%%%%%%%%%%%
\subsection{Metafinite Structures}
\label{sub:MetafiniteStructures}
%%%%%%%%%%%%%%%%%%%%%%%%%%%%%%%%%%%%%%%%%%%%%%%

Metafinite structures may be thought of as finite structures $\A$ that are given finite access to a potentially infinite structure $\rr$ by a finite set $\W$ of weight functions $W \: A^{k} \to R$. 
They were first used to generalize Fagin's theorem from $\NP$ to $\NP(\R)$ in \cite{gradel1995descriptive}, and their general theory was developed in~\cite{graedel1998metafinite}. 
Just as $\rr$-machines separate the finite and discrete operations of index registers from the potentially infinite and continuous operations of $\rr$-registers, metafinite structures separate the finite and discrete aspect of a structure from the potentially infinite and continuous aspects of a structure. 
Good examples to hold in mind as we delve into the specifics are graphs with real number weights on the edges, relational databases with a numerical domain that supports arithmetic operations on that domain, or more generally any kind finite structure with numerical attributes.  

In order to talk about metafinite structures, we need a notion of metafinite vocabularies. Just as a metafinite structure has a clean separation between a finite structure $\A$ and a potentially infinite structure $\rr$ that interact via a finite set of weight functions $\W$, a metafinite vocabulary will consist of three separate vocabularies.

\begin{definition}
\label{def:MetafiniteVocabulary}
    A \textit{metafinite vocabulary} is a triple $\MVoc = (\primary(\MVoc),\secondary(\MVoc),\weight(\MVoc))$ where
    \begin{enumerate}
        \item the \textit{primary vocabulary} $\primary(\MVoc)$ is a finite vocabulary,
        \item the \textit{secondary vocabulary} $\secondary(\MVoc)$ is a potentially infinite vocabulary,
        \item the \textit{weight vocabulary} $\weight(\MVoc)$ is a finite set of function symbols $W$, each with a nonnegative integer $k_W \in \N$ called the arity of $W$. 
    \end{enumerate} 
\end{definition}

We say that $\Voc$ is a \textit{metafinite vocabulary over $\rr$} whenever the secondary vocabulary of $\Voc$ is the vocabulary of $\rr$. In line with our assumption about first-order structures, we will assume that the secondary vocabulary of $\Voc$ contains two constant symbols $0$ and $1$, but we will not assume that the primary vocabulary has these constants. We will also assume that both the primary and secondary vocabularies contain a symbol for equality, although we will continue to omit this symbol when we specify vocabularies.

\begin{definition}
\label{def:MetafiniteRStructure}
    A \textit{metafinite structure} is a triple $\D = (\A,\rr,\W)$ where
    \begin{enumerate}
        \item the \textit{primary part} of $\D$ is a finite structure $\A$ in a finite vocabulary $\Voc(\A)$,
        \item the \textit{secondary part} of $\D$ is a potentially infinite structure $\rr$ in a potentially infinite vocabulary $\Voc(\rr)$,
        \item the set of \textit{weight functions} of $\D$ is a finite set $\W$ of functions $W \: A^{k_W} \to R$.  
    \end{enumerate}
    We say that $\D$ is \textit{Boolean} if the image of each weight function $W \in \W$ is contained in $\{0,1\}$ so that $W \:A^{k_W} \to \{0,1\}\subseteq R$.
\end{definition}

The \textit{size} of a metafinite structure $\D$, denoted by $|\D|$, is the cardinality of the universe $A$ of its primary part $\A$. The \textit{vacabulary} of $\D$ is the triple $\Voc(\D) = (\Voc(\A),\Voc(\rr),\Voc(W))$, where $\Voc(W)$ is a set containing exactly one function symbol of arity $k_w$ for each $w \in W$. In practice, we will conflate the function $W \: A^{k_W} \to R$ with the symbol for $w$. Whenever the secondary part of $\D$ is $\rr$, then we say that $\D$ is an \textit{$\rr$-structure}. For a given metafinite vocabulary $\MVoc$ over $\rr$, let $\struct(\rr,\Voc)$ be the class of $\rr$-structures in this vocabulary.

\begin{example}
    A simple class of metafinite structures is obtained by letting $\A = (V,E)$ be a finite graph, letting $\rr = \R$ be the ordered ring of real numbers, and letting $W$ contain a single binary function $w \: V^2 \to \R$ that assigns $0$ to all pairs $(v_1,v_2)$ that are not related by $E$. The resulting $\R$-structure is a finite graph with edges that have real number weights. Conversely, every graph with real weights can be interpreted as an \R-structure.
\end{example}

\begin{example}
\label{example:StringsAreMetafiniteStructures}
    Another simple class of $\rr$-structures is obtained by letting $\A = (\{1, \dots, n\})$ be an initial segment of the positive integers and letting $W$ contain a single unary function $X \: \{1,\dots,n\} \to R$. The resulting $\rr$-structure induces the string
    $(X_1, \dots, X_n) \in R^n$ such that $X_i = X(i)$. 
    Conversely, every string $(X_1,\dots,x_n) \in R^n$ induces an $\rr$-structure in this same class, where $X$ is the unary function $X \: \{1,\dots,n\} \to R$ such that $X(i) = X_i.$ From this we can see that elements of $R^*$ are in bijection with $\rr$-structures of a certain class.
\end{example}

In order to treat $\rr$-structures in any vocabulary as inputs to $\rr$-machines, we must encode these structures as strings in $R^*$. Any encoding of $\rr$-structures as strings imposes a total ordering on the universe of the primary part of these structures. 
Conversely, choosing a total ordering on the primary part of $\rr$-structures is an essential step towards devising an encoding scheme.

% \till{The following definition is a bit confusing to me.}
% \jeremy{I should change this definition to remove ambiguity between the symbol $\leq$ and its interpretation in a structure.}

\begin{definition}
    For every finite structure $\A$, let $\ordering(\A)$ denote a \textit{choice of total ordering} $\leq$ on the universe of $\A$. We say that a finite structure $\A$ is \textit{ordered} if $\ordering(\A)$ is a relation of $\A$. Note that this means $\leq$ is in the vocabulary of $\A$. If $\ordering(\A)$ is not a relation of $\A$, then we say that $\A$ is \textit{unordered}. We say that a metafinite structure $\D$ is ordered if its primary part is ordered, and otherwise we say it is unordered.
\end{definition}

\begin{remark}
    The existence of the choice function $\ordering(\textup{ })$ can be guaranteed either by invoking the axiom of global choice or by restricting our attention to metafinite structures whose primary part has a universe that is inherently ordered, as is the case with initial segments of $\N$. The only difference between ordered and unordered structures is that ordered structures have the chosen order on their universe in their vocabulary. 
\end{remark}

% \jeremy{I should put this paragraph into the definition of encodings somehow, for easy reference and clarity}

Given our choice of total ordering $\leq$ on $\A$, a weight function $W \: A^k \to R$ of an $\rr$-structure $\D$ may be encoded as a string $\code(w) = (r_1,\dots,r_m) \in R^m$ where $m = |A^k|$ and $r_i$ is the value of $w$ on the $i$th tuple of $A^k$ under the lexicographic ordering on $A^k$ induced $\leq$. Using this encoding of weight functions of $\D$, we devise a method of encoding the entirety of $\D$ by pushing all of the constants, functions, and relations in the primary part of $\D$ into the set of weight functions of $\D$, thereby converting $\D$ into a metafinite algebra.

\begin{definition}
    A \textit{metafinite algebra} is a metafinite structure in which the primary part $\A$ has an empty vocabulary. In other words, $\A$ is just a set $A$ without any constants, functions, or relations. 
\end{definition}

Every $\rr$-structure $\D = (\A,\rr,\W)$ may be converted into a metafinite algebra in the following way. First, transform each constant and function of $\A$ into a relation by viewing a constant as a $0$-ary function and recalling that a function, as a set of ordered pairs, is simply a functional relation. Then encode each relation $P \subseteq A^k$ of $\A$ by its characteristic function $\chi_P \: A^k \to \{0,1\}$ before adding $\chi_P$ to $W$ and removing the constant, function, or relation corresponding to $\chi_P$ from the primary part of $\D$. The result is a metafinite algebra $\D' = (\A',\rr,\W')$ over $\rr$ that contains the same information as $\D$. 

% \till{I have my trouble with this type of encoding.
% I think you should have a blank symbol.
% Otherwise, you do not know where code(w1) ends and code(w2) ends. }

% \jeremy{I should explain here why we can understand $\code(\D)$ without separators. Essentially because we have a fixed vocabulary $\Voc$ we know the arity of all finitely many weight functions, and implicitly we have an ordering on these weight functions $W_1,\dots,W_N$...}

\begin{definition}
\label{def:EncodingMetafiniteStructures}
    The \textit{encoding} of an $\rr$-structure $\D=(\A,\rr,\W)$ is the string $\code(\D) \in R^*$ obtained in the following way. First convert $\D$ into a metafinite algebra $\D' = (\A',\rr,\W')$ with $\W' = \{W_1,\dots,W_N\}$. Then let $\code(\D) = (\code(W_1), \dots, \code(W_n)) \in R^*$. 
\end{definition}

We do not include the code for the equality relation in the code for a metafinite structure, even though we assume it is always a relation of the primary vocabulary. This means that every string $X \in R^*$ is of the form $\code(\D)$ for some $\rr$-structure $\D$ from the class described in Example \ref{example:StringsAreMetafiniteStructures}. Thus, every subset of $R^*$ is a set of metafinite structures in a fixed vocabulary, allowing us to recast decision problems in these terms without loss of generality.

\begin{definition}
\label{def:DecisionProblemofMetafiniteStructures}
    A \textit{decision problem of $\rr$-structures} in vocabulary $\Voc$ is a subset $S \subseteq \struct_\rr(\Voc)$. If $\mathcal{C}$ is a complexity class over $\rr$ then we say that $S$ belongs to $\mathcal{C}$ if $\{\code(\D) : \D \in S\}$ is in $\mathcal{C}$. 
\end{definition}

This equivalence between decision problems over $\rr$ and decision problems of $\rr$-structures allows us to rephrase the basic notions of complexity theory over $\rr$ in terms of $\rr$-structures. 
%\till{I feel that whenever you say \rr-structure then you really mean metafinite \rr structure. No? Maybe let's discuss when you have time.}
In particular, we may view the complexity class $\P(\rr)$ as the class of all decision problems of $\rr$-structures $S$ in any vocabulary such that membership in $S$ may be tested in polynomial time. 

When rephrasing the definition of $\NP(\rr)$ in terms of $\rr$-structures, we replace witness strings $Y \in R^n$ with witness weight functions $Y \: A^k \to R$, where $|A^k| = n$. Indeed, relative to a choice of ordering on $A$, we may view the function $Y$ as the string $Y$, and vice versa, under our encoding scheme. Defining the size $|Y|$ of the function $Y$ as the cardinality of its domain $A^k$, we see that $|Y|$ is the length of $Y$ when $Y$ is viewed as a string. 

%\jeremy{Need to update this definition to reflect the fact that we are using a pairing function now, rather than simple concatenation, as well as the fact that we are using witnesses of exactly length $q(n)$ rather than $\leq q(n)$.}

The expansion of an $\rr$-structure $\D$ with the weight function $Y$ is denoted by $(\D,Y)$. Note that $\code((\D,Y)) = \code(\D)\code(Y)$, so expanding $\D$ with $Y$ corresponds to concatenation. Given this translation, we say that a decision problem of $\rr$-structures $L$ in vocabulary $\Voc$ is in $\NP(\rr)$ if and only if there is a positive integer $q$ and a decision problem of $\rr$-structures $S$ in vocabulary $(\primary(\MVoc), \secondary(\MVoc), \weight(\MVoc) \cup \{Y_1,\dots,Y_m\})$ such that $S \in \P(\rr)$ and

$$L = \{\D \in \struct(\MVoc) : (\exists W \: A^q \to R) \ (\D,W) \in S\}.$$

We may characterize decision problems of $\rr$-structures in $\Sigma_k\rr$ in the same way, with an alternating string of quantified witness weight functions.

% \begin{remark}
%     We allow multiple weight functions $Y_1,\dots,Y_m$ to encode a single string witness $Y$ because the length of $Y$ may not be of the form $A^k$ for any $k$. This is also why we allow weight functions of arity zero: then the length of $\code(Y_1)\dots\code(Y_m)$ may be any natural number.
% \end{remark}

\subsection{Metafinite Logic}
\label{sub:MetafiniteLogic}

% \jeremy{Is this the appropriate place for such a long historical overview, or is it better that some of this goes in the section where we prove Fagin's theorem over $\rr$?}

Fagin's theorem shows us that existential second-order logic captures the complexity class $\NP$ over finite structures \cite{fagin1974generalized}\cite{libkin2004elements}. It does this by first viewing the class $\NP$ as a class of decision problems of finite structures, and then showing that a set $L$ of finite structures is in $\NP$ if and only if there is a sentence $\varphi_L$ of existential second-order logic such that $L = \{\A : \A \satisfies \varphi_L\}$. 

The essential idea in the proof of Fagin's theorem is that an accepting computation sequence of a Turing machine can be described in first-order logic with an appropriate relational vocabulary. Passing to existential second-order logic allows us to existentially quantify over these relations, thereby removing the requirement that they are in our vocabulary, as well as to express the existence of a witness $Y$ for the membership of a finite structure $\A$ in a given decision problem $L$. 

Gr{\''a}del and Meer established an analogous result in \cite{gradel1995descriptive} by showing that existential second-order logic over $\R$ captures $\NP(\R)$ over $\R$-structures. The essential idea in their proof is similar to the essential idea of Fagin's theorem: The existence of a witness and an accepting computation sequence of an $\R$-machine can be expressed in a sentence of existential second-order logic over $\R$. 

Just as $\R$-machines have two sorts of registers---index registers and $\R$-registers---the logic that Gr{\''a}del and Meer used is a many-sorted second-order logic with two sorts---a sort of point terms to describe the index registers, and a sort of weight terms to describe the $\R$-registers. Such a logic requires structures with a domain to interpret each sort of term, hence the invention of $\R$-structures wherein the primary part interprets the point terms and the secondary part interprets the weight terms. The requirement that the primary part of an $\R$-structure is finite reflects both the fact that input strings to $\R$-machines are finite, as well as the fact that terminating $\R$-machine computations are finite.

Metafinite logic was developed by Gr{\''a}del and Gurevich in \cite{graedel1998metafinite} as a generalization of the logic of Gr{\''a}del and Meer that allows us to replace the structure $\R$ with any other structure $\rr$. It is a restriction of many-sorted second-order logic that is tailored to express properties of metafinite structures, as we shall see. For a given metafinite structure $\D=(\A,\rr,\W)$, we have two domains, $A$ and $R$, and so we have two sorts of terms---a sort of point terms, which refer to elements of the primary part $A$, and a sort of weight terms, which refer to elements of the secondary part $R$. 

\begin{definition}[First-Order Terms]
\label{def:MetafiniteLogicFirstOrderTerms}
    Fix a set $\Var = \{x_1,x_2,x_3,\dots\}$ of first-order variables that range only over the primary part of a metafinite structure. The \textit{first-order terms} of a metafinite vocabulary $\Voc = (\primary(\MVoc),\secondary(\MVoc),\weight(\MVoc))$ are as follows:
    \begin{enumerate}
        \item The set of \textit{point terms} is the closure of $\Var$ and the constants of $\primary(\MVoc)$ under applications of functions from $\primary(\MVoc)$. In other words, point terms are all the terms in the primary vocabulary.
        \item Every constant of $\secondary(\MVoc)$ is a \textit{weight term}, but note that elements of $\Var$ are not weight terms. 
        \item If $p_1, \dots, p_k$ are point terms and $w$ is a $k$-ary function symbol of $\weight(\MVoc)$, then $w(p_1,\dots,p_k)$ is a weight term. In other words, applying a weight function to point terms results in a weight term.
        \item If $t_1, \dots t_k$ are weight terms and $f$ is a $k$-ary function symbol of $\secondary(\MVoc)$, then $f(t_1,\dots,t_k)$ is a weight term. In other words, weight terms are closed under the application of functions in the secondary vocabulary. 
    \end{enumerate} 
\end{definition}

Suppose $\D = (\A,\rr,\W)$ is a metafinite structure with vocabulary $\Voc(\D)$. For a given point term $p$, if the set of variables occurring in $p$ is a subset of $\{x_1,\dots,x_n\}$, then $p$ induces a function $p : A^n \to A$. Similarly, for a given weight term $t$, if the set of variables occurring in $t$ is a subset of $\{x_1,\dots,x_n\}$, then $t$ induces a function $t : A^n \to R$. For any tuple $(a_1,\dots,a_n) \in A^n$, $p(a_1,\dots,a_n)$ is computed by replacing each variable $x_i$ of $p$ with $a_i$ and recursively applying the functions of $\Voc(\A)$ that occur in $p$, while $t(a_1,\dots,a_n)$ is computed by recursively applying the functions of $\Voc(\A)$, $\Voc(\rr)$, and $W$ that occur in $t$. 

\begin{definition}[First-Order Formulas]
\label{def:MetafiniteLogicFirstOrderFormulas}
    \textit{First-order atomic formulas} in vocabulary $\MVoc$ are of the form $P(p_1,\dots,p_k)$ or $Q(t_1,\dots,t_k)$,  where $P$ and $Q$ are relation symbols of arity $k$ from $\primary(\MVoc)$ and $\secondary(\MVoc)$, respectively, and where $p_1,\dots,p_k$ are point terms and $t_1,\dots,t_k$ are weight terms. The set $\FO(\MVoc)$ of \textit{first-order formulas} in vocabulary $\MVoc$ is the closure of the set of first-order atomic formulas in vocabulary $\Voc$ under Boolean operations $\{\neg,\vee,\wedge\}$ and quantification $\{\forall, \exists\}$ over first-order variables. The set $\FO(\rr)$ of first-order formulas over $\rr$ is the union of all $\FO(\Voc)$ such that $\secondary(\MVoc) = \Voc(\rr).$ That is, $$\FO(\rr) = \bigcup \{\FO(\Voc) : \secondary(\MVoc) = \Voc(\rr)\}.$$
\end{definition}

Recall that a first-order variable ranges only over the universe $A$ of the primary part $\A$ of a metafinite structure $\D$. This accords with our usual understanding of first-order logic on the structure $\A$. Second-order logic on the structure $\A$ also allows quantification over $k$-ary relations on the universe of $\A$, for each $k \in \N$, allowing us to create formulas stating, for example, that there is a binary relation on $A$ that is a linear ordering on $A$. Since each $k$-ary relation $P$ on $A$ may be encoded by its characteristic function $\chi_R \: A^k \to \{0,1\}$ and added to the set of weight functions for a metafinite structure with primary part $\A$, we can see that allowing quantification over the set $W$ subsumes second-order logic over the structure $\A$.

By allowing quantification over weight functions, second-order metafinite logic extends first-order metafinite logic. In order to allow quantification over weight functions of arity $k$, we introduce second-order variables of arity $k$, for each nonnegative integer $k \in \N$.

\begin{definition}[Second-Order Terms]
\label{def:MetafiniteLogicSecondOrderTerms}
    For each nonnegative integer $k$, fix a set $\Var_k = \{X_1^k,X_2^k,X_3^k,\dots\}$ second-order variables of arity $k$ that range over the $k$-ary weight functions of a metafinite structure. The \textit{second-order terms} of a metafinite vocabulary $\MVoc$ include all first-order terms of $\Voc$. In addition, if $t_1,\dots,t_k$ are second-order weight terms and $X^k$ is a second-order variable of arity $k$, then $X^k(t_1,\dots,t_k)$ is a weight term.  
\end{definition}

For a given second-order weight term $t$ in the vocabulary of $\D$, if the set of second-order variables occurring in $t$ is a subset of $\{X_1^{k_1}, \dots, X_m^{k_m}\}$, then any tuple $(w_1, \dots, w_m)$ of weight functions of the appropriate arity induce a first-order term $t(w_1, \dots,w_m)$ in the vocabulary of $\D$ expanded with symbols for each of the weight functions, which in turn induces a function $t(W_1,\dots,W_N)\:A^n \to A$. We sometimes write $t(W_1,\dots,W_m,a_1,\dots,a_n)$ for $t(W_1,\dots,W_m)(a_1,\dots,a_n)$. 

\begin{definition}[Second-Order Formulas]
\label{def:MetafiniteLogicSecondOrderFormulas}
    \textit{Second-order atomic formulas} in vocabulary $\Voc$ are defined just as first-order atomic formulas are defined, except weight terms are now allowed to be second order. The set of \textit{second-order formulas} in vocabulary $\Voc$ is the closure of the set of second-order atomic formulas in vocabulary $\Voc$ under Boolean operations $\{\neg, \vee,\wedge\}$ and quantification $\{\forall,\exists\}$ over both first-order and second-order variables.  The set $\SO(\Voc)$ of \textit{second-order formulas} in vocabulary $\Voc$ is the closure of the set of second-order atomic formulas in vocabulary $\Voc$ under Boolean operations $\{\neg,\vee,\wedge\}$ and quantification $\{\forall, \exists\}$ over first-order variables. The set $\SO(\rr)$ of first-order formulas over $\rr$ is the union of all $\SO(\Voc)$ such that $\secondary(\MVoc) = \Voc(\rr).$ That is, $$\SO(\rr) = \bigcup \{\SO(\Voc) : \secondary(\MVoc) = \Voc(\rr)\}$$
\end{definition}

\begin{remark}
    Note that a second-order variable of arity $0$ ranges over weight functions $r \: A^0 \to R$, which are essentially the same as elements of $R$. This means that quantifying over second-order variables subsumes quantifying over the secondary part of metafinite structures. It is essential to Lemma \ref{lemma:FirstOrderMetafiniteEvaluationPolynomialTime} that first-order metafinite logic only allows quantification over the primary part of metafinite structures, which we achieve by only allowing quantification over first-order variables.
\end{remark}

A variable of $\varphi$ is \textit{free} if it is not bound by a quantifier. We write  $\varphi(X_1^{k_1},\dots,X_m^{k_m},x_1,\dots,x_n)$ to emphasize the the free variables of $\varphi$ are a subset of $\{X_1^{k_1},\dots,X_m^{k_m},x_1,\dots,x_n\}$, but we may still simply write $\varphi$ even if $\varphi$ has free variables. If $\varphi$ has no free variables, then $\varphi$ is a \textit{sentence}. 

\begin{definition}
\label{def:MetafiniteLogicSecondOrderSemantics}
    Let $\varphi$ be a second-order formula in the vocabulary of a metafinite structure $\D = (\A,\rr,\W)$ such that the free variables of $\varphi$ are a subset of $\{X_1^{k_1},\dots,X_m^{k_m},x_1,\dots,x_n\}$. Given weight functions $W_1,\dots,W_m \in W$ of the appropriate arity, and elements $a_1,\dots,a_n \in A$, the satisfaction relation $\D \satisfies \varphi(W_1,\dots,W_N,a_1,\dots,a_n)$ is defined recursively as follows. 
    \begin{enumerate}
        \item If $\varphi$ is the atomic formula $P(p_1,\dots,p_k)$, then $\D \satisfies \varphi(W_1,\dots,W_m,a_1,\dots,a_n)$ if and only if $$(p_1(a_1,\dots,a_n), \dots, p_k(a_1,\dots,a_n)) \in P^\A.$$
        \item If $\varphi$ is the atomic formula $Q(t_1,\dots,t_k)$ then $\D \satisfies \varphi(W_1,\dots,W_m,a_1,\dots,a_n)$ if and only if $$(t_1(W_1,\dots,W_m,a_1,\dots,a_n),\dots,t_k(W_1,\dots,W_m,a_1,\dots,a_n)) \in Q^\rr.$$
        \item If $\varphi$ is a Boolean combination of $\psi_1$ and $\psi_2$, then the satisfaction relation is defined in the usual way. For example, if $\varphi$ is $\neg \psi$, then $\D \satisfies \varphi(W_1,\dots,W_m,a_1,\dots,a_n)$ if and only if $$\D \not\satisfies \psi(W_1,\dots,W_m,a_1,\dots,a_n).$$
        \item If $\varphi$ results from quantifying a first-order variable in $\psi$, then the satisfaction relation is defined by quantifying of $A$. For example, if $\varphi$ is $\forall x_{n+1} \psi$, then $\D \satisfies \varphi(W_1,\dots,W_m,a_1,\dots,a_n)$ if and only if $$\D \satisfies \psi(W_1,\dots,W_m,a_1,\dots,a_n,a)\text{ for all } a \in A.$$
        \item If $\varphi$ results from quantifying a second-order variable of arity $k$ in $\psi$, then the satisfaction relation is defined by quantifying over weight functions of arity $k$ in $W$. For example, if $\varphi$ is $\forall X_{m+1}^{k} \psi$, then $\D \satisfies \varphi(W_1,\dots,W_m,a_1,\dots,a_n)$ if and only if $$\D \satisfies \psi(W_1,\dots,W_m,W_{m+1},a_1,\dots,a_n) \textup{ for all } W_{m+1} \: A^k \to R .$$ 
    \end{enumerate}
\end{definition}

\begin{definition}
\label{def:MetafiniteLogicallyEquivalentFormulas}
    Two formulas $\varphi$ and $\psi$ are \textit{logically equivalent} whenever they are satisfied by exactly the same structures. That is $\varphi$ is logically equivalent to $\psi$ whenever, for all structures $\A$, $$\D \satisfies \varphi \text{ if and only if }\D\satisfies \psi.$$ 
\end{definition}

\begin{definition}
\label{def:MetafiniteLogicSecondOrderPrenexNormalForm}
    A formula $\varphi$ is in \textit{prenex normal form} if it is of the form $$Q_1 X_1^{k_1} \dots Q_N X_N^{k_N} Q_1 x_1 \dots Q_n x_n \psi,$$ where $Q_i$ are quantifiers, $X_i^{k_i}$ and $x_j$ are second and first-order variables, respectively, and $\psi$ is quantifier free.
\end{definition}

The inductive proof that every second-order formula $\varphi$ is semantically equivalent to some second-order formula $\varphi'$ in prenex normal form is essentially equivalent to the proof of the analogous result for first-order logic, except that the induction uses two new conversion rules that can be found in \cite{libkin2004elements} on page 115.

\begin{theorem}
    For every formula $\varphi$ there is a formula $\varphi'$ in prenex normal form such that $\varphi$ is semantically equivalent to $\varphi'$.  \qed
\end{theorem}

\begin{definition}[Second-Order Fragments]
\label{def:MetafiniteLogicSecondOrderFragments}
    For each $k \in \N$, the \textit{$k$th existential fragment} of $\SO(\rr)$ is the set $\Sigma_k\SO(\rr)$ of all sentences in $\SO(\rr)$ of the form $$Q_{1,1} X_{1,1} \dots Q_{1,m_1} X_{1,m_1} \dots Q_{N,1} X_{N,1} \dots Q_{N,m_N} X_{N,m_N}Q_1x_1 \dots Q_n x_n \psi,$$ where the quantifier blocks $Q_i = (Q_{i,1},\dots,Q_{i,m_i})$ are all the same quantifier, starting with an existential block $Q_1$ and alternating until reaching the prenex normal form first-order formula  $Q_1x_1\dots Q_nx_n\psi.$ In particular $\Sigma_1\SO(\rr) = \Sigma_1\SO(\rr)$ is \textit{existential second-order logic over $\rr$}. The \textit{$k$th universal fragment} of $\SO(\rr)$ is the set $\Pi_k\SO(\rr)$ that is defined by the same condition, except the first quantifier block is universal. In particular $\Pi\SO(\rr)=\Pi_1\SO(\rr).$
\end{definition}

\begin{definition}[Second-Order Boolean Fragments]
\label{def:MetafiniteLogicSecondOrderBooleanFragments}
    For each $k \in \N$, the \textit{$k$th existential Boolean fragment} of $\SO(\rr)$ is the set $\exists_k\SO(\rr)$ of all sentences $\varphi$ in $\Sigma_k\SO(\rr)$ such that the only constants from $\rr$ occurring in $\varphi$ are 0 and 1. In particular $\exists\SO(\rr) = \exists_1\SO(\rr).$ The \textit{$k$th universal Boolean fragment} of $\SO(\rr)$ is the set $\forall_k\SO(\rr)$ of all sentences $\varphi$ whose only constants from $\rr$ are $0$ and $1$. In particular $\forall\SO(\rr) = \forall_1\SO(\rr).$
\end{definition}

% \jeremy{Need to elaborate on this.}
Note that we are justified in calling these fragments Boolean because they may be encoded as Boolean strings and tested with Turing machines.

\subsection{Describing Metafinite Structures} 

The \textit{model-checking problem} asks for a given sentence $\varphi$ in given logic $\mathcal{L}$ and a given structure $\A$, does $\A \satisfies \varphi$? It is via this problem that we will relate logic and computational complexity. 

\begin{definition}
    Let $\mathcal{K}$ be a complexity class and $\mathcal{L}$ be a logic. The \textit{data complexity} of $\mathcal{L}$ is $\mathcal{K}$ if for every sentence $\varphi$ of $\mathcal{L}$, $$\{\code(\D) : \D \satisfies \varphi\} \in \mathcal{K}.$$ 
\end{definition}

% \begin{definition}
%     A property $\mathcal{P}$...
% \end{definition}

\begin{definition}[Logics Capturing Complexity Classes]
\label{def:LogicCapturesComplexityClassOnStructures}
    Let $\mathcal{K}$ be a complexity class, $\mathcal{L}$ be a logic, and $\mathcal{C}$ be a class of metafinite structures. We say that \textit{$\mathcal{L}$ captures $\mathcal{K}$ on $\mathsf{C}$} if the following two things hold:
    \begin{enumerate}
        \item For all sentences $\varphi$ of $\mathcal{L},$ the decision problem $\{\D \in \mathsf{C} : \D \satisfies \varphi\}$ is in $\mathcal{K}.$
        \item For every property $\mathcal{P}$ of structures from $\mathsf{C}$ that can be tested with complexity $\mathcal{K}$, there is a sentence $\varphi_\mathcal{P}$ of $\mathcal{L}$ such that $\A \satisfies \varphi_\mathcal{P}$ if and only if $\A$ has property $\mathcal{P}$, for each $\A \in \mathsf{C}$.
    \end{enumerate}
\end{definition}

% \begin{definition}
%     Let $\mathcal{L}$ be a logic, $\mathcal{C}$ be a complexity class over $\rr$, and $\mathcal{K}$ a class of $\rr$-structures. We say that $\mathcal{L}$ \textit{captures} $\mathcal{C}$ on $\mathcal{K}$ if two things are true:
%     \begin{enumerate}
%         \item For every sentence $\varphi$ of $\mathcal{L}$, $\{\D \in \mathcal{K} : \D \satisfies \varphi\}$ belongs to $\mathcal{C}$.
%         \item For every $S \subseteq $
%     \end{enumerate}
% \end{definition}

%%%%%%%%%%%%%%%%%%%%%%%%%%%%%%%%%%%%%%%%%%%%%%%%
\section{Cook-Levin Theorem Over a Structure}
\label{sec:cook-levin}
%%%%%%%%%%%%%%%%%%%%%%%%%%%%%%%%%%%%%%%%%%%%%%%%

% \jeremy{Need to come back to say what we do in this section precisely.}
In this section, under weak assumptions on the structure $\rr$, we will show that $\SAT(\rr) = \Sigma_1\SAT(\rr)$ is $\NP(\rr)$-complete, and we will generalize this result to the higher levels of the polynomial hierarchy over $\rr$. Specifically, we will show that $\Sigma_k \SAT(\rr)$ is complete for $\Sigma_k \rr$, and $\Pi_k\SAT(\rr)$ is complete for $\Pi_k\rr$ under polynomial-time $\rr$-machine reductions, for each $k \in \N$ with $k \geq 1$. Moreover, we will show that $\exists_k\SAT(\rr)$ is complete for $\exists_k\rr$ and $\forall_k\SAT(\rr)$ is complete for $\forall_k\rr$ under polynomial-time Turing machine reductions, for all $k \in \N$ with $k \geq 1$. 
We assume throughout this section that $\rr$ is a bipointed structure.

%%%%%%%%%%%%%%%%%%%%%%%%%%%%%%%%%%%%%%%%%%%%%%%%%%%%
\subsection{$\SAT(\rr)$ is in $\NP(\rr)$}
\label{sub:SATR-NPR-Membership}
%%%%%%%%%%%%%%%%%%%%%%%%%%%%%%%%%%%%%%%%%%%%%%%%%%%%

In order to show that $\SAT(\rr)$ is in $\NP(\rr)$, it is sufficient to show that given a quantifier-free formula $\varphi(x_1,\dots,x_n)$ and a tuple $(r_1,\dots,r_n) \in R^n$, we can decide whether or not $\rr \satisfies \varphi(r_1,\dots,r_n)$ using an $\rr$-machine that runs in time polynomial in the size of $\varphi$. Given the recursive construction of $\varphi$, the recursive algorithm $\evaluate$ naturally suggests itself as a decision procedure.

\begin{definition}
\label{def:AlgorithmEvaluate}
    The algorithm $\evaluate$ takes two inputs: a quantifier-free formula $\varphi$ in $\Voc(\rr)$, and a tuple $\overline{r} = (r_1,\dots,r_n) \in R^n$ such that $n$ is greater than or equal to the index of any variable occurring in $\varphi$. The algorithm then recursively decides if $\rr \satisfies \varphi(\overline{r})$ as follows
    \begin{enumerate}
        \item if $\varphi$ is the atomic formula $t_1 = t_2$, then $\evaluate$ returns 1 if $t_1^\rr(\overline{r}) = t_2^\rr(\overline{r})$, and returns 0 otherwise,
        \item If $\varphi$ is the atomic formula $P(t_1,\dots,t_k)$, then $\evaluate$ returns 1 if $(t_1^\rr(\overline{r}),\dots,t_k^\rr(\overline{r})) \in P^\rr$, and returns 0 otherwise.
    \end{enumerate}
    This is the base case of the algorithm. There are three recursive cases corresponding to the Boolean connectives $\{\neg,\wedge,\vee\}$ that accord with the meaning of these connectives: 
    \begin{enumerate}
    \setcounter{enumi}{2}
        \item if $\varphi$ is $\neg \psi$, then $\evaluate$ returns 1 if $\evaluate(\psi,\overline{r}) = 0$, and returns 0 otherwise,
        \item if $\varphi$ is $\psi_1 \wedge \psi_2$, then $\evaluate$ returns 1 if $\evaluate(\psi_1,\overline{r}) = 1$ and $\evaluate(\psi_2,\overline{r})=1$, and returns 0 otherwise,
        \item if $\varphi$ is $\psi_1 \vee \psi_2$, then $\evaluate$ returns 1 if $\evaluate(\psi_1,\overline{r}) = 1$ or $\evaluate(\psi_2,\overline{r})=1$, and returns 0 otherwise.
    \end{enumerate} 
\end{definition}

We argue that whenever $\Voc(\rr)$ is of finite type, then there is an $\rr$-machine $\M$ implementing $\evaluate$ and that $\M$ runs in polynomial time. 
% \hans{I find this confusing. What if $\Voc(\rr)$ is of finite type is not of finite type? The section doesn't seem to specify this? And should we maybe better write: Without explicitly constructing ...
% When I read below, we see that in some cases the machine does not exist. I think the situation, what we precisely want to prove in this section, with what assumptions should be made very clear at the start of the section. In general, I think it is best if such a result is stated in the form of a lemma or theorem.}
In order for $\M$ to implement \evaluate, $\M$ needs to do three things:

% \jeremy{I need to emphasize here that it is not just parsing formulas that causes a problem: first-order logic can be encoded as a context-free grammar, which can be parsed in polynomial time by a Turing machine. What is impossible is encoding the formulas in the vocabulary of $\rr$ in such a way that an $\rr$-machine is able to parse these formulas and use the syntax tree of this formula to evaluate the truth of this formula on some given interpretation of its free variables. This is really because there are structures for which it is impossible to encode the functions and relations of a structure into strings in a way that an $\rr$-machine can take the code of this function/relation and some input and evaluate the result of this function/relation on this input.}
    \begin{enumerate}
        \item $\M$ needs to parse encodings of formulas in order to both ($i$) determine if a string is a valid encoding of some formula and ($ii$) convert the encoding of that formula into a syntax tree for that formula in order to determine which case of $\evaluate$ to apply to this formula. For example, if $\varphi$ is $\psi_1 \vee \psi_2$, then $\M$ needs to be able to use some (yet to be specified) definition of $\code(\varphi) \in R^*$ to determine that $\vee$ was the last connective applied in the construction of $\varphi$ so that it can apply the $\vee$ case of $\evaluate$ to $\varphi$.
        \item $\M$ needs to implement recursion in order to recursively apply \evaluate in the cases where $\varphi$ is not an atomic formula.
        \item $\M$ needs to determine whether or not $\rr$ satisfies atomic formulas $t_1 = t_2$ and $P(t_1,\dots,t_k)$  when the free variables of this formula have been given values according to $(r_1,\dots,r_n)$.
    \end{enumerate} 

Programming $\M$ to accomplish (1) and (2) may be thought of as compiling \evaluate into the machine language of $\rr$-programs. It is plausible that $\rr$-machines should be able to accomplish (3) because they are able to execute equality tests and tests on relations $P$ in $\Voc(\rr)$, however, there are infinite type vocabularies for which this is impossible. Examples include the uncountably infinite vocabulary $\Voc(\R_\textup{lin})$ for the structure $\R_\textup{lin} = (\R,0,1,-,+,(\scalar{r} : r \in \R))$ and the countably infinite vocabulary $\Voc(\N_\textup{sc})$ for the structure $\N_\textup{sc} = (\N,0,1,(\scalar{n} : n \in \text{Prime} \subseteq \N))$. 

This follows from results of Hemmerling in 
\cite{Hemmerling98a,hemmerling1996computability}, which show that if $\rr$ is one of these structures, then it is impossible to encode the symbols of $\Voc(\rr)$ into strings in $R^*$ in such a way that an $\rr$-machine can extract the function or relation of $\rr$ from its code. In other words, there is no \textit{effective encoding} of $\rr$. This implies that it is also impossible to encode formulas of $\Voc(\rr)$ into strings in $R^*$ in such a way that an $\rr$-machine can evaluate, creating an obstacle to accomplishing (3). 
% For example, if our formula is $\varphi \equiv f(x_1) = x_2$ and $\overline{r} = (r_1,r_2)$ then there is no definition of $\code(\varphi) \in R^*$ that allows an $\rr$-machine to 

%\jeremy{I don't think this is strictly a consequence of the lack of an effective encoding. This just says that there is no way for a machine to execute the $\evaluate$ algorithm. It could be that there is another algorithm verifying instances of $\SAT(\N_\textup{sc}).$ I should also include the note that $\R_\textup{lin}$ has a complete problem, although it is not $\SAT$ and we don't know if $\SAT$ is complete.}
As a consequence, it is unlikely that  $\SAT(\N_{\text{sc}})$ is in $\NP(\N_\text{sc})$ simply because there is no way to effectively encode $\SAT(\N_\text{sc})$ as strings in $\N^*$. We know even more in the case of $\N_\text{sc}$ because a result of Ga\ss ner implies that there is no complete problem for $\NP(\N_\textup{sc})$ \cite{gassner1997np}.

However, Hemmerling also establishes in \cite{Hemmerling98a} that whenever $\Voc(\rr)$ is of finite type, then $\rr$ is effectively encodable. This is true even when $\rr$ has uncountably many constants because the only operation associated to a constant is writing that constant, and thus a constant may stand as a code for itself. We use this result to devise an encoding of formulas of $\Voc(\rr)$ of finite type such that an $\rr$-machine can parse these encodings in polynomial time.

First we enumerate the function symbols $\Fun(\rr) = \{f_1, \dots, f_{|\Fun|}\}$ and relation symbols $\Rel(\rr) = \{P_1,\dots,P_{|\Rel|}\}$ of $\Voc(\rr)$. The grammar of first-order logic is context sensitive because function and relation symbols have a specified arity. Despite this, we may encode first-order logic in a context-free grammar by including the arity of function and relation symbols as an explicit part of their codes, as described in Exercise 5.2.8 of \cite{gallier2015logic}, where we replace the unary encoding of arity with a binary encoding.

% For example, this grammar encodes a $k$-ary function symbol $f_n$ as $fI^k\$I^n\$$, and the production rules of this grammar enforce that the encoding of $k$ terms must follow the encoding of $f_n$.

The result is a context-free grammar 
% (specifically an SLR(1) grammar)
that can be parsed in polynomial time by a Turing machine \cite{sipser1996introduction}. Since $\rr$-machines can simulate Turing machines in polynomial time by Theorem \ref{thm:RMachinesSimulateBooleanOneTapeTuringMachinesWithConstantOverhead}, this grammar can also be parsed by an $\rr$-machine in polynomial time. Furthermore, assuming that $\Voc(\rr)$ is of finite type implies that $\rr$ is effectively encodable, so an $\rr$-machine can extract the functions and relations occurring in $\varphi$ in constant time after it parses the grammatical structure of the encoding of $\varphi$ in polynomial time. Thus, whenever $\Voc(\rr)$ is of finite type, we can encode formulas in $\Voc(\rr)$ in such a way that they can be parsed in polynomial time by an $\rr$-machine. 

\begin{definition}
\label{def:EncodingFormulasOfVocabularyR}
    For any bipointed structure of finite type $\rr$, we fix an encoding of $\Voc(\rr)$-formulas $\varphi$, denoted by $\code(\varphi) \in R^*$, such that $\code(\varphi)$ can be parsed by an $\rr$-machine in time polynomial in the length of $\code(\varphi)$, which we denote as $|\code(\varphi)|$ or $|\varphi|$. We assume that $|\varphi|$ is polynomial in the maximum of the number of symbols occurring in $\varphi$ and the greatest index of any symbol occurring in $\varphi$. Furthermore, we assume that the code of each function and relation of $\Voc(\rr)$, as well as that of each variable, connective, and quantifier, is a Boolean string in $\{0,1\}^*$. 
\end{definition}

\begin{remark}
    Our encoding is such that $\code(\varphi) \in \{0,1\}^* \subseteq R^*$ if and only if the only constants occurring in $\varphi$ are 0 and 1.
\end{remark}

Using this encoding, we can program an $\rr$-machine $\M$ to (1) parse encodings of formulas in polynomial time. We may program $\M$ to (2) recursively apply $\evaluate$ by implementing recursion using standard techniques, such as looping back to an earlier step in the program of $\M$ via branching instructions. In order to (3) compute the base case of $\evaluate$ for an atomic formula $t_1 = t_2$ or $P(t_1,\dots,t_k)$ and a tuple $(r_1,\dots,r_n)$, $\M$ can first compute the values $t_i^\rr(r_1,\dots,r_n)$ by writing the constant symbols occurring in $t_i$ before recursively applying the functions occurring in $t_i$ to these constants and values from $(r_1,\dots,r_n)$. The machine $\M$ is able to do because each of these constants and functions is an operation of $\rr$ that $\M$ can execute via $\rr$-instructions. Similarly, $\M$ is able to determine if $t_1^\rr(\overline{r})=t_2^\rr(\overline{r})$ or if $(t^\rr_1(\overline{r}),\dots,t^\rr_k(\overline{r})) \in P^\rr$ because $\M$ can test equality and any relation of $\rr$. All that remains to be seen is that $\M$ runs in polynomial time. 

\begin{lemma}
\label{lemma:evaluate runs in polynomial time}
    If $\rr$ is a bipointed structure of finite type, then there is an $\rr$-machine $\M$ that implements the algorithm $\evaluate$ and runs in time polynomial in $|\varphi|$ on any input $(\varphi,(r_1,\dots,r_n))$, where $\varphi$ is a quantifier-free $\Voc(\rr)$-formula, and $(r_1,\dots,r_n) \in R^n$.
\end{lemma}
\begin{proof}
    Suppose $\M$ receives input $( \varphi,(r_1,\dots,r_n))$. As $\M$ parses $\varphi$ in polynomial time, $\M$ can verify that the largest index of a variable occurring in $\varphi$ is at most $n$, returning 0 if this is not the case. If $\varphi$ is the atomic formula $t_1=t_2$, it suffices to show that $\M$ can compute the values $t_i^\rr(r_1,\dots,r_n) \in R$ in time polynomial in $|t_1=t_2|$ because, given these values, $\M$ can test if they are equal in constant time. A similar argument applies in the case that $\varphi$ is the atomic formula $P(t_1,\dots,t_k)$, so we proceed to show that $\M$ can compute $t_i^\rr(r_1,\dots,r_n)$ in time polynomial in $|\varphi|$, for any term $t_i$ occurring in any formula $\varphi$.
    
    %If $\varphi$ is the atomic formula $P(t_1,\dots,t_k)$, it suffices to show that $\M$ can compute the values $t_i(r_1,\dots,r_n) \in R$ in time polynomial in $|P(t_1,\dots,t_k)|$ because, given these values, $\M$ can test if $(t_1(r_1,\dots,r_n),\dots,t_k(r_1,\dots,r_n)) \in P$ in constant time. 

    Each constant occurring in a term $t_i$ is included in $\code(t_i)$, which is part of the input $(\code(\varphi),(r_1,\dots,r_n))$. As they are part of the input, $\M$ does not need to write these constants. Similarly, the codes of the functions occurring in $t_i$ are part of the input, and these codes can be extracted from the input in polynomial time because we have encoded $\varphi$ in a context-free grammar, which may be parsed in polynomial time by a Turing machine that $\M$ can simulate. 
    
    By our assumption that $\rr$ is of finite type, there is a constant-time effective encoding of $\rr$, so the codes of the functions occurring in $t_i$ may be used to apply these functions to the constants occurring in $t_i$ as well as the values from $(r_1,\dots,r_n)$. Thus, $\M$ can compute $t_i^\rr(r_1,\dots,r_n)$ in time polynomial in $|t_i| \leq |\varphi|$, for any formula $\varphi$. If $\varphi$ is an atomic formula, then the above implies that $\M$ can decide in polynomial time if $\rr \satisfies \varphi(r_1,\dots,r_n)$. 
    
    %Writing a constant and evaluating a function can be executed in one time step by $\M$, so $\M$ can compute the value $t_i(r_1,\dots,r_n)$ in time polynomial in the number of occurrences of constants and functions in $t_i$, which is at most $|t_i|$. This implies that $\M$ can compute $t_i(r_1,\dots,r_n)$ in time polynomial in $|t_i|<|P(t_1,\dots,t_k)|$. 
    
    If $\varphi$ is not atomic, then $\M$ will recursively call itself on inputs of the form $(\psi,(r_1,\dots,r_n))$, where $\psi$ is a subformula of $\varphi$. To see that $\M$ runs in polynomial time when $\varphi$ is not atomic, it suffices to show that the number of recursive calls that $\M$ makes to itself is polynomial in $|\varphi|$. This is the case because each recursive call that $\M$ makes corresponds to a Boolean connective from $\{\neg,\wedge,\vee\}$ occurring in $\varphi$.
\end{proof}

\begin{theorem}
\label{thm:SAT_R-NP_R-Membership}
    If $\rr$ is of finite type, then $\SAT(\rr)$ is in $\NP(\rr)$.
\end{theorem}

\begin{proof}
    Let $S$ be the set of pairs $(\exists \overline{x} \varphi(\overline{x}),\overline{r})$ such that $\rr \satisfies \varphi(\overline{r})$, where $\varphi(\overline{x})$ is a quantifier-free $\Voc(\rr)$-formula, and $\overline{r} \in R^n$ and $n = |\exists \overline{x} \varphi(\overline{x})|$. By Lemma \ref{lemma:evaluate runs in polynomial time}, $S \in \P(\rr)$. Furthermore, the length of any reasonable encoding of $(r_1, \dots,r_n)$ will be at most linear in $|\exists x_1 \dots \exists x_n \varphi(x_1, \dots,x_n)|$. Given the fact that $\exists x_1 \dots \exists x_n \varphi(x_1,\dots,x_n) \in \SAT(\rr)$ if and only if there is some $(r_1, \dots,r_n) \in R^n$ such that $(\exists x_1 \dots \exists x_n \varphi(x_1,\dots,x_n),(r_1,\dots,r_n)) \in S$, we can see that $\SAT(\rr) \in \NP(\rr)$.
\end{proof}

% \hans{I think we should tell at the start of the section that we are going to prove this. Either in words, or as a restatable. In particular, the finite type should be mentioned in the beginning of the section.}

%%%%%%%%%%%%%%%%%%%%%%%%%%%%%%%%%%%%%%%%%%%%%%%%
\subsection{$\SAT(\rr)$ is $\NP(\rr)$-hard}
\label{sub:SATRIsNPR-hard}
%%%%%%%%%%%%%%%%%%%%%%%%%%%%%%%%%%%%%%%%%%%%%%%%

In what follows, we prove in Theorem \ref{theorem: SAT_R is NP_R hard} that whenever $\rr$ is a bipointed structure of finite type that has all constants,
%\hans{maybe make ''all'' more explicit?} 
then $\SAT(\rr)$ is $\NP(\rr)$-hard. For such structures, it is also the case that $\SAT(\rr)$ is in $\NP(\rr)$, as established in Theorem \ref{thm:SAT_R-NP_R-Membership}. Together, these results imply that for such structures, $\SAT(\rr)$ is $\NP(\rr)$-complete, as expressed in Theorem \ref{thm:CookLevin}.

The result of \Cref{thm:CookLevin} is similar to results in \cite{blum1989theory} and \cite{megiddo1990general} but it is more general because it applies to structures other than rings. A similar result is established in \cite{goode1994accessible} for the same class of structures that Theorem \ref{thm:CookLevin} applies to, but the proof therein uses polynomial-time constructible circuit families. The proof we give here has the advantage that it uses machines in a way that is quite close to the proof of the Boolean Cook-Levin theorem, as in \cite{megiddo1990general}, and it applies to all (bipointed) structures of finite type that have all constants, as in \cite{goode1994accessible}.

\begin{theorem}[$\SAT(\rr)$ is $\NP(\rr)$-hard \cite{goode1994accessible}]
\label{theorem: SAT_R is NP_R hard}
    If $\rr$ is bipointed, of finite type, and has all constants, then $\textup{SAT}(\rr)$ is $\textup{NP}(\rr)$-hard. 
\end{theorem}
\begin{proof}
    Fix some $L \in \text{NP}(\rr)$. By definition, there is some polynomial $q$ and some polynomial-time $\rr$-machine $\M$ such that $$ L = \{v  \in R^* : (\exists w \in R^{q(|v|)}) \ \M(v,w)=1 \}.$$ Let $n = |v|$ be the length of $v$, and let $m$ be an integer such that $\M$ halts in fewer than $n^m$ steps on every input $(v,w)$ such that $|w| = q(|v|)$. Such an $m$ exists because there is a polynomial $p \: \N \to \N$ such that $\M$ runs in time $p$, so on inputs of the form $(v,w)$, $\M$ will run in time $p(2(|v|+|w|)) = p(2(n+q(n)))$, which is again polynomial in $n$. 
    
    On input $(v,w)$, the zeroth index register is initialized with value $|(v,w)|$, while all other index register are initialized with 0. Because $\M$ can only increment the contents of index register by one with every time step, the largest possible value of any index register during the computation of $\M$ on input $(v,w)$ is $|(v,w)|+n^m$. Without loss of generality, we assume that $m$ is large enough so that $n^m$ is greater than the largest value of the index registers during a computation of $\M$ on any input $(v,w)$ of the above form.
    
    We build a quantifier-free formula $\varphi_v$ in the vocabulary of $\rr$ such that $\varphi_v$ is satisfiable in $\rr$ if and only if $v \in L$. The formula $\varphi_v$ will encode the sequence of configurations of $\M$ induced by any input of the form $(v,w)$. Recall that a configuration of $\M$ is a tuple $(\ell,\overline{n},\overline{r}) \in \{0,\dots,\ell_\program\} \times \N^{k_\program} \times R^\N$. Since $\M$ uses at most $n^m$ $\rr$-registers, a configuration of $\M$ is uniquely determined by a tuple $$(\ell,\overline{n},\overline{r}) \in \{0,\dots,\ell_\program\} \times \{0,\dots,n^m-1\}^{k_\program} \times R^{n^m}.$$ To encode the configuration $(\ell,\overline{n},\overline{r})$ in a subformula of $\varphi_v$, we will use atomic subformulas of the form $S_{p,t} = r$, $I_{j,p,t} = r$, or $Z_{i,t} = r$ where $S_{p,t},$ $I_{j,p,t}$, and $Z_{i,t}$ are variables, and $r$ is an element of $R$.
    
    The structure $\rr$ may not contain a copy of the integers, but because we always assume that $\rr$ is bipointed we know $\rr$ has two constants $0$ and $1$, so we may encode the possible contents $\{0,\dots,n^m-1\}$ of the index registers as binary strings $b_1 \dots b_{\lfloor \log(n^m) \rfloor +1}$ of length $\lfloor \log(n^m) \rfloor +1$. We use $\lfloor \log(n^m)\rfloor $ variables $\{I_{j,1,t}, \dots , I_{j,\lfloor \log(n^m) \rfloor +1 ,t}\}$ to encode the value of $j$th index register at time $t$. If the value of the $j$th index register is $n_j$ at time $t$, and if the binary representation of $n_j$ is $b_{1} \dots b_{\lfloor \log(n^m) \rfloor +1}$, then we encode this as $$\Index_j(n_j,t) \equiv \bigwedge_{p=1}^{\lfloor \log(n^m) \rfloor +1} I_{j,p,t} = b_p,$$ which is a formula in the vocabulary of $\rr$ because $\rr$ has $0$ and $1$ as constants. Similarly, if $\M$ executes instruction $\ell$ at time $t$, and if the binary representation of $\ell$ is $b_1 \dots b_{\lfloor \log(n^m) \rfloor +1}$, then we encode this as $$\Instruction(\ell,t) \equiv \bigwedge_{p=1}^{\lfloor \log(n^m) \rfloor +1} S_{p,t} = b_p.$$ In contrast to the case for the contents of the index registers and the instruction, we may encode the fact that the $i$th $\rr$-register of $\M$ contains $r$ at time $t$ using a single variable, as $$Z_{i,t} = r.$$ In this way, variables encode the following information:
    \begin{itemize}
        \item $\{S_{p,t} : 1 \leq p \leq \lfloor \log(n^m) \rfloor +1\}$ encode the instruction $\ell$ at time $t$,
        \item $\{I_{j,p,t} : 0 \leq j \leq k_\program-1 \text{ and } 1 \leq p \leq \lfloor \log(n^m) \rfloor +1\}$ encode the contents of the $k_\program$  index registers at time $t$,
        \item $\{Z_{i,t} : 0 \leq i \leq n^m-1 \}$ encode the contents of the $n^m$ $\rr$-registers of $\M$ at time $t$.
    \end{itemize}

    The formula $\varphi_v$ will be a conjunction of three subformulas, $\varphi_v \equiv \start_v \land \update \land \accept$. The subformula $\start_v$ is itself the conjunction of subformulas that partially encode the starting configuration of $\M$ on any input $(v,w)$. Recall that we encode pairs $(v,w)$ as strings of the form $(0,v_1,0,\dots,0,v_n,1,w_1,0,\dots,0,w_{q(n)})$, so the initial configuration of $\M$ on input $(v,w)$ is the tuple $$\conf_0((v,w)) = (0,(2n+2q(n),0,\dots,0),(0,v_1,0,\dots,0,v_n,1,w_1,0,\dots,0,w_{q(n)},0,\dots,0)).$$ 
    
    The first entry $0$ of $\conf_0((v,w))$ expresses that $\M$ executes instruction $0 : \mathsf{instruction}_0$ at time $t=0$. We express this with the subformula $\Instruction(0,0)$ defined above. The second entry $(2n+2q(n),0,\dots,0)$ of $\conf_0((v,w))$ expresses that the zeroth index register is initialized to hold $|(v,w)|$ at time $t=0$, while all other index registers are initialized to $0$. We express this with the subformula $$\Index_0(2n+2q(n),0) \wedge
         \left( \bigwedge_{j = 1}^{k_\program-1} \Index_j(0,0) \right).$$

    We do not want to encode a particular witness $w$ into $\varphi_v$ because we want $\varphi_v$ to encode a computation of $\M$ on \textit{any} potential witness. This means that we do not know specific values for $w_1, \dots, w_{q(n)}$ in the third entry of $\conf_0((v,w))$, but we do know that $(0,v_1,0,\dots,0,v_n,1)$ is an initial segment of this entry, and that every remaining element in this string with an even index is $0$. We express this with the subformula $$\Input_v \equiv \left( \bigwedge_{i = 1}^n Z_{2i-1,0} = v_i \right) \wedge \left( \bigwedge_{i=0}^{n-1} Z_{2i,0} = 0 \right) \wedge Z_{2n,0} = 1 \wedge \left( \bigwedge_{i=n+1}^{\lfloor (n^m-1)/2 \rfloor} Z_{2i,0 } = 0 \right).$$ 
    
    Note that $\Input_v$ does not constrain each of the variables in $\{Z_{i,0} : 0 \leq i \leq n^m-1\}$. Furthermore, all variables remain free in $\Input_v,$ and they will remain free in $\varphi_v$. Also note that $\Input_v$ is a formula in the vocabulary of $\rr$ because we assumed that $\rr$ has all constants, which allows us to use $v_1,\dots,v_n \in R$ as constants. We combine the above subformulas into one subformula $\start_v$ partially encoding $\conf_0((v,w)),$ while leaving $w$ undetermined, as 

    $$\start_v \equiv \Instruction(0,0) \wedge \Index_0(2n+2q(n),0) \wedge
         \left( \bigwedge_{j = 1}^{k_\program-1} \Index_j(0,0) \right) \wedge 
         \Input_v.$$

    We encode the way in which $\M$ modifies its registers based on instruction $\ell$ at time $t$ with a family of formulas $\update(\ell,t)$, where $\ell$ and $t$ are parameters that determine the specific formula $\update(\ell,t)$. This means that $\ell$ and $t$ are not variables of $\update(\ell,t)$. However $\update(\ell,t)$ will always be a statement about the variables whose time index is either $t$ or $t+1$. Each instance of $\update(\ell,t)$ includes subformulas stating that if a register is unmodified by $\M$ at time $t$, then at time $t+1$ it contains the same value that it did at time $t$. In what follows, $i,i',$ and $t$ range over $\{0,\dots,n^m-1\}$, $j,j'$ range over $\{0,\dots,k_\program-1\}$, and $p$ ranges over $\{1, \dots, \lfloor \log(n^m) \rfloor +1\}$. The definition of $\update(\ell,t)$ is based on the possible instructions in $\program$ and their meanings as detailed in \Cref{tab:RInstructions} and \Cref{def:TransitionRelationOfRMachines}.

\begin{enumerate}

        \item If the $\ell$th instruction of $\program$ is $\ell : Z \lbr i \rbr := c$, then 
        $$
        \begin{array}{rcl}
            \update(\ell,t) & \equiv & \Instruction(\ell+1,t+1)  \\
            & & \\
            & \wedge & Z_{i,t+1} = c  \\
            & & \\
            & \wedge & \bigwedge_{i' \in \{0,\dots,n^m-1\} \setminus \{i\}} \left( Z_{i',t+1} = Z_{i',t} \right)  \\
            & & \\
            & \wedge & \bigwedge_{j=0}^{k_\program-1} \bigwedge_{p=1}^{\lfloor  \log(n^m) \rfloor} \left( I_{j,p,t+1} = I_{j,p,t}  \right)
        \end{array}$$

        \item If the $\ell$th instruction of $\program$ is $\ell: Z \lbr i_0 \rbr := f(Z \lbr i_1 \rbr, \dots, Z \lbr i_k \rbr)$, then 
        $$
        \begin{array}{rcl}
             \update(\ell,t) & \equiv  & \Instruction(\ell+1,t+1)  \\
             &  & \\ 
             & \wedge & Z_{i_0,t+1} = f(Z_{{i_1},t}, \dots, Z_{i_k,t})  \\
             &  & \\
             & \wedge & \bigwedge_{i' \in \{0,\dots,n^m-1\} \setminus \{i_0\}} Z_{i',t+1} = Z_{i',t} \\
             & & \\
             & \wedge & \bigwedge_{j=0}^{k_\program-1} \bigwedge_{p=1}^{\lfloor \log(n^m) \rfloor +1} I_{j,p,t+1} = I_{j,p,t} 
        \end{array}$$

        \item If the $\ell$th instruction of $\program$ is $ \ell : \ifthenbranch{Z \lbr i_1 \rbr = Z \lbr i_2 \rbr}{\ell_1}{\ell_2},$ then 
        $$
        \begin{array}{rcl}
            \update(\ell,t) &  \equiv & \Big( Z_{i_1,t} = Z_{i_2,t} \to \Instruction(\ell_1,t+1) \Big)   \\
            & & \\
            & \wedge & \Big( \neg Z_{i_1,t} \neq Z_{i_2,t} \to \Instruction(\ell_2,t+1) \Big) \\
            & & \\
            & \wedge &  \Big(\bigwedge_{i \in \{0,\dots,n^m-1\}} Z_{i,t+1} = Z_{i,t} \Big)  \\
            & & \\
            &  \wedge &  \Big(\bigwedge_{j=0}^{k_\program-1} \bigwedge_{p=1}^{\lfloor \log(n^m) \rfloor +1} I_{j,p,t+1} = I_{j,p,t} \Big)
        \end{array} 
        $$

        \item If the $\ell$th instruction of $\program$ is $ \ell : \ifthenbranch{P(Z \lbr i_1 \rbr, \dots, Z \lbr i_k \rbr)}{\ell_1}{\ell_2},$ then 
        $$
        \begin{array}{rcl}
            \update(\ell,t) &  \equiv & \Big( P(Z_{i_1,t},\dots,Z_{i_k,t}) \to \Instruction(\ell_1,t+1) \Big)   \\
            & & \\
            & \wedge & \Big( \neg P(Z_{i_1,t},\dots,Z_{i_k,t}) \to \Instruction(\ell_2,t+1) \Big) \\
            & & \\
            & \wedge &  \Big(\bigwedge_{i \in \{0,\dots,n^m-1\}} Z_{i,t+1} = Z_{i,t} \Big)  \\
            & & \\
            &  \wedge &  \Big(\bigwedge_{j=0}^{k_\program-1} \bigwedge_{p=1}^{\lfloor \log(n^m) \rfloor +1} I_{j,p,t+1} = I_{j,p,t} \Big)
        \end{array} 
        $$

        \item If the $\ell$th instruction of $\program$ is $\ell : Z \lbr I \lbr j \rbr \rbr := Z \lbr I \lbr j' \rbr \rbr$, then 
        $$\begin{array}{rcl}
            \update(\ell,t) & \equiv & \Instruction(\ell+1,t+1) \\ 
            & & \\
            & \wedge & \bigwedge_{(n_j,n_{j'}) \in \{0,\dots,n^m-1\}^2} \Big[ (\Index_j(n_j,t) \wedge \Index_{j'}(n_{j'},t)) \to  \\
            & & \\
             & & \Big[\Big( Z_{n_j,t+1} = Z_{n_{j'},t} \Big)  \wedge \\
             & & \\
             & & \left( \bigwedge_{i' \in \{0,\dots,n^m-1\} \setminus \{n_j\}} Z_{i,t+1} = Z_{i,t} \right)  \wedge \\
             & & \\
             &  & \left( \bigwedge_{j=0}^{k_\program} \bigwedge_{p=1}^{\lfloor \log(n^m) \rfloor +1} I_{j,p,t+1} = I_{j,p,t} \right) \Big]\Big]
        \end{array}$$

        \item If the $\ell$th instruction of $\program$ is $\ell : \ifthenbranch{I \lbr j \rbr = I \lbr j' \rbr}{\ell_1}{\ell_2}$, then 
        $$
        \begin{array}{rcl}
            \update(\ell,t) & \equiv & \bigwedge_{(n_j,n_{j'}) \in \{0,\dots,n^m-1\}^2} \Big[ (\Index_j(n_j,t) \land \Index_{j'}(n_{j'},t)) \to \\
            & & \\
            & & \Big[ \Big( \left( \bigwedge_{p=1}^{\lfloor \log(n^m) \rfloor +1} I_{j,p,t} = I_{j',p,t} \right) \to \Instruction(\ell_1,t+1) \Big) \wedge \\
            & & \\
            & & \Big( \neg \left( \bigwedge_{p=1}^{\lfloor \log(n^m) \rfloor +1} I_{j,p,t} = I_{j',p,t} \right) \to \Instruction(\ell_2,t+1) \Big) \Big] \Big] \\
            & & \\
            & \wedge &  \Big(\bigwedge_{i \in \{0,\dots,n^m-1\}} Z_{i,t+1} = Z_{i,t} \Big)  \\
            & & \\
            & \wedge &  \Big(\bigwedge_{j=0}^{k_\program-1} \bigwedge_{p=1}^{\lfloor \log(n^m) \rfloor +1} I_{j,p,t+1} = I_{j,p,t} \Big)
        \end{array}$$

        \item If the $\ell$th instruction of $\program$ is $\ell : I \lbr j \rbr := I \lbr j \rbr + 1$, then 
        $$
        \begin{array}{rcl}
            \update(\ell,t) & \equiv &  \Instruction(\ell+1,t+1) \\
            & & \\ 
            & \wedge & \bigwedge_{n_j \in \{0,\dots,n^m-2\}} \Big( \Index_j(n_j,t) \to \Index_j(n_j+1,t+1) \Big)\\
            & & \\
            & \wedge &  \Big(\bigwedge_{i \in \{0,\dots,n^m-1\}} Z_{i,t+1} = Z_{i,t} \Big)  \\
           & & \\
           & \wedge &  \Big( \bigwedge_{j' \in \{0 \dots k_\program-1\}\setminus \{j\}} \bigwedge_{p=1}^{\lfloor \log(n^m) \rfloor +1} I_{j',p,t+1} = I_{j',p,t} \Big)
        \end{array}$$

        \item If the $\ell$th instruction of $\program$ is $\ell : I \lbr j \rbr := I \lbr j \rbr \monus 1$, then 
        $$
        \begin{array}{rcl}
            \update(\ell,t) & \equiv &  \Instruction(\ell+1,t+1) \\
            & & \\ 
            & \wedge & \bigwedge_{n_j \in \{0,\dots,n^m-1\}} \Big( \Index_j(n_j,t) \to \Index_j(n_j \monus 1,t+1) \Big)\\
            & & \\
            & \wedge &  \Big(\bigwedge_{i \in \{0,\dots,n^m-1\}} Z_{i,t+1} = Z_{i,t} \Big)  \\
           & & \\
           & \wedge &  \Big( \bigwedge_{j' \in \{0 \dots k_\program-1\}\setminus \{j\}} \bigwedge_{p=1}^{\lfloor \log(n^m) \rfloor +1} I_{j',p,t+1} = I_{j',p,t} \Big)
        \end{array}$$

        \item If the $\ell$th instruction of $\program$ is $\ell_\program : \mathsf{stop}$ and $t \leq n^m-1$, then 
        $$
        \begin{array}{rcl}
            \update(\ell,t) & \equiv & \Instruction(\ell_\program,t+1)  \\
             & & \\
             & \wedge &  \Big(\bigwedge_{i} Z_{i,t+1} = Z_{i,t} \Big) \\
             & & \\
             & \wedge & \Big( \bigwedge_{j=0}^{k_\program-1}\bigwedge_{p=1}^{\lfloor \log(n^m) \rfloor +1} I_{j,p,t+1} = I_{j,p,t} \Big)
        \end{array}$$
    \end{enumerate}

Letting $\update \equiv \bigwedge_{\ell = 0}^{\ell_\program} \bigwedge_{t=0}^{n^m-1} \update(\ell,t)$ and $\accept \equiv \Index_0(1,n^m) \wedge Z_{0,n^m} = 1,$ we define the formula $\varphi_v$ as $$\varphi_v(\overline{x}) \equiv \start_v(\overline{x}) \wedge \update(\overline{x}) \wedge \accept(\overline{x}),$$ where $\overline{x}$ denotes the tuple of all variables in $\varphi_v$.  We now show that $v \in L$ if and only if $\rr \satisfies \exists\overline{x}(\varphi_v(\overline{x}))$. 

Suppose $v \in L$. Then there is some $w \in R^{q(n)}$ such that $\M(v,w) = 1$. This means that the computation $\conf_0((v,w)) \to \conf_1 \to \dots \to \conf_N$ of $\M$ on input $(v,w)$ ends with an accepting configuration, which is a configuration of the form $\conf_N = (\ell_\program,(1,n_1,\dots,n_{k_\program}),(1,r_1,r_2,\dots,r_{n^m-1}))$ with $N <n^m$. We use this sequence of configurations to define values in $R$ for each variable of $\varphi_v$. For each $0 \leq t \leq N$, we have a configuration $\conf_t$ given by the computation. For each $ N < t \leq n^m-1$, let $\conf_t = \conf_N$. Thus, we have one configuration $\conf_t$ for each $t$ with $0 \leq t \leq n^m-1$. 

Suppose $\conf_t = (\ell, (n_0,\dots,n_{k_\program}),(r_0,\dots,r_{n^m-1})).$We use $\conf_t$ to assign values to all variables of $\varphi_v$ indexed by $t$, including $S_{p,t}$, $I_{j,p,t}$ and $Z_{i,t}$. The binary representation of $\ell$ determines values for the variables $S_{p,t}$, the binary representation of $n_j$ determines values for $I_{j,p,t}$, and $r_i$ determines a value for $Z_{i,t}$. Let $\overline{r}$ be a tuple of elements in $R$ that matches the variable tuple $\overline{x}$ in such a way that $x_i$ is assigned value $r_i$ according to the above scheme. Since $\varphi_v$ says precisely that this tuple $\overline{r}$ of values encodes an accepting computation, we can see that $\rr \satisfies \varphi_v(\overline{r})$, which implies that $\rr \satisfies \exists \overline{x}(\varphi_v(\overline{x})).$

Conversely, suppose $\rr \satisfies \exists \overline{x}(\varphi_v(\overline{x})).$ Then there is some tuple $\overline{r}$ of values in $R$ such that $\rr \satisfies \varphi_v(\overline{r}).$ In particular, $\rr \satisfies \start_v(\overline{r}) \wedge \update(\overline{r})$, so the values in $\rr$ encode the computation of $\M$ on the input $(v,w)$, where $w_i$ is determined by the vales assigned to the variable $Z_{2n+2i-1,0}$ for $i \in \{1,\dots,q(n)\}$. These variables determine $w$ because of the form of the initial configuration $\conf_0((v,w))$ of $\M$ on any input, as described above. Since it is also the case that $\rr \satisfies \accept(\overline{r})$, it must be that $\M(v,w)=1$, which implies that $v \in L.$

Finally, we verify that we can build $\exists\overline{x}(\varphi_v(\overline{x}))$ from $v$, $q$, and $\M$ using an $\rr$-machine that runs in polynomial time in the length of $v$. It is sufficient to note that the number of atomic formulas in $\varphi_v$ is polynomial in the length of $v$. Encoding $\varphi_v$ as an element of $R^*$, each such atomic formula can be written down by an $\rr$-machine in polynomial time, which means that the string in $R^*$ that encodes $\exists\overline{x}(\varphi_v(\overline{x}))$ may be written down in polynomial time. Thus, $L$ reduces to $\SAT(\rr)$ in polynomial time, from which we conclude that $\SAT(\rr)$ is $\NP(\rr)$-hard. 
\end{proof}

Taken together, \Cref{thm:SAT_R-NP_R-Membership} and \Cref{theorem: SAT_R is NP_R hard} imply the following.

\begin{restatable}[Cook-Levin Analogue~\cite{goode1994accessible}]{theorem}{SATRIsNPRcomplete}
\label{thm:CookLevin}
Let \rr be a bipointed structure of finite type with all constants. 
Then $\SAT(\rr)$ is $\NP(\rr)$-complete under polynomial-time $\rr$-machine reductions.
\end{restatable}

%\SATRIsNPRcomplete*

% \hans{Are we making sufficiently clear which results are new, and which results we give a new proof? And, if we give a new proof for an existing result, what is the value of doing that?}

%%%%%%%%%%%%%%%%%%%%%%%%%%%%%%%%%%%%%%%%%%%%%%%%
\subsection{$\exists\SAT(\rr)$ is $\exists \rr$-Complete}
\label{sub:BooleanSATIsBPNPRHard}
%%%%%%%%%%%%%%%%%%%%%%%%%%%%%%%%%%%%%%%%%%%%%%%%

% \begin{definition}
%     Let $\Sigma_k\textup{SAT}^0(\rr)$ be the set of sentences in $\Sigma_k\textup{SAT}(\rr)$ in which the only constants that occur are $0$ and $1$. In particular $\SAT^0(\rr) = \Sigma_1\SAT^0(\rr).$
% \end{definition}

% \begin{example}
%     If $\rr = (\R,0,1,+,-,\times,/,=,<)$, then $\textup{SAT}_\rr^0 = \textup{ETR}_\R^\mathbb{Q}$. 
% \end{example}

% \begin{example}
%     If $\rr = (\R,0,1,+,-,\times,=,<)$, then $\textup{SAT}_\rr^0 = \textup{ETR}_\R^\mathbb{Z}$
% \end{example}

% \begin{example}
%     If $\rr = (\R,0,1,+,-,\times,=)$, then $\textup{SAT}_\rr^0 = \textup{HN}_\R^\mathbb{Z}$
% \end{example}

% \begin{theorem}
% \label{theorem: SAT^0_R is BP(NP^0_R)-complete}
%     If $\rr$ is a structure that has a polynomial-time binary encoding with decidable codes, then $\textup{SAT}_\rr^0$ is $\textup{BP}(\textup{NP}_\rr^0)$-complete under $\mathsf{FP}$-reductions.
% \end{theorem}

\begin{restatable}[Boolean Cook-Levin Analogue]{corollary}{BooleanCookLevin}
\label{thm:BooleanCookLevin}
    Let $\rr$ be a bipointed structure of finite type. Then $\exists \SAT(\rr)$ is $\exists \rr$-complete under polynomial-time Turing machine reductions.
\end{restatable}

\begin{proof}
    Each sentence $\varphi \in \exists\SAT(\rr)$ contains no constants other than 0 or 1. According to our encoding scheme in \Cref{def:EncodingFormulasOfVocabularyR}, this implies that $\code(\varphi) \in \{0,1\}^*$ is a Boolean string, so we have $\exists\SAT(\rr) \subseteq \{0,1\}^*.$ By looking at the algorithm $\evaluate$, we can see that when inputs are restricted to formulas without constants other than 0 or 1, then $\evaluate$ may be implemented by an $\rr$-machine without constants other than 0 or 1. Thus, by an argument analogous to \Cref{thm:SAT_R-NP_R-Membership}, we can see that $\exists\SAT(\rr) \in \exists\rr$.

    To see that $\exists\SAT(\rr)$ is $\exists\rr$-hard, fix some decision problem $L \in \exists\rr$ and some arbitrary $v \in \{0,1\}^*$, and consider the reduction constructed in the proof of \Cref{theorem: SAT_R is NP_R hard} that $\SAT(\rr)$ is $\NP(\rr)$-hard. There are only two ways that elements of $R$ might be introduced into $\varphi_v$ as constants:
    \begin{enumerate}
        \item $v$ is hard-coded into $\varphi_v$ with the use of constants, and
        \item the machine constants of $\M$ are encoded into $\varphi_v$ using constants.
    \end{enumerate}
    Since we assume that $v \in \{0,1\}^*$ and that $\M$ has no machine constants other than 0 and 1, we see that the only constants occurring in $\varphi_v$ are $0$ and $1$. This fact, together with the argument in the proof of Theorem \ref{theorem: SAT_R is NP_R hard}, implies that $v \in L$ if and only if $\varphi_v \in \exists\SAT(\rr)$. 

    The reduction $v \mapsto \varphi_v$ is surprisingly simple in that an $\rr$-machine implementing this reduction does not need access to any of the functions or relations of $\rr$: it only needs to be able to read and write elements of $R$. When $L \in \exists\rr$, we can see from the above that even this ability is not required because both $v$ and $\code(\varphi_v)$ are binary strings, and reducing $v$ to $\varphi_v$ only depends on a binary encoding of $\M$, which exists because $\M$ has no constants other than 0 or 1. Thus, this reduction can be carried out by a Turing machine, and the same complexity analysis shows that it runs in polynomial time.

\end{proof}

\subsection{Complete Problems for $\Sigma_k\rr$ and $\exists_k\rr$}
\label{sub:HigherPolynomialCookLevin}
%%%%%%%%%%%%%%%%%%%%%%%%%%%%%%%%%%%%%%%%%%%
In this subsection we will show that $\Sigma_k \SAT(\rr)$ is $\Sigma_k\rr$-complete and $\Pi_k\SAT(\rr)$ is $\Pi_k \rr$-complete by induction on $k \in \N$. The proof we give essentially requires the use of constants from the universe of $\rr$, but the reduction we construct does not introduce constants not already in the language being reduced, so it is also the case that $\exists_k\SAT(\rr)$ is $\exists_k\rr$-complete and $\forall_k \SAT(\rr)$ is $\forall_k\rr$-complete under many-one polynomial-time reductions with a Turing machine. 

Although one can show that $\Pi_1\SAT(\rr)$ is $\coNP(\rr) = \Pi_1\rr$-complete as an easy corollary of Theorem \ref{thm:CookLevin}, it is instructive to give a direct proof that $\Pi_1\SAT(\rr)$ is $\coNP(\rr)$-hard because this reduction provides intuition for the reductions we will construct for higher levels of the hierarchy.

\begin{theorem}
    $\Pi_1\SAT(\rr)$ is $\coNP(\rr)$-hard.
\end{theorem}
\begin{proof}
    Suppose $L \in \coNP(\rr)$. Then there is some polynomial $q$ and some polynomial-time $\rr$-machine $\M$ such that for all $n \in \N$ and all $v \in R^n$, $$v \in L \text{ if and only if } (\forall w \in R^{q(n)})\ \M(v,w) = 1.$$ Let each of $\start_v$, $\update$, and $\accept$ be formulas in the vocabulary of $\rr$ that are constructed as in the proof of Theorem \ref{theorem: SAT_R is NP_R hard}, and let $\varphi_v$ be the formula $$\varphi_v \equiv (\start_v \wedge \update) \to \accept.$$ 
    
    Given the definition of $\Input_v,$ the variables in $\varphi_v$ that encode a potential witness string $w$ loaded into the $\rr$-registers at time zero are the variables $Z_{2n+1,0},Z_{2n+3,0},Z_{2n+5,0},\dots,Z_{2n+2q(n)-1,0}.$ We write $\forall Z$ to denote the quantifier block $$\forall Z_{2n+1,0} \forall Z_{2n+3,0} \forall Z_{2n+5,0} \dots \forall Z_{2n+2q(n)-1,0}.$$ With this definition in mind, we define the $\Pi_k$ sentence $\Phi_v$ in the vocabulary of $\rr$ as $$\Phi_v \equiv \forall Z \forall \overline{y}(\varphi_v(Z,\overline{y})),$$ where $\overline{y}$ is the tuple of all variables occurring in $\varphi_v$ except those occurring in $Z$. Since the number of atomic subformulas occurring in $\varphi_v$ is polynomial in $n = |v|$, there is an $\rr$-machine that constructs $\Phi_v$ from $v$, $q$ and $\M$ in polynomial time in $v$. Thus, it is sufficient to show that $v \in L$ if and only if $\rr \satisfies \Phi_v$.

    Suppose $v \in L.$ Then $(\forall w \in R^{q(n)}) \ \M(v,w) = 1.$ Let $w$ be an arbitrary string in $R^{q(n)}$ that assigns values to the variables occurring in $Z,$ and let $\overline{r}$ be any tuple that assigns some arbitrary element $r_i$ of $R$ to each variable $y_i$ occurring in $\overline{y}.$ If $\rr \satisfies (\start_v \wedge \update)(w,\overline{r}),$ then we know that $(w,\overline{r})$ encodes the computation of $\M$ on input $(v,w).$ Since $\M(v,w)=1$ by our assumption that $v \in L,$ we can see from the definition of $\accept$ that $\rr \satisfies \accept(w,\overline{r}).$ As $w$ and $\overline{r}$ were arbitrary, it must be that for all $w \in R^{q(n)}$ and all $\overline{r},$ $\rr \satisfies \varphi_v(w,\overline{r})$. Thus, $\rr \satisfies \Phi_v.$

    Conversely, suppose $\rr \satisfies \Phi_v.$ Then $\rr \satisfies \forall Z \forall \overline{y}(\varphi_v(Z,\overline{y})).$ Any string $w \in R^{q(n)}$ determines values for the variables occurring in  $Z$, while the computation $\conf_0((v,w)) \to \dots \to \conf_N$ of $\M$ on input $(v,w)$ determines values $\overline{r}$ for the remaining variables $\overline{y}$ of $\varphi_v$, as in the proof of Theorem \ref{theorem: SAT_R is NP_R hard}. Because $(w,\overline{r})$ encodes the computation of $\M$ on input $(v,w)$, we can see that $\rr \satisfies (\start_v \wedge \update)(w,\overline{r})$. Then $\rr \satisfies \accept(w,\overline{r})$ by our assumption that $\rr \satisfies \Phi_v$, which implies that $\M(v,w)=1$. Thus, $v \in L$.
\end{proof}

The main difference between the proof that $\Sigma_1\SAT(\rr)$ is $\NP(\rr)$-hard and the proof that $\Pi_1\SAT(\rr)$ is $\coNP(\rr)$-hard is that in the former proof, it sufficed to find a single tuple of elements in $R$ that encodes an accepting computation of $\M$, while in the latter proof we want to show something about all tuples of $R$ of a certain length. In general, such a tuple will not encode a computation of $\M$, let alone an accepting computation, so we introduce the implication $(\start_v \wedge \update) \to \accept$ to restrict our attention to those tuples that do encode a computation. 

This will be important when showing that $\Sigma_k\SAT(\rr)$ is $\Sigma_k\rr$-hard in the case that $k$ is even. To see this, suppose $L \in \Sigma_k\rr.$ Then there is some polynomial $q$ and some polynomial-time $\rr$-machine $\M$ such that for all $n \in \N$ and all $v \in R^n,$ $$v \in L \iff (\exists w_1 \in R^{q(n)})(\forall w_2 \in R^{q(n)}) \dots (\forall w_k \in R^{q(n)}) \M(v,w_1,\dots,w_k) = 1.$$ We want to construct a formula $\varphi_v$ such that $$v \in L \iff \rr \satisfies \exists Z_1 \forall Z_2 \dots \forall Z_k \forall \overline{y} \  \varphi_{v}(Z_1,\dots,Z_k,\overline{y}),$$ where the variables $Z_i$ encode a potential witness strings $w_i$, and the variables $\overline{y}$ encode the rest of the computation of $\M$ on input $(v,w_1,\dots,w_k).$ We need to quantify the variables $\overline{y}$ to be quantified universally, matching the universal quantification over $Z_k$, in order avoid another quantifier alternation, thereby ensuring the resulting sentence to be a $\Pi_k$ sentence. So we will use a formula of the form $(\start_v \wedge \update) \to \accept$ whenever the last quantifier $Q_k$ is universal. 

In addition, when constructing $\varphi_v,$ we need to modify $\start_v$ to build a formula $\start_{v,k}$ stating that $\M$ starts with an input of the form $(v,w_1,\dots,w_k)$ in its registers. Towards this end, let $m$ be such that $\M$ halts in at most $n^m$ steps on any input of the form $(v,w_1,\dots,w_k)$ with each $w_i \in R^{q(n)}$. Furthermore, assume that $n^m$ is greater than the largest value attained by any index register during a computation of $\M$ on any input of the above form, and assume that $n^m$ is greater than the length $\ell_\program+1$ of the program of $\M.$

Recall that our pairing function represents the tuple $(v,w_1,\dots,w_k)$ as the string $$(0,v_1,0,\dots,0,v_n,1,w_{1,1},0,\dots,0,w_{1,q(n)},1,\dots ,1,w_{k,1},0,\dots,0,w_{k,q(n)}),$$ where a 1 in an even-index position (starting from index $i=0$) demarcates the end of one entry and the beginning of another. Thus, upon receiving an input of the form $(v,w_1,\dots,w_k),$ $\M$ will initialize $\rr$-registers $Z \lbr 0 \rbr, \dots , Z \lbr n^m-1 \rbr$  with the values  $$(0,v_1,0,\dots,0,v_n,1,w_{1,1},0,\dots,0,w_{1,q(n)},1,\dots ,1,w_{k,1},0,\dots,0,w_{k,q(n)},0,\dots,0).$$ Let $\Input_{v,k}$ be the formula constructed as $\Input_v$ from the proof of Theorem \ref{theorem: SAT_R is NP_R hard}, except replace the clause $Z_{2n,0}=1$ with the clause $\bigwedge_{j=0}^{k-1}(Z_{2(n+jq(n)),0}=1)$ expressing the fact that at time zero we now have $k$ witness strings $w_1,\dots,w_k$ that need to be demarcated, rather than just one.

Similarly, on any input of the above form, $\M$ will initialize index registers $I \lbr 0 \rbr ,\dots, I \lbr k_\program -1 \rbr$ with values $$(2(n+kq(n)),0,\dots,0),$$ where the zeroth index register contains the length of the input. Let $\start_{v,k}$ be the formula constructed as in the proof of Theorem $\ref{theorem: SAT_R is NP_R hard}$, except modify $\Input_v$ as described above and replace the clause $\Index_0(2n+2q(n),0)$ with the clause $\Index_0(2(n+kq(n)),0)$ expressing the fact that at time zero the zeroth index register should contain the length of an input of the above form $(v,w_1,\dots,w_k)$ with $k$ witness strings.

With this definition of $\start_{v,k}$ in mind, let $\update$ and $\accept$ be the formulas constructed in the proof of Theorem \ref{theorem: SAT_R is NP_R hard}, and define the formula $\varphi_{v,k}$ follows.
If $k$ is even, then $$\varphi_{v,k} \equiv (\start_{v,k} \wedge \update) \to \accept,$$ and if $k$ is odd, then $$\varphi_{v,k} \equiv \start_{v,k} \wedge \update \wedge \accept.$$ 
Given the definition of $\Input_{v,k}$, the variables in $\varphi_{v,k}$ that encode a potential witness string $w_j$ loaded into the $\rr$-registers at time zero are those variables $Z_{i,0}$ such that $i$ is odd and $2(n+(j-1)q(n)) \leq i \leq 2(n+jq(n))-1$. Letting $a(j) = 2(n+(j-1)q(n)),$ we can see that these variables are $Z_{a(j)+1,0}, Z_{a(j)+3,0},\dots, Z_{a(j)+2q(n)-1,0}.$ For any quantifier $Q_j \in \{\exists,\forall\},$ we write $Q_jZ_j$ to denote the quantifier block $$Q_jZ_{a(j)+1,0}Q_jZ_{a(j)+3,0} \dots Q_jZ_{a(j)+q(n)-1,0}.$$ With this abbreviation in mind, define the $\Sigma_k$ sentence $\Phi_{v,k}$ in the vocabulary of $\rr$ as $$\Phi_v \equiv \exists Z_1 \forall Z_2 \dots Q Z_k Q \overline{y} (\varphi_v(Z_1,\dots,Z_k,\overline{y})),$$ where the quantifier blocks $Q_jZ_j$ alternate and  $\overline{y}$ is the tuple of all variables occurring in $\varphi$ except those occurring in $Z_1,\dots,Z_k$. Noting that $\Phi_{v,k}$ depends on the $L \in \Sigma_k$ from which we build $\varphi_{v,k},$ we will show for all $L \in \Sigma_k \rr$ that $v \in L$ if and only if $\rr \satisfies \Phi_{v,k}$ by induction on $k$.

\begin{lemma}[Encoding Computations in First Order Formulas]
\label{lemma:BaseCaseHigherLevelsHardnessInduction}
    Let $\M$ be an arbitrary polynomial-time $\rr$-machine, $q$ be an arbitrary polynomial, and $v$ be an arbitrary element of $R^n$ for any $n \in \N.$ Then for all $k \in \N$ and for all $(w_1,\dots,w_k) \in (R^{q(n)})^k,$ the following are equivalent:
    \begin{enumerate}
        \item $\M(v,w_1,\dots,w_k) = 1$%\\
        %Hans: the newlines give too much vertical whitespace
        \item $\rr \satisfies \exists \overline{y} \ \varphi_{v,k}(w_1,\dots,w_k,\overline{y})$ %\\
        \item $\rr \satisfies \forall \overline{y} \ \widetilde{\varphi}_{v,k}(w_1,\dots,w_k,\overline{y})$
    \end{enumerate}
\end{lemma}
\begin{proof}
    We prove that $(1)$ is equivalent to $(2).$ The proof that $(1)$ is equivalent to $(3)$ is similar. Suppose $\M(v,w_1,\dots,w_k)=1.$ Then the computation $\conf_0((v,w_1,\dots,w_k)) \to \dots \to \conf_N$ of $\M$ on input $(v,w_1,\dots,w_k)$ determines values $\overline{r}$ for the variables $\overline{y}$ as in the proof of Theorem \ref{theorem: SAT_R is NP_R hard}. Since $\overline{r}$ encodes this computation and since this is an accepting computation, we see that $\rr \satisfies \varphi_{v,k}(w_1,\dots,w_k,\overline{r}).$ Thus, $\rr \satisfies \exists \overline{y} \ \varphi_{v,k}(w_1,\dots,w_k,\overline{y}).$

    Conversely, suppose $\rr \satisfies \exists \overline{y} \ \varphi_{v,k}(w_1,\dots,w_k,\overline{y}).$ Then there are values $\overline{r}$ in $R$ for the variables $\overline{y}$ such that $\rr \satisfies \varphi_{v,k}(w_1,\dots,w_k,\overline{r}).$ From the definition of $\varphi_{v,k},$ we can see that this means that $\overline{r}$ encodes the computation of $\M$ on input $(v,w_1,\dots,w_k).$ Since $\rr \satisfies \accept(w_1,\dots,w_k,\overline{r}),$ it must be that this is an accepting computation. Thus, $\M(v,w_1,\dots,w_k)=1.$ 
\end{proof} 

Note that the above lemma implies \Cref{lemma:EncodingComputationFOFormula} from the main body of this paper. At this point, the reader might be convinced that hardness for each level of the polynomial hierarchy has been established: since the condition (1) is equivalent to the condition (2) and to the condition (3) for all strings $w_1,\dots,w_k,$ one may simply place the appropriate alternating sequence of quantifiers before each condition to see that they remain equivalent. Indeed, the above lemma shoes the \textit{logical equivalence} of all three statements, and so we may use these statements interchangeably by a metatheorem of first-order logic called the replacement theorem \cite{kleene2002mathematical}. Without invoking the replacement theorem, it would be necessary to carry out an inductive proof of a more general statement with unquantified strings as parameters. The replacement theorem is an easier way to arrive at the following theorem.

\begin{theorem}
    For all $k \in \N,$ $\Sigma_k\SAT(\rr)$ is $\Sigma_k\rr$-hard and $\Pi_k\SAT(\rr)$ is $\Pi_k\rr$-hard.
\end{theorem}   
\begin{proof}
    Suppose $L \in \Sigma_k\rr$. For any $v \in R^n$, let $\varphi_{v,k}$ be the formula depending on $L$ and $v$ defined above, and let $\Phi_{v,k}$ be the $\Sigma_k$ sentence in the vocabulary of $\rr$ defined by $$\Phi_{v,k} \equiv  \exists Z_1 \forall Z_2 \dots Q Z_k Q \overline{y} \  \varphi_{v,0,k}(Z_1,\dots,Z_k,\overline{y}).$$ Then by Lemma \ref{lemma:BaseCaseHigherLevelsHardnessInduction} together with the logical replacement theorem of \cite{kleene2002mathematical}, $v \in L$ if and only if $\rr \satisfies \Phi_{v,k}.$ Since the number of atomic subformulas in $\varphi_{v,0,k}$ is polynomial in $|v|,$ an $\rr$-machine can construct $\Phi_{v,k}$ in polynomial time from $v$ and $L$. Thus $v \mapsto \Phi_{v,k}$ is a polynomial-time reduction from $L$ to $\Sigma_k\SAT(\rr),$ so we see that $\Sigma_k\SAT(\rr)$ is $\Sigma_k\rr$ hard. Similarly, $\Pi_k\SAT(\rr)$ is $\Pi_k\rr$-hard.
\end{proof}

We also have an easy proof of membership: from the correctness of the $\evaluate$ algorithm and its polynomial running time, established above and in \Cref{lemma:evaluate runs in polynomial time}, respectively, together with the replacement theorem, which allows us to prepend the appropriate string of quantifiers while retaining logical equivalence, we can derive the following.

\begin{theorem}
    For all $k \in \N$, $\Sigma_k\SAT(\rr)$ is in $\Sigma_k\rr$ and $\Pi_k\SAT(\rr)$ is in $\Pi_k\rr$.
\end{theorem}
\begin{proof}
    Recall from Lemma \ref{lemma:evaluate runs in polynomial time} that the algorithm $\evaluate$ runs in polynomial time. Take some $\Sigma_k$ sentence $\Phi \equiv \exists \overline{x}_1 \forall \overline{x}_2 \dots Q_k \overline{x}_k \varphi(\overline{x}_1,\dots,\overline{x}_k)$ in the vocabulary of $\rr$. Let $n = |\Phi|$ be the length of the encoding of $\Phi$ and let $q(n) = n$ be the constant polynomial. Each quantifier block $Q_i\overline{x}_i$ is at most as long as $n$, and so any $\overline{r}_i \in R^n$ determines an interpretation for each variable in $\overline{x}_i$. Then by the correctness of the $\evaluate$ algorithm, together with the logical replacement theorem \cite{kleene2002mathematical}, we can see that $\Phi \in \Sigma_k\SAT(\rr)$ if and only if  $$(\exists \overline{r}_1 \in R^n) (\forall \overline{r}_2 \in R^n) \dots (Q_k \overline{r}_k \in R^n) \ \evaluate(\varphi,\overline{r}_1,\dots,\overline{r}_n) = 1.$$ Thus, $\Sigma_k\SAT(\rr) \in \Sigma_k\rr$ and similarly $\Pi_k\SAT(\rr) \in \Pi_k\rr$.
\end{proof}

From these two theorems, we have the following theorem as a corollary.

\HigherCookLevin*

Small modifications to the proofs above give the following theorem via an analysis of what happens when we restrict to Boolean inputs and machines without constants other than 0 or 1, just as in the proof in \Cref{sub:BooleanSATIsBPNPRHard} of \Cref{thm:BooleanCookLevin} that $\exists\SAT(\rr)$ is $\exists\rr$-complete. 

\HigherBooleanCookLevin*

% \begin{conjecture}
%     If $\rr$ can be interpreted in $\mathcal{S}$ with polynomial length formulas, then $\textup{SAT}_\rr^0$ Turing-reduces to $\textup{SAT}_\mathcal{S}^0$. 
% \end{conjecture}

% \begin{theorem}
%     If evaluating quantifier-free formulas in the vocabulary of $\rr$ can be done by a Turing machine in polynomial time, then $\text{BP}(\text{SAT}_\rr)$ is Turing-complete for $\text{BP}(\text{NP}^0_\rr)$. 
% \end{theorem}

%%%%%%%%%%%%%%%%%%%%%%%%%%%%%%%%%%%%%%%
\section{Fagin's Theorem Over a Structure}
\label{sec:Fagin'sTheorem}
%%%%%%%%%%%%%%%%%%%%%%%%%%%%%%%%%%%%%%%
In this section, we show that $\Sigma_1\SO(\rr)$ captures $\NP(\rr)$ for any bipointed $\rr$. Notably, we do not need to assume that $\rr$ is of finite type nor that $\rr$ has all constants. According to \Cref{def:LogicCapturesComplexityClassOnStructures}, showing that $\Sigma_1\SO(\rr)$ captures $\NP(\rr)$ requires showing two things: (1) if $\Phi$ is a $\Sigma_1\SO(\rr)$ sentence, then $\{\D : \D \satisfies \Phi\} \in \NP(\rr)$, and (2) if  $L$ is a decision problem of $\rr$-structures such that $L \in \NP(\rr)$, then there is some $\Sigma_1\SO(\rr)$ sentence $\Phi$ such that $L = \{\D : \D \satisfies \Phi\}$. Subsection \ref{sub:DataComplexityESO(R)} establishes point (1) in \Cref{thm:DataComplexitySigmaSOR}, and Subsection \ref{sub:ESO(R)CapturesNP(R)} shows point (2) in Theorem \ref{thm:SigmaSORDescribesNPR}, thereby establishing \Cref{thm:Fagin}. In Subsection \ref{sub:BooleanESO^0(R)capturesBP^0(NP(R))} we modify the proofs of the above theorems to prove that $\exists\SO(\rr)$ captures $\exists\rr$ over Boolean $\rr$ structures, thereby establishing \Cref{thm:BooleanFagin}, and in Subsection \ref{sub:SOCapturesPH} we extend these results to the Polynomial and Boolean hierarchies, thereby establishing \Cref{thm:HigherFagin} and \Cref{thm:HigherBooleanFagin}.
%%%%%%%%%%%%%%%%%%%%%%%%%%%%%%%%%%%%%%%%%%%
\subsection{Data Complexity of $\Sigma_1\SO(\rr)$}
\label{sub:DataComplexityESO(R)}
%%%%%%%%%%%%%%%%%%%%%%%%%%%%%%%%%%%%%%%%%%%

In this subsection we will show that for any metafinite vocabulary $\MVoc$ over $\rr$ and any sentence $\Phi$ of $\Sigma_1\SO(\MVoc)$, the decision problem of $\rr$-structures $\{ \D \in \struct(\MVoc) : \D \satisfies \Phi\}$ belongs to $\NP(\rr)$. In contrast to the argument showing that $\SAT(\rr)$ is in $\NP(\rr)$ in \Cref{thm:SAT_R-NP_R-Membership}, we do not need to assume that $\rr$ is of finite type. Towards this end, fix a metafinite vocabulary $\Voc = (\primary(\MVoc),\secondary(\MVoc),\weight(\MVoc))$ over $\rr$, by which we mean that $\secondary(\MVoc) = \Voc(\rr)$.

Note that the basic operations in the primary part $\A$ of an $\rr$-structure $\D = (\A,\rr,\W)$ in this vocabulary are not basic operations of the structure $\rr$, which means that an $\rr$-machine cannot execute them in unit time as it does the operations of $\rr$. However, given our knowledge of the fixed vocabulary $\Voc$, we can program an $\rr$-machine to simulate the operations of the primary part of $\D$ in time polynomial in $|\D|=|A|$ when it is given $\code(\D)$ as input. 

We must first determine an encoding of the elements in the primary universe of $\D$ in order to feed them as inputs to an $\rr$-machine because, in general, the universe of $\A$ is not a subset of the universe of $\rr$. Recall that we have a chosen total ordering $\ordering(\A)$ on the primary universe of $\D$, which orders $A$ as $\{a'_1,\dots,a'_{|\D|}\}.$ Given $a \in A$, we let $\code(a) = \bin(i)$, where $a = a'_i$ and $\bin(i)$ is the binary representation of $i$. 

Using any standard encoding of tuples, we extend this encoding of $A$ to encodings of tuples $(a_1,\dots,a_k) \in A^k$ as binary strings $\code(a_1,\dots,a_k) \in \{0,1\}^*$. Since $\code(a)$ directly references the place of $a$ in the total order on $A$, we can program an $\rr$-machine to determine the place of a tuple $(a_1,\dots,a_k)$ in the induced lexicographic order on $A^k$, based on $\code(a_1,\dots,a_k)$. Furthermore, for a fixed $k$, an $\rr$-machine can do this in time polynomial in $|\D|$ via $k$ applications of binary search. 

%\jeremy{Wouldn't this require us to know $|\D|$? We could do this by cycling through $A$ once and identifying the largest element. This would also be easier than the polynomial evaluation method I suggested below. This is similar to identifying an element of maximal rank in the proof of Fagin's theorem, but different because we want to compute it now, rather than define it.}

Suppose $P$ is a $k$-ary relation of the primary part of $\D$. Given $\code(\D)$ and $\code(a_1,\dots,a_k)$ as inputs, we can program an $\rr$-machine $\M$ to check if $(a_1,\dots,a_k) \in P$ in the following way. First, $\M$ determines the place $i$ of $(a_1,\dots,a_k)$ in the lexicographic order on $A^k$. Second, using our knowledge of $\Voc$, we program $\M$ to use $|\code(\D)|$ to determine $|\D|$, and from this to identify the substring $\code(P)$ of $\code(\D)$. Third, $\M$ examines $\code(P)$ to determine if the $i$th bit of $\code(P)$ is one. 

The length of $\code(\D)$ is polynomial in $|\D|$, where this polynomial is based on the arity of the symbols in $\Voc$. The second step of the above algorithm can be accomplished by repeatedly evaluating this polynomial on values $n \leq |\code(\D)|$ until the result is greater than or equal to $|\code(\D)|$. Each evaluation can be done in time linear in $|\D|$, so all of the second step can be done in time polynomial in $|\D|$. In the worst case, the third step would need to scan the whole of $\code(\D)$, which means that it can be done in time polynomial in $|\D|$. Thus, evaluating if $(a_1,\dots,a_k) \in P$ can be done by a polynomial-time $\rr$-machine.

Similarly, any other operation of the primary part of $\D$ and any weight function of $\D$ can be simulated by a polynomial time $\rr$-machine that receives $\code(\D)$ as an input. The ability of $\rr$-machines to simulate the operations of $\rr$-structures in polynomial time is the basis for the following lemma.

\begin{lemma}
\label{lemma:FirstOrderMetafiniteEvaluationPolynomialTime}
    For any metafinite vocabulary $\MVoc$ over $\rr$, if $\varphi$ is a sentence of first-order metafinite logic in this vocabulary, then $\{\D \in \struct(\MVoc) : \D \satisfies \varphi\} \in \P(\rr)$.
\end{lemma}
\begin{proof}
    We first prove a more general statement: For any first-order formula $\varphi$ in prenex normal form $$Q_1x_1 \dots Q_kx_k \psi(x_1,\dots,x_k,x_{k+1},\dots,x_s),$$ for any metafinite structure $\D = (\A,\rr,\W)$ with $|A|=n$, and for any $(a_{k+1},\dots,a_s) \in A^{s-k}$, we can determine if $$\D \satisfies Q_1x_1 \dots Q_kx_k\psi(x_1,\dots,x_k,a_{k+1},\dots,a_s)$$ in time polynomial in $n$. We proceed by induction on the number of quantifiers in the formula. If there are no quantifiers, then our formula is simply $\psi(x_1,\dots,x_s)$. Given some $(a_1,\dots,a_s) \in A^s$, to decide if $\D \satisfies \psi(a_1,\dots,a_s)$, we only need to compute the value of $\psi$ on the input $(a_1,\dots,a_s)$. As in the above discussion, this can be done in polynomial time in the size of $\D$. %\jeremy{This does depend on the size of $\D$ as explained in the discussion above. We should simply use the above discussion to say that evaluating $\D \satisfies \psi(a_1,\dots,a_s)$ can be done in polynomial time. Can it actually be done in linear time? Then our inductive hypothesis, which we should state explicitly, should be that we can evaluate a formula with $k$ quantifiers can be done in $O(n^k)$ time.} 
    To determine if $$\D \satisfies Q_1x_1 \dots Q_kx_kQ_{k+1}x_{k+1}\psi(x_1,\dots,x_k,x_{k+1},a_{k+2}\dots,a_s),$$ it suffices to check, for all $a_{k+1} \in A$, whether or not $$\D \satisfies Q_1x_1 \dots Q_kx_k\psi(x_1,\dots,x_k,a_{k+1},\dots,a_s).$$ By our inductive hypothesis, each of these checks can be done in polynomial time. Since this means making $|A|=n$ checks, in total we again need only polynomial time, where the exponent of this polynomial has been increased by one. This completes our inductive argument. 
    
    Any sentence $\varphi$ is logically equivalent to a sentence $\varphi'$ in prenex normal form, which means that deciding if $\D \satisfies \varphi$ is equivalent to deciding if $\D \satisfies \varphi'$. By the above, this may be done in polynomial time.
\end{proof}

\begin{theorem}
\label{thm:DataComplexitySigmaSOR}
    For any metafinite vocabulary $\MVoc$ over $\rr$, if $\Phi$ is a sentence of $\Sigma_1\SO(\MVoc)$, then $\{\D \in \struct(\MVoc) : \D \satisfies \Phi\} \in \NP(\rr).$
\end{theorem}
\begin{proof}
    Any $\Sigma_1\SO(\MVoc)$ sentence $\Phi$ is of the form $\exists w_1 \dots \exists w_h \varphi(w_1,\dots,w_h)$, where $w_1,\dots,w_h$ are second-order variables. Suppose $\D = (\A,\rr,\W)$ is a metafinite $\Voc$-structure. Given any interpretation of each $w_i$ as a weight functions $w_i \: A^k \to R$, the formula $\varphi(w_1,\dots,w_k)$ is a first-order sentence in vocabulary $\Voc \cup \{w_1,\dots,w_h\} = (\primary(\MVoc),\secondary(\MVoc),\weight(\MVoc) \cup \{w_1,\dots,w_h\}).$ Writing $(w_1,\dots,w_h)$ as $\overline{w}$, we can see that by \Cref{lemma:FirstOrderMetafiniteEvaluationPolynomialTime} applied to this vocabulary, $$\{(\D,\overline{w}) : (\D,\overline{w}) \satisfies \varphi(\overline{w})\} \in \P(\rr).$$ Thus, $\{\D : \D \satisfies \Phi\} = \{\D : \exists \overline{w} (\D,\overline{w}) \satisfies \varphi(\overline{w})\} \in \NP(\rr).$
\end{proof}

%%%%%%%%%%%%%%%%%%%%%%%%%%%%%%%%%%%%%%%%%%%
\subsection{$\Sigma_1\SO(\rr)$ Captures $\NP(\rr)$}
\label{sub:ESO(R)CapturesNP(R)}
%%%%%%%%%%%%%%%%%%%%%%%%%%%%%%%%%%%%%%%%%%%
In this subsection, we show that if $L$ is any decision problem of $\rr$-structures such that $L \in \NP(\rr)$, then there is some $\Sigma_1\SO(\rr)$ sentence $\Phi$ such that $L = \{\D : \D \satisfies \Phi\}.$ In order to do this, we take some $\rr$-machine $\M$ such that $\M$ verifies $L$, and we construct $\Phi$ so that it encodes the computation of $\M$ in the sense that $\D \satisfies \Phi$ if and only if there is some $Y$ such that $\M(\D,Y) = 1.$

In order to build $\Phi$ in this way, we need to encode the starting configuration of $\M$ on input $(\D,Y)$ into a formula, as we do in \Cref{lemma:FirstOrderFormulaEncodingStartingConfiguration}. In contrast to the argument showing that $\SAT(\rr)$ is $\NP(\rr)$-hard in \Cref{theorem: SAT_R is NP_R hard}, we do not need to assume that $\rr$ has all constants. 

%\jeremy{Need to rephrase this sentence...}
We then show our main result in  \Cref{thm:SigmaSORDescribesNPR}, which, together with the result of  \Cref{thm:DataComplexitySigmaSOR} stating that the data complexity of $\Sigma_1\SO(\rr)$ is $\NP(\rr)$, implies \Cref{thm:Fagin}.

% \begin{lemma}
% \label{lemma:ArithmeticIsFirstOrderDefinable}
%     For any metafinite vocabulary $\MVoc$, there are first-order formulas $\orderform(E)$, $\sumform_m(E,S)$, and $\prodform_m(E,S,P)$ such that for all $\MVoc$-structures $\D = (\A,\rr,\W)$ and for all positive integers $m$,
%     \begin{enumerate}
%         \item $(\D,E) \satisfies \orderform(E)$ if and only if $E\:A^2 \to R$ encodes a linear order $E \subseteq A^2$ on $A$,
%         \item $(\D,E,S) \satisfies \orderform(E) \wedge \sumform_m(E,S)$ if and only if $S\:A^{3m} \to R$ encodes the graph of the summation function $S \subseteq A^{3m}$ on $A^m$ lexicographically ordered according to $E$,
%         \item $(\D,E,S,P) \satisfies \orderform(E) \wedge \sumform_m(E,S) \wedge \prodform_m(E,S,P)$ if and only if $P\:A^{3m} \to R$ encodes the product function $P \subseteq A^{3m}$ on $A^m$ lexicographically ordered according to $E$.
%     \end{enumerate}
% \end{lemma}

\begin{lemma}
\label{lemma:ArithmeticIsFirstOrderDefinable}
    For any metafinite vocabulary $\MVoc$, there are first-order formulas $\orderform(E)$, $\sumform_m(E,S)$, and $\prodform_m(E,S,P)$ such that for all $\MVoc$-structures $\D = (\A,\rr,\W)$ and for all positive integers $m$, $$(\D,E,S,P) \satisfies \orderform(E) \wedge \sumform(E,S) \wedge \prodform(E,S,P)$$ if and only if each of the following are true:
    \begin{enumerate}
        \item $E\:A^2 \to R$ encodes a linear order $E \subseteq A^2$ on $A$,
        \item $S\:A^{3m} \to R$ encodes the graph of the summation function $S \subseteq A^{3m}$ on $A^m$ lexicographically ordered according to $E$, and
        \item $P\:A^{3m} \to R$ encodes the graph of the product function $P \subseteq A^{3m}$ on $A^m$ lexicographically ordered according to $E$.
    \end{enumerate}
\end{lemma}
\begin{proof}
    For each of the above relations $F \in \{E,S,P\}$, we identify a relation $F \subseteq A^k$ with its characteristic function $F\:A^k \to \{0,1\} \subseteq R$. The formula $\orderform(E)$ is simply the conjunction of the usual axioms for linearly ordered sets using $E$ as the relation symbol. 
    
    Let $\overline{x}$, $\overline{y}$, and $\overline{z}$ be $m$-tuples of variables. Given ordering $E$, the induced lexicographic order on $A^m$ is definable with a first-order formula using the relation $E$. We abbreviate this formula as $E_m(\overline{x},\overline{y})$. Similarly, the least element in $A^m$ is definable by a formula $\texttt{zero}(\overline{x})$, the maximum element is definable by a formula $\texttt{max}(\overline{x})$, and the immediate successor relation is definable by a formula $\texttt{successor}(\overline{x},\overline{y})$ expressing that $\overline{y}$ is the immediate successor of $\overline{x}$. We define $\texttt{successor}$ in such a way that the maximum element of $A^m$ is the immediate successor of itself, so that $\texttt{successor}$ defines a total function on the finite set $A^m$.

    Given these formulas, we encode the usual recursive definition of addition into a formula $\sumform(E,S)$, which is then the conjunction of a clause for the base case and a clause for the recursive case of this definition. The clause corresponding to the base case is $$ \forall \overline{x} \forall \overline{z}  (\texttt{zero}(\overline{z}) \to S(\overline{x},\overline{z},\overline{x})).$$ The clause corresponding to the recursive case is $$\forall \overline{x}\forall \overline{y} \forall \overline{z} \forall \overline{y'} \forall \overline{z'}\Big[ \Big(S(\overline{x},\overline{y},\overline{z}) \wedge \texttt{successor}(\overline{y},\overline{y'}) \wedge \texttt{successor}(\overline{z},\overline{z'})\Big) \to S(\overline{x},\overline{y'},\overline{z'})\Big].$$ Similarly, we can encode the usual recursive definition of multiplication in terms of addition into a formula $\prodform(E,S,P)$, which is the conjunction a clause corresponding to the base case and a clause corresponding to the recursive case of this definition.
\end{proof}

\begin{lemma}
\label{lemma:FirstOrderFormulaEncodingStartingConfiguration}
    For any metafinite vocabulary $\MVoc$ over $\rr$ and for any positive integer $m$ such that $m \geq \sum_j k_j$, the sum of arities $k_j$ of all symbols in $\primary(\MVoc) \cup \weight(\MVoc)$,  there is first-order formula $\Input(E,S,P,Z)$ in the expanded vocabulary $\MVoc \cup \{E,S,P,Z\}$ such that for all $\MVoc$-structures $\D = (\A,\rr,\W)$, $$(\D,E,S,P,Z) \satisfies \Input(E,S,P,Z)$$ 
    if and only if each of the following are true:
    \begin{enumerate}
        \item $E \subseteq A^2$ is a linear order on $A$,
        \item $S \subseteq A^{3m}$ is the graph of the summation function on $A^m$ lexicographically ordered according to $E$
        \item $P \subseteq A^{3m}$ is the graph of the product function on $A^m$ lexicographically ordered according to $E$, and
        \item $Z\:A^m \to R$ encodes the string $\code_E(\D)0\dots 0$, which we express by writing $Z = \code_E(\D)0 \dots 0$.
    \end{enumerate}
\end{lemma}
\begin{proof}
    \Cref{lemma:ArithmeticIsFirstOrderDefinable} deals with the first three conditions, so we will focus on the fourth condition by constructing a formula $\bigwedge_{j=1}^{h+1} \Input_j(E,S,P,Z)$ expressing the fourth condition. We will then take the conjunction of this formula with the previously constructed formulas to arrive at $$\Input(E,S,P,Z) \equiv \orderform(E) \wedge \sumform_m(E,S) \wedge \prodform_m(E,S,P) \wedge \bigwedge_{j=1}^{h+1}\Input_j(E,S,P,Z).$$

    Assume that $\MVoc$ is a vocabulary of metafinite algebras, meaning that $\primary(\MVoc)$ is empty and $\weight(\MVoc)=\{w_1,\dots,w_j,\dots,w_h\}$, where $j \in \{1,\dots,h\}$. The heart of the idea is illustrated with this case, and the proof in the case where $\primary(\MVoc)$ is nonempty may be obtained with a simple modification of the construction presented below. For any metafinite structure $\D = (A,\{w_1,\dots,w_j,\dots,w_h\})$ of vocabulary $\MVoc$ and cardinality $|A| = n$, the code of $\D$ relative to the ordering $E$ on $A$ is $$\code_E(\D) = \code_E(w_1) \dots \code_E(w_j) \dots \code_E(w_h) \in R^*.$$ Recall from Definition \ref{def:EncodingMetafiniteStructures} that for each weight function $w_j \: A^{k_j} \to R$, the code of $w_j$ is the string $\code_E(w_j) = (r_1,\dots,r_{n^{k_j}}) \in R^{n^{k_j}}$ where $r_i$ is the value of $w_j$ applied to the $i$th tuple $(p_1,\dots,p_{k_j})$ of $A^{k_j}$ according to the lexicographic order on $A^{k_j}$ induced by $E$.  

    In what follows, we will analyze the string $\code_E(\D)$ in terms of the order relation $\leq$ on the natural numbers that index entries in this string, as well as in terms of the arithmetic operations of addition $+$ and multiplication $\times$ on the natural numbers. Using the relations $E$, $S$, and $P$ that encode these operations, we will be able to use our analysis of $\code(\D)$ to construct a first order formula expressing that $Z = \code_E(\D)0 \dots 0$. 
    
    We think of $\overline{p} = (p_1,\dots,p_m) \in A^m$ as the $n$-ary representation of $\sum_{i=1}^m p_i \times n^{m-i} \in \{0,\dots,n^m-1\}$, thereby identifying $A^m$ with $\{0,\dots,n^m-1\}$. In the same way, we identify each $A^{k_j}$ with $\{0,\dots, n^{k_j}-1\}.$ According to this $n$-ary representation, $\code_E(w_j)$ is the substring of $\code_E(\D)$ indexed by tuples $(p_1,\dots,p_m)$ such that $$\sum_{i=1}^{j-1} n^{k_i} \leq \sum_{i=1}^m p_i \times n^{m-i} \leq \left(\sum_{i=1}^j n^{k_i} \right)-1.$$ Keeping in mind that the length of $\code_E(w_j)$ is $n^{k_j}$, we can see that for every $\overline{p} = (p_1,\dots,p_m)$ in the above range, there is a unique $ \overline{q} = (q_1,\dots,q_{k_j}) \in A^{k_j}$ such that $\overline{p} = \sum_{i=1}^{j-1}n^{k_i} + \overline{q},$ by which we mean that $$\sum_{i=1}^m p_i \times n^{m-i} = \sum_{i=1}^{j-1} n^{k_i} + \sum_{i=1}^{k_j} q_i \times n^{k_j-1}.$$ Viewing the function $Z \: A^m \to R$ as a string in $R^{n^m}$, we can see that $Z$ agrees with $\code_E(\D)$ on the substring $\code_E(w_j)$ if and only if $Z(\overline{p}) = w_j(\overline{q})$ for all $\overline{p}$ in the range from $\sum_{i=1}^{j-1}n^{k_i}$ to $\left(\sum_{i=1}^j n^{k_i} \right)-1$ and for the unique $\overline{p}$ such that $\overline{p} = \sum_{i=1}^{j-1}n^{k_i} + \overline{q}$.

    We now proceed to express the above conditions in a first-order formula. By Lemma \ref{lemma:ArithmeticIsFirstOrderDefinable}, each of $\leq$, $+$ and $\times$ are first-order definable using the weight function symbols $E$, $S$ and $P$ encoding the corresponding operations in $A^m$ ordered lexicographically by $E$. This implies that there is a first-order formula $\texttt{range}_j(E,S,P,p_1,\dots,p_m)$ in vocabulary $\MVoc \cup \{E,S,P\}$ expressing that $ \sum_{i=1}^{j-1} \leq \overline{p} \leq \sum_{i=1}^j n^{k_i}$.   Similarly, there is a first-order formula $\texttt{offset}_j(E,S,P,p_1,\dots,p_m,q_1,\dots,q_{k_j})$ in this same vocabulary expressing that $\overline{q} = \sum_{i=1}^{j-1}n^{k_i} + \overline{q}.$ Given these formulas, we can express that $Z$ agrees with $\code_E(\D)$ on the substring $\code_E(w_j)$ with the formula 
    $$\begin{array}{rcl}
        \Input_j(E,S,P,Z) & \equiv & \forall p_1 \dots \forall p_m \Big[ \texttt{range}_j(E,S,P,p_1,\dots,p_m) \to \\
        & & \\
        & & \exists q_1 \dots \exists q_{k_j} \Big[ \texttt{offset}_j(E,S,P,p_1,\dots,p_m,q_1,\dots,q_{k_j}) \wedge \\
        & & \\
        & & \hspace{1.75cm} Z(p_1,\dots,p_m) = w_j(q_1,\dots,q_{k_j})\Big] \Big].  \\
    \end{array}$$

    Similarly, there is a first-order formula $\texttt{range}_{h+1}(E,S,P,p_1,\dots,p_m)$ expressing that $\left(\sum_{i=1}^h n^{k_i}\right)-1 \leq \overline{p}$, so we can express that $Z$ ends with zeros beyond $\code_E(\D)$ with the formula $$\Input_{h+1}(E,S,P,Z) \equiv \forall p_1 \dots \forall p_m \Big[ \texttt{range}_{h+1}(E,S,P,p_1,\dots,p_m) \to Z(p_1,\dots,p_m) = 0 \Big].$$ We define the formula $\Input(E,S,P,Z)$ $\Input$ as the conjunction $$\Input(E,S,P,Z) \equiv \orderform(E) \wedge \sumform_m(E,S) \wedge \prodform_m(E,S,P) \wedge \bigwedge_{j=1}^{h+1}\Input_j(E,S,P,Z).$$ The first three formulas guarantee that we can formally express the above arithmetic with the weight functions $E$, $S$, and $P$, and the fourth formula guarantees that $Z = \code_E(\D)0\dots0.$
\end{proof}

\begin{theorem}
\label{thm:SigmaSORDescribesNPR}
    For every metafinite vocabulary $\MVoc$ over $\rr$, if $L$ is a decision problem of $\rr$-structures such that $L \in \NP(\rr)$, then there is a sentence $\Phi$ of $\Sigma_1\SO(\MVoc)$ such that $$L = \{\D \in \struct(\MVoc)  :  \D \satisfies \Phi(L)\}.$$
\end{theorem}
\begin{proof}
    Since $L \in \NP(\rr)$, there is some positive integer $k$ and some polynomial-time $\rr$-machine $\M$ such that $$L = \{\D \in \struct(\MVoc) : (\exists Y \: A^k \to R) \ \M(\D,Y)=1 \}.$$
    %If $\D \in \NP(\rr)$, then there must be a certificate $Y$, such that $(\D, Y) \in \P(\rr)$.We will first show that for any problem $(\D, Y)$, we can construct a first-order formula that decides whether $(\D, Y) \in \P(\rr)$.
    Let $n = |\D| = |A|$ be the size of the $\rr$-structure $\D = (\A,\rr,\W)$ of vocabulary $\Voc$, and let $m$ be a natural number such that $\M$ stops on $(\D, Y)$ after fewer than $n^m$ steps for all $Y\:A^k \to R$. Furthermore, assume that $n^m$ is larger than the value attained by any index register of $\M$ during a computation on any input $(\D,Y)$ of the above form. Lastly, assume that $m$ is larger than the sum of the arities of all symbols in $\primary(\MVoc)\cup\weight(\MVoc)$ so that \Cref{lemma:FirstOrderFormulaEncodingStartingConfiguration} applies.

    %We aim to encode the computation of the machine $\M$ on any input $(\D,Y)$ using a formula of $\Sigma_1\SO(\rr)$. In order to do this, we will build a first-order formula with another introduce a number of relations symbols that will be interpreted in the structure $(\D,Y)$, including $E$...
    
    % Intuitively, let $\bar{t}$ be some number $0 \leq \bar{t} \leq n^m-1$ that represents the time step of $M$.
    % Since the machine terminates after $n^m$ steps, the machine's contents will never change after time $n^m-1$.
    % Similarly, let $\bar{p}$ represent the index of some register of $M$.
    % As $M$ runs for at most $n^m$ timesteps, the head can never reach an index of $n^m$ or beyond, showing that $0 \leq \bar{p} \leq n^m-1$.
    % Finally, with $j$ we represent the instructions that might happen at a timestep.
    % Clearly, the set of instructions is polynomial, as seen in \Cref{tab:RInstructions}.

    We assume that we have access to some linear order $E \subseteq A^2$ on $\A$ that we will existentially quantify over at the end of our construction. The linear order $E$ induces a lexicographic order on $A^m$, and we think of $\overline{p} = (p_1,\dots,p_m) \in A^m$ as the $n$-ary representation of $\sum_{i=1}^m p_i \times n^{m-i} \in \{0,\dots,n^m-1\}$, thereby identifying $A^m$ with $\{0,\dots,n^m-1\}$.  
    % there is a relation $\lO$ on $\A$ such that $\lO$ is a linear order of $\A = \{ \alpha_1, \alpha_2, \dots, \alpha_n \}$.
    % Given $\lO$, we define $\lOL$ as a lexicographical order on $\A^m$, i.e., $(\alpha_{i_1}, \alpha_{i_2}, \dots, \alpha_{i_m}) < (\alpha_{j_1}, \alpha_{j_2}, \dots, \alpha_{j_m})$ iff there exists a $1 \leq k$ such that $\alpha_{i_k} < \alpha_{j_k}$ and for all $1 \leq l < k$, $\alpha_{i_l} = \alpha_{j_l}$.
    % Then, we identify $\A^m$ with $\{ 0, \dots, n^m-1 \}$ using $\lOL$.
    % Due to this, we may think of m-tuples $\bar{t}=(t_1, \dots, t_m)$ (and $\bar{p} = (p_1, \dots, p_m)$) as ranging over $\{ 0, \dots, n^m-1 \}$.
    Notably, by Lemma \ref{lemma:ArithmeticIsFirstOrderDefinable}, there are first-order formulas that define the minimal element in this order, as well as the successor operation, addition, and multiplication. We represent each of these operations by simply writing $\bar{t} = 0$, $\overline{t} = \overline{u} + 1$, $\bar{t} = \bar{u} + \bar{v}$, and $\bar{t} = \bar{u} \times \bar{v}$, but formally they should be interpreted as the corresponding statement in terms of the relations $E$, $S$, and $P$. Using the elements of $A^m$, we will encode the computation of $\M$ on input $(\D,Y)$ in the following way:
    \begin{itemize}
        \item We introduce the relation $C \subseteq A^{2m}$ that we will constrain with formulas expressing that $C(\overline{u},\overline{t})$ is true in $(\D,Y)$ if and only if $\M$ on input $(\D,Y)$ executes instruction $\overline{u}$ at time $\overline{t}$. We choose the letter $C$ to stand for ``command''.
        \item We introduce the relations $I_0,\dots,I_{k_\program-1}$, where $k_\program-1$ is the maximum index register accessed by the program $\program$ of $\M$. For each $i$, we will constrain $I_i \subseteq A^{2m}$ with formulas expressing that $I_i(\overline{u},\overline{t})$ is true in $(\D,Y)$ if and only if the $i$th index register of $\M$ on input $(\D,Y)$ contains $\overline{u}$ at time $\overline{t}$.
        \item We introduce the weight function $Z \: A^{2m} \to R$ that we will constrain with formulas expressing that $Z(\overline{p},\overline{t}) = r$ is true in $(\D,Y)$ if and only if the $\overline{p}$th $\rr$-register of $\M$ on input $(\D,Y)$ contains $r$ at time $\overline{t}$.
    \end{itemize} 
    
    % Next, we introduce the relation symbols $\textsc{inst}$ $\Instruction$ and $\textsc{reg}_1, \dots, \textsc{reg}_k$.
    % We define $\textsc{inst}(\bar{t}, \bar{u})$ as a Boolean value whether $M$ executes instruction $\bar{u}$ at time $\bar{t}$. 
    % Furthermore, we define $\textsc{reg}_i(\bar{t}, \bar{u})$ as a Boolean value whether the $i$'th register of $M$ contains the value $\bar{u}$ at time $\bar{t}$.

    % Finally, we use one function $Z : \A^m \rightarrow \rr$ that determines which value the $\bar{p}$'th $\rr$-register contains at time $\bar{t}$, i.e. $Z(\bar{t}, \bar{p}) = r$, where $r$ is the value in the $\bar{p}$'th $\rr$-register at time $\bar{t}$.

    Now our goal is to build first-order formulas that correctly constrain the relations $C, I_0,\dots,I_{k_\program-1}$ and the weight function $Z$. We begin by constructing a first-order formula $\start(E,S,P,Z)$ that encodes the starting configuration of $\M$ on input $(D,Y)$. On this input $\M$ starts by executing instruction $0$, which we express with the formula $C(\overline{0},\overline{0}).$ 

    We need to constrain $I_0$ to represent the fact that $\M$ starts with $|\code(D,Y)|$ in its zeroth index register. Assuming without loss of generality that we have converted $(\D,Y)$ into a metafinite algebra $(A,\{w_1,\dots,w_h\})$ as described in \Cref{def:EncodingMetafiniteStructures}, the length of  $\code((\D,Y))$ is $\sum_{i=1}^h n^{k_i}$. The $n$-ary representation of $n^{k_i}$ is an element of $(p_1,\dots,p_m) \in A^m$ such that all $p_j = 1$ when $m-j=k_i$, and $p_j=0$ for all other $j$. 
    %\jeremy{Should I really use the $p$ here? Though it is a ``point'' term, it might be more clear to use $a$.} 
    Since we can express $0$ and $1$ as the minimum element of $A$ and its successor, we can express the $n$-ary representation of $n^{k_i}$ in a first-order formula, which implies that we can express $\sum_{i=1}^h n^{k_i}$ in a first-order formula. Note that we can express the above summation because we can express addition, Lemma  \ref{lemma:ArithmeticIsFirstOrderDefinable}, and because $h$ and each $k_i$ are fixed by the vocabulary $\MVoc$. 
    
    Thus, we constrain $I_0$ with the formula $\exists \overline{u}\left(\overline{u} = \sum_{i=1}^h n^{k_i} \wedge I_0(\overline{0},\overline{u})\right)$, which expresses that $\M$ starts with $|\code((\D,Y))| = \sum_{i=1}^h n^{k_i}$ in its zeroth index register when it receives input $(\D,Y)$. Furthermore, we constrain $I_1,\dots,I_{k_\program-1}$ with the formula $\bigwedge_{i=1}^{k_\program-1} I_i(\overline{0},\overline{0})$ expressing that $\M$ starts with $0$ in all other index registers.
    
    By Lemma \ref{lemma:FirstOrderFormulaEncodingStartingConfiguration} applied to vocabulary $\MVoc \cup \{Y\}$, there is a first-order formula $\texttt{input}(E,S,P,Z)$ that is true in $(\D,Y)$ if and only if the function $\overline{p} \mapsto Z(\overline{0},\overline{p})$ encodes the string $\code((\D,Y))0 \dots 0$, which constrains $Z$ to correctly represent the fact that at time $\overline{0}$, $\M$ starts with the code of $(\D,Y)$ loaded into its $\rr$-registers, followed by zeros. We define the first-order formula $\start(E,S,P,Z)$ as the conjunction of all of the above formulas encoding the starting configuration of $\M$ on input $(\D,Y).$ That is, $$\start(E,S,P,Z) \equiv C(\overline{0},\overline{0}) \wedge \exists \overline{u}\left(\overline{u} = \sum_{i=1}^h n^{k_i} \wedge I_0(\overline{0},\overline{u})\right) \wedge \bigwedge_{i=1}^{k_\program-1} I_i(\overline{0},\overline{0}) \wedge \Input(E,S,P,Z).$$ 
    
    Next, we encode the way $\M$ modifies its registers based on instruction $\ell$ at time $t$ with the family of formulas $\update_\ell$, where $\ell$ is a parameter determining the specific form of this formula, and $\overline{u}$, $\overline{p}_j$, and $\overline{t}$ are $m$-tuples of first-order variables ranging over $A^m$. We let $\ell$ range over $\{0,\dots,\ell_\program\}$, where $\ell_\program$ is the label of the last instruction in $\program$, and we base the definition of $\update_\ell$ on the possible instructions in  $\program$ as well as their meanings, as detailed in \Cref{tab:RInstructions} and \Cref{def:TransitionRelationOfRMachines}.
    
    Without loss of generality, we assume that $\ell_\program$ and all indices occurring in $\program$ are smaller than $n^m$. Note that any fixed value $i$ of $\{0,\dots,n^m-1\}$ is first-order definable as the $i$th successor of the minimum element of $A^m$. In the following formulas, an equality such as $\overline{x} = i$ should be interpreted as the formula stating that $\overline{x}$ is the $i$th element of $A^m.$

\begin{enumerate}
        \item If the $\ell$th instruction of $\program$ is $\ell : Z \lbr i \rbr := c$, then 
        $$
        \begin{array}{rcl}
            \update_\ell & \equiv & \forall \overline{u}\forall\overline{p}\forall\overline{t} \Big[ \Big(\overline{u} = \ell \wedge \overline{p} = i \wedge C(\overline{u},\overline{t}) \Big) \to  Z(\overline{p},\overline{t}+1) = c \Big].  \\
        \end{array}$$

% This is a more full description of what update should be, but I think it obscures the idea.
        % $$
        % \begin{array}{rcll}
        %     \update_\ell & \equiv & \forall \overline{u}\forall\overline{t}\forall\overline{p}\forall\overline{p}\forall\overline{q}  & \Big[ \Big(\overline{u} = \ell \wedge \overline{p} = i \wedge \overline{q} \neq i C(\overline{u},\overline{t}) \Big) \to \\
        %     & & & \\
        %     & & & \ \ C(\overline{u}+1,\overline{t}+1)\wedge \\
        %     & & & \\
        %     & & & \ \ \bigwedge_{i=0}^{k_\program-1} I_i(\overline{p},\overline{t}) \to I_i(\overline{p},\overline{t}+1) \wedge \\
        %     & & & \\
        %     & & & \ \ Z(\overline{p},\overline{t}+1) = c \wedge\\
        %     & & & \\
        %     & & & \ \ Z(\overline{q},\overline{t}+1) = Z(\overline{q},\overline{t}) \Big]
        % \end{array}$$

        \item If the $\ell$th instruction of $\program$ is $\ell: Z \lbr i_0 \rbr := f(Z \lbr i_1 \rbr, \dots, Z \lbr i_k \rbr)$, then 
        $$
        \begin{array}{rcl}
             \update_\ell & \equiv  &\forall \overline{u}\forall\overline{p}_0\dots\forall\overline{p}_k\forall\overline{t}\Big[\Big(\overline{u} = \ell \wedge \bigwedge_{j=0}^k \overline{p}_j = i_j \wedge C(\overline{u},\overline{t}) \Big) \to \\
             &  & \\ 
             &  & Z(\overline{p}_0,\overline{t}+1) = f(Z(\overline{p}_1,\overline{t}), \dots, Z(\overline{p}_k,\overline{t}))\Big]. \\
        \end{array}$$

        % \item If the $\ell$th instruction of $\program$ is $ \ell : \ifthenbranch{P(Z \lbr i_1 \rbr, \dots, Z \lbr i_k \rbr)}{\ell_1}{\ell_2},$ then
        % $$
        % \begin{array}{rcl}
        %     \update_\ell & \equiv & \forall \overline{u}_0\forall\overline{u}_1\forall\overline{u}_2\forall\overline{p}_0\dots\forall\overline{p}_k\forall\overline{t}\Big[\Big(\bigwedge_{j=0}^2\overline{u} = \ell \wedge \bigwedge_{j=0}^k \overline{p}_j = i_j \wedge \Instruction(\overline{u}_0,\overline{t}) \Big) \to  \\
        %     & & \\
        %     & & \forall \overline{p}_1 \dots \forall \overline{p}_{k_\program} \Big[\Big<\Index_1(\overline{p}_1, \overline{t}) \wedge \dots \wedge \Index_{k_\program}(\overline{p}_{k_\program}, \overline{t})\Big> \to    \\
        %     & & \\
        %     & & \Big<\Big(P(Z(\overline{p}_1,\overline{t}) \dots, Z(\overline{p}_k,\overline{t})) \rightarrow \Instruction(\overline{u}_1, \overline{t}+1)\Big) \wedge    \\
        %     & & \\
        %     & & \Big(\neg P(Z(\overline{p}_1,\overline{t}) \dots, Z(\overline{p}_k,\overline{t})) \rightarrow \Instruction(\overline{u}_2, \overline{t}+1)\Big)\Big>\Big]\Big].
        % \end{array} 
        % $$

        \item If the $\ell$th instruction of $\program$ is $ \ell : \ifthenbranch{Z \lbr i_1 \rbr = Z \lbr i_2 \rbr}{\ell_1}{\ell_2},$ then
        $$
        \begin{array}{rcl}
            \update_\ell & \equiv & \forall \overline{u}_0\forall\overline{u}_1\forall\overline{u}_2\forall\overline{p}_0\dots\forall\overline{p}_k\forall\overline{t}\Big[\Big(\bigwedge_{j=0}^2\overline{u} = \ell \wedge \bigwedge_{j=0}^k \overline{p}_j = i_j \wedge C(\overline{u}_0,\overline{t}) \Big) \to  \\
            & & \\
            & & \Big[\Big(Z(\overline{p}_1,\overline{t}) = Z(\overline{p}_2,\overline{t}) \rightarrow C(\overline{u}_1, \overline{t}+1)\Big) \wedge    \\
            & & \\
            & & \Big(\neg (Z(\overline{p}_1,\overline{t}) = Z(\overline{p}_2,\overline{t})) \rightarrow C(\overline{u}_2, \overline{t}+1)\Big)\Big]\Big].
        \end{array} 
        $$

        \item If the $\ell$th instruction of $\program$ is $ \ell : \ifthenbranch{P(Z \lbr i_1 \rbr, \dots, Z \lbr i_k \rbr)}{\ell_1}{\ell_2},$ then
        $$
        \begin{array}{rcl}
            \update_\ell & \equiv & \forall \overline{u}_0\forall\overline{u}_1\forall\overline{u}_2\forall\overline{p}_0\dots\forall\overline{p}_k\forall\overline{t}\Big[\Big(\bigwedge_{j=0}^2\overline{u} = \ell \wedge \bigwedge_{j=0}^k \overline{p}_j = i_j \wedge C(\overline{u}_0,\overline{t}) \Big) \to  \\
            & & \\
            & & \Big[\Big(P(Z(\overline{p}_1,\overline{t}) \dots, Z(\overline{p}_k,\overline{t})) \rightarrow C(\overline{u}_1, \overline{t}+1)\Big) \wedge    \\
            & & \\
            & & \Big(\neg P(Z(\overline{p}_1,\overline{t}) \dots, Z(\overline{p}_k,\overline{t})) \rightarrow C(\overline{u}_2, \overline{t}+1)\Big)\Big]\Big].
        \end{array} 
        $$
        
        % \item If the $\ell$th instruction of $\program$ is $\ell : Z \lbr I \lbr i_1 \rbr \rbr := Z \lbr I \lbr i_2 \rbr \rbr$, then for all $t$
        % $$\begin{array}{rcl}
        %     \update_\ell & \equiv & \forall\overline{u}\forall\overline{p}_1\forall\overline{p}_2\forall\overline{t}\Big[\overline{u}=\ell\wedge\overline{p}_1=i_1\wedge\overline{p}_2=i_2\wedge\Instruction(\overline{u},\overline{t}) \to \\ 
        %     & & \\
        %     & & \forall \overline{p}_1\forall \overline{p}_2\Big[\Big( \Index_j(\overline{p}_1, \overline{t}) \wedge \Index_{j'}(\overline{p}_2, \overline{t}) \Big) \to Z(\overline{p}_1,\overline{t}+1) = Z(\overline{p}_2,\overline{t})\Big]\Big] \\ 
        % \end{array}$$

        \item If the $\ell$th instruction of $\program$ is $\ell : Z \lbr I \lbr i_1 \rbr \rbr := Z \lbr I \lbr i_2 \rbr \rbr$, then for all $t$
        $$\begin{array}{rcl}
            \update_\ell & \equiv & \forall\overline{u}\forall\overline{t}\Big[\overline{u}=\ell\wedge C(\overline{u},\overline{t}) \to \\ 
            & & \\
            & & \forall \overline{p}_1\forall \overline{p}_2\Big[\Big( I_{i_1}(\overline{p}_1, \overline{t}) \wedge I_{i_2}(\overline{p}_2, \overline{t}) \Big) \to Z(\overline{p}_1,\overline{t}+1) = Z(\overline{p}_2,\overline{t})\Big]\Big] \\ 
        \end{array}$$
        
        % \item If the $\ell$th instruction of $\program$ is $\ell : \ifthenbranch{I \lbr i_1 \rbr = I \lbr i_2 \rbr}{\ell_1}{\ell_2}$, then for all $t$
        % $$
        % \begin{array}{rcl}
        %     \update_\ell & \equiv & \forall\overline{u}_0\forall\overline{u}_1\forall\overline{u}_2\forall\overline{t}\Big[\Big(\overline{u}_0=\ell \wedge \overline{u}_1 = \ell_1 \overline{u}_2 = \ell_2 \wedge  \Instruction(\overline{u}_0, \overline{t})\Big) \to \\ & & \\
        %     & & \Big( \exists c( I_j(c, t) \wedge I_k(c, t)) \rightarrow \Instruction(l_1, t+1) \\ & & \\
        %     & \wedge & \neg (\exists c ( I_j(c, t) \wedge I_k(c, t)) \rightarrow \Instruction(l_2, t+1) )\Big).
        % \end{array}$$

        \item If the $\ell$th instruction of $\program$ is $\ell : \ifthenbranch{I \lbr i_1 \rbr = I \lbr i_2 \rbr}{\ell_1}{\ell_2}$, then for all $t$
        $$
        \begin{array}{rcl}
            \update_\ell & \equiv & \forall\overline{u}_0\forall\overline{u}_1\forall\overline{u}_2\forall\overline{t}\Big[\Big(\overline{u}_0=\ell \wedge \overline{u}_1 = \ell_1 \wedge  \overline{u}_2 = \ell_2 \wedge  C(\overline{u}_0, \overline{t})\Big) \to \\ & & \\
            & & \forall\overline{p}_1\forall\overline{p}_2\Big[\Big(I_{i_1}(\overline{p}_1,\overline{t}) \wedge I_{i_2}(\overline{p}_2,\overline{t}) \wedge \overline{p}_1 = \overline{p}_2 \to C(\overline{u}_1,\overline{t}+1)\Big) \wedge  \\ & & \\
            & &  \hspace{1cm}\Big(I_{i_1}(\overline{p}_1,\overline{t}) \wedge I_{i_2}(\overline{p}_2,\overline{t}) \wedge \neg(\overline{p}_1 = \overline{p}_2) \to C(\overline{u}_1,\overline{t}+1)\Big)\Big]\Big].
            
        \end{array}$$

        \item If the $\ell$th instruction of $\program$ is $\ell : I \lbr i \rbr := I \lbr i \rbr + 1$, then 
        $$
        \begin{array}{rcl}
            \update_\ell & \equiv &  \forall\overline{u}\forall\overline{t}\Big[\Big(\overline{u} = \ell \wedge C(\overline{u},\overline{t}) \to \\
            & & \\ 
            & & \forall\overline{p}\Big[I_i(\overline{p},\overline{t}) \to I_i(\overline{p}+1,\overline{t}+1)\Big]\Big].
        \end{array}$$

        \item If the $\ell$th instruction of $\program$ is $\ell : I \lbr i \rbr := I \lbr i \rbr \monus 1$, then 
        $$
        \begin{array}{rcl}
            \update_\ell & \equiv &  \forall\overline{u}\forall\overline{t}\Big[\Big(\overline{u} = \ell \wedge C(\overline{u},\overline{t}) \to \\
            & & \\ 
            & & \forall\overline{p}\Big[I_i(\overline{p},\overline{t}) \to I_i(\overline{p} \monus 1,\overline{t}+1)\Big]\Big].
        \end{array}$$

        \item If the $\ell$th instruction of $\program$ is $\ell_\program : \mathsf{stop}$, then 
        $$
        \begin{array}{rcl}
            \update_\ell & \equiv & \forall\overline{u}\forall\overline{p}_0 \dots \forall\overline{p}_{k_\program-1}\forall\overline{t}\Big[\Big(\overline{u}=\ell \wedge C(\overline{u},\overline{t}) \wedge \bigwedge_{i=0}^{k_\program-1} I_i(\overline{p}_i,\overline{t}) \Big) \to \\
            & & \\
            & & \Big(C(\overline{u},\overline{t}+1) \wedge Z(\overline{p},\overline{t})=Z(\overline{p},\overline{t}) \wedge \bigwedge_{i=0}^{k_\program-1}I_i(\overline{p}_i,\overline{t}+1)\Big)\Big].
        \end{array}$$
\end{enumerate}

In each case defining $\update_\ell$, we have additional clauses stating that every register that is not modified by the instruction remains unchanged at time $\overline{t}+1$, and that if instruction $\ell$ is not a branching instruction, then $\ell+1$ is the next instruction executed. In particular, the last case of the definition $\update_\ell$ asserts that if $\M$ reaches the instruction $\ell_\program : \mathsf{stop}$ at time $\overline{t}$, then the contents of the registers of $\M$ will not change at any time in between time $\overline{t}$ and time $n^m-1$. We define the formula $\update$ as the conjunction $$\update \equiv \update_0 \wedge \dots \wedge \update_{\ell_\program}.$$

Lastly, we must encode the statement that $\M$ accepts the input $(\D,Y)$. By definition, $\M$ accepts its input when it stops with $1$ in its zeroth index register $I \lbr 0 \rbr$, indicating that the length of the output is 1, and 1 in its zeroth $\rr$ register $Z\lbr 0 \rbr$, giving the output. We express this with the formula $$\accept \equiv I_0(1,n^m-1) \wedge Z(1,n^m-1)=1.$$ 

The formula $\varphi_\M \equiv \start \wedge \update \wedge \accept$ is a first-order sentence in vocabulary $$\MVoc \cup \{Y,C,I_0,\dots,I_{k_\program-1},E,S,P,Z\},$$ which means that $$\Phi_\M \equiv \exists Y \exists C \exists I_0 \dots \exists I_{k_\program-1} \exists Z \exists E \exists S \exists P \ \varphi_\M(W,C,I_0,\dots,I_{k_\program-1},E,S,P,Z)$$ is an existential second-order sentence in vocabulary $\MVoc$. We now show that $\D \in L$ if and only if $\D \satisfies \Phi_\M$.

Suppose $\D \in L$. Then there is some $Y \: A^k \to R$ such that $\M(\D,Y) = 1$. We may use the computation of $\conf_0(\D,Y) \to \conf_1 \to \dots \to \conf_{n^m-1}$ of $\M$ on input $(\D,Y)$ to determine relations $C$ and $I_0, \dots, I_{k_\program-1}$. For example, we let $C \subseteq A^{2m}$ be the relation that contains $(\overline{u},\overline{t}) \in A^{2m}$ such that $\overline{u}$ is the $n$-ary representation of the instruction label $\ell$ that $\M$ executes at time $\overline{t}$ on input $(\D,Y)$. We let $I_0,\dots, I_{k_\program-1}$ be determined by the computation of $\M$ on input $(\D,Y)$ in a similar way. 

When $(\D,Y)$ is fed to $\M$ as an input, we implicitly assume there is some chosen linear ordering $E \subseteq A^2$ on the primary universe of $\D$. This is so that we can define $\code_E((\D,Y))$ as a string in $R^*$, as discussed in \Cref{def:EncodingMetafiniteStructures}. Let $E$ be this chosen linear ordering, and let $S \subseteq A^{3m}$ and $P \subseteq A^{3m}$ be the graphs of addition and multiplication in $A^m$ lexicographically ordered according to $E$. Finally, the computation of $\M$ on input $(\D,Y)$ determines a weight function $Z \: A^{2m} \to R$  by defining $Z(\overline{p},\overline{t})$ to be the value $r \in R$ that $\M$ holds in its $\overline{p}$th $\rr$-register at time $\overline{t}$. 

Since the subformulas of $\varphi_\M$ constraining these relations and weight functions say precisely that they encode an accepting computation of $\M$ on input $(\D,Y)$, we can see that $$(\D,Y,C,I_0,\dots,I_{k_\program-1},E,S,P,Z) \satisfies \varphi_\M(W,C,I_0,\dots,I_{k_\program-1},E,S,P,Z),$$ and this implies that $\D \satisfies \Phi_\M$.

Conversely, suppose $\D \satisfies \Phi_\M$. Then there are relations and weight functions such that $$(\D,Y,C,I_0,\dots,I_{k_\program-1},E,S,P,Z) \satisfies \varphi_\M(W,C,I_0,\dots,I_{k_\program-1},E,S,P,Z).$$ In particular, $$(\D,Y,C,I_0,\dots,I_{k_\program-1},E,S,P,Z) \satisfies (\start \wedge \update)(W,C,I_0,\dots,I_{k_\program-1},E,S,P,Z),$$ so these relations and weight functions encode the computation of $\M$ on input $(\D,Y)$. Since it is also the case that $$(\D,Y,C,I_0,\dots,I_{k_\program-1},E,S,P,Z) \satisfies \accept(W,C,I_0,\dots,I_{k_\program-1},E,S,P,Z),$$ it must be that $\M$ accepts input $(\D,Y)$. Thus, $\D \in L$.
\end{proof}

\Cref{thm:SigmaSORDescribesNPR} together with \Cref{thm:DataComplexitySigmaSOR} gives us the following corollary. 

\begin{restatable}[Fagin's Analogue]{theorem}{Fagin}
\label{thm:Fagin}
    Let $\rr$ be a bipointed structure. Then $\Sigma_1\SO(\rr)$ captures $\NP(\rr)$ on metafinite structures over $\rr$.
\end{restatable}

\begin{example}
\label{example:SOFormulaDescribingSAT(R)}
    \Cref{thm:SAT_R-NP_R-Membership} tells us that if $\rr$ is of finite type, then $\SAT(\rr) \in \NP(\rr)$. We will construct a sentence $\Phi$ of $\Sigma_1\SO(\rr)$ such that $\D \satisfies \Phi$ if and only if $\D$ encodes a quantifier-free $\Voc(\rr)$-formula $\psi$ such that $\rr \satisfies \exists \overline{x} \psi(\overline{x})$. Note that since $\Voc(\rr) = (\Con,\Fun,\Rel)$ is of finite type, there are finitely many function and relation symbols in this vocabulary. The metafinite vocabulary $\MVoc$ that we use has a primary vocabulary that consists of the relation symbols
    $$\{\text{Var}, \text{Con}, \text{Term}, \text{Subform}, \text{Form}\} \cup \{F_f : f \in \Fun\} \cup \{Q_P : P \in \Rel\} \cup  \{\text{Equal}\} \cup \{\text{Not}, \text{And}, \text{Or}\}.$$
    These relation symbols will specify the syntactic structure of $\psi$. The metafinite vocabulary $\MVoc$ also has a weight function symbol $C$ that will specify the constants occurring in $\psi$. The $\rr$-structure $\D_\psi = (\A,\rr,\W)$ encoding $\psi$ will have primary universe $A = \{1,\dots,n\}$, which has one element for each term and subformula occurring in $\psi$. Essentially, $\D_\psi$ will encode the syntax tree of $\psi$ in the following way:
    \begin{enumerate}
        \item $\text{Var}(i)$ indicates that $i$ represents a variable occurring in $\psi$,
        \item $\text{Con}(i)$ indicates that $i$ represents a constant occurring in $\psi$,
        \item $\text{Term}(i)$ indicates that $i$ represents a term occurring in $\psi$,
        \item $\text{Subform}(i)$ indicates that $i$ represents a subformula of $\psi$,
        \item $\text{Form}(i)$ indicates that $i$ represents the formula $\psi$,
        \item $F_f(i,j)$ indicates that term $j$ results from applying $f$ to term $i$,
        \item $Q_P(i_1,\dots,i_k,j)$ indicates that subformula $j$ results from applying relation symbol $P$ to terms $i_1,\dots,i_k$,
        \item $\text{Equal}(i_1,i_2,j)$ indicates that subformula $j$ results from applying $=$ to terms $i_1$ and $i_2$,
        \item $\text{Not}(i,j)$ indicates that subformula $j$ results from applying $\neg$ to subformula $i$,
        \item $\text{And}(i_1,i_2,j)$ indicates that subformula $j$ results from applying $\wedge$ to subformulas $i_1$ and $i_2,$
        \item $\text{Or}(i_1,i_2,j)$ indicates that subformula $j$ results from applying $\vee$ to subformulas $i_1$ and $i_2.$
    \end{enumerate}
    For example, if $\psi$ is the formula $(f(g(x)) = c) \wedge P(y,y)$, then we may represent $\psi$ with the following enumerated syntax tree:

    \begin{center}
    \Tree [.$8:(f(g(x))=c)\wedge P(y,y)$ [.$6:f(g(x))=c$ [.$5:f(g(x))$ [.$4:g(x)$ $1:x$ ] ] [.$3:c$ ] ]
    [.$7:P(y,y)$ $2:y$ ] ]
    \end{center}
    We then encode this syntax tree into an $\rr$-structure $\D_\psi$ by specifying the following information:
    
        \begin{enumerate}
            \item $A = \{1,\dots,8\}$
            \item $\text{Var} = \{1,2\}$
            \item $\text{Con} = \{3\}$
            \item $\text{Term} = \{1,2,3,4,5\}$
            \item $\text{Subform} = \{6,7,8\}$
            \item $\text{Form} = \{8\}$
            \item $F_g = \{(1,4)\}$
            \item $F_f = \{(4,5)\}$
            \item $Q_P = \{(2,2,7)\}$
            \item $\text{Equal} = \{(5,3,6),(6,7,8)\}$,
            \item $C \: A \to R$ is the weight function such that $C(i) = \begin{cases}
                c & \text{ if } i = 3 \\
                0 & \text{ if } i \neq 3.
            \end{cases}$
        \end{enumerate}
    The $\MVoc$-sentence $\Phi$ we construct will consist of two parts: first, a subformula $\valid$ asserting that an $\rr$-structure $\D$ encodes a valid $\Voc(\rr)$-formula, and second, a subformula $\satisfied(X)$ asserting that $X \: A \to R$ specifies $\overline{r}$ such that $\rr \satisfies \psi(\overline{r})$. Let $M$ be the maximum arity of a symbol in $\Fun \cup \Rel$ of $\Voc(\rr)$. Then $\valid$ is the formula whose prefix is $\forall i_1 \dots \forall i_M \forall i \forall j \exists t$ and whose matrix is the conjunction of the following formulas:
    \begin{enumerate}
        \item $(\text{Var}(i) \vee \text{Con}(i)) \to  \text{Term}(i)$
        \item $\text{Term}(i) \fromto \neg \text{Subform}(i)$
        \item $\text{Form}(i) \to \text{Subform}(i)$
        \item $\text{Form}(t)$
        \item $(\text{Form}(i_1) \wedge \text{Form}(i_2)) \to i_1 = i_2$
        \item $\bigwedge_{f \in \Fun} F_f(i_1,\dots,i_{k_f},j) \to (\bigwedge_{\ell = 1}^{k_f} \text{Term}(i_\ell) \wedge \text{Term}(j))$ 
        \item $\bigwedge_{P \in \Rel} Q_P(i_1,\dots,i_{k_P},j) \to (\bigwedge_{\ell = 1}^{k_P} \text{Term}(i_\ell) \wedge \text{Subform}(j))$
        \item $\text{Equal}(i_1,i_2,j) \to (\text{Term}(i_1) \wedge \text{Term}(i_2) \wedge \text{Subform}(j))$
        \item $\text{Not}(i,j) \to (\text{Subform}(i) \wedge \text{Subform}(j))$
        \item $\text{And}(i_1,i_2,j) \to (\text{Subform}(i_1) \wedge \text{Subform}(i_2) \wedge \text{Subform}(j))$
        \item $\text{Or}(i_1,i_2,j) \to (\text{Subform}(i_1) \wedge \text{Subform}(i_2) \wedge \text{Subform}(j))$
    \end{enumerate}
    The subformula $\satisfied(X)$ has a free second-order variable $X$ ranging over weight functions $X \: A \to R$ that specifies values $\overline{r}$ to assign to the variables $\overline{x}$ of $\psi$, as well as the value $t^\rr(\overline{r}) \in R$ of each term $t$ that occurs in $\psi$ and the truth value $\psi'(\overline{r}) \in \{0,1\}$ of each subformula $\psi'$ occurring in $\psi$. The subformula $\satisfied(X)$ asserts that $X$ correctly computes the value of terms and the truth value of subformulas occurring in $\psi$ by using the prefix $\forall i_1 \dots \forall i_M \forall i \forall j$ and the matrix resulting from the conjunction of the following formulas:
    \begin{enumerate}
        \item $\text{Con}(i) \to X(i) = C(i)$
        \item $\bigwedge_{f \in \Fun} F_f(i_1,\dots,i_{k_f},j) \to X(j) = f(X(i_1),\dots,X(i_{k_f}))$
        \item $\text{Subform}(i) \to (X(i) = 0 \vee X(i) = 1)$
        \item $\bigwedge_{P \in \Rel} Q_P(i_1,\dots,i_{k_P},j) \to (X(j) = 1 \fromto P(X(i_1),\dots,X(i_{k_P})))$
        \item $\text{Equal}(i_1,i_2,j) \to (X(j) = 1 \fromto X(i_1) = X(i_2))$
        \item $\text{Not}(i,j) \to (X(j) = 1 \fromto X(i) = 0)$
        \item $\text{And}(i_1,i_2,j) \to (X(j) = 1 \fromto X(i_1) = 1 \wedge X(i_2) = 1)$
        \item $\text{Or}(i_1,i_2,j) \to (X(j) = 1 \fromto X(i_1) = 1 \vee X(i_2) = 1)$
    \end{enumerate}
    The sentence of $\Sigma_1\SO(\rr)$ expressing that $\D$ encodes some $\psi(\overline{x}) \in \SAT(\rr)$ is then $$\Phi \equiv \exists X \ \valid \wedge \satisfied(X).$$
    \end{example}

%%%%%%%%%%%%%%%%%%%%%%%%%%%%%%%%%%%%%%%%%%%
\subsection{$\exists\SO(\rr)$ Captures $\exists \rr$}
\label{sub:BooleanESO^0(R)capturesBP^0(NP(R))}
%%%%%%%%%%%%%%%%%%%%%%%%%%%%%%%%%%%%%%%%%%%

In this subsection, we prove \Cref{thm:BooleanFagin}, which says that $\exists\SO(\rr)$ captures $\exists\rr$ over Boolean $\rr$-structures. According to \Cref{def:LogicCapturesComplexityClassOnStructures}, showing this requires showing two things for any metafinite vocabulary $\MVoc$ over $\rr$: (1) If $\Phi$ is an $\exists\SO(\Voc)$ sentence, then $\{\D \in \Bstruct(\Voc) : \D \satisfies \Phi\} \in \exists\rr$, and (2) If $L$ is a decision problem of Boolean $\rr$-structures in vocabulary $\Voc$ such that $L \in \exists\rr$, then there is some $\exists\SO(\Voc)$ sentence $\Phi$ such that $L = \{\D \in \Bstruct(\Voc) : \D \satisfies \Phi\}.$ The proofs of these statements are simple modifications of the corresponding statements for $\Sigma_1\SO(\rr)$ and $\NP(\rr)$.

\begin{lemma}
\label{lemma:BooleanFirstOrderMetafiniteEvaluationPolynomialTime}
    For any metafinite vocabulary $\MVoc$ over $\rr$, if $\varphi$ is a sentence of first-order metafinite logic in this vocabulary such that the only constants from $\rr$ that occur in $\varphi$ are 0 and 1, then $\{\D \in \Bstruct(\Voc) : \D \satisfies \varphi\} \in P(\rr)^0.$
\end{lemma}
\begin{proof}
    Assume without loss of generality that $\varphi$ is in prenex normal form, and recall that $\P(\rr)^0$ is the class of decision problems that have a decision algorithm implemented on a polynomial-time $\rr$ machine with no machine constants other than 0 and 1.
    From \Cref{lemma:FirstOrderMetafiniteEvaluationPolynomialTime}, we know that we can decide if $\D \satisfies \varphi$ in polynomial time with an $\rr$-machine $\M$. In order for $\M$ to execute the algorithm described in \Cref{lemma:FirstOrderMetafiniteEvaluationPolynomialTime}, only the constants appearing in $\varphi$ need to be machine constants of $\M$. Since only $0$ and $1$ occur in $\varphi$, we can see that $\M$ is constant free. 
\end{proof}

\begin{theorem}
\label{thm:DataComplexityExistsSOR}
    For any metafinite vocabulary $\MVoc$, if $\Phi$ is a sentence of $\exists\SO(\MVoc)$, then $$\{\D \in \Bstruct(\Voc) : \D \satisfies \Phi\} \in \exists\rr.$$
\end{theorem}
\begin{proof}
    Since we restrict $L = \{\D \in \Bstruct(\MVoc) : \D \satisfies \Phi\}$ to Boolean $\rr$-structures, if we can show that $L \in \NP(\rr)^0,$ we then will have shown that $L \in \BP(\NP(\rr)^0) = \exists\rr$. Any $\exists\SO(\MVoc)$ sentence $\Phi$ is of the form $\exists X_1 \dots \exists X_h \varphi(X_1,\dots,X_h)$, where $X_1,\dots,X_h$ are second-order variables and $\varphi(W_1,\dots,W_h)$ is a first-order formula in vocabulary $\MVoc \cup \{X_1,\dots,X_h\}$, expanded by regarding the variable $X_i$ as a weight function symbol. Given any interpretation of each $X_i$ as a Boolean weight function $W_i \: A^k \to \{0,1\} \subseteq R$, \Cref{lemma:BooleanFirstOrderMetafiniteEvaluationPolynomialTime} tells us that $$\{(\D,W_1,\dots,W_h) : (\D,W_1,\dots,W_h) \satisfies \varphi(W_1,\dots,W_h)\} \in P(\rr)^0.$$ Thus, $L = \{\D : \exists W_1 \dots \exists W_h (\D,W_1,\dots,W_h) \satisfies \varphi(W_1,\dots,W_h)\} \in \NP(\rr)^0.$ This argument is essentially the same as the argument occurring in the proof of \Cref{thm:DataComplexitySigmaSOR}, except using \Cref{lemma:BooleanFirstOrderMetafiniteEvaluationPolynomialTime} instead of \Cref{lemma:FirstOrderMetafiniteEvaluationPolynomialTime}.
\end{proof}

\begin{theorem}
\label{thm:ExistsSORDescribesER}
    For every metafinite vocabulary $\MVoc$ over $\rr$, if $L$ is a decision problem of Boolean $\rr$-structures such that $L \in \exists\rr$, then there is a sentence $\Phi$ of $\exists\SO(\Voc)$ such that $$L = \{\D \in \Bstruct(\Voc) : \D \satisfies \Phi\}.$$
\end{theorem}
\begin{proof}
    Since $L \in \NP(\rr)$, there is some positive integer $k$ and some polynomial-time $\rr$-machine $\M$, with no machine constants other than 0 and 1, such that $$L = \{\D \in \Bstruct(\Voc) : (\exists Y \: A^k \to R) \ \M(\D,Y) = 1\}.$$ With the same method as in the proof of \Cref{thm:SigmaSORDescribesNPR}, we can construct a $\Sigma_1\SO(\MVoc)$ sentence $\Phi$ such that there is some $Y\:A^k \to \R$ for which $\M(D,Y) = 1$ if and only if $D \satisfies \exists Y \Phi_\M(Y)$. It remains to show that our method of construction results in $\Phi_\M$ having no constants from $\rr$ other than 0 and 1 when the same is true of $\M$. The only place where constants from $\rr$ are potentially introduced into $\Phi_\M$ is in case 1 of the definition of $\update_\ell$, for instructions in the program $\program$ of $\M$ of the form $\ell : Z \lbr i \rbr := c$. Since $\M$ uses only the constants 0 and 1, we know that any $c$ occurring in an $\update_\ell$ of this form is either 0 or 1. Thus, $\exists Y \Phi_\M(Y)$ is an $\exists \SO(\Voc)$ formula.
\end{proof}

Theorems \ref{thm:DataComplexityExistsSOR} and \ref{thm:ExistsSORDescribesER} together give us the following corollary.

\begin{restatable}[Boolean Fagin's Analogue]{corollary}{BooleanFagin}
\label{thm:BooleanFagin}
    Let $\rr$ be a bipointed structure. Then $\exists\SO(\rr)$ captures $\exists \rr$ on Boolean metafinite structures over $\rr$.
\end{restatable}

%%%%%%%%%%%%%%%%%%%%%%%%%%%%%%%%%%%%%%%%%%%
\subsection{$\Sigma_k\SO(\rr)$ Captures $\Sigma_k\rr$ and $\exists_k\SO(\rr)$ Captures $\exists_k\rr$}
\label{sub:SOCapturesPH}
%%%%%%%%%%%%%%%%%%%%%%%%%%%%%%%%%%%%%%%%%%%

In this subsection, we prove \Cref{thm:HigherFagin} and \Cref{thm:HigherBooleanFagin}. A common way to prove that the $k$th existential fragment of second-order logic $\Sigma_k\SO$ captures the $k$th existential level of the polynomial hierarchy $\Sigma_k$ is to use the oracle characterization of $\Sigma_k$ and proceed by induction. This proof method generalizes to establish \Cref{thm:HigherFagin} that $\Sigma_k\SO(\rr)$ captures $\Sigma_k\rr$, but we cannot modify it to get a proof of \Cref{thm:HigherBooleanFagin} that $\exists_k\SO(\rr)$ captures $\exists_k\rr$.

To get an intuition for what goes wrong, suppose that we have already proven that $\Sigma_k\SO(\rr)$ captures $\Sigma_k\rr$ and we now want to show that $\Sigma_{k+1}\SO(\rr)$ captures $\Sigma_{k+1}\rr = \NP(\rr)^{\Sigma_k\rr}.$ For any $L \in \NP(\rr)^{\Sigma_k\rr}$, there is a polynomial-time verification $\rr$-machine $\M$ with access to some oracle $S \in \Sigma_k\rr$ such that $\M$ verifies $L$. By our inductive hypothesis, there is a sentence $\Psi$ of $\Sigma_k\SO(\rr)$ such that $\D \in S$ if and only if $\D \satisfies \Psi$. With a suitable modification to $\Phi_\M$ as constructed in the proof of \Cref{thm:SigmaSORDescribesNPR}, we can express any oracle calls that $\M$ makes to $S$ with the sentence $\Psi$ placed inside of $\Phi_\M$, and bringing the second-order quantifiers to the front of $\Phi_\M$ results in one more quantifier alternation. Thus, $\Sigma_{k+1}\SO(\rr)$ captures $\Sigma_{k+1}\rr$, completing our inductive argument. 

However, if we try to adapt this proof to the Boolean analogue, the sentence $\Psi$ causes a problem. Even though restricting $\M$ to have no constants other than 0 and 1 means that the same is true of $\Phi_\M$, in general $\Psi$ may have constants other than 0 and 1 because it describes a machine $N$ that may have constants other than 0 or 1, which follows from the definition of $\Sigma_k\rr$. If we could replace $\Sigma_k\rr$ with $\exists_k\rr$, then our problem would be solved, but it is unlikely that $\exists \rr^{\Sigma_k\rr} = \exists\rr^{\exists_k\rr}$ in general. Since this simpler proof technique does not work, we will directly adapt the proof that $\Sigma_1\SO(\rr)$ captures $\NP(\rr)$ to show that $\Sigma_k\SO(\rr)$ captures $\Sigma_k\rr$.

Towards this end, fix a metafinite vocabulary $\MVoc$ over $\rr$, and let $\M$ be a polynomial time $\rr$-machine. Construct the formulas $\update$ and $\accept$ as in the proof of \Cref{thm:SigmaSORDescribesNPR}, but modify the formula $\start$ constructed therein to become a family of formulas $\start_{j,k}$, which results from building the subformula $\Input$ for the vocabulary $$\Voc \cup \{X_1,\dots,X_j,Y_1,\dots,Y_k\}$$ via the same method used in  \Cref{lemma:FirstOrderFormulaEncodingStartingConfiguration}. (Previously we had built $\Input$ specifically for the vocabulary $\Voc \cup\{Y\}$.)
%\hans{The following sentence is long; maybe break it in two or more smaller pieces.}
%We need to prove a slightly more complicated statement than we might expect in order to proceed by induction on $k$, so following \Cref{definition:ReducingFormulaForHigherLevels} and the surrounding discussion, 
Define the family of formulas $\varphi_{\M,k}$ and $\widetilde{\varphi}_{\M,k}$ for all $k \in \N$ as follows. 
%If $k$ is even, then 
% $$\begin{array}{rcl}
%     \varphi_{\M,k} & \equiv & (\start_{\M,k} \wedge \update) \to \accept \\
%     \widetilde{\varphi}_{\M,k} & \equiv & \ \start_{\M,k} \wedge \update \wedge \accept.
% \end{array}$$
% If $k$ is odd, then 
$$\begin{array}{rcl}
    \varphi_{\M,k} & \equiv & \ \start_{\M,k} \wedge \update \wedge \accept \\
    \widetilde{\varphi}_{\M,k} & \equiv & (\start_{\M,k} \wedge \update) \to \accept.
\end{array}$$ 

The heart of the idea is captured in the following lemma. 
\begin{lemma}[Encoding Computations in Second-Order Logic]
\label{lemma:HigherFaginEncoding}
    For all $\rr$-structures $\D$ in vocabulary $\Voc$ and for all interpretations of $Y_i$ as weight functions $Y_i \: A^q \to R$, the construction of these formulas implies that the following are equivalent:
\begin{enumerate}
    \item $\M(\D,Y_1,\dots,Y_k) = 1$
    \item $\D \satisfies \varphi_{\M,k}(Y_1,\dots,Y_k)$
    \item $\D \satisfies \widetilde{\varphi}_{\M,k}(Y_1,\dots,Y_k).$
\end{enumerate}
\end{lemma}
\begin{proof}
    The proof follows essentially the same argument as that given in the proof of Fagin's theorem over $\rr$, shown in \Cref{thm:Fagin}.
\end{proof}

%\jeremy{I should update the formulation of this statement to include $Y$'s in $\varphi$, and possibly change them to $W$'s. I also need to quantify over these formulas...}
Note that the above lemma implies \Cref{lemma:EncodingComputationSOFormula} from the main body of this paper. Given this, it is intuitively clear that we can stick the appropriate alternating sequence of quantifiers in front of these statements to get the desired result. Indeed, this is justified by the replacement theorem \cite{kleene2002mathematical}, which says we may use logically equivalent statements interchangeably. We just need the following theorem before we arrive at our main result.

\begin{theorem}
\label{thm:DataComplexitySigmaKSOR}
    For any metafinite vocabulary $\MVoc$ over $\rr$, and for any $k \in \N$ if $\Phi$ is a sentence of $\Sigma_k\SO(\MVoc)$, then $\{\D \in \struct(\MVoc) : \D \satisfies \Phi\} \in \Sigma_k\rr$, and if $\widetilde{\Phi}$ is a sentence of $\Pi_k\SO(\rr)$, then $\{\D \in \struct(\MVoc) : \D \satisfies \widetilde{\Phi}\} \in \Pi_k\rr$.
\end{theorem}
\begin{proof}
    If $\Phi \in \Sigma_k\SO(\rr)$, then $\Phi$ is of the form $(\exists \overline{Y}_1) (\forall \overline{Y}_2) \dots (Q \overline{Y}_k) \varphi$, and it is intuitively clear that $\{\D : \exists \overline{Y}_1\forall \overline{Y}_2 \dots Q \overline{Y}_k (\D,\overline{Y}_1,\dots,\overline{Y}_k) \satisfies \varphi\} \in \Sigma_k\rr$ because we can test if structures satisfy first-order formulas by \Cref{lemma:FirstOrderMetafiniteEvaluationPolynomialTime}.
\end{proof}

As a corollary of \Cref{lemma:HigherFaginEncoding} and \Cref{thm:DataComplexitySigmaKSOR}, we have the desired theorem.

\HigherFagin*

The same sort of analysis as in \Cref{thm:DataComplexityExistsSOR} and \Cref{thm:ExistsSORDescribesER} can be generalized to give us the following corollary.

\HigherBooleanFagin*

%%%%%%%%%%%%%%%%%%%%%%%%%%%%%%%%%%%%%%%
\section{Oracles and Hierarchies}
\label{sec:OracleMachinePolynomialHierarchy}
%%%%%%%%%%%%%%%%%%%%%%%%%%%%%%%%%%%%%%%

%%%%%%%%%%%%%%%%%%%%%%%%%%%%%%%%%%%%%%%
\subsection{Oracles for $\Sigma_k\rr$}
%%%%%%%%%%%%%%%%%%%%%%%%%%%%%%%%%%%%%%%

%\jeremy{I need to update this section to use the $q(n)$ witness definition, rather than the $\leq q(n)$ definition.}

We begin with some technical lemmas that simplify the main result.

\begin{lemma}
\label{lemma:Pi_kRSubseteqP(R)^Sigma_kR}
    For all $k \in \N$, $\Pi_k\rr \subseteq \P(\rr)^{\Sigma_k\rr}.$
\end{lemma}
\begin{proof}
    Suppose $L \in \Pi_k\rr$. Then $\overline{L} \in \Sigma_k \rr$. Let $\M$ be the machine with a $\Sigma_k\rr$ oracle that on input $v$ returns 0 if $v \in \overline{L}$ and 1 if $v \not\in \overline{L}.$ Since $\M$ runs in polynomial time and decides $L$, we can see that $L \in \P(\rr)^{\Sigma_k\rr}.$  
\end{proof}

\begin{lemma}
\label{lemma:P(R)Sigmak+1IsSubsetSigmaPiP(R)Sigmak}

    For all $k \in \N$, $\P(\rr)^{\Sigma_{k+1}\rr} \subseteq \Sigma \Pi   \left(\P(\rr)^{\Sigma_k\rr}\right).$
\end{lemma}
\begin{proof}
    Suppose $L \in P(\rr)^{\Sigma_{k+1}\rr}$, let $\M$ be a polynomial time $\rr$-machine with a $Q$ oracle, for some $Q \in \Sigma_{k+1}\rr$, such that $\M$ decides $L$. Take some arbitrary element $v$ of $R^*$. On input $v$, $\M$ will make finitely many queries $u_1,\dots,u_p$ by executing oracle query instructions, and each oracle query instruction is of the form $$\ifthenbranch{u_i \in Q}{\ell_{(i,1)}}{\ell_{(i,0)}},$$ where $\ell_{(i,1)}$ and $\ell_{(i,0)}$ are numbers labeling instructions in the program of $\M$. We can encode the answer that the oracle returns after query $u_i$ in a single bit $b_i$ such that 
    $$
    b_i = \begin{cases}
        1 & \text{ if } u_i \in Q \\
        0 & \text{ if } u_i \not\in Q.
    \end{cases}
    $$
    
    If we were given the number $p$ of queries that $\M$ will make on input $v$, as well as the answers $b_1,\dots,b_p$ to those queries, then we could simulate $\M$ on input $v$ with an $\rr$-machine that does not have a $Q$ oracle in the following way: Simulate $\M$ until it makes its first oracle query. If $b_1=1$, then continue to simulate $\M$ starting from instruction $\ell_{(1,1)}$. If $b_1=0$, then continue to simulate $\M$ starting from instructions $\ell_{(1,0)}$. Continue to simulate $\M$ in this manner, using bit $b_i$ to decide which way to branch upon query $i$, until $\M$ halts. This allows us to rephrase membership in $L$ in terms of this answer sequence as follows: 
    $$
    {
    \renewcommand{\arraystretch}{1.5}
    \begin{array}{rcl}
        v \in L & \iff & \exists (p,(b_1,\dots,b_p))  \\
         & & [\text{On input $v$, if $\M$ branches according to answers $b_1,\dots,b_p,$} \\
         & & \text{then $\M$ will make queries $u_1,\dots,u_p$ before accepting $v$}] \wedge \\
         & & [\bigwedge_{i=1}^p (b_i=1 \implies u_i \in Q] \wedge \\
         & & [\bigwedge_{i=1}^p (b_i=0 \implies u_i \not\in Q].
    \end{array}
    }
    $$

    Of course, to know that our simulation of $\M$ on $v$ returns the correct result, we must know that the answers $b_1, \dots,b_p$ are correct, which is what the last two clauses of the above condition guarantee. It is unlikely that any $\rr$-machine is able to determine whether or not $u_i \in Q$ in polynomial time without using an oracle. However, because $Q \in \Sigma_{k+1}\rr = \Sigma \left(\Pi_k\rr\right),$ there is a simpler oracle $S \in \Pi_k\rr$ that we can use to make this determination with the help of witness strings. In particular, $S$ can tell us about $u_i$ via the following equivalences:

    % Each of the queries $\varphi_i$ is of the form $\exists x_i \psi_i(x_i)$ where $\exists x_i$ stands for the quantifier block $\exists x_{(i,1)}\dots\exists x_{(i,n_{i})},$ and $\psi_i$ is a $\Pi_k$ formula in the vocabulary of $\rr$ that may now have free variables after removing the $\exists$ quantifiers. Each $w_i \in R^{n_i}$ induces a $\Pi_k$ sentence $\psi_i(w_i)$ that results from replacing the free variables of $\psi_i$ with the values specified in $w_i = (w_{(i,1)},\dots,w_{(i,n_i)})$. This is a sentence in the vocabulary of $\rr$ because we assume that $\rr$ has all constants, so in particular it has the constants occurring in $w_i$. 

    % Given this, we can see that $\varphi_i \in \Sigma_{k+1}\SAT(\rr)$ if and only if $(\exists w_i \in R^{n_i})\, \psi_i(w_i) \in \Pi_k\SAT(\rr)$. In order to directly use a $\Sigma_k\SAT(\rr)$ oracle, we will replace the $\Pi_k$ formula $\psi_i$ with a $\Sigma_k$ formula. The formula $\neg \psi_i$ is logically equivalent to the $\Sigma_k$ formula $\overline{\psi}_i$ that is obtained by pushing the negation operator to the front of the quantifier free portion of $\psi_i$ and changing each quantifier from $\forall$ to $\exists$ and vice versa. Noting that $\rr \satisfies \psi_i(w_i)$ if and only if $\rr \not\satisfies \neg \psi(w_i)$ if and only if $\rr \not\satisfies \overline{\psi}_i(w_i)$, we can see the following equivalences:
    $$
    \begin{array}{rcl}
        u_i \in Q & \textup{ if and only if } & (\exists w_i \in R^{n_i}) \, (u_i,w_i) \in S,\\
        & & \\
        u_i \not\in Q & \textup{ if and only if } & (\forall w'_i \in R^{n_i}) \, (u_i,w'_i) \not\in S.
    \end{array}
    $$

    The implication $b_i = 1 \implies u_i \in Q$ partially guaranteeing the correctness of answer $b_i$ is then equivalent to $(\exists w_i \in R^{n_i}) \, [b_i=1 \implies  (u_i,w_i) \in S].$ Note that pulling the existential quantifier to the front of the implication, as we have done, is logically equivalent to leaving it in front of the statement $(u_i,w_i) \in S$. Thus, we can rephrase the condition determining membership in $L$ as follows:
    $$
    {
    \renewcommand{\arraystretch}{1.5}
    \begin{array}{rcl}
        v \in L & \iff & \exists (p,(b_1,\dots,b_p),(w_1,\dots,w_p)) \forall(w'_1,\dots,w'_p)  \\

         & & [\text{On input $v$, if $\M$ branches according to answers $b_1,\dots,b_p,$} \\

         & & \text{then $\M$ will make queries $u_1,\dots,u_p$ before accepting $v$}] \wedge \\

         & & [\bigwedge_{i=1}^p (b_i=1 \implies (u_i,w_i) \not\in \overline{S}] \wedge \\

         & & [\bigwedge_{i=1}^p (b_i=0 \implies (u_i,w'_i) \in \overline{S}].
    \end{array}
    }
    $$

    Since $\M$ runs in polynomial time, the number $p$ of queries that $\M$ makes on input $v$ will be polynomial in $|v|$. Similarly, the length of the code of each query $u_i$ will be polynomial in $|v|$, as will the length of any potential witness $w_i$ for $u_i$. Taken together, this means that the length of the encoding of $(p,(b_1,\dots,b_p),(w_1,\dots,w_p))$ will be polynomial in $|v|$, as will the length of the encoding of $(w'_1,\dots,w'_p)$. Thus, to show that $L \in \Sigma\Pi\left(\P(\rr)^{\Sigma_k(\rr)}\right),$ it suffices to show that the quantifier-free portion of the above condition can be checked in polynomial time by an $\rr$-machine $N$ with an $S$ oracle.

    The machine $N$ rejects any input that is not of the form $$(v,(p,(b_1,\dots,b_p),(w_1,\dots,w_p)),(w'_1,\dots,w'_p)).$$ Given an input of the above form, $N$ will simulate $\M$ on input $v$ until $\M$ makes its first query $u_1$. If $b_1=1$, then $N$ uses its oracle to check if $(u_1,w_1) \in S$. If this is not the case, then $N$ rejects its input. If this is the case, then $N$ branches according to $b_1$ by proceeding to simulate $\M$ staring from instruction $\ell_{(1,1)}$. If $b_1 = 0$, then $N$ proceeds similarly, except it uses its oracle to check if $(u_1,w'_1) \in S$ before rejecting or continuing to simulate $\M$ from instruction $\ell_{(1,0)}$.
    
    The machine $N$ continues to simulate $\M$ in this manner, checking at the $i$th query if answer $b_i$ is correct, relative to the witness strings $w_i$ and $w'_i$, before rejecting or branching according to $b_i$. If $\M$ accepts $v$, then $N$ accepts $(v,(p,(b_1,\dots,b_p),(w_1,\dots,w_p)),(w'_1,\dots,w'_p))$, and otherwise $N$ rejects this input. Everything that $N$ does in addition to simulating $\M$ can be done in polynomial time in $|v|$. Since $\M$ runs in polynomial time in $|v|$, simulating $\M$ by branching according to answers $b_1,\dots,b_p$ can also be done in polynomial time. Thus, $N$ runs in polynomial time.
\end{proof}

\OraclePolynomialHierarchyCharacterization*
\begin{proof}
    We proceed by induction on $k$. Recall that because we define $\NP(\rr)$ and $\coNP(\rr)$ in terms of witnesses rather than nondeterministic machines, we have $$\NP(\rr)^{\Sigma_k\rr} = \Sigma \left(\P(\rr)^{\Sigma_k\rr}\right) \text{ and } \coNP(\rr)^{\Sigma_k\rr} = \Pi \left( \P(\rr)^{\Sigma_k\rr} \right).$$ By definition we have the equality $\Sigma_0\rr = \P(\rr)$. Thus, $\NP(\rr)^{\Sigma_0\rr} = \NP(\rr)^{\P(\rr)} = \NP(\rr)$ and similarly $\coNP(\rr)^{\Sigma_0\rr} = \coNP(\rr),$ thereby establishing our base case.

    Assume that $\Sigma_k \rr = \Sigma \left( \P(\rr)^{\Sigma_{k-1}\rr}\right)$ and $\Pi_k \rr = \Pi \left( \P(\rr)^{\Sigma_{k-1}\rr}\right)$. Suppose $L \in \Sigma_{k+1}\rr$. Since $\Sigma_{k+1}\rr = \Sigma \left(\Pi_k\rr\right)$, we know there is some polynomial $q$ and some $S \in \Pi_k\rr$ such that for all $v \in R^*$, $$v \in L \text{ if and only if } \left(\exists w \in R^{q(|v|)}\right) (v,w) \in S.$$ Since $S \in \Pi_k\rr$, the inclusion $\Pi_k\rr \subseteq \P(\rr)^{\Sigma_k\rr}$ established in Lemma \ref{lemma:Pi_kRSubseteqP(R)^Sigma_kR} implies that $S \in \P(\rr)^{\Sigma_k\rr}$. Thus, using the same polynomial $q$ to bound the length of potential witnesses, we have $L \in \Sigma\left(\P(\rr)^{\Sigma_k\rr}\right)$, thereby establishing the inclusion $\Sigma_{k+1}\rr \subseteq \Sigma\left(\P(\rr)^{\Sigma_k\rr}\right).$ 

    Conversely, suppose $L \in \Sigma\left(\P(\rr)^{\Sigma_k\rr}\right)$. Then there is some polynomial $q$ and some $S \in \P(\rr)^{\Sigma_k\rr}$ such that for all $v \in R^*$, $$v \in L \text{ if and only if } \left(\exists w \in R^{q(|v|)}\right) (v,w) \in S.$$ Since $S \in \P(\rr)^{\Sigma_k\rr}$, the inclusion $\P(\rr)^{\Sigma_k\rr} \subseteq \Sigma\Pi \left( \P(\rr)^{\Sigma_{k-1}\rr} \right)$ established in Lemma \ref{lemma:P(R)Sigmak+1IsSubsetSigmaPiP(R)Sigmak} implies that $S \in \Sigma\Pi \left( \P(\rr)^{\Sigma_{k-1}\rr} \right).$ By our inductive hypothesis, we know $\Pi \left( \P(\rr)^{\Sigma_{k-1}\rr} \right) = \Pi_k\rr$, so $S \in \Sigma\left( \Pi_k\rr \right)$, and this implies that $L \in \Sigma \Sigma \left(\Pi_k\rr \right)$. It is clear that $\Sigma \Sigma \left(\Pi_k\rr \right) = \Sigma \left(\Pi_k\rr \right)$, so we can see that $L \in \Sigma \left(\Pi_k\rr \right) = \Sigma_{k+1}\rr$, thereby establishing the inclusion $\Sigma\left(\P(\rr)^{\Sigma_k\SAT(\rr)}\right) \subseteq \Sigma_{k+1}\rr$.
\end{proof}

%%%%%%%%%%%%%%%%%%%%%%%%%%%%%%%%%%%%%%%
\subsection{Oracles for $\exists_k\rr$}
%%%%%%%%%%%%%%%%%%%%%%%%%%%%%%%%%%%%%%%

Recall that we define $\exists\rr = \BP^0(\NP(\rr))$. That is, $\exists\rr$ is the class of decision problems $L \subseteq \{0,1\}^*$ for which there is a polynomial-time verification algorithm, implemented on a constant-free $\rr$-machine that only accepts Boolean inputs, using witness strings $w \in R^*$. The meaning of $\exists\rr^{\Sigma_k\rr^0}$ is then the class of decision problems $L \subseteq \{0,1\}$ where the above mentioned $\rr$-machine has an oracle for some $Q \in \Sigma_k\rr^0$. 

We can alter the above lemmas and theorem to derive \Cref{thm:OracleBooleanHierarchyCharacterization} below. Specifically, we modify \Cref{lemma:Pi_kRSubseteqP(R)^Sigma_kR} to show that $\Pi_k\rr^0 \subseteq (\P(\rr)^0)^{\Sigma_k\rr^0}$, and we alter \Cref{lemma:P(R)Sigmak+1IsSubsetSigmaPiP(R)Sigmak} to show that $(\P(\rr)^0)^{\Sigma_k\rr^0} \subseteq \Sigma\Pi((\P(\rr)^0)^{\Sigma_{k-1}\rr^0}).$ Using these altered lemmas, we may then use essential the same argument as in the proof of \Cref{thm:OraclePolynomialHierarchyCharacterization} to derive the following corollary.

% That is, the meaning of $\exists\rr^{\Sigma_k\rr^0}$ is $$\exists\rr^{\Sigma_k\rr^0} = \BP^0(\NP(\rr))^{\Sigma_k\rr^0} = \BP^0(\NP(\rr)^{\Sigma_k\rr^0}).$$ Indeed, this is the only way to interpret $\exists\rr^{\Sigma_k\rr^0}$ because raising a complexity class $\mathcal{C}$ to an oracle class $\mathcal{O}$ only makes sense if $\mathcal{C}$ is defined in terms of machines. From this observation, the following theorem is almost an immediate consequence of \Cref{thm:OraclePolynomialHierarchyCharacterization}.

\OracleBooleanHierarchyCharacterization*

This corollary tells us that it is unlikely that we can replace the non-Boolean oracle class $\Sigma_k\rr^0$ with the Boolean oracle class $\exists_k\rr$, as one might expect from a straightforward analogy with the classical polynomial hierarchy. For example, while $\NP^{\NP} = \Sigma_2$ and $\NP(\rr)^{\NP(\rr)} = \Sigma_2\rr$, \Cref{thm:OracleBooleanHierarchyCharacterization} tells us that $\exists\rr^{\NP(\rr)^0} = \exists_2\rr$, so it is unlikely that $\exists\rr^{\exists\rr} = \exists_2\rr$. It is instructive to reflect on why our proof requires the constant-free, non-Boolean oracle class $\Sigma_k\rr^0$. 

Already for $k=0$, in order to show that $\exists\rr^{\Sigma_0\rr^0} = \exists\rr$, we need to use a constant-free machine to simulate a constant-free machine with an oracle $Q \in \Sigma_0\rr^0=\P(\rr)^0$. This is possible only because $Q$ is itself decidable with a constant-free machine. Otherwise, we could hide a constant in the oracle $Q$, making this simulation impossible. We can see this from the observation that $\BP(\P(\rr)) \subseteq \exists\rr^{\P(\rr)}$, and in general there are problems in $\BP(\P(\rr))$ that cannot be decided with a constant-free machine. For more detail on hiding constants, see the warning below \Cref{def:ConstantFreeBooleanPart} and \cite{CK95}. Thus, we need to take the constant-free part of $\Sigma_k\rr$ as an oracle class.

We then want to show for $k=1$ that $\exists\rr^{\NP(\rr)^0} = \exists_2\rr$. To show the inclusion $\exists\rr^{\NP(\rr)^0} \subseteq \exists_2\rr$, it is again important that we use the constant-free part of $\NP(\rr)$ so that we are not able to hide constants in oracles. To show the inclusion $\exists_2\rr \subseteq \exists\rr^{\NP(\rr)^0}$, it is important that we use a non-Boolean oracle class because witness strings can be non-Boolean.

In more detail, suppose $L \in \exists_2\rr$. Then $L \subseteq \{0,1\}^*$ and there is a constant-free, polynomial-time machine $\M$ such that for all $v \in \{0,1\}^*$, $$v \in L \text{ if and only if } (\exists w_1 \in R^*)(\forall w_2 \in R^*) \M(v,w_1,w_2)=1.$$ Looking only at the universally quantified witness strings $w_2 \in R^*$, we can see that $Q = \{u \in R^* : (\forall w_2 \in R^*) \M(u,w_2)=1\} \in \coNP(\rr)^0$. Furthermore, $v \in L$ if and only if there is some $w_1 \in R^*$ such that $(u,w_1) \in Q$. Because a constant-free machine with an oracle for $\overline{Q} = R^* \setminus Q \in \NP(\rr)^0$ can decide $Q$ in polynomial time, we see that $L \in \exists\rr^{\NP(\rr)^0}$. In general, the oracle $\overline{Q}$ will be non-Boolean because the witness strings $w_1, w_2 \in R^*$ will be non-Boolean. Thus, we cannot generally take the Boolean part of $\NP(\rr)^0$, nor of $\Sigma_k\rr^0$, as our oracle class.

\end{document}